\documentclass[openright,11pt,twoside,singlespace]{citthesis}

\newcommand{\citepeg}[1]{\citep[{e.g.,}][]{#1}}
\newcommand{\citepcf}[1]{\citep[{see}\phantom{}][]{#1}}
\newcommand{\citesee}[1]{\citep[{see}\phantom{}][]{#1}}
\newcommand{\rha}[0]{\rightarrow}
\def\ale{\mathrel{\hbox{\rlap{\hbox{\lower4pt\hbox{$\sim$}}}\hbox{$<$}}}}
\def\age{\mathrel{\hbox{\rlap{\hbox{\lower4pt\hbox{$\sim$}}}\hbox{$>$}}}}
\def\degr{\hbox{$^\circ$}}
\def\arcmin{\hbox{$^\prime$}}
\def\arcsec{\hbox{$^{\prime\prime}$}}
\def\fd{\hbox{$\phantom{.}\!\!^{\rm d}$}}
\def\fh{\hbox{$\phantom{.}\!\!^{\rm h}$}}
\def\fm{\hbox{$\phantom{.}\!\!^{\rm m}$}}
\def\fs{\hbox{$\phantom{.}\!\!^{\rm s}$}}

\def\sfr{\hbox{$M_\odot$ yr$^{-1}$}}
\def\flux{\hbox{erg s$^{-1}$ cm$^{-2}$}}

\def\gsim{\mathrel{\hbox{\rlap{\lower.55ex \hbox {$\sim$}}
                   \kern-.3em \raise.4ex \hbox{$>$}}}}
\def\lsim{\mathrel{\hbox{\rlap{\lower.55ex \hbox {$\sim$}}
                   \kern-.3em \raise.4ex \hbox{$<$}}}}

\newcommand\nodata{ ~$\cdots$~ }%
\newcommand\ion[2]{#1$\;${\small\rmfamily\@Roman{#2}}\relax}%

\def\periodafterfoot{\hskip -8pt.\hskip 11pt}
\def\commaafterfoot{\hskip -8pt,\hskip 8pt}
\def\nadaafterfoot{\hskip -3 pt}
\def\semiafterfoot{\hskip -5pt;\hskip 4pt}
\def\negafterfoot{\hskip -3 pt$\!$}

\newcommand{\secfootnote}[2]{
     \renewcommand{\thefootnote}{\fnsymbol{footnote}}
     \footnotetext[#1]{#2}
     \renewcommand{\thefootnote}{\arabic{footnote}}
} 
\def\secfootdag{2}

\newcommand{\secauthor}[1]{
	{ \center \normalsize \sc {#1} \rm \endcenter }
}

\newcommand{\secaffil}[1]{%
	\vspace{-2ex}
	{\center \footnotesize {#1} \endcenter}
	\normalfont\normalsize
}%

\newcommand{\secaffils}[1]{%
	\vspace{-4ex}
	{\center  \footnotesize {#1}  \endcenter}
	\normalfont\normalsize
}%

\def\eg{{e.g.,~}}
\def\ie{{i.e.,~}}

\def\hst{{\it HST}}

\usepackage{spacing}
\usepackage{natbib}
\usepackage{appendix,amsbsy}
\usepackage{fancyhdr}
\usepackage[medium]{mytitlesec}
\bibpunct{(}{)}{;}{a}{}{,}
\titleformat{\chapter}[display]
 {\normalfont\Large\filcenter\sffamily}
 {\titlerule[1pt]
   \vspace{1pt}
   \titlerule
   \vspace{1pc}
   \Large\MakeUppercase{\chaptername} \thechapter}
 {1pc}
 {\titlerule
  \vspace{1pc}
  \Huge}

\titleformat{\section}[frame]
 {\normalfont}
 {\filright\footnotesize\enspace\sffamily  SECTION \thesection\enspace}
 {4pt}
 {\Large\sffamily\filcenter}

\begin{document}

\title{Toward an Understanding of the\\Progenitors 
of Gamma-Ray Bursts}
\author{Joshua Simon Bloom}		%
\degree{Doctor of Philosophy}		
\degreemonth{June}			%
\degreeyear{2002}			%
\thesisdate{1 April 2002}		%
\department{Department of Astronomy and Astrophysics}
\techreportnumber{MA-xx}		
\maketitle				
\copyrightpage				
\newpage				

\begin{acknowledgementspage}

I first encountered my future thesis adviser, Professor Shri Kulkarni,
at his colloquium on soft gamma-ray repeaters in 1993.  I was a
sophomore at Harvard and had just decided to concentrate in astronomy
and physics for my undergraduate degree. I remember that afternoon
vividly: his talk was explosive and Shri was ebullient and animated.
In that one hour, he managed, with great gusto, to capture the essence
of the field and left me intensely excited about a subject that I had
known little about. I know now that such a talent is rare.

Shri is indeed a rarity.  He is at once enthusiastic, brusque,
contradictory, logical, forward-thinking, generous, unfathomably
demanding, focused, multi-plexed, socio-pol\-itic\-ally charged,
insightful, and unwaveringly pragmatic. While he is no ordinary
exemplar for a graduate student, I could not have been luckier nor
happier to have had such a person as a Ph.D.~adviser. He constantly
challenged me to attain technical excellence and strive for a deeper
clarity in my work. He fostered my curiosity and enabled me, as both a
person and a scientist, to shine. I cannot thank him enough for all
that he has done for me.

I have been so enriched by interactions with so many over the past
five years. From the amazing folks at the Palomar and Keck
observatories to the X-ray and $\gamma$-ray satellite builders to
co-authors whom I have never met face-to-face, this thesis reflects
the strengths and dedication of literally hundreds of people.
Dr.~Dale Frail, who I had the great fortune to work with on no less
than 29 published refereed papers, has been one of the most important
influences in my development.  I will forever consider Dale a mentor,
colleague, and a good friend.

Dale and Shri's original little 1996 afterglow duo blossomed into no
less than a full-fledged team of a researchers. I am indebted to all
of the past and existing members of the ``Caltech--NRAO--CARA GRB
Collaboration'' who, without fail, graciously shared their talents and
energies with me. The individuals who contributed to my thesis work
are either co-authors on the respective published papers or explicitly
mentioned in chapter-specific acknowledgments. Here, I particularly
thank Prof.~S.~George Djorgovski, Re'em Sari, Prof.~Jules Halpern
(Columbia University), Alan Diercks, Chris Clemens, Steve Odewahn,
Paul Price, Edo Berger, and Dan Reichart. Our collective adventures in
the outer-limits of astronomy have been a source of immense pleasure
and gratification.

My past and current officemates have been each extraordinarily
influential. I thank Ken Banas, Ben Oppenheimer, Alice Shapley, Kurt
Adelberger, and Rob Simcoe for their friendship, support, and
knowledge.  Over our frequent coffee and sushi breaks, I spent
hundreds of hours talking and sharing ideas with Pieter van Dokkum and
Dan Stern. I learned a great deal from Pieter and Dan about galaxies,
cosmology, and data reductions; I hold them in the highest regard both
personally and professionally.  I shall never be able to fully
articulate just how important my peers and colleagues have been.

For the past three years, I have been supported by a Fannie and John
Hertz Foundation Graduate Fellowship. Each year, the Fellowship is
supposed to be awarded to ``America's most promising technical
talents...who can be expected to have the greatest impact on the
application of the physical sciences to human problems during the next
half-century.''  I would never profess to be such a person but I
certainly will strive attain that lofty goal. Indeed, it has been an
honor to serve as Hertz Fellow.

I heartily thank my thesis committee members: Profs.~Fiona Harrison
(chair), Shri, George, Marc Kamionkowski, Nick Scoville.  From the
onset to the end, they consistently and thoughtfully encouraged me to
build this thesis with an eye on the big picture. I thank the staff in
the Astronomy department for their support and Josep Paredes at the
University of Barcelona for his hospitality.

Even with such stimulating work environments, no one could be (or, at
least, should be) truly fulfilled without close friendships. In this
respect, I am very pleased to also count among my confidants and
partners in pursuit of happy and full life: Michelle Brent, Josh
Eisner, Gordon Squires, Ken B., Ben O., Geoff Criqui, John Ciorciari,
Jared Bush, Jamie Miller, Jessica Rovello, Steve Agular, and Mark
Galassi.  These are the people who helped sustain me and who were
always there through my highs and lows.

This thesis is dedicated to all my friends and those six people that
are closest to my heart, without whom thesis would not have been
possible nor worthwhile. Anna, Bec, Mom, Dad, Nanny, and Pop-pop,
thank you.

\end{acknowledgementspage}

\begin{abstractpage}

\begin{spacing}{1.4}

The various possibilities for the origin (``progenitors'') of
gamma-ray bursts (GRBs) manifest in differing observable properties.
Through deep spectroscopic and high-resolution imaging observations of
some GRB hosts, I demonstrate that well-localized long-duration GRBs
are connected with otherwise normal star-forming galaxies at moderate
redshifts of order unity. Using high-mass binary stellar population
synthesis models, I quantify the expected spatial extent around
galaxies of coalescing neutron stars, one of the leading contenders
for GRB progenitors. I then test this scenario by examining the offset
distribution of GRBs about their apparent hosts making extensive use
of ground-based optical data from Keck and Palomar and space-based
imaging from the {\it Hubble Space Telescope}.  The offset
distribution appears to be inconsistent with the coalescing neutron
star binary hypothesis (and, similarly, black-hole--neutron star
coalescences); instead, the distribution is statistically consistent
with a population of progenitors that closely traces the ultra-violet
light of galaxies.  This is naturally explained by bursts which
originate from the collapse of massive stars (``collapsars'').  This
claim is further supported by the unambiguous detections of
intermediate-time (approximately three weeks after the bursts)
emission ``bumps'' which appear substantially more red than the
afterglows themselves. I claim that these bumps could originate from
supernovae that occur at approximately the same time as the associated
GRB; if true, GRB 980326 and GRB 011121 provide strong observational
evidence connecting cosmological GRBs to high-redshift supernovae and
implicate massive stars as the progenitors of at least some
long-duration GRBs. Regardless of the true physical origin of these
bumps, it appears that all viable alternative models of these bumps
(such as dust scattering of the afterglow light) require a substantial
amount of circumburst matter that is distributed as a wind-stratified
medium; this too, implicates massive stars.  Also suggested herein are
some future observations which could further solidify or refute the
supernova claim.  In addition to the observational and modeling work,
I also constructed the {\it Jacobs Camera} (JCAM), a dual-beam optical
camera for the Palomar 200--inch Telescope designed to follow-up rapid
GRB localizations.

\end{spacing}

\end{abstractpage}

\begin{singlespace}
\tableofcontents			
\listoffigures				
\listoftables				
\end{singlespace}

\starttext				

\begin{prefacepage}
\addcontentsline{toc}{section}{Preface}
\label{chap:preface}

\begin{singlespace}
\begin{quote}
 {But the reason so many of you live, work and study here is
that there are so many more questions yet to be answered...And so I
wonder,...Are we alone in the universe? {\bf What causes gamma ray
bursts?}  What makes up the missing mass of the universe? What's in
those black holes, anyway? And maybe the biggest question of all: How
in the wide world can you add \$3 billion in market capitalization
simply by adding .com to the end of a name?}
\begin{flushright}
William Jefferson Clinton\\
42nd President of the United States of America\\
{\it Science and Technology Policy Speech} \\
21 January 2000, Caltech
\end{flushright}
\end{quote}
\end{singlespace}

These ponderances, spoken just before the bursting of the Internet
``bubble,'' invigorated me. Rarely, I suspect, does a sitting
president publicly ask the question that is the central endeavor of
one's PhD thesis. But, given the topic and the timing---gamma-ray
bursts, during one of the most enlightening periods of understanding
of the phenomenon---perhaps we should not be surprised that this
happened.  Of course, the cynical view is that the President's speech
writers had scoured the Caltech web site before his visit and
encountered a number of the astronomy press releases that had been
generated here over the years, drawing up a number of relevant
questions for a Caltech-specific audience. The idealist view, one that
is perhaps more comforting to accept, is that the question of what
makes gamma-ray bursts really is on the minds not just of a few
astronomers but on a much larger audience.

For those that are familiar with the phenomenon, this public appeal of
gamma-ray bursts is almost assured---GRB descriptions are, after all,
awash in superlatives. We now know, for instance, that GRBs are one of
the {\it brightest} events in the universe, briefly reaching
luminosities comparable to the integrated luminosity of a few hundred
thousand galaxies. The bursts are some of the {\it rarest}
well-studied transient events (only a few per galaxy per 10 million
years) and probably represent the violent death and/or birth of the
{\it most dense} objects known, namely black holes and neutron
stars. Because of the extreme densities and accelerations of the masses
involved in triggering a GRB, the events leading up to a GRB explosion
holds the greatest promise of impulsively releasing as yet undetected
gravitational waves.

That this central question---What causes gamma-ray bursts?---can even
be asked now with a straight face is nothing short of remarkable.
Not too long ago---pre-1997, to be precise---no one knew for sure from
where GRBs originated. Was the origin of the bursts that of primordial
anti-matter comets smashing into the Oort cloud \citep[distance of
scale of 100 pc;][]{der96}\footnotemark\footnotetext{1 parsec = 3.085
$\times 10^{18}$ cm or 3.27 light-years}\negafterfoot\ or the
re-connection of super-conducting cosmic strings \citep[distance of
scale of 10$^{10}$ pc;][]{pac88} or some place more conventional (such
as in our Galaxy or more distant galaxies)?

This remarkable ignorance, let alone the lack of any solid connection
between GRBs and other known astrophysical entities, persisted for 29
years and 10 months after the first detection of a GRB.  Then, in May
1997, shortly following the detection of the first long-lived emission
following a GRB (``afterglow''), finally the first step was taken. An
optical spectrum of the afterglow obtained at the Keck telescopes
revealed the burst to have originated from at least a redshift of
\hbox{$z = 0.835$}, proving that at least one GRB originated from
``cosmological'' distances.

This was the state of affairs when I began my PhD thesis at Caltech in
the fall of 1997---just one GRB was known to be of a cosmological
origin and a total of two afterglows had been discovered.  My early
interest in afterglow follow-up work stemmed from the belief that the
intense study of afterglows would surely lead to physical insights
about the nature of the afterglows themselves and GRB progenitors.
Indeed, with over 30 GRB afterglows discovered and studied in the past
five years, the community has solidified GRBs as of a cosmological
origin and learned a great deal about the physical processes
underlying afterglows.  My thesis, however, focuses on the later issue, uncovering the progenitors.

The ultimate conclusion of this work---that most GRBs probably arise
from the death of massive stars rather than the coalescences of massive
compact binary stars---was rather unexpected since the predominant
view in 1997 was that the latter objects were likely responsible for
such bursts.  The connection between GRBs and massive stars
represents, as \citet{vkw00} have also pointed out, a harmonious and
beautiful full-circle revelation: the first theory of gamma-ray bursts,
posited by \citet{col68} before GRBs had even be discovered, suggested
that GRBs could arise during a supernova explosion.

\bigskip

\noindent {\it Notes on the contents of this thesis}: Thanks to the
relative ignorance of the physics of GRBs and the small wavelength
regime in which GRBs and their aftermath had been observed (X-rays to
GeV gamma-rays), the introduction to a PhD thesis ten or even five
years ago could realistically have captured the sum-total knowledge of
the phenomena for the reader.  Yet, after 5 heady years in the
afterglow era, the information explosion\footnotemark\footnotetext{As
of mid-2001, GRBs entered the literature at a rate of 1.5 per day,
50\% higher than the rate in 1994 \citep{hurley02}.}\nadaafterfoot\
would surely require hundreds of pages for a comprehensive exposition
of GRBs (think of writing a comprehensive review on galaxy evolution
or supernovae).

This is not my intention with the scope of the following summary
chapter; instead, I give a brief overview of the observations and
current theoretical understanding of the
phenomena\footnotemark\footnotetext{The first section of chapter
\ref{chap:intro} contains some of the text in the published version of
chapter \ref{chap:offset}. The introduction of chapter
\ref{chap:offset} is commensurately abridged.}\periodafterfoot Then I
present some of the more salient topics related to progenitors and
their host galaxies, drawing specific attention to the work in body of
the thesis.  For a more thorough review of the current state of the
field, the reader is referred to \citet{fm95} (GRB science pre-1997),
\citet{pir99} (fireball and afterglow physics), \citet{vkw00}
(afterglow observations), and \citet{fwh99} (progenitor models).

The next eight chapters were prepared for a total of five different
journals, each, as such, with differing presentation and referencing
styles.  The audience for each article varied as well---the {\it
Nature} article (chapter \ref{chap:sn-grb}), for instance, was geared
to a general science audience, while the {\it PASP} article (chapter
\ref{chap:jcam}) was written primarily for an astronomical
instrumentation audience.  In the interest of continuity, I have
homogenized the referencing style and nomenclature from chapter to
chapter.

In some chapters, I have also included some additional text and
figures that were necessarily cut from the published article due to
space constraints.  In chapter \ref{chap:sgrbs}, for instance, I
include example $\gamma$-ray light curves of possible supernova--GRBs,
and in chapter \ref{chap:sn-grb} I provide an expanded explanation of
some of the data reduction methods.  Looking back, some ideas in the
chapters proved to be less salient (and correct!)  than others and so
there was the temptation to jettison the chaff. Since this thesis is
comprised mostly of published articles that themselves (should)
reflect the progression of the field, I have, however, tried to steer
clear of constructing such a revisionist history. As such, all of the
``new'' ideas augmented to the published versions appeared first in
submitted versions of the paper. Chapter \ref{chap:epi} is designated
as forum for redresses and epilogues to the body of this thesis.

\end{prefacepage}

\cleardoublepage
\pagestyle{fancy}
\fancyhead{} 
\renewcommand{\chaptermark}[1]{
 \markboth{\chaptername \ \thechapter}{}} 

\fancyhead[LE,RO]{\slshape {\footnotesize \rightmark}} 
\fancyhead[LO,RE]{\slshape {\footnotesize \leftmark}, p.~\thepage} 
\fancyfoot[C]{\phantom{~}}

\chapter{Introduction and Summary}
\label{chap:intro}

\section{History, Phenomenolgy, and Afterglows}
\label{sec:hist}

\label{sec:intro-phenom}

Gamma-ray bursts, otherwise extinguished by the Earth's atmosphere,
were discovered serendipitously \citep{kso73,sko74} by space-based US
satellites designed to insure compliance with the Limited Nuclear Test
Ban Treaty (signed July 25, 1963) by searching for the $\gamma$-ray
emission that accompanies nuclear weapons testing.  In gross
properties, the 23 bursts presented in the discovery papers were not
unlike the some 3000 observed to date by the many GRB-specific
satellites that followed. One notable sub-class of the gamma-ray burst
(GRB) phenomenon are the so-called ``Soft Gamma-Ray Repeaters'' (SGRs)
which, since the late-1970s, have been definitively associated with
highly-magnetized isolated neutron stars (magnetars) in our Galaxy or
the Large Magellanic Cloud (LMC) (see
\citealt{hard01} for review). 

The duration of ``classic'' GRBs (i.e., those GRBs which are not
SGRs), observationally determined as the time that the flux exceeds
some threshold above the sky background level, ranges from a few
milliseconds to thousands of seconds
\citepeg{kkb+96}.  The peak of the spectral energy distribution of
bursts falls in the range of $\sim$50--1000 keV \citep{mpp+95} but
both ends of this range likely exist due to the trigger inefficiencies
of GRB satellites (see, e.g., fig.~3 of \citealt{lp99}; although see
\citealt{bra98a})\footnotemark\footnotetext{Until very recently, a
burst was not considered a GRB until its energy spectrum peaked above
$\sim30$ keV.  It is the adherence to this definition of GRBs that may
be restricting a deeper insight into the nature of GRBs. If a burst is
found to peak at lower-energies (a so-called X-ray Flash; XRF), but
retains most of the other properties of classic GRBs, then it may
simply be a GRB at a high redshift (one of remaining holy-grails of
observational GRB astronomy).  \citet{hzkc01} has recently written a
nice admonition to the community about classification of such
phenomena.}\periodafterfoot

GRBs seem to occur at random times and from random locations in the
sky, about 4 times per day at current detector thresholds. Though it
was known that the sky distribution of GRBs appeared roughly isotropic
for years, the {\it Burst and Transient Source Experiment} (BATSE),
the prime workhorse of GRB astronomy in the early- to mid-1990s,
placed the strongest constraints showing GRBs to be isotropically
distributed to within a high degree of confidence (\citealt{mfw+92};
see also table 8 of \citealt{pmp+99}).

Though, by 1995, no classical GRB had been definitively connected with
any other astrophysical entity, the observed isotropy was taken by
many as evidence for a cosmological progenitor origin. The paucity of
faint bursts relative to the number expected if the bursts originated
homogeneously in Euclidean space also served as evidence for a
cosmological origin \citepeg{feh+93,fb95}. During the ``Great Debate''
on the distance scale to GRBs \citep{nem95}, \citet{pac95} argued for
the cosmological origin of GRBs based primarily on these two points
while \citet{lam95} explained how the same data were consistent with a
galactic progenitor origin\footnotemark\footnotetext{That two such
contrasting views on the same data existed should remind us of the
quote from Antonio in the \underline{Merchant of Venice}, Act I,iii:
``Mark you this, Bassanio, The devil can cite Scripture for his
purpose.''}\periodafterfoot I attended the Debate and thought that
D.~Lamb made a persuasive argument for Galactic scenarios despite the
small theoretical parameter space then allowed by the isotropy and
brightness distribution observations.

The main impedance to progress was the difficulty of localizing bursts
to an accuracy high enough to unequivocally associate an individual
GRB with some other astrophysical entity. Several concerted efforts
were made to find counterparts
\citepeg{scd+87,vkr94,vhj95,fk95,gre95}, but we now know that such
surveys were either too shallow or too delayed in time to catch
rapidly fading GRB counterparts. Aside from the hope that GRBs would
produce/induce emission at some other frequency, some counterpart
searches were strongly motivated by (prescient) theoretical
predictions for the existence of lower-frequency counterparts
\citep{pr93,mr93a,katz94a,mrp94}. These ``afterglow'' models were a 
natural consequence of the (also prescient) predictions for
cosmological GRBs from highly-relativistic outflow of a
low-baryon-loaded fireball
\citep{pac86,goo86,mr92,mr93}.

In large measure the localization problem was due to both the
transient nature of the phenomena and the fact that the incident
direction of $\gamma$-rays are difficult to pinpoint with a single
detector; for example, the typical 1-$\sigma$ uncertainty in the
location of a GRB using the {\it Burst and Transient Source
Experiment} (BATSE) was $4$--$8$ degree in radius \citep{bpk+99}. The
Interplanetary Network \citep[IPN; see][]{cbb+99} localized GRBs using
burst arrival times at several spacecrafts throughout the solar system
and provided accurate localizations (3 $\sigma$ localizations of
$\sim$few to hundreds $\times$ sq.~arcmin) to ground-based observers;
however, the localizations were reported with large time delays (days
to months after the GRB).

The crucial breakthrough came in early 1997, shortly following the
launch of the \hbox{{\it BeppoSAX}} satellite \citep{bbp+97}. On-board
instruments
\citep{fcd+97,jmb+97} were used to rapidly localize the prompt and
long-lived hard X-ray emission of the GRB of 28 February 1997 (GRB
970228) to a 3 $\sigma$ accuracy of 3 arcmin (radius) and relay the
location to ground-based observers in a matter of hours.  Fading X-ray
\citep{cfh+97} and optical \citep{vgg+97} emission (afterglow) associated 
with GRB 970228 were discovered. Ground-based observers noted
\citep{mkd+97,vgg+97} a faint nebulosity in the vicinity of the
optical transient (OT) afterglow. Subsequent {\it Hubble Space
Telescope} (HST) imaging resolved the nebulosity
\citep{slp+97} and showed that the morphology was indicative of a
distant galaxy \citep{slp+97}.  We now know the redshift of this
faint, blue galaxy is $z=0.695$ (chapter \ref{chap:grb970228},
\citealt{bdk01}).

The next prompt localization of a GRB yielded the first measured
distance to a GRB through optical absorption spectroscopy: GRB 970508
occurred from a redshift \hbox{$z \ge 0.835$} \citep{mdk+97}.  The
first radio afterglow was detected from GRB 970508 which, through
observations of scintillation, led to the robust inference of
super-luminal motion of the GRB ejecta (\citealt{fkn+97}; see below).
These measurements (along with the dozen other redshifts now
associated with individual GRBs) have effectively ended the distance
scale debate and solidified GRBs as one of the most energetic
phenomena known
\citepcf{kbb+00,fks+01}. One of the most remarkable aspects of
GRB science \citepeg{wrm97} was the enormous success, as evidenced by
the observations of GRB 970228 and GRB 970508, of the theoretical
predictions for the existence and behavior of GRB afterglows
\citep{pr93,mr97a,viet97}.

\begin{figure*}[tbp]
\centerline{\psfig{figure=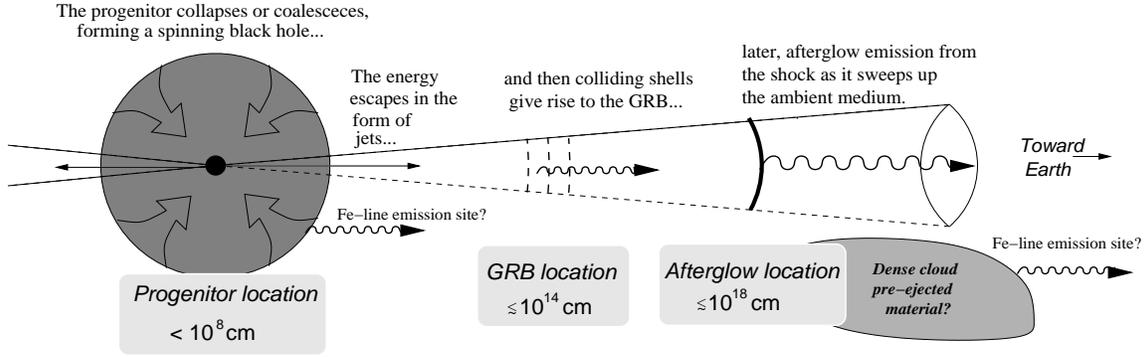,width=6.0in,angle=0}}
\caption[Anatomy of a gamma-ray burst explosion]{
Anatomy of a gamma-ray burst explosion.  The dark circle represents
the newly formed spinning black hole at the center of an imploding
star (or merging compact binary system).  The long-lived afterglow
emission that we see arises from the swept-up material; in this
material, relativistic electrons radiate sychrotron light in an
amplified magnetic field.  Due to the extreme velocity of the jet, the
whole sequence of events is compressed in time as viewed from Earth.}
\label{fig:intro-explo}
\end{figure*}

\begin{figure*}[p]
\centerline{\psfig{file=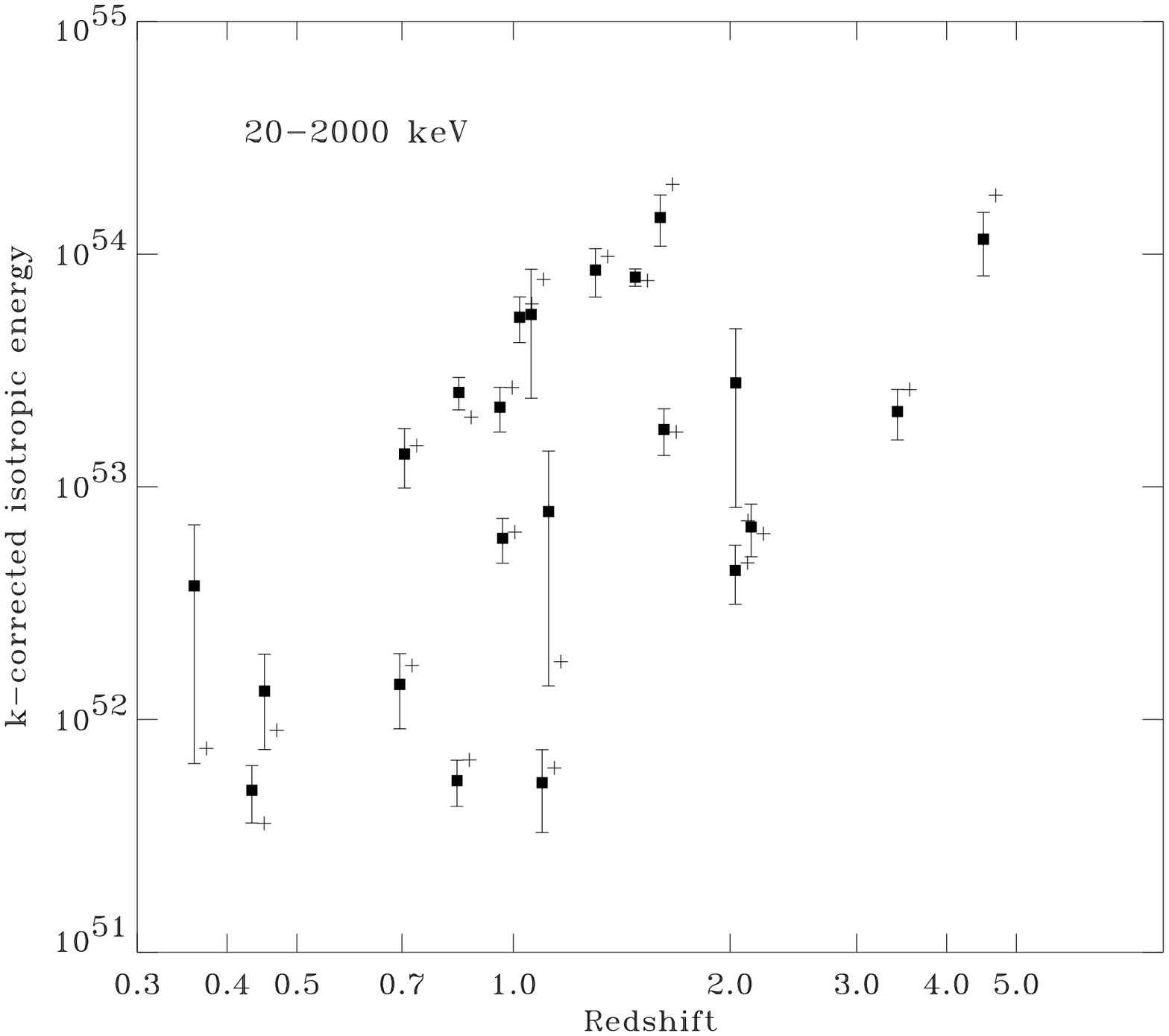,width=4.3in,angle=0}}
\centerline{\psfig{file=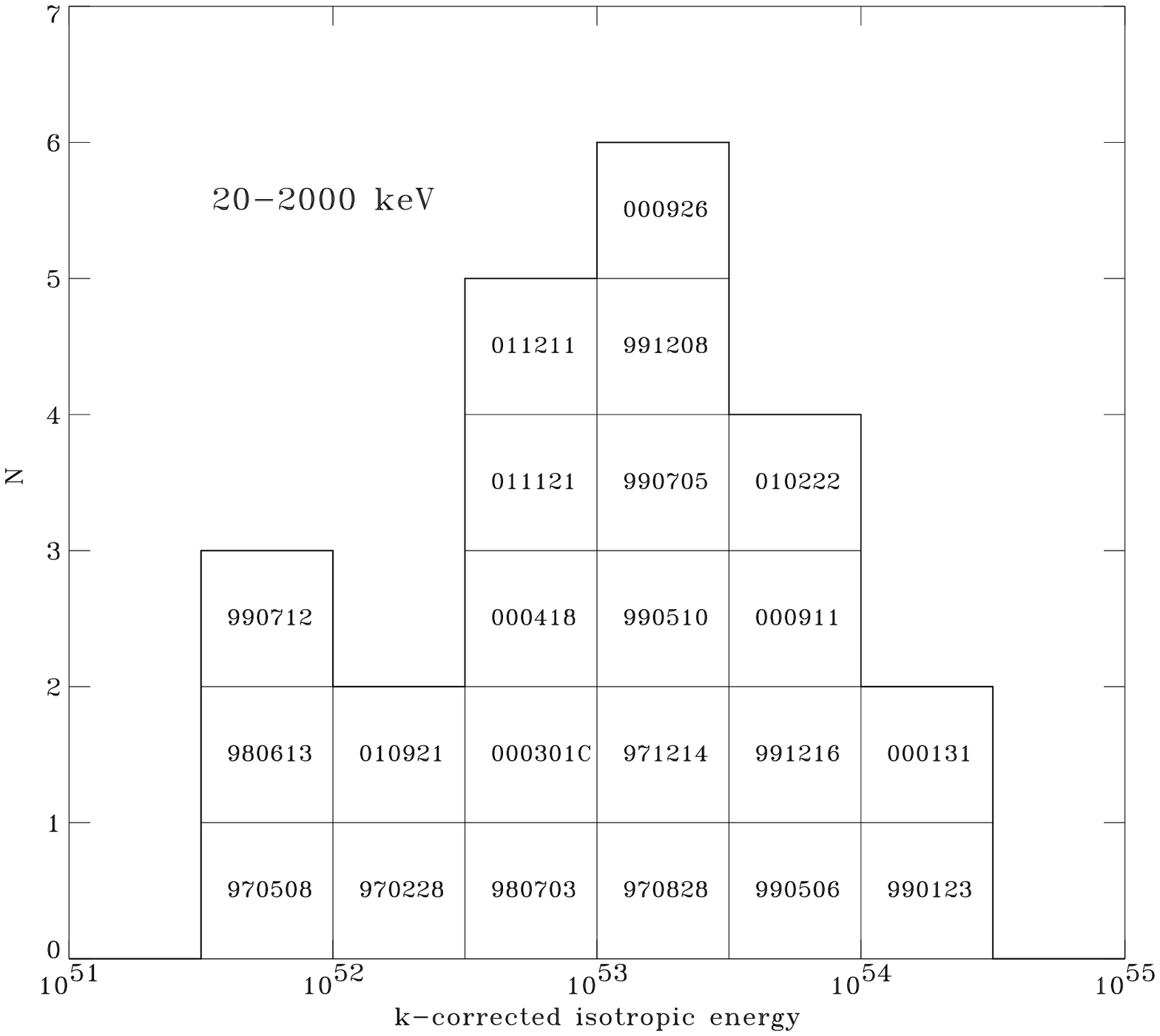,width=3.2in,angle=0}
\psfig{file=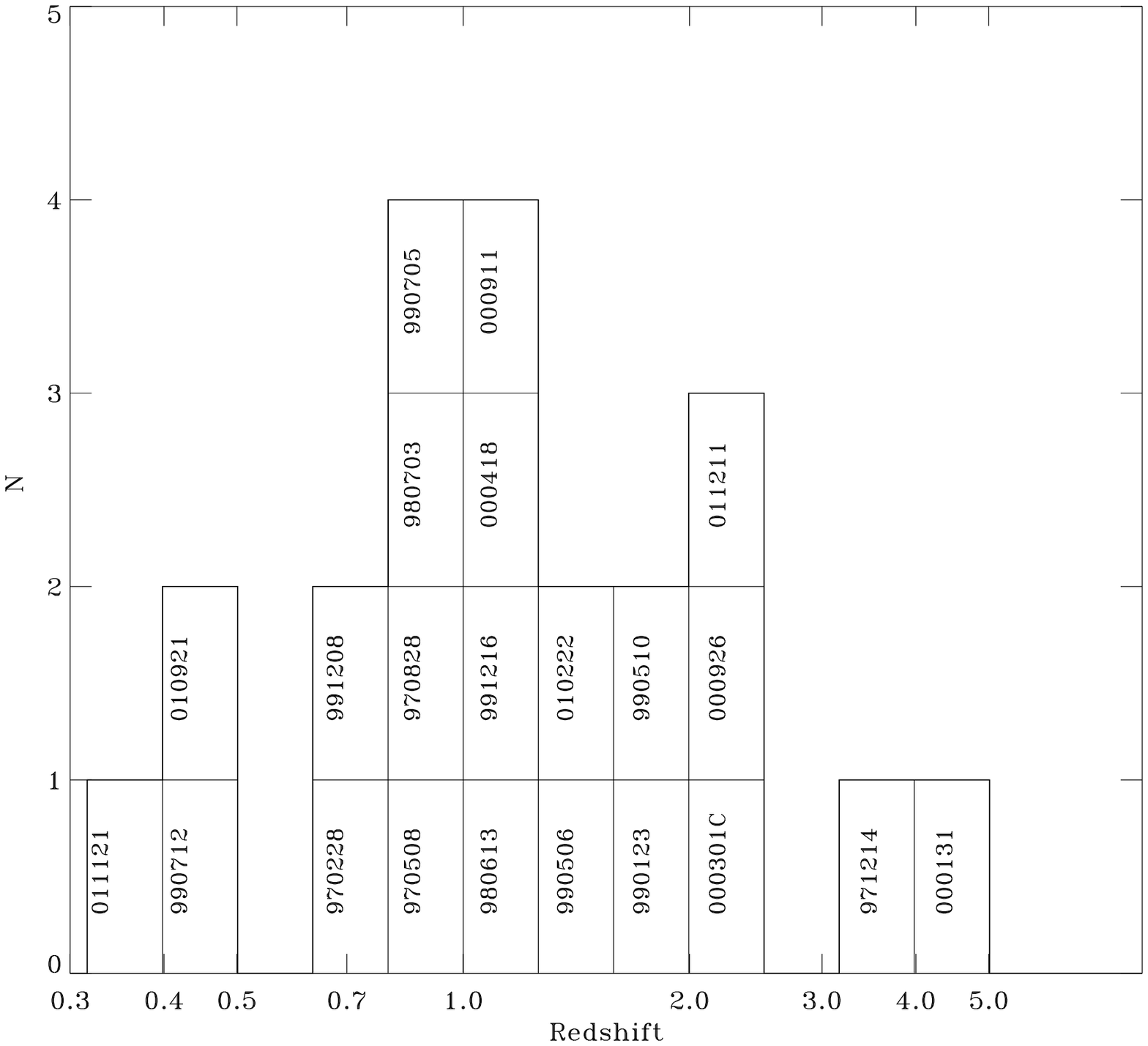,width=3.2in,angle=0}}
\caption[The redshift and energy distributions of the 22 cosmological
GRBs with known redshift]{The redshift and energy distributions of the
22 cosmological GRBs with known redshift. At top, the $k$-corrected
prompt energy release versus redshift in the restframe 20--2000 keV
bandpass assuming isotropic emission.  The 2 $\sigma$ error bars are
shown as well as the derived energies if no cosmological
$k$-correction is applied (denoted as a cross ``$+$''). At bottom
left, the histogram of $k$-corrected GRBs energies. At bottom right,
the observed redshift distribution of GRBs with measured
redshifts. Note that while this observed distribution is reflective of
the true GRB rate as a function of redshift, this is not, in general,
the true GRB rate: for example, owing to a lack of strong star
formation lines at observer-frame optical wavebands, there is a strong
selection against finding emission-line redshifts in the redshift
range $1.7
\ale z \ale 2.5$. Figures adapted and updated from
\citet{bfs01}.}
\label{fig:intro-energy}
\end{figure*}

The cosmological nature of GRBs now frames our basic understanding of
the physics of GRB phenomena.
The general energetics are well-constrained: given the observed
fluences and redshifts, approximately $10^{51}$--$10^{53}$ erg in
$\gamma$-ray radiation is released in a matter of a few seconds in
every GRB (see fig.~\ref{fig:intro-energy}). The GRB variability
timescale suggests that this energy is quickly deposited by a
``central engine'' in a small volume of space (radius $r \ale 1000$
km) and is essentially optically thick to $\gamma$-ray radiation at
early times.  This opaque fireball of energy (see below) then expands
adiabatically and relativistically until the $\gamma$-ray radiation
can escape; there, the GRB is thought to arise from the interaction of
internal shocks initiated by the central engine \citepeg{frrw99}.

The short variability timescale of a GRB and high total energy release
would tend to imply that the optical depth to pair-production at the
explosion site is exceedingly high ($\tau_{\gamma\gamma} \age
10^{12}$) yet GRB spectra are optically thin.  It was recognized in
the mid-1980s that this so called ``compactness problem'' can be
avoided by invoking relativistic motion
\citep{goo86,pac86}. If the surface of emission of $\gamma$-rays is
moving toward the observer at a bulk Lorentz factor $\Gamma$, then the
optical depth to pair-production is reduced due to two effects. First,
the fraction of photons that can pair-produce in the frame of the
moving surface is lower than inferred by an outside observer (by
$\Gamma^2$ for a flat spectrum), who measures a blue-shifted
spectrum. Second, special relativistic effects allow for the emission
radius to be larger by $\Gamma^2$ than the variability estimate (about
$10^{14}$ cm rather than 10$^8$ cm).  By requiring that
$\tau_{\gamma\gamma} < 1$, the source of GRB emission must be moving
with $\Gamma \age 100$ at the time of the GRB.

The coupling of the fireball energy to any entrained baryons will tend
to stall the outward expansion of the flow.  Specifically, if the
total energy in the fireball is $E_0$ ($\approx 10^{51}$ erg), then the
total amount of baryonic mass allowed in the flow is $M = E_0/\Gamma
c^2 \ale 10^{-5} M_\odot$. The elegant solution to the compactness
problem, then, places an important constraint on the nature of GRB
ejecta (see also \citealt{pir99}): the fireball must be nearly devoid
of baryons.

After the GRB, the relativistic blastwave continues its outward
expansion and begins to sweep up the ambient medium.  Taking $n = 1$
cm$^{-3}$ to be the density of the surrounding medium, the blastwave
begins to slow considerably by a radius $R \approx (E_0 / 4 m_{\rm H}
c^2
\Gamma^2)^{1/3} \approx 10^{16}$--10$^{17}$ cm and some of the kinetic 
energy is then converted into internal motion within the shock
\citep{mr93}. Here $m_H$ is the mass of a hydrogen atom. The transient afterglow phenomenon, thought to arise at
this radius, is likely due to synchrotron radiation arising from the
interaction of the relativistic ejecta and the ambient medium
surrounding the burst site
\citep[see][for reviews]{vkw00,kbb+00,dfk+01a}.

Indirectly, the initial $\Gamma$ of some GRBs have been constrained in
the context of the compactness problem and early-time observations of
afterglows \citepeg{mlr93,sp99b,rf00,sr02}. By noting an abrupt
quenching of scintillation behavior a few days after GRB 970508, 
\citet{fkn+97} showed that the afterglow emitting region must have grown 
with apparent superluminal speeds, observationally solidifying
relativistic motion as a fundamental property of GRBs.
 
There have now been tentative detections of transient X-ray line
features in five GRB afterglows \citep[e.g., 970508 and
970828][]{pcf+99,yno+99}. The most convincing detection so far comes
from observations of the afterglow of GRB 991216 \citep{pggs+00}.
Individually, the observational significance of the line detections
are marginal, but on the whole there appears to be a good case for line
emission features in the afterglow of some GRBs. If so, there must
exist dense matter in the vicinity of the explosion
\citepeg{wmkr00,vpps99,lcg99a}.  \citet{mr01} have also suggested that
the Fe lines may be produced if some of the waste energy from the
central engine heats up a ``bubble'' of matter from the progenitor, a
natural consequence of jets propagating in a dense stellar interior.
The bubble breaks out from the progenitor remnant on a timescale of
hours to days.  Figure \ref{fig:intro-explo} depicts the relative
locations of the suggested sources of the X-ray line emission from a
generalized progenitor.

Like most other high-energy phenomena, it is now widely accepted that
gamma-ray burst emission is collimated (or ``jetted'').
Observationally, jetting should be manifested as a (variable)
polarization signal in the afterglow \citep{gl99,sar99}; such
signatures have been detected in some GRB afterglows
\citep{clg+99,wvg+99,bl00,rwv+00}. The observed evolution of GRB
afterglows, often showing a break in the light curves from 0.5--50
days, also appear to conform to basic predictions from the dynamics of
jetted outflows \citep{rho99,sph99}.  Aside from relaxing the overall
energy requirements, the establishment of jetting in GRBs also implies
that the true rate of GRBs in the universe is substantially higher
than previously believed (by a factor of $\sim$550, \citealt{fks+01}).

\section{Proposed GRB Progenitor Scenarios}
\label{sec:intro-prog}

While the GRB emission and the afterglow phenomenon of long-duration
bursts are now reasonably well-understood, one large outstanding
question remains: what makes a gamma-ray burst?  Specifically, what are
the astrophysical objects, the ``progenitors,'' which produce GRBs?
Those theoretical progenitor scenarios which have remained feasible in
the afterglow era are principally constrained by the following
considerations:

\begin{singlespace}
\begin{itemize}

\item The implied (isotropic) energy release in $\gamma$-rays are typically
10$^{-3}$--$10^{-1}$ times the rest-mass energy of the Sun.  The
estimated efficiency of conversion of the initial input energy (either
Poynting flux or baryonic matter) to $\gamma$-rays ranges from
$\sim$1\% \citepeg{kum99} to as much as $\sim$60\% \citepeg{ks01};
therefore, the best-guess estimate of the total energy release
(including neutrino and gravitational-wave losses) is roughly
comparable to the rest-mass energy of one solar mass.
\medskip

\item The GRB variability timescale (few ms) observed implies that the 
energy deposition takes place in a small region of space (radius of $c
\times 1$ ms $\approx 300$ km).  The range in total burst durations
suggest that the central engine must live for less than one second and
up to thousands of seconds.
\medskip

\item The inferred rate of GRB occurrence (table \ref{tab:rates}) and the 
lack of burst repetition \citepeg{hmp+98} suggest that GRB events are
rare and catastrophically destroy the individual progenitors.

\end{itemize}
\end{singlespace}

The progenitor scenarios which most naturally explain these
observables fall in to three broad classes---the coalescence of binary
compact stellar remnants, the explosion of a massive star
(``collapsar''), and the accretion-induced collapse of a
differentially rotating compact object \citep[``DRACO'';][]{kr98}.  An
active galactic nucleus (AGN) origin is another possibility, whereby a
main-sequence (MS) or white-dwarf (WD) is tidally disrupted near a
super-massive black hole.  In such scenarios, however, the variability
timescale still requires the energy source to be stellar-mass objects
\citep{car92,cw99}.  

\begin{figure*}[tbp]
\centerline{\psfig{file=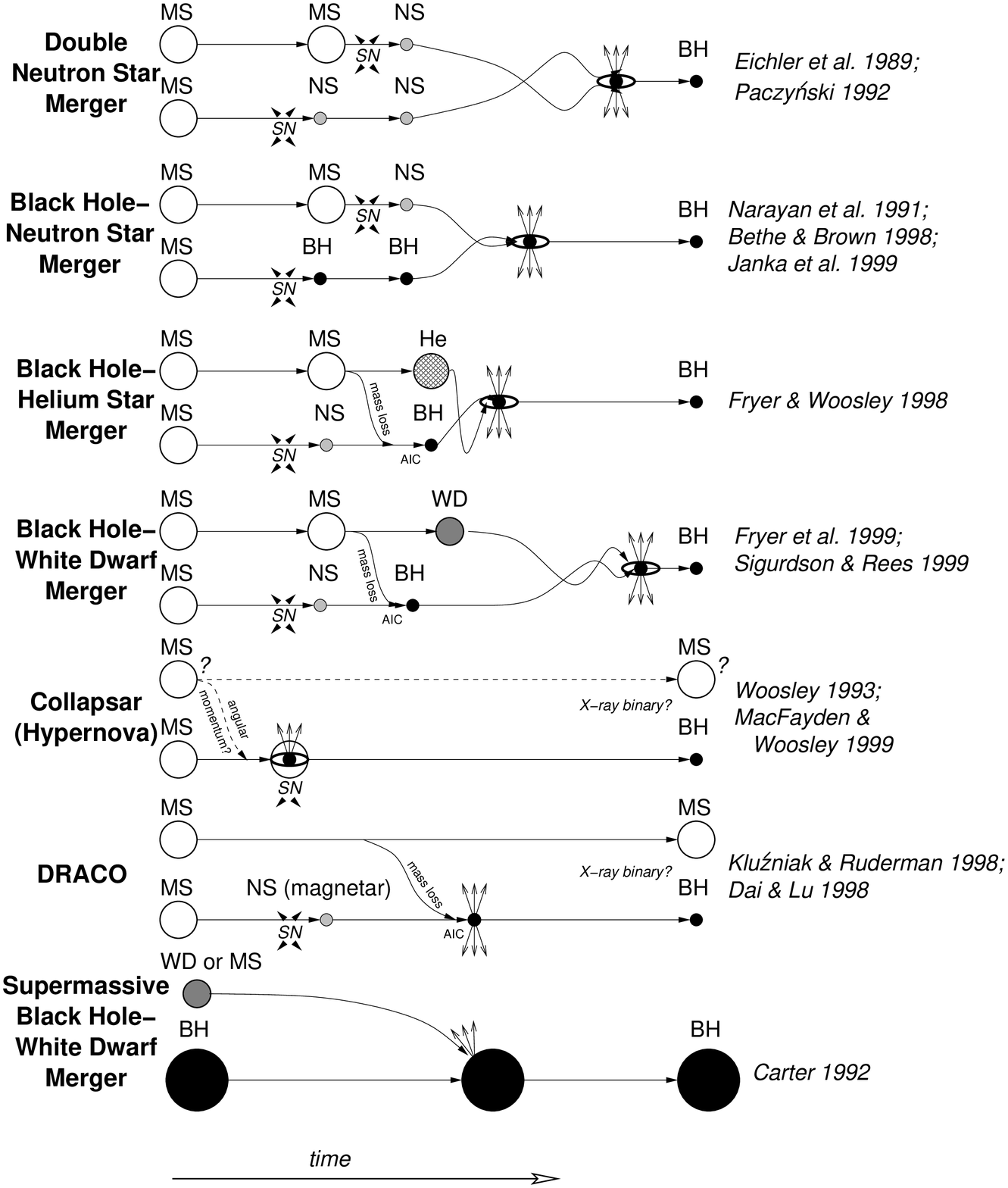,width=6.0in,angle=0}}
\caption[Schematic of plausible theoretical scenarios for the
progenitors of classic GRBs]{Schematic of plausible theoretical
scenarios for the progenitors of classic gamma-ray bursts.  In merger
scenarios, the primary star (more massive at ZAMS) is depicted as the
bottom component.  The dominant production channel for each scenario
is shown.  The (rough) relative in-spiral time due to gravitational
radiation for the four scenarios at top are shown (e.g., BH--He
mergers occur, in general, much more rapidly than NS--NS or NS--WD
mergers).  AIC = ``accretion-induced collapse''; SN = supernova
explosion. \nocite{dl98,jer+99}}
\label{fig:prog-mod}
\end{figure*}

Here, I briefly summarize the popular progenitor models and refer the
reader to \citet{fwh99} for a more in-depth review of the black-hole
accretion disk progenitors models.  Figure \ref{fig:prog-mod} depicts
a schematic compilation of the major progenitor scenarios for classic
GRBs along with references for the various theoretical treatments of
each scenario. The time sequence for the (supposed) predominant
production channel for each family of progenitor scenarios is shown,
although there are variants for each family that could plausibly
produce the same trigger.  For instance, only binary progenitors for
merging scenarios are depicted, but some mergers may occur after
stellar capture in dense cluster cores \citepeg{sr97}.

A spinning BH is formed in both the collapsar and the merging remnant
class of progenitors.  The debris, either from the stellar core of the
collapsar or a tidally disrupted neutron star, forms a temporary
accretion disk (or ``torus'') which then falls into the BH releasing a
fraction of gravitational potential energy of the matter. In this
general picture \citep[see][for a review]{ree99}, the lifetime of the
accretion disk accounts for the duration of the GRB and the
light-crossing time of the BH accounts for the variability timescale.
The GRB is powered by the energy extracted either from the spin energy
of the hole or from the gravitational energy of the in-falling
matter.
\begin{deluxetable}{lcr}
\singlespace
\tablecolumns{3} 
\tablewidth{0in} 
\tablecaption{Estimated Rates of GRBs and Plausible Progenitors\label{tab:rates}}
\tabletypesize{\small}
\tablehead{
\colhead{Progenitor/} & \multicolumn{1}{c}{Rate (yr$^{-1}$ Gpc$^{-3}$)} & Ref. \\
\colhead{Phenomenon}       & \colhead{local rate ($z=0$) }}
\startdata
NS--NS                      & 80 & \citet{phi91}   \\
BH--NS                      & 10--300 & \citet{fwh99}  \\
BH--WD                      & 10 & \citet{fwh99}  \\
BH--He                      & 1000 & \citet{fwh99}  \\
Type Ib/Ic                  & 6$\times 10^{4}$ & \citet{phi91} \\ 
GRBs       & 0.5\tablenotemark{a} &\citet{sch01,wbbn98} \\
           & 250\tablenotemark{b} & \citet{fks+01} 
\enddata

\tablenotetext{a}{Rates not including beaming.  Assumes that GRBs follow the star-formation rate in the universe.}

\tablenotetext{b}{Rates including the effects of beaming.}

\tablecomments{Aside from the rate of Type Ib/Ic SNe events, the rates of GRBs and possible GRB progenitors are uncertain by at least a factor of two. In the case of NS--NS and BH--NS mergers, the true rates probably are uncertain by at least an order of magnitude \citepeg{knst01}.  All of the progenitor scenarios listed closely scale with the rate of star formation; therefore, the rates at redshift of $z=1$ are a factor of $\sim$ten higher than locally.}

\end{deluxetable}

The coalescing compact binary class \citep{pac86,goo86,eic+89} was
favored before the first redshift determination because the existence
of coalescence events of a double neutron star binaries (NS--NS) was
observationally assured: at least a few known NS--NS systems in our
Galaxy (e.g., PSR 1913+16, PSR 1534+12) will merge in a Hubble time
thanks to the gravitational radiation of the binary orbital angular
momentum
\citepcf{tay94}.  Further, the best estimate of the rate of NS--NS
coalescence in the Universe \citepeg{phi91,npp92} was comparable to an
estimate of the GRB rate \citepeg{feh93}\footnotemark\footnotetext{The
latter estimate assumed a constant bursting rate as a function of
redshift and that the faintest bursts only had been detected to
redshift of unity.  These two assumptions, which proved to be
incorrect, upwardly biased the local unbeamed estimate by $\sim$150
\citep{wbbn98}.}\periodafterfoot Recently, stellar evolution models
have suggested that black hole--neutron star binaries (BH--NS) may be
formed at rates comparable to or even higher than NS--NS binaries
\citepeg{bb98}, though no such systems have yet been observed.  There
are other merging remnant binaries which may form GRBs, notably
merging black hole--white dwarf (BH--WD) binaries
\citep{ty94,fwhd99} and black hole--helium star binaries (BH--He)
\citep{fw98}. Table \ref{tab:rates} provides a summary of the various 
rates estimates of some of these GRB progenitors.

The collapsar class is comprised of a rotating massive Wolf-Rayet
star, either isolated or in a binary system, whose iron core
subsequently collapses directly to form a black hole. The basic
picture of a collapsar (or ``failed Type Ib supernova,'' as it was
often called) was pioneered by \citet{woo93}.  To avoid baryon loading
the progenitor star should have lost most, if not all, of its extended
gas envelope of hydrogen by the time of collapse.  The progenitors of
collapsars---likely Wolf-Rayet stars---are then closely related to the
progenitors of hydrogen-deficient supernova, namely type Ib/Ic
supernovae \citepcf{mw99}. 

As can be seen in table \ref{tab:rates}, clearly not all type Ib/Ic
supernovae can be accompanied by a GRB, even if beaming is taken into
account. Perhaps one distinguishing difference is that high angular
momentum is necessary in collapsars. High angular momentum
centrifugally supports a transient torus around the BH, fostering an
extended timescale ($\sim$tens of seconds) for mass-energy injection.
Further, angular momentum creates a natural rotation axis that sets up
large density gradients which then allow for the expanding blastwave
to reach relativistic speeds. The efficiency of energy conversion is
also helped around a spinning BH for two reasons.  First, the
innermost stable orbit around a Kerr BH is smaller than for a
non-rotating BH, allowing for more gravitational potential of the
accretion torus to be tapped.  Second, rotational energy extraction
from the hole becomes possible via the Blandford-Znajek process.
Given the benefits of high angular momentum in the collapsar scenario,
it is thus reasonable to suggest that collapsars might be more readily
formed in close binary systems.

The accretion-induced collapse scenarios posit that the energy for a
gamma-ray burst is stored in the rotation and/or magnetic field of
a compact object. \citet{uso92} first suggested that cosmological
GRBs could be powered by dipole radiation from a magnetized accreting
white dwarf which collapses to form a neutron star.
\citet{kr98} suggested a variant to this by noting that 
differential rotation is temporarily induced when an object collapses
due to accretion (either a white dwarf collapsing to a neutron star or
a neutron star collapsing to a black hole). During the
differential-rotation stage, a buoyant magnetic field dissipates the
rotational energy and could create a brief episodic burst of
electromagnetic energy.  Though the timescales for the energy
dissipation by such a mechanism appear to be plausible for GRBs, for
many years DRACO and magnetar models were not favored since the
energy reservoir was considerably less than one solar mass.  However,
after the relatively recent firm establishment of jetting in GRBs,
thus reducing the overall energy requirements, such models might now
be more viable than previously believed.

\section{Summary of the Thesis: Constraining GRB Progenitors}

Like supernovae, the progenitors of gamma-ray bursts have become a
subject of intense interest and study for many decades. But unlike
observed supernovae, GRBs are more distant and occur much less often.
This implies that the systematic study of GRB progenitors must
progress without the aid of nearby examples, where by assumption the
study of such examples could be done with high photometric and
astrometric precision.  Indeed, unlike as has already been seen with a
few supernovae \citep[e.g.,~SN 1987a as a
blue-supergiant;][]{wm87,wpp+89}, there is little hope that we will
ever have a resolved pre-discovery image of a GRB progenitor.

We are thus forced to uncover the progenitors of GRBs by indirect
means, and it is the thrust of my thesis work to do so.  
This thesis is separated into three parts corresponding to different
aspects of attack on the GRB progenitor question, progressing from the
very large scale (galaxies and locations within galaxies) to the very
small scale (the stellar scale). In the first part, entitled ``The
Large-scale Environments of GRBs,'' I posit that observations the host
galaxies of GRBs place important constraints on the nature of GRB
progenitors.  I show in chapter \ref{chap:nsns} that it is possible to
distinguish between various progenitor models by observing the
distribution of GRBs around their host galaxies. In chapter
\ref{chap:offset}, I present a comprehensive observational study of
the distribution of GRBs around galaxies which, in the context of
chapter \ref{chap:nsns}, provides one of the strongest constraints
to-date on the nature of the progenitors.

In part two of this thesis, entitled ``The GRB/Supernova Connection,''
I present the first observational evidence for bright
intermediate-time emission in a GRB afterglow.  I interpret this light
curve ``bump'' as due to a supernova which occurred contemporaneously
with a distant GRB.  Then, in light of the observations of a very
nearby supernova with a probable connection to a GRB, I propose a new
sub-classification of GRBs based on the physical underpinnings of a
true GRB-supernova connection.  In part three of this thesis, entitled
``An instrument to study the small-scale environments of GRBs,'' I
describe the design and construction of the {\it Jacobs Camera} (JCAM)
for the Palomar 200 inch Telescope.

\subsection{Progenitor clues from the large-scale environments}
\label{sec:sum-largescale}

\noindent{\bf The largest scale:} Studies of GRBs on the gigaparsec scale 
offer some big clues to the nature of the progenitors.  First and
foremost, of course, is the detection of redshifted absorption lines
which immediately renders all Galactic models for GRBs untenable.  The
relatively moderate redshift distribution of GRBs
(fig.~\ref{fig:intro-energy}) also suggests that some of the more
exotic progenitor scenarios at high redshifts, such as bursts from
super-conducting cosmic strings \citep{pac88}, are incorrect.  The
remainder of the viable models all posit a progenitor birthsite (though
not necessarily explosion site) near to and as part of the stellar
mass of galaxies.

The connection between GRBs and stars is borne out by the observed
redshift distribution (fig.~\ref{fig:intro-energy}). Qualitatively,
the rate of GRBs appears to peak around redshift of unity, just as the
inferred star-formation rate (SFR) in the universe (see \citealt{pm01}
for a recent review in the context of GRBs).  The most massive stars
(collapsars) explode soon ($\ale 10^{7}$ yr) after zero-age main
sequence (ZAMS) whereas merging neutron stars require a median time to
merge of \hbox{$\sim 2$--$10 \times 10^{8}$~yr} since ZAMS (e.g.,
\citealt{phi91}; \citealt{npp92}; \citealt{pzs96}; chapter \ref{chap:nsns},
\citealt{bsp99}).  In principle, therefore, due to the significant time 
from ZAMS to the mergers of NS--NS and BH--NS binaries, such merging
remnants should produce GRBs at preferentially {\it lower} redshift
than collapsars and promptly bursting binaries (BH--He); I quantified
this ``redshift offset'' of delayed merging binaries in chapter
\ref{chap:nsns} (see also \citealt{fwh99}). There are, unfortunately, 
a number of biases in the observed GRB sample (and universal SFR
measurements themselves!)  that preclude such a quantitative
comparison to solidify the stellar-origin connection and distinguish
between progenitors\footnotemark\footnotetext{Higher redshift
afterglows should be, in general, more dim and then systematically be
observed less frequently. Very high redshift bursts ($z \age 6$) would
escape detection at optical wavelengths due to blanketing from the
Lyman $\alpha$ forest.  Moreover, the window for redshift discovery is
not uniform for all redshifts; this is especially true for redshifts
discovered by emission spectroscopy of associated host galaxies. So in
practice, distinguishing the GRB($z$) rate from the SFR($z$) rate is
exceedingly difficult \citepeg{bn00} without tens if not hundreds more
GRB redshift measurements (see fig.~\ref{fig:sfr}).}\periodafterfoot
To date, no one has adequately accounted for all of the observational
biases to determine the true, underlying rate; as more uniformly
detected redshifts become possible with {\it
Swift}\footnotemark\footnotetext{The {\it Swift} satellite is a GRB
MIDEX mission scheduled for launch in late 2003.  The satellite will
localize GRBs to sub-arcsecond resolution just seconds after trigger
at a rate of about 2--3 per week. The on-board optical telescope will
be used for high-quality photometry and spectroscopy of the
afterglows. See \citet{bar00} and \citet{bhn+00} for
details.}\commaafterfoot $\!\!$though, such biases may one day be
quantifiable.

More uniformly selected are the associated host galaxies of GRBs and
thereby as a sample may be more informative than
redshifts. Observer-frame $R$-band magnitudes of GRB hosts appear, for
instance, to be in rough agreement with the hypothesis that the GRB
rate follows the SFR of the universe
\citep{mm98,hf99}.  The GRB hosts themselves, too, appear to be a fair
representation of the luminosity function of the general field
population \citep{dkb+01}; that is, GRB hosts do not appear to be
extraordinarily bright nor faint.  By inspection of the host images in
figure \ref{fig:offset1}, $\sim$50\% of the hosts could be classified
as a type irregular, peculiar, or merger, consistent with the fraction
from redshift unity galaxies studies \citep{res00,lal00}.
Furthermore, the distribution of unobscured star-formation, as proxied
by the equivalent width of [O II] $\lambda$3727, appears to follow the
[O II] distribution of galaxies at $z\sim1$ \citep{dkb+01}.

\noindent{\bf The galaxy scale:} The ability to compare the gross 
properties of GRB hosts with other samples is, of course, possible
only after studies on the individual hosts themselves. Chapters 
\ref{chap:grb970508}--\ref{chap:grb970228} represent some of the first 
moderately detailed observations of GRB host galaxies.  Chapter
\ref{chap:grb970228} (on the host of GRB 970228) is the most recent 
of these studies and shows, I believe, the power of combining detailed
spectroscopy with high-resolution imaging to arrive at a rather
complete picture of the nature of individual GRB hosts. In that
chapter, I strongly refuted a claim (that was based on photometry of
the host alone) that the host was extraordinary when compared to other
galaxies at similar redshifts \citep{ftp+99}.

Before the detailed modeling of light curves were used to constrain
the nature of supernovae progenitors, the location of supernovae in
and around galaxies provided important clues to the nature of the
progenitors \citepeg{rea53,jl63}.  For instance, only Type Ia
supernovae have been found in elliptical galaxies naturally leading to
the idea that the progenitor population can be quite old whereas the
progenitors of Type II and Type Ibc are likely to be closely related
to recent star formation \citep[see][for review]{vand92}.  Further, in
late-type galaxies, Type Ibc and Type II supernovae appear to be
systematically closer to HII star forming regions than Type Ia
supernovae \citepeg{btf94}.  This is taken as strong evidence that the
progenitors of Type Ibc and Type II SNe are massive stars
\citepcf{fil97}.

Similarly, given the delayed time to merge, the instantaneous rate of
GRBs from binary mergers is more a function of the integrated (as
opposed to instantaneous) star formation rate in its parent galaxy.
So if GRBs arise from the death of massive stars we do not expect
early-type (i.e., elliptical and S0) host galaxies, whereas GRBs from
merging remnants could occur in such galaxies. Indeed, no elliptical
host galaxy has yet been uncovered.

More important, independent of galaxy type, the locations of GRBs
within (or outside) galaxies provide a measurable signal to help
distinguish between progenitor scenarios. Massive stellar explosions
occur very near their birth-site, likely in active HII star-forming
regions, since the time since ZAMS is so small.  BH--He binaries will
merge quickly and so are also expected to be located near star-forming
regions \citep{fwh99}.  In stark contrast, as explored in chapter
\ref{chap:nsns}, NS--NS (and NS--BH) binaries merge far from their
birthsite. These stellar remnant progenitors will merge after at least
one of the binary members has undergone a supernova.  Each supernova
is thought to impart a substantial ``kick'' on the resulting neutron
star \citepeg{hp97}; for those binary systems which survive both
supernovae explosions, the center--of--mass of the remnant binary
itself will receive a velocity boost on the order of a few hundred km
s$^{-1}$ \citepeg{bp95}.  That is, NS--NS or NS--BH binaries will be
ejected from their birthsite.  The gradual angular momentum loss in
the binary due to gravitational radiation causes the binary to
coalesce/merge which then leads to a GRB.  The time to merge ($\sim
10^{6}$--$10^{9}$ yr) depends on the masses of the remnants and binary
orbit parameters.  Population synthesis models have all shown that
roughly one-third to one-half of NS--NS and BH--NS binary mergers will
occur beyond 10 kpc in projection from the centers of their hosts
\citep{bsp99,fwh99}.  The distribution of merger sites depends
sensitively on the gravitational potential of the host and the
(radial) distribution of massive star birth sites.

The realization that the offset of GRBs from their host galaxies could
be one of the cleanest observational tests for the progenitors
motivated the work in chapter \ref{chap:offset}. There, by observing
the distribution of 20 cosmological GRBs about their hosts, I find
good evidence for a progenitor population which follows the UV light
of their host galaxies.

\subsection{An instrument to study the small-scale environments of GRBs}
\label{sec:sum-jcam}

GRBs and afterglows both influence and are influenced by the immediate
surroundings.  As noted in \S \ref{sec:hist}, the afterglow stage
begins at distance of $\sim10^{14}$ cm from the explosion site and at
such a distance the impact of the progenitor should be felt.  For
instance, \citet{wl99} noted that the mass-loss history from a
collapsar progenitor is likely to be complex and non-axisymmetric.
The resulting density inhomogeneities would result in small-scale
variations of GRB light curves about an otherwise smooth synchrotron
shock afterglow.  Though the measurement of such features is clearly
of interest, after a few hours the r.m.s.~fluxuations on timescales of
minutes are less than 5\%.  Practically, this implies that the
imprints of the small-scale environments cannot be measured by small
aperture telescopes.

To this end, we built the Jacobs Camera instrument
(JCAM\footnotemark\footnotetext{This instrument was privately funded
by a donation from M.~Jacobs.}\nadaafterfoot) for the Hale 200-inch
telescope at Palomar Observatory.  The instrument---a dual CCD imaging
camera with very quick readout---has just ended its commissioning
phase and we hope to make the instrument public within a year. Since
JCAM is permanently mounted at the East Arm, we can begin observing
simultaneously in two optical bandpasses in 5--20 minutes from
trigger. I helped to find funding for the instrument, design and
integrate the electronics hardware, design and implement the software
control systems, integrate the instrument and the telescope, and
perform the necessary calibrations and maintenance.  I have been
solely responsible for the commissioning of JCAM.

Chapter \ref{chap:jcam} is a version of the JCAM data paper which has
been submitted to {\it PASP} for publication.  There, I argue that
rapid observations of GRB afterglows can yield important clues to the
nature of the progenitors.  By mapping the evolution of the afterglow
in time and in broadband colors, we hope to be able to directly probe
density inhomogeneities (or lack thereof) on scales of $10^{14}$ cm or
smaller and constrain the initial Lorentz factor of GRBs (see \S
\ref{sec:motivation}).

\subsubsection{Wherefore another optical imaging instrument?}

The past decade saw the advent of large-format CCDs and IR arrays for
astronomical imaging, fostered by a wide range of scientific
objectives (e.g.,~surveys to detect weak lensing, microlensing events,
near-Earth asteroids, planetary transits, or Lyman-break galaxies) and
only recently made practical by the rapid fall in detector cost and,
most important, data storage. At Palomar Observatory alone three new
large-format instruments have emerged as the dominant paradigm for new
instrumentation at the site---on the Hale 200 inch, the {\it Large
Format Camera} (LFC\footnotemark\footnotetext{$5 \times 10^7$ pix,
0.12 deg$^2$; \raggedright{\tt
http://www.astro.caltech.edu/$\sim$ras/lfc/lfc.html}}\negafterfoot)
for optical imaging and the {\it Wide Field Infrared Camera}
(WIRC\footnotemark\footnotetext{$4 \times 10^6$ pix, 0.02
deg$^2$; \raggedright {\tt
http://isc.astro.cornell.edu/$\sim$don/wirc.html}}\negafterfoot) for
infrared imaging; on the Oschin 48 inch Schmidt, the {\it Near-Earth
Asteroid Tracking} system (NEAT\footnotemark\footnotetext{$5 \times
10^7$ pix, 3.75 deg$^2$; \raggedright {\tt
http://neat.jpl.nasa.gov/neatoschincam.htm}}\negafterfoot).  Despite
the advantages of a large field, costs and data rates from
large-format cameras are still formidable. The LFC, for example, costs
over \$500 k in hardware alone, and, on a typical night generates over
8 GB of data.  Given that the mounting and cooling of the prime-focus
instruments requires over 24 hrs, such instruments are clearly not
well-suited for unanticipated transient follow-up.

On the 200 inch telescope and on the Keck telescopes, the primary
instrument tends to change every several days depending on lunar phase
and the observers' science. An informal survey of the Keck I 2001B
schedule shows that only 45\% of the nights were scheduled with an
imager suitable for deep optical follow-up.  The other prime
disadvantages of large-telescope optical instrumentation are the
rather large full-frame readout times (e.g.,~70 sec for the LFC, 145
sec for COSMIC), large data rates, and the inability to observe many
bands simultaneously.

JCAM is an entirely different approach to instrumentation at the
Palomar. The scientific intent in the construction of JCAM, described
more fully in \S \ref{sec:motivation}, was to provide quick access to
multi-color optical imaging of GRB afterglows. The instrument is now
continuously mounted at the East Arm f/16 focus of the Hale 200 inch
and can be accessed by the installation of the Coud\'e mirror and
secondary mirror in the light path (see \S \ref{sec:jcam-ops}).  One
advantage of a small number of pixels per image ($2.6 \times 10^5$) is
that data transfer rates, even over T1-line quality connections, are
manageable; this has allowed us to observe with JCAM remotely over the
Internet, the first such instrument at Palomar Observatory.

\subsection{The stellar scale: connection to supernovae}
\label{sec:sum-grbsn}

Over the span of just one month in 1998, two GRBs (980326 and 980425)
were detected that continue to influence our understanding of GRB
progenitors, in particular the connection of GRBs to supernovae and
hence to massive stars.  GRB 980425 was associated with a nearby
peculiar supernovae SN 1998bw at a distance of 39.1 Mpc, suggesting
the burst had an extraordinarily low gamma-ray energy output (8
$\times 10^{47}$ erg) compared with the energetics of other GRBs
(chapter \ref{chap:sgrbs}). GRB 980326, as described in chapter
\ref{chap:sn-grb}, was the first cosmological GRB for which an
associated supernova was found.

\subsubsection{A sub-class of under-luminous GRBs produced by
supernovae}
 
In the localization error box of the {\it BeppoSAX} WFC,
\citet{gvv+98} discovered a Type Ib/Ic supernova which was later
recognized as more energetic compared to type Ib/Ic supernovae as
measured from optical expansion velocities \citep{imn+98}. Though, on
a purely phenomenological basis, the chance probability for a spurious
association between SN 1998bw and GRB 980425 appeared to be small
\citep{gvvk+99}, there was some ambiguity as to the whether a {\it
bone fide} X-ray afterglow was detected at a position inconsistent
with the SN \citep{paa+99}. Regardless of the true physical connection
with GRB 980425, we recognized that the prompt radio emission detected
from SN 1998bw necessitated relativistic shock propagation during the
initial period of the explosion \citep{kbf+98a,kfw+98}---the first
evidence for a relativistic shock in a supernova \citep{wkf99}.  That
the young shock likely contained enough energy ($\age$ 10$^{49}$ erg)
to power a (weak) GRB strengthened the connection between the two
phenomena on a {\it physical} basis. Indeed, it is now widely accepted
that the physical association is real \citepeg{whe01,sal01}.

Chapter \ref{chap:sgrbs} was written against the backdrop of this
exciting possibility of a physical connection between a supernova and
a gamma-ray burst. There, I realized that the physical model put forth
by \citet{kbf+98a} held some predictions about the general properties
of the resulting supernovae and accompanying GRB (of course, GRB
980425 and SN 1998bw, naturally exhibited all of these
properties). Could this be a newly discovered subclass of GRBs?  I
asked whether any of the known GRBs and supernovae fit the proposed
properties of what I call S-GRBs (supernova GRBs).  Despite
provocative speculations by a number of other authors about a few
other GRB/SN connections, I unfortunately (but not unexpectedly) found
no convincing evidence for another plausible association.

Given the lack of a believable association of any other SN/GRB pair, I
then sought to place constraints on the frequency of such S-GRBs
finding that at most a few percent of such bursts comprise the known
GRB sample; figure \ref{fig:sn-6707} shows some examples of possible
S-GRBs based solely on the similarity of the light curves with GRB
980425. 

\subsubsection{GRB 980326: the first connection of cosmological GRBs 
to supernovae}

GRB 980326 was one the softest and faintest GRBs localized by {\it
BeppoSAX}, but was otherwise unremarkable as a GRB. In the first few
days, the afterglow exhibited a rapid decline seen in many other GRBs.
As had become standard practice, about 30 days after the burst we
imaged the GRB field and obtained a long integration spectrum of the
supposed host galaxy. The idea was to find the redshift of the GRB
through emission spectroscopy.  Though continuum was detected, no
obvious lines were seen in the ``host'' spectrum.  Seeking to refine
the astrometry so as to improve the blind-offset spectroscopy
observations, we re-observed the field of GRB 980326 eight months
after the burst and to our surprise the ``host'' was gone, having
faded by at least a factor of 10 in flux from our intermediate-time
imaging and spectroscopic detections. Two years later, a faint host
was finally detected at the afterglow position using {\it HST} (see \S
\ref{subsec:980326}).

A redress of the ``host'' hypothesis was clearly warranted, leading us to
conclude that we had seen something unusual in the month after GRB
980326.  Our conclusions (as well as observations) are presented in
chapter \ref{chap:sn-grb}.  There I discuss how the data are
consistent  with the presence of an underlying supernova which peaked in
brightness around the same time as our intermediate-time observations.
Moreover, I found that the supernova component was consistent with the
peak flux of SN 1998bw if the the supernova had occurred at a redshift
of unity. A number of alternative physical explanations had been put
forth that might have explained the intermediate-time ``bump'' (e.g.,~dust
heating and re-radiation, delayed energy input from a long lasting
remnant, thermal expansion after a merger of a binary neutron star).
Yet, as noted in \S \ref{sec:transient}, all such interpretations
failed to explain either the timescale of the bump or the colors of
the bump.

Soon after this chapter was submitted, a re-analysis of the optical
photometry of GRB 970228 showed the light curve to be consistent with
a supernova component \citep{rei99,gtv+00}.  Unfortunately, the flux
of the afterglow of GRB 970228 was of a brightness comparable to the
SN component, and so the significance of the detection of the bump is
greatly diminished relative to that in the GRB 980326. Unlike with GRB
980326, however, GRB 970228 showed photometric evidence for a bump in
three bandpasses and showed a roll-over in a broadband spectrum
consistent with that of a type Ib/Ic SN \citep{rei01b}. Most
important, the redshift of GRB 970228 was known (see chapter
\ref{chap:grb970228}), which removed the peak-brightness---redshift
degeneracy that existed in GRB 980326.

\subsubsection{GRB 011121: multi-color observations of a supernova-like component}

GRB 011121 holds the record as the cosmological GRB with the lowest
known redshift ($z = 0.36$). For this reason, the peak in any
associated supernova component was expected to be at $\sim 7000$ \AA,
squarely in the observer-frame optical bands.  Starting about two
weeks after the burst, we triggered a series of {\it HST} observations
of the source with the hope of constraining the nature of any
intermediate-time emission. As discussed in chapter
\ref{chap:grb011121}, such emission is seen as a clear excess above
that expected from the early afterglow.

Indeed, chapter \ref{chap:grb011121} builds upon the previous emission
bump observations in 980326 and 970228, showing excess at four
different epochs in as many as five optical filters per epoch. The
expected spectral roll-over beyond 7000 \AA, if the source of emission
was a supernova, is seen. Further, the characteristic rise and decay
of core-collapsed supernovae on timescales of weeks, is also seen in
the emission bump. I argue that the physical origin of the emission
bump is from a supernova which occurred at nearly the same time as the
GRB itself.

\mypart{The Large-scale Environments of GRBs}{The Large-scale Environments of Gamma-Ray Bursts}{}

\chapter[The Spatial
Distribution of Coalescing Neutron Star Binaries: \\ Implications for
Gamma-Ray Bursts]{The Spatial Distribution of Coalescing Neutron Star
Binaries: Implications for Gamma-Ray Bursts$^\dag$}
\label{chap:nsns}

\secfootnote{\secfootdag}{A version of this chapter was first published 
in {\it The Monthly Notices of the Royal Astronomical Society}, 305,
p.~763--768 (May 1999).}

\secauthor{Joshua S.~Bloom$^{1}$ and Steinn Sigurdsson$^2$, and Onno R.~Pols$^{2,3}$}

\secaffils{$^1$ California Institute of Technology, MS 105-24,
Pasadena, CA 91106 USA}

\secaffils{$^2$ Institute of Astronomy, Madingley Road, Cambridge, 
                CB3 0HA, England}

\secaffils{$^3$ Instituto de Astrof\'{\i}sica de Canarias, 
		c/ Via L\'actea s/n, E-38200 La Laguna, Tenerife, Spain}

\begin{abstract}

We find the distribution of coalescence times, birthrates, spatial
velocities, and subsequent radial offsets of coalescing neutron stars
(NSs) in various galactic potentials accounting for large asymmetric
kicks introduced during a supernovae.  The birthrates of bound NS--NS
binaries are quite sensitive to the magnitude of the kick velocities
but are, nevertheless, similar ($\sim 10$ per Galaxy per Myr) to
previous population synthesis studies.  The distribution of merger
times since zero-age main sequence is, however, relatively insensitive
to the choice of kick velocities.  With a median merger time of $\sim
10^8$ yr, we find that compact binaries should closely trace the star
formation rate in the Universe.  In a range of plausible galactic
potentials (with $M_{\rm galaxy} \age 3\times 10^{10} M_\odot$) the
median radial offset of a NS--NS merger is less than 10 kpc. At a
redshift of $z=1$ (with $H_0 = 65$ km s$^{-1}$ Mpc$^{-1}$ and $\Omega
= 0.2$), this means that half the coalescences should occur within
$\sim 1.3$ arcsec from the host galaxy.  In all but the most shallow
potentials, 90 percent of NS--NS binaries merge within 30 kpc of
the host.  We find that although the spatial distribution of
coalescing neutron star binaries is consistent with the close spatial
association of known optical afterglows of gamma-ray bursts (GRBs)
with faint galaxies, a non-negligible fraction ($\sim 15$ percent) of
GRBs should occur well outside ($\age 30$ kpc) dwarf galaxy hosts.
Extinction due to dust in the host, projection of offsets, and a range
in interstellar medium densities confound the true distribution of
NS--NS mergers around galaxies with an observable set of optical
transients/galaxy offsets.
\end{abstract}

\section{Introduction}

The discovery of an X-ray afterglow \citep{cfp+97a} by BeppoSAX (Costa
\citep{bbp+97} and subsequently an optical transient
associated with gamma-ray burst (GRB) 970228 \citet{vgg+97} led to the
confirmation of the cosmological nature of GRBs \citet{mdk+97}.  The
broadband optical afterglow has been modeled relatively successfully
\citep{mr93a,wrm97,wax97a,wkf98} as consistent with an expanding
relativistic fireball
\citep{rm94,pr93,katz94a,mr97a,viet97,spn98,rm98}.  Still, very little
is known about the nature of the progenitors of GRBs, and, for that
matter, their hosts.  Broadband fluence measures and the known
redshifts of some bursts implies a minimum (isotropic) energy budget
for GRBs of $\sim 10^{52-53}$ ergs (\citealt{mdk+97};
\citealt{kdr+98}; see also \citealt{bfs01} and fig.~\ref{fig:intro-energy}).  
The log $N$-log $P$ brightness distribution, the observed rate, $N$,
of bursts above some flux, $P$, versus flux, indicates a paucity of
dim events from that expected in a homogeneous, Euclidean space.  With
assumptions of a cosmology, source evolution and degree of anisotropy
of emission, the log $N$-log $P$ has been modeled to find a global
bursting rate.  Assuming the bursts are non-evolving standard candles
\citet{fb95} found $\sim 1$ burst event per galaxy per Myr
(GEM) to be consistent with the observed log $N$-log $P$.  More
recently, \citet{wbbn98} (see also, \citealt{tot97}; \citealt{lpp97})
found the same data consistent with GRBs as standard candles assuming
the bursting rate traces the star-formation rate (SFR) in the
Universe; such a distribution implies a local burst rate of $\sim
0.001$ GEM and a standard peak luminosity of $L_0 = 8.3
\times 10^{51}$ erg s$^{-1}$ \citep{wbbn98}.

Given the energetics, burst rate and implied fluences, the
coalescence, or merger, of two bound neutron stars (NSs) is the
leading mechanism whereby gamma-ray bursts are thought to arise
\citep{pac86,goo86,eic+89,npp92}.  One quantifiable prediction of the
NS--NS merger hypothesis is the spatial distribution of GRBs (and GRB
afterglow) with respect to their host galaxies. Conventional wisdom,
using the relatively long--lived Hulse-Taylor binary pulsar as a
model, is that such mergers can occur quite far ($\age 100$ kpc)
outside of a host galaxy.  Observed pulsar (PSR) binaries with a NS
companion provide the only direct constraints on such populations,
but the observations are biased both toward long lived systems, and
systems that are close to the Galactic plane.

The merger rate of NS--NS binaries has been discussed both in the
context of gravitational wave-detection and GRBs 
\citepeg{phi91,nps91,npp92,ty94,lpp+95}. 
Recently \citet{fbb98,lpp97,pzs96} studied the effect of asymmetric
kicks on birthrates of NS--NS binaries, but did not quantify the
spatial distribution of such binaries around their host galaxies.
\citet{ty94} discussed the spatial distribution of
NS--NS mergers but neglected asymmetric kicks and the effect of a
galactic potential in their simulations.  Only \citet{pzy98} have
discussed the maximum travel distance of merging neutron stars
including asymmetric supernovae kicks.

It is certainly of interest to find the rate of NS--NS coalescences
{\it ab initio} from population synthesis of a stellar
population. This provides an estimate of beaming of GRBs, assuming
they are due to NS--NS mergers, and hence an estimate of probable
frequency of gravitational wave sources, providing a complementary
rate estimate to those of \citet{phi91} and \citet{nps91}, which are
based on long lived NS-PSR pairs only and are very conservative. It
also provides an estimate of how the NS--NS mergers trace the
cosmological star formation rate (SFR) of the Universe, if mean
formation rates and binarity of high mass stars are independent of
star formation environments such as metallicity.

Here we concentrate on estimating the spatial distribution of
coalescing NS--NS binaries around galaxies.  To do so, both the system
velocity and the interval between formation of the neutron star binary
and the merger through gravitational radiation is found by simulation
of binary systems in which two supernovae occur.  We explore the
effects of different asymmetric kick amplitudes, and the resultant
birthrates and spatial distribution of coalescing NS--NS binaries born
in different galactic potentials.

In section 2 we briefly outline the prescription for our Monte Carlo
code to simulate bound binary pairs from an initial population of
binaries by including the effect of asymmetric supernovae kicks.  In
section 3 we outline the integration method of NS--NS pairs in various
galactic potentials.  Section 4 highlights the birthrates and spatial
distributions inferred from the simulations.  Section 5 concludes by
discussion the implications and predictions for gamma-ray burst
studies.

\section{Neutron Star Binary Population Synthesis}

We used a modified version of the code created for binary evolution by
Pols \citep{pm94} taking into account the evolution of eccentricity
through tidal interaction and mass transfer before the first and
second supernova, and allowing for an asymmetric kick to both NSs
during supernovae. The reader is referred to \citet{pm94} for a more
detailed discussion account of the binary evolution code.

\subsection{Initial conditions and binary evolution}

In general, the evolution of a binary is determined by the initial
masses of the two stars ($m_1$, $m_2$), the initial semi-major axis
($a_o$) and the initial eccentricity ($e_o$) of the binary at zero-age
main sequence (ZAMS).  We construct Monte Carlo ensembles of high-mass
protobinary systems (with primary masses between $4 M_\odot$ and $100
M_\odot$) by drawing from an initial distribution of each of the four
parameters as prescribed and motivated in \citet{pzv96}.  We treat
mass transfer and common-envelope (CE) phases of evolution as in
\citet{pm94}. CE evolution is treated as a spiral-in process; we use a
value of $\alpha=1$ for the efficiency parameter of conversion of
orbital energy into envelope potential energy; see equation [17] of
\citet{pm94}.  We treat circularization of an initially eccentric
orbit as in \citet{pzv96}.

During detached phases of evolution we assume that mass accreted by
the companion is negligible so that $a M_{\rm tot}$ = constant.  Mass
lost by the binary system in each successive time step results in a
change in eccentricity according to the sudden mass loss equations
\citep[see, for example, eqns.~A.21 and A.24 of][]{wb96}. 
We ignore the effect of gravitational radiation and magnetic braking
in the early stages of binary evolution.

The simple approximation of the 4-parameter distribution function,
albeit rather {\it ad hoc}, appears to adequately reproduce the
observed population of lower mass stars in clusters \citepeg{pm94}.
The effect on the distribution of NS--NS binaries after the second
supernova by variation of the 4-parameter space is certainty of
interest, but we have used the canonical values.  A fair level of
robustness is noted in that varying the limits of the initial
distributions of $a_o$ and $e_o$ does not the change the implied
birthrates of bound NS binaries nearly as much as plausible variations
in the asymmetric kick distribution.  This effect was noted in 
\citet{pzs96} and \citet{pzv96}.

\subsection{Asymmetric supernovae kicks}

Several authors \citepeg{pac90,no90,ll94,cc97} have sought to
constrain the distribution of an asymmetric kick velocities from
observations of isolated pulsars which are the presumed by-products of
type II supernovae.  Even careful modeling of the selection effects in
observing such pulsars has yielded derived mean velocities that differ
by nearly an order of magnitude.  It is important here to use a good
estimate for the actual physical impulse (the ``kick velocity'') the
neutron stars receive on formation. The observed distribution of
pulsar velocities does not reflect the kick distribution directly as
it includes the Blaauw kick \citep{bla61} from those pulsars formed in
binaries, and selection effects on observing both the high and low
speed tail of the pulsar population \citepeg{hbwv97}. \citet{hp97}
found that the observed distribution is adequately fit by a Maxwellian
velocity distribution with $\sigma_{\rm kick} = 190$ km s$^{-1}$
(which corresponds to a 3-D mean velocity of ~300 km s$^{-1}$).  Since
it is not clear that pulsar observations require a more complicated
kick-velocity distribution, we chose to adopt a Maxwellian but vary
the value of $\sigma_{\rm kick}$.

When a member of the binary undergoes a supernova, we assume the
resulting NS receives a velocity kick, $v_k$, drawn from this
distribution.  Although the direction of this kick may be coupled to
the orientation of the binary plane, we choose a kick with a random
spatial direction, since there is no known correlation between the
kick direction or magnitude and the binary parameters.

If $\alpha$ is the angle between the velocity kick and the orbital
plane and $v$ is relative velocity vector of the two stars, then,
following earlier formulae
\citepeg{pzv96,wb96}, the new-semi major axis of the binary is
\begin{equation}
a' = \left( \frac{2}{r} - \frac{v^2 + v_k^2 + 2 v v_k \cos \alpha}
	{{\rm G}({\rm M}_{\rm NS} + {\rm M}_2)}\right)^{-1}
\label{eq:a}
\end{equation}
where $r$ is the instantaneous distance between the two stars before
SN, $M_2$ is the mass of the companion (which may already be a NS),
and $M_{\rm NS}=1.4 M_{\odot}$ is the mass of the resulting neutron
star.  We neglect the effects of supernova-shell accretion on the mass
of the companion star.  If $a'$ is positive, the new eccentricity is
\begin{equation}
e' = \left[1 - \frac{| \vec r \times \vec v_r|^2}{a'{\rm G}({\rm
M}_{\rm NS} + {\rm M}_2)}\right]^{1/2}~,
\label{eq:e}
\end{equation}
where the resultant relative velocity is $\vec v_r = \vec v + \vec
v_k$. Assuming the kick directions between successive SN are
independent, the resulting kick to the bound system (whose magnitude
is given by equation 2.10 of \citealt{bp95}) is added in quadrature to
the initial system velocity to give the system velocity ($v_{\rm
sys}$).
 
To produce 1082 bound NS--NS binaries with a \citeauthor{hp97} kick
velocity distribution and initial conditions described above, we
follow the evolution of 9.7 million main sequence binaries which
produce a total of $\sim$ 1 million neutron stars through supernovae.
Assuming a supernova rate of 1 per 40 years \citep{tls94} and 40\%
binary fraction \citep[as in][]{pzs96}, we find an implied birthrate
of NS--NS binaries by computing the number of binaries with SN type II
per year and multiply by the ratio of bound NS--NS systems to SN type
II as found in the simulations.  We neglect the (presumed small)
contribution of other formation channels (e.g., three-body
interactions) to the overall birthrate of NS--NS binaries.  The
implied birthrate of NS--NS binaries from various kick-velocity
magnitudes are given in table 2.

\section{Evolution of Binaries Systems in a Galactic Potential}

The large-scale dynamics of stellar objects are dominated by the halo
gravitational potential while the initial distribution of stellar
objects is characterized by a disk scale length.  We take the disk
scale and halo scale to vary independently in our galactic models.  We
assume that the NS--NS binaries are born in an exponential stellar
disk, with birthplace drawn randomly from mass distribution of the
disk. The initial velocity is the local circular velocity
(characterized but the halo) plus $v_{\rm sys}$ added with a random
orientation.

We then integrate the motion of the binary in the galactic potential
assuming a \citet{her90} halo; we ignore the contribution of the disk
to the potential.  We assume scale lengths for the disk and halo: the
disk scale ($r_{\rm disk}$) determines the disk distribution, the halo
scale length ($r_{\rm break}$) and circular velocity ($v_{\rm circ}$)
determine the halo mass (see table 1).  The movement of the NS--NS
binaries on long time-scales is sensitive primarily to the depth of
the galactic potential (here assumed to be halo dominated) and how
quickly it falls off at large radii.  Assuming isothermal halos
instead of Hernquist profiles would decrease the fraction of NS--NS
pairs that move to large galactocentric radii, but the differences in
distribution are dominated by the true depth of the halo potentials in
which the stars form rather than their density profiles at large
radii.

\begin{figure}
\centerline{\psfig{file=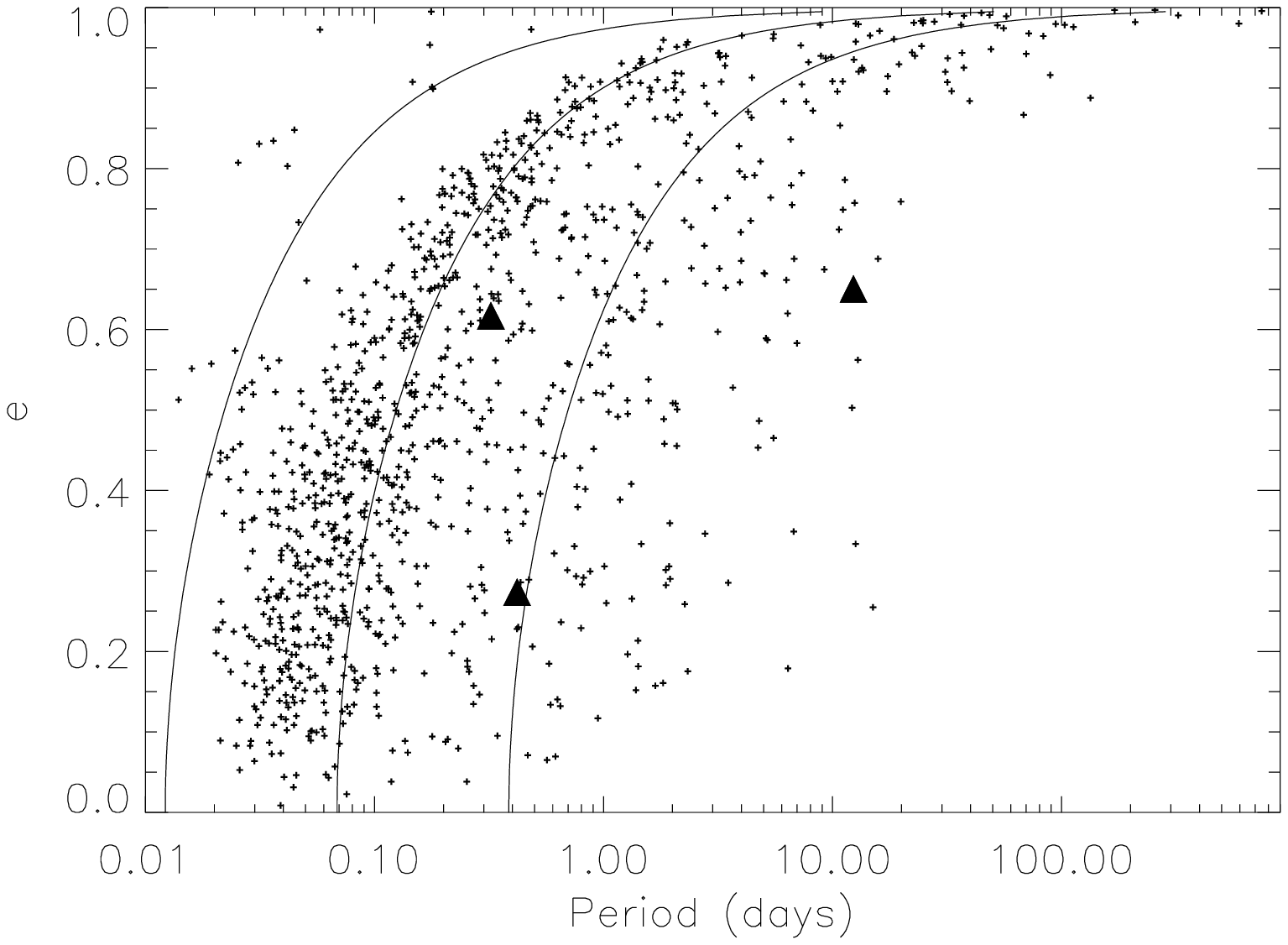,width=5in}}
\caption[The distribution of orbital parameters after the second 
supernovae for bound NS--NS pairs]{The distribution of orbital
parameters (period and eccentricity) after the second supernovae for
bound NS--NS pairs. From left to right are lines of constant merger
time after second SN ($10^6$, $10^8$, $10^{10}$ yrs). The parameters
of the observed NS pairs 1913+16 \citep{tw89}, 1534+12 \citep{wol91},
and 2303+46 \citep{td88} are marked with triangles.  With an
observational bias toward long-lived systems, clearly the observed
PSR-NS systems are not indicative of the true NS binary distribution.}
\label{fig:ae}
\end{figure} 
We use a symplectic leapfrog integrator to advance the binary in the
galactic potential, and a simple iteration scheme to evolve the
semi-major axis and eccentricity of the binary as gravitational
radiation drives $a$ and $e$ to zero, assuming the orbit averaged
quadrupole dominated approximation \citep{pet64}.  The integration is
continued until either $1.5 \times 10^{10}$ years have passed (no
merger in Hubble time) or the characteristic time to merger is short
compared to the dynamical time scale of the binary in the halo (i.e.,
the binary will not move any further before it merges).  We then
record the 3-D position of the binary relative to the presumptive
parent galaxy and the time since formation.

\section{Results}

\subsection{Orbital parameter distribution after the second supernova}

Figure \ref{fig:ae} shows the distribution of orbital parameters
(semi-major axis and period) after the second supernova for bound
NS--NS pairs for the \citet{hp97} kick distribution ($\sigma_{\rm
kick} = 190$ km s$^{-1}$). As found previously \citepeg{pzs96}, bound
systems tend to follow lines of constant merger time.  The density of
systems in figure \ref{fig:ae} can be taken as the probability density
of finding a NS--NS binaries directly after the second supernova.  In
time, the shorter-lived systems (higher $e$ and shorter period) merge
due to gravitational radiation.  Thus, at any given time after a burst
of star-formation there is an observational bias toward finding
longer-lived systems.  In addition, there is a large observational
bias against finding short period binaries \citep{jk91}.  That the
observed PSR-NS systems lie in the region of parameter space with low
initial probability is explained by these effects.  The time-dependent
probability evolution has been discussed and quantified in detail by
\citet{pzy98}.
\begin{figure}
\centerline{\psfig{file=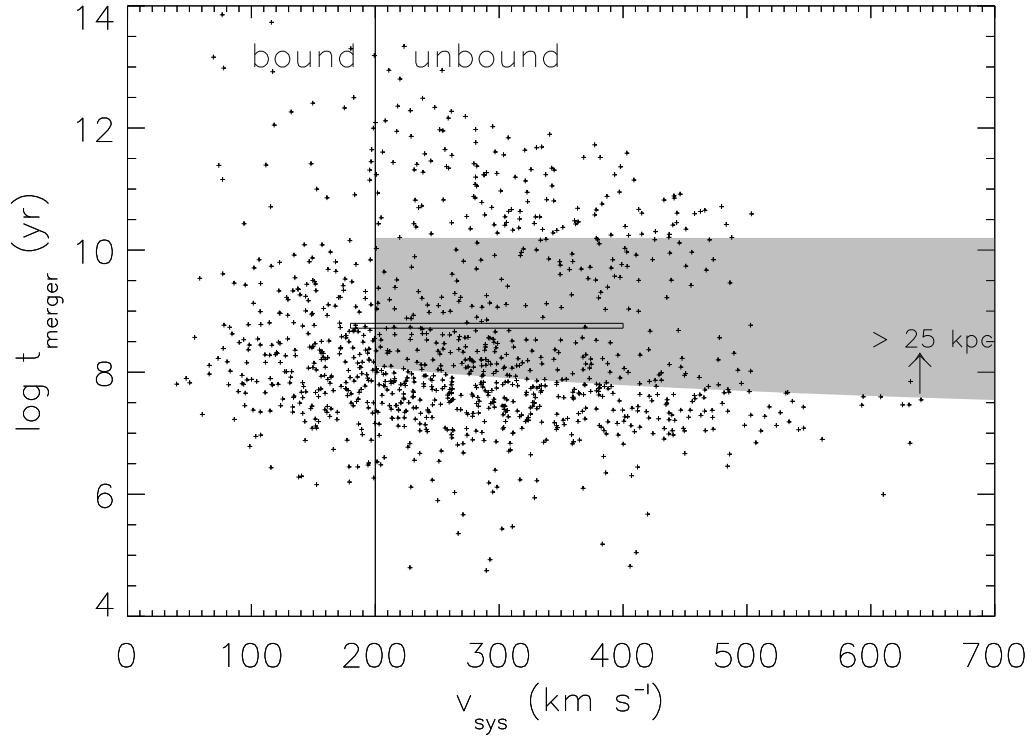,width=6in}}
\caption[The distribution of merger times after second supernovae as
a function of system velocity]{The distribution of merger times after
second supernovae as a function of system velocity.  Left of the
vertical line, all pairs created are gravitationally bound to an
under-massive host ($3 \times 10^{10} M_\odot$) at the disk scale
radius.  Of the pairs that are unbound, only the pairs in the shaded
region could travel more than $\sim 25$ kpc (linearly) from their
birthplace and merge within a Hubble time ($\ale 1.5\times 10^{10}$
yrs). Since the spatial velocity of observed NS binaries includes both
the initial circular velocity of the system and the system velocity
due to kicks from each supernova, the true system velocities are
highly uncertain.  For comparison, though, we denote the range of
accepted kick velocities of PSR 1913+16 with a long rectangle (the
merger time is much better constrained than that depicted); this
illustrates a general agreement of the system velocity of PSR 1913+16
and the modeled distribution of bound NS binaries.  The slightly
longer merger time of PSR 1913+16 than expected from the density of
systems in this parameter space is explained in section 4.1 of the
text.}
\label{fig:tv}
\end{figure} 
Figure \ref{fig:tv} shows the distribution of merger times as a
function of system velocity. A majority of systems merge in $\sim
10^8$ yr spread over system velocities of $50$ -- $500$ km s$^{-1}$.
A subclass of systems have spatial velocities and merger times that
are anti-correlated.

\begin{deluxetable}{lccccccc}
\tablewidth{0pt}
\tablecaption{The Spatial Distribution of Coalescing Neutron Star Binaries in
Various Galactic Potentials\label{table:sd}}
\tablehead{
 & \multicolumn{5}{c}{Galaxy parameters} & \multicolumn{2}{c}{Coalescence Distance} \\
Run & $v_{\rm circ}$  & $r_{\rm break}$ & $r_{\rm
  disk}$  &  $M$  & L & $d_{\rm median}$ & $d_{\rm avg}$  \\
    & (km s$^{-1}$) &  (kpc) & (kpc) & ($10^{11} M_{\odot}$) &  & (kpc)  & (kpc)
}

\startdata

 a &  100 &  1  & 1 & 0.092 & $\ale  0.05 L_{*}$ & 4.3 & 66.2 \cr
 b &  100 &  3  & 1 & 0.278 & $\simeq  0.1 L_{*}$ & 4.0 & 50.1 \cr 
 c &  100 &  3  & 3 & 0.278 & $\simeq  0.1 L_{*}$ & 8.7 & 68.8 \cr
 d &  150 &  3  & 1 & 0.625 & $\simeq  0.5 L_{*}$ & 3.1 & 24.8 \cr
 e &  150 &  3  & 3 & 0.625 & $\simeq  0.5 L_{*}$ & 7.7 & 54.1 \cr
 f &  225 &  3  & 3 &  1.41 & $\simeq  1 L_{*}$  & 6.0  & 29.9 \cr
 g &  225 &  3  & 1 &  1.41 & $\simeq  1 L_{*}$  &  2.3 & 7.1 \cr
 h &  225 &  5  & 3 &  2.34 & $\simeq  2 L_{*}$  &  6.0 & 21.4 \cr
 i &  225 &  5  & 5 &  2.34 & $\simeq  2 L_{*}$  &  9.9 & 30.2 \cr

\enddata

\tablecomments{Though the average distance from center a
pair travels before coalescence ($d_{\rm avg}$) generally decreases
with increasing galactic mass, the median distance ($d_{\rm median}$)
scales with disk radius ($r_{\rm disk}$).}
\end{deluxetable}

\subsection{Coalescence/birth rates}

We have explored the consequences of different kick strengths on the
birthrates of NS--NS binaries. Table \ref{table:rate} summarizes these
results.

Earlier work \citepeg{sut78,dc87,vwb90,wvv92,bp95} in which asymmetric
kicks were incorporated with a single NS component binary (as in
LMXBs, HMXBs) noted a decrease in birthrate with increased kick
magnitude.  \citet{pzs96} and \citet{lpp97} found a similar effect on
the bound NS pair birthrates.  \citet{lip97} provides a good review
of the expected rates. Clearly, the birthrate of NS--NS binaries is
also sensitive to the total SN type II rate (which is observationally
constrained to no better than a factor of two, and theoretically
depends both on the uncertain high mass end of the initial mass
function and the total star formation at high redshift), and is also
sensitive to the fraction of high mass stars in binaries with high
mass secondaries.

We concentrate our discussion of NS--NS binary birthrates to galactic
systems for which the SN type II is fairly well-known (such as in the
Galaxy). It is important to note, however, that the SN type II rate
may be quite high in low surface-brightness and dwarf galaxies
\citepeg{bf96}. This would subsequently lead to a higher
NS--NS birthrates in such systems than a simple mass scaling to rates
derived for the Galaxy.

Recently \citet{vl96} find (observationally) the birthrate of NS--NS
binaries to be 8 Myr$^{-1}$. \citet{lpp97} find between 100 and 330
events per Myr in simulations.  \citet{pzs96} found birthrates
anywhere from 9 to 384 Myr$^{-1}$ depending mostly on the choice of
asymmetric kick strength in their models.  We note that our derived
birthrate of $\sim 3$ Myr$^{-1}$ for high $\sigma_v = 270$ km s$^{-1}$
is comparable to those found \citeauthor{pzs96} (particularly model
``ck'') with an average 3-D kick velocity of $450$ km s$^{-1}$.  Also,
for low velocity kicks ($\sigma_{\rm kick} = 95$ km s$^{-1}$) our
birthrates approach that of Portegies Zwart \& Spreeuw models with no
asymmetric kicks.

The discrepancies between this and other work, therefore, we believe,
are largely due to the choices of supernovae kick distributions and
strengths. That the absolute birthrate varies by an order of magnitude
depending on the binary evolution code and asymmetric kick
distributions used in different studies hints at the uncertainty in
the {\it ab initio} knowledge of the true birthrates.
\begin{deluxetable}{cccc}
\tablewidth{0pt}
\tablecaption{
The Bound NS--NS Binary Birthrate and Merger Time Properties
as a Function of Supernova Kick Strength \label{table:rate}}
\tablehead{ \colhead{$\sigma_{\rm kick}$} (km s$^{-1}$) & \colhead{Birthrate (Myr $^{-1}$)} & \colhead{$\tau_{\rm median}$ (yr)} & \colhead{$\tau_{\rm avg}^a$ (yr)}}

\startdata

 95  &  49 & $1.4 \times 10^8$  & $9.4 \times 10^8$ \cr
 190 &  10 & $7.0 \times 10^7$ &  $8.0 \times 10^8$ \cr
 270 &  3  & $5.5 \times 10^7$  & $7.0 \times 10^8$ \cr

\enddata

\tablecomments{$^a$Average merger time of pairs merging in less than
  $1.5\times 10^{10}$ years. A Maxwellian distribution
characterized by a velocity dispersion ($\sigma_{\rm kick}$) is
assumed.}

\end{deluxetable}

\subsection{Spatial distribution}

\begin{figure}
\centerline{\psfig{file=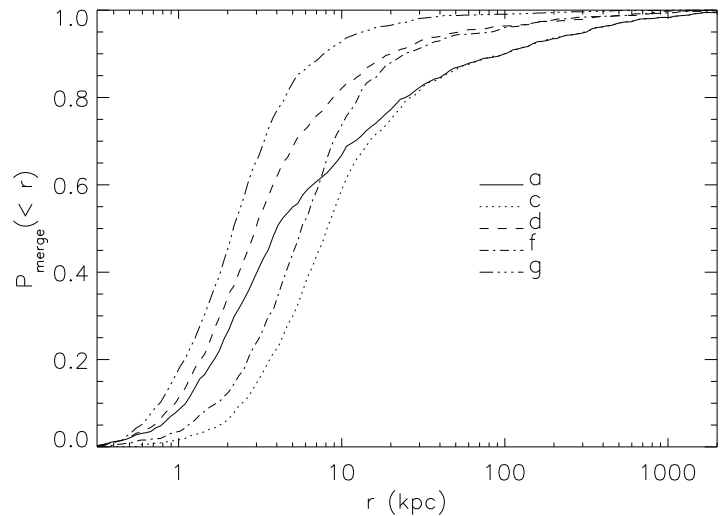,width=4.2in}}
\caption[The radial distribution of coalescing neutron stars around
galaxies of various potentials]{The radial distribution of coalescing
neutron stars around galaxies of various potentials.  The letters
refer to runs in table 1.  In all scenarios, at least 50\% of the
mergers occur within 10 kpc of the host galaxy.  The wider radial
distribution of in the under-luminous galaxy scenarios (a,c) reflects
the smaller gravitational potential of under-luminous galaxies.}
\label{fig:face}
\end{figure}

Approximately half of the NS--NS binaries merge within $\sim 10^8$
years after the second SN; this merger time is relatively quick on
the timescale of star-formation. In addition, half of the pairs coalesce
within a few kpc of their birthplace and within 10 kpc of the galactic
center (see figure \ref{fig:face}) {\it regardless} of the potential
strength of the host galaxy.  As shown in figure \ref{fig:face},
galaxies with $M_{\rm galaxy} > 10^{10} M_\odot$ ($L \age 0.1 L_*$),
90 (95) percent of the NS--NS mergers will occur within 30 (50) kpc of
the host.
In the least massive dwarf galaxies with $M_{\rm galaxy} \simeq 9
\times 10^{9} M_\odot$ ($\ale 0.1 L_*$), 50 (90, 95) percent of mergers
occur within $\sim 10$ ($100$, $300$) kpc of the host (see figure
\ref{fig:face}).  So, for example, assuming a Hubble constant of $H_0
= 65$ km s$^{-1}$ Mpc$^{-1}$ and $\Omega = 0.2$, we find that 90 (95)
percent of NS binaries born in dwarf galaxies at redshift $z=1$ will
merge within $\sim 12.7$ arcsec ($\sim 38.2$ arcsec) of the host
galaxy.  These angular offsets can be considered the extreme of the
expected radial distribution since the potentials are weakest and we
have not included the effect of projection.  We would expect 50 (90,
95) percent of the mergers near non-dwarf galaxies to occur within $\sim
1.3$ (3.8, 6.4) arcsec from their host at $z=1$ for the cosmology
assumed above.

Given the agreement of our orbital parameter distribution (figure 1)
and velocity distribution (figure 2) with that of \citet{pzy98}, the
discrepancy between the derived spatial distribution (see figure 8 of
\citeauthor{pzy98}) is likely due to our use of a galactic
potential in the model.  This inclusion of a potential naturally keeps
merging NSs more concentrated toward the galactic center than without
the effect.

\section{Discussion}


Although the NS--NS birthrate decreases with increased velocity kick,
the distribution of merger time and system velocity is not affected
strongly by our choice of kick distributions. Rather, the shapes of
the orbital and velocity distributions (figures 1 and 2) are closely
connected with the pre-SN orbital velocity, which is itself connected
simply with the evolution and masses. That is, bound NS binaries come
from a range of parameters which give high orbital velocities in the
pre-second SN system.  The orbital parameters (and merger time
distribution) of binaries which survive the second SN are not
sensitive to the exact kick-velocity distribution.  We suspect this
may be because bound systems can only originate from a parameter space
where the kick magnitude and orientation are tuned for the pre-second
SN orbital parameters.  The overall fraction of systems that remain
bound {\it is} sensitive to the kick distribution insofar as the kick
distribution determines how many kicks are in the appropriate range of
parameter space.
\begin{figure}
\centerline{\psfig{file=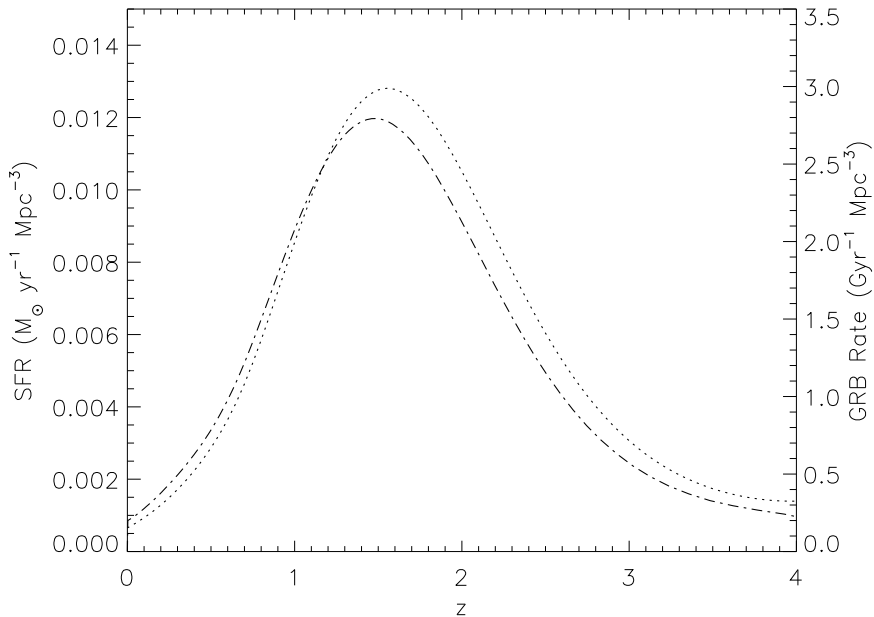,width=5in}}
\caption[NS--NS merger rate dependence on redshift]{NS--NS merger rate dependence on redshift. The dotted line
is a reproduction of the SFR from \citet{mad97} with corresponding
units on the left--hand axis.  The SFR curve as seen as a lower limit
to the true star-formation history since dust may obscure a large
fraction SFR regions in galaxies. The right hand axis is the
(unobscured) GRB rate if the bursts arise from the merger of two
NS--NS assuming a merger time distribution found in the present study
(dot--dashed line).  Both the SFR and merger rate are in co-moving
units (assuming $H_0 = 50$ km s$^{-1}$ Mpc$^{-1}$).  The normalization
of the burst rate is taken from \citet{wbbn98}.}
\label{fig:sfr}
\end{figure} 

Since NS--NS binaries are formed rapidly (with an average time since
ZAMS of $\sim 22$ million years) and the median merger time is of
order one hundred million years regardless of the kick velocity
distribution (see table 2), the rate of NS--NS mergers should closely
trace the star formation rate.  In the context of gamma-ray bursts,
where merging NSs are seen as the canonical production mechanism, this
result implies that the GRB merger rate should evolve proportionally
to the star formation rate (see figure \ref{fig:sfr}; see also
\citealt{bpzy98}).  Indeed, several studies
\citep{tot97,lpp97,wbbn98} have consistently fit the GRB log $N$--log $P$ 
curve to a model which assumes such a rate density evolution.

If indeed gamma-ray bursts arise from the coalescence of neutron star
binaries, then we confirm that GRBs should trace the star formation
rate in the Universe; thus most GRBs should have redshifts near the
peak in star formation \citep[currently believed to be $1 \ale z \ale
2$;][]{mpd98} although the observed distribution may be skewed to
lower redshifts by obscuration at high redshift
\citepeg{hsd+98}. Determination of the distribution of X-ray and
optical counterparts to GRBs may help constrain the true cosmological
star formation history, though the observations of GRB counterparts
are vulnerable to some of the same extinction selection effects that
complicate determination of high redshift star formation rates.
Figure \ref{fig:sfr} illustrates the redshift dependence of the GRB
rate assuming the bursts arise in NS--NS mergers.

The minimum required local (isotropic) bursting rate of 0.025 galactic
event per Myr \citep{wbbn98} is consistent with our birthrate results
(table 2) assuming a beaming fraction of 1/10--1/100 for the gamma ray
emission and our canonical values for the type II supernova rate and
supernova binary fraction.  The effects of beaming should be observed
in both the light curves of GRB afterglow and in deep transient
searches \citepeg{wl98}.

In the case of GRB 970508, \citet{bdkf98} (chapter
\ref{chap:grb970508}) and \citet{cgb+98} \citepcf{nbs+97}
found that the host is an under-luminous dwarf galaxy; the close
spatial connection (offset $< 1$'') of the OT with the galaxy is then
a case (albeit weak) against the NS--NS merger hypothesis as the {\it
a priori} probability is $\ale 20\%$ (figure
\ref{fig:face}).  \citet{pac98b} first pointed out that the close
spatial association with a dwarf galaxy is a case against the NS--NS
merger hypothesis.  Certainly more transients are required to rule
against the NS--NS merger hypothesis; we note, however, that dust
obscuration and projection effects may severely bias the sample (see
above discussion).

The verdict on the reconciliation of the expected radial distribution
of NS--NS mergers with hosts of known GRBs is still
out. \citet{slp+97} found the optical transient associated with GRB
970228 to be slightly offset from the center of a dim galaxy, but
without a redshift it is still unclear as to the the true luminosity
of the host and thus the expected offset of the OT in the NS--NS
merger hypothesis.  Similarly, small or negligible offsets of GRB
afterglows with faint galaxies has been found in other GRBs.
\citet{kdr+98} found the redshift of the
purported host galaxy of GRB 971214 to be $z=3.4$ implying the host is
$L \age L_*$; the expected offsets of NS--NS mergers around massive
galaxies (figure
\ref{fig:face}, models d through i) is then consistent with their
finding of an OT offset $\simeq$ 0.5 kpc.

A few well-established offsets cannot tell us what is the true
distribution of GRBs around host galaxies.  As more OTs are
discovered, we will hopefully build up a large sample to statistically
test the offsets.  Fortunately, the unobscured afterglow emission
strength is coupled with the density $n$ of the surrounding
interstellar medium (ISM) with intensity scaling as $\sqrt n$
\citep{bmr93,mrw98b}; however, high ISM densities tracing dust will 
tend to obscure rest-frame UV and optical emission from the transient.
In the absence of strong absorption from the surrounding medium,
transients of GRBs are preferential found close to where they are
born, in the disk.  However, dust obscuration and projection effects
severely complicate determination of the true offset of OTs from their
host galaxy.  Furthermore, identification of the host with a GRB
becomes increasingly difficult with distances beyond a few light radii
($\sim 10$ kpc) of galaxies.

If all afterglows, especially those where little to no absorption is
implied, are found more highly concentrated than predicted in figure
\ref{fig:face}, the NS--NS merger hypothesis would lose favor to
models which keep progenitors more central to their host. GRBs as
events associated with single massive stars such as microquasars
\citep{pac98b} or failed type Ib SN \citep{woo93} could be
possible.  Alternatively, one may consider neutron star--black hole
(BH) binaries as the progenitors of GRBs \citep{moc93}.  Most black
hole X-ray binaries have low-spatial velocities (although Nova Sco has
$v_{\rm sys} \simeq 100$ km s$^{-1}$; see \citealt{bps95}) so NS--BH
binaries should have system velocities $\sim 3$ to 10 times smaller
than NS--NS binaries.  One would expect NS--BH systems to be borne
with higher eccentricities than NS--NS systems leading to quicker
merger. Moreover, NS--BH binaries are more massive than NS--NS
binaries and merger time due to gravitational radiation scales
strongly with mass. Thus the attraction is that NS--BH mergers would
be preferentially closer to their host and their merger rate might be
small enough so as to require no beaming. Alternatively, gamma-ray
bursts could arise from several of these plausible progenitor models
and still be consistent with basic relativistic fireball models.

\section{Conclusions}

A reconciliation with the expected distribution of presumed
progenitors of GRBs and observed transient/host offsets is clearly
required.  We find for all plausible galactic potentials that the
median radial offset of a NS--NS merger is less than 10 kpc.  And in
all but the most shallow potentials, ninety percent of NS--NS binaries
merge within 30 kpc of the host. At a redshift of $z=1$ (with $H_0 =
65$ km s$^{-1}$ Mpc$^{-1}$ and $\Omega = 0.2$), this means that ninety
percent the coalescences should occur within $\sim 4$ arcsec from the
host galaxy.  Although the expected spatial distribution of coalescing
neutron star binaries found herein is consistent with the close
spatial association of known optical afterglows of gamma-ray bursts
with faint galaxies, a non-negligible fraction ($\sim 15$\%) of GRBs
{\it should} occur well outside dwarf galaxy hosts if the NS--NS
hypothesis is correct.  Otherwise, other models which keep progenitors
closer to their host (e.g.,~BH--NS mergers, microquasars, or failed
SN type Ib SNe) would be preferred.

As all the progenitor models mentioned are connected with high-mass
stars, the true GRB afterglow rate as a function of redshift should
trace the star-formation rate in the universe.  However, environmental
effects, such as dust obscuration, may severely bias the estimate of
the true offset distribution.  Even in the NS--NS models where
progenitors have a natural mechanism to achieve high spatial
velocities, most will be closely connected spatially to their
host. Redshifts derived from absorption in the afterglow spectra
should be nearly always that of the nearest galaxy \citep{bsw+97}.
Rapid burst follow-up ($\ale$ 1 hr), with spectra taken while the
optical transients are bright should confirm some form of absorption
from the host galaxy.

We have confirmed the strong dependence of birthrate of NS--NS
binaries on kick velocity distribution and found the independence of
the orbital parameters after the second supernova (and hence merger
times and spatial velocity) on the choice of kicks.  The methodology
herein can be extended to include formation scenarios of black
holes. This could provide improved merger rate estimates for LIGO
sources, and estimate the relative contribution of coalescences
between neutron stars and low mass black holes to the event
rate. Detailed modeling of the Milky Way potential would also allow
predictions for the distribution of NS--PSR binaries observable in the
Milky Way, which would provide an independent test of the assumptions
made in these models.

\acknowledgments

It is a pleasure to thank Peter M\'esz\'aros, Melvyn Davies, Gerald
Brown, Hans Bethe, Ralph Wijers, Peter Eggleton, Sterl Phinney, Peter
Goldreich, Brad Schaefer, and Martin Rees for helpful insight at
various stages of this work. We especially thank Simon Portegies Zwart
as referee. JSB thanks the Hershel Smith Harvard Fellowship for
funding. SS acknowledges the support of the European Union through
a Marie Curie Individual Fellowship.

\chapter[The Host Galaxy of GRB 970508]{The Host Galaxy of GRB 970508$^\dag$}
\label{chap:grb970508}

\vspace{-1cm}

\secauthor{J.~S.~Bloom, S.~G.~Djorgovski, S.~R.~Kulkarni}
\secaffil{California Institute of Technology, Palomar Observatory,
105-24, Pasadena, CA 91125}
\secauthor{D.~A.~Frail}
\secaffil{National Radio Astronomy Observatory, P.~O.~Box O, Socorro, NM
87801}

\secfootnote{\secfootdag}{A version of this chapter was first published 
in {\it The Astrophysical Journal Letters}, 518, p.~L1--L4 (1999).}

\medskip

\begin{abstract}

We present late-time imaging and spectroscopic observations of the
optical transient (OT) and the host galaxy of GRB 970508.  Imaging
observations roughly 200 and 300 days after the burst provide
unambiguous evidence for the flattening of the light-curve.  The
spectroscopic observations reveal two persistent features which we
identify with [O II] $\lambda\lambda$3727 \AA\ and [Ne III]
$\lambda$3869 \AA\ at a redshift of $z = 0.835$ --- the same redshift
as the absorption system seen when the transient was bright.  The OT
was coincident with the underlying galaxy to better than 370
milliarcsec or a projected radial separation of less than 2.7 kpc.
The luminosity of the [O II] line implies a minimum star formation
rate of $\age 1 ~M_\odot ~{\rm yr}^{-1}$.  In our assumed cosmology,
the implied restframe absolute magnitude is $M_B = -18.55$, or $L_{B}
= 0.12 L_*$.  This object, the likely host of GRB 970508, can thus be
characterized as an actively star-forming dwarf galaxy.  The close
spatial connection between this dwarf galaxy and the OT requires that
at least some fraction of progenitors be not ejected in even the
weakest galactic potentials.

\end{abstract}

\section{Introduction}

After an initial brightening lasting $\sim$1.5 days, the optical
transient (OT) of GRB 970508 faded with a nearly pure power-law slope
by 5 mag over $\sim$100 days \citepeg{ggv+98a,gcm+98,skz+98}.
Indications of a flattening in the light curve \citep{pjg+98} were
confirmed independently by \citet{bkdf98b}, \citet{ctg+98}, and
\citet{skz+98}. Recently, \citet{zsb98b} fit the $BVRI$ light curves 
of the OT $+$ host and found the broadband spectrum of the presumed
host galaxy.

The existence of an [O II] emission line at the absorption system
redshift \citep{mccb97} was taken as evidence for an underlying, dim
galaxy host. After {\it Hubble Space Telescope} (\hst) images
revealed the point-source nature of the light, several groups 
\citepeg{fbp97,pfb+98a,nbs+97} 
suggested that the source responsible for the [O II] emission must be
a very faint ($R$ $>$ 25 mag), compact (less than 1\arcsec) dwarf galaxy
at $z$=0.835 nearly coincident on the sky with the transient. These
predictions are largely confirmed in the present study.

In this Letter, we report on the results of deep imaging and
spectroscopy of the host galaxy of GRB 970508 obtained at the 10 m
Keck II telescope.

\section{Observations and Analysis}

Imaging and spectroscopic observations were obtained using the Low
Resolution Imaging Spectrograph (LRIS) \citep{occ+95} on the \hbox{10
m} Keck II Telescope on Mauna Kea, Hawaii. The log of the observations
is presented in table \ref{tab:grb970508obs}. All nights were
photometric.  The imaging data were reduced in the standard manner.

To follow the light-curve behavior of the OT + host over $\sim$300
days from the time of the burst, we chose to tie the photometric zero
point to a previous study \citep{skz+98} which predicted late-time
magnitudes based on early (less than 100 days) power-law behavior in
several bandpasses. This photometric tie provides an internally
consistent data set for our purposes. Other studies of the light curve
include \citet{ggv+98a} and \citet{pfb+98a}. V.~Sokolov (1998, private
communication) provided magnitudes of eight ``tertiary" field stars ($R
= 18.7$--23 mag) as reference, since the four secondary comparison
stars \citep{skz+98} were saturated in all of our images. The zero
points were determined through a least-squares fit and have
conservative errors of \hbox{$\mathstrut{\sigma_{{B}}}$ = 0.05}
and
\hbox{$\mathstrut{\sigma_{{R}}}$ = 0.01} mag.

For our spectroscopic observations, we used a 300 lines mm$^{-1}$
grating, which gives a typical resolution of $\approx$15 \AA\ and a
wavelength range from approximately 3900 to 8900 \AA. The
spectroscopic standards G191B2B \citep{msb88} and HD 19445 \citep{og83}
were used to flux-calibrate the data of October and November,
respectively. Spectra were obtained with the slit position angle at
51\degr\ in order to observe both the host galaxy of GRB 970508 and g1
\citesee{dmk+97}. This angle was always close to the
parallactic angle, and the wavelength-dependent slit losses are not
important for the discussion below. Internal consistency implied by
measurements of independent standards implies an uncertainty of less
than 20\% in the flux zero-point calibration. Exposures of arc lamps
were used for the wavelength calibration, with a resulting r.m.s.\
uncertainty of about 0.3 \AA\ and possible systematic errors of the
same order, due to the instrument flexure.

\section{Results}

Table \ref{tab:grb970508obs} gives a summary of the derived magnitudes
at the position of the OT and, as a comparison, the extrapolated
magnitudes from a pure power-law decay fit by \citet{skz+98}. The OT +
host is brighter by greater than 0.8 mag in both the $B$- and $R$-band,
leading to the obvious conclusion that the transient has faded to
reveal a constant source. We used a Levenberg-Marquardt $\chi^2$
minimization method to fit a power-law flux (OT) plus constant flux
(galaxy) to the $B$ and $R$ light curves using data compiled in
\citet{skz+98}:
\begin{equation}
f_{\rm total} = f_{0}t^{-\alpha} + f_{\rm gal},
\end{equation}
where t is the time since the burst measured in days. The quantities
$f_0$ and $f_{\rm gal}$ are the normalization of the flux of the
transient and the persistent flux of the underlying galaxy,
respectively. We find \hbox{$B\mathstrut{_{{\rm gal}}}$ = 26.77 $\pm$ 0.35
mag}, \hbox{$R\mathstrut{_{{\rm gal}}}$ = 25.72 $\pm$ 0.20 mag},
\hbox{$B\mathstrut{_{0}}$ = 19.60 $\pm$ 0.04 mag}, \hbox{$R\mathstrut{_{0}}$ =
18.79 $\pm$ 0.03 mag}, with values for the decay parameters of
\hbox{$\mathstrut{\alpha_{{B}}}$ = -1.31 $\pm$ 0.03} and
\hbox{$\mathstrut{\alpha_{{R}}}$ = -1.27 $\pm$ 0.02}. Note that
the power-law decline did not start until day $\sim$ 1.6, so the
normalizations, $B_0$ and $R_0$, do not actually correspond to the
true flux of the transient on day 1.

\begin{figure*}[tp]
\centerline{\psfig{file=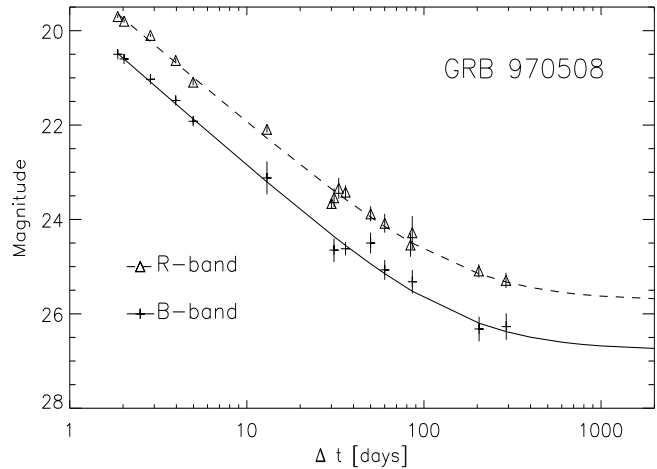,height=2.7in,width=3.8in}}
\caption[Light curve of the optical transient of GRB 970508]{Light curve of the optical transient
of GRB 970508.  Both $R$- (dashed; triangles) and $B$-band (solid;
crosses) data were compiled and transformed to a single photometric
system by \citet{skz+98} (see references therein).  The latest two
data points on each light curve are from this chapter.  The constant
flux of the underlying galaxy (the purported host) dominates the light
at late times.}
\label{fig:grb970508ltcurve}
\end{figure*}

To search for any potential offset of the OT and the galaxy, we used
an early image of the gamma-ray burst (GRB) field obtained on COSMIC
at the Palomar 200 inch telescope on 1997 May 13.6 UT while the
transient was still bright \citep[$R$ $\approx$ 20 mag;
see][]{dmk+97}. Assuming the power-law behavior continued, the light
at the transient position is now dominated by the galaxy, with the
transient contributing less than 30\% to the total flux (see
figure~\ref{fig:grb970508ltcurve}).

We registered the Keck LRIS and the P200 COSMIC $R$-band (300 s) images
by matching 33 relatively bright ($R <$ 23 mag) objects in a 4\arcmin\
$\times$ 4\arcmin\ field surrounding the GRB transient. The coordinate
transformation between the two images accounted for pixel scale,
rotating, translation, and higher order distortion. The r.m.s.\ of the
transformed star positions (including both axes) was $\sigma$ = 0.56
LRIS pixels (=0\arcsec.121). We find the angular separation of the OT
and the galaxy to be less than 0.814 pixels (=0\arcsec.175), which
includes the error of the transformation and centering errors of the
objects themselves. The galaxy is thus found well within 1.7 pixels =
0\arcsec.37 (3 $\sigma$) of the OT.

\begin{figure*}[tbh]
\centerline{\psfig{file=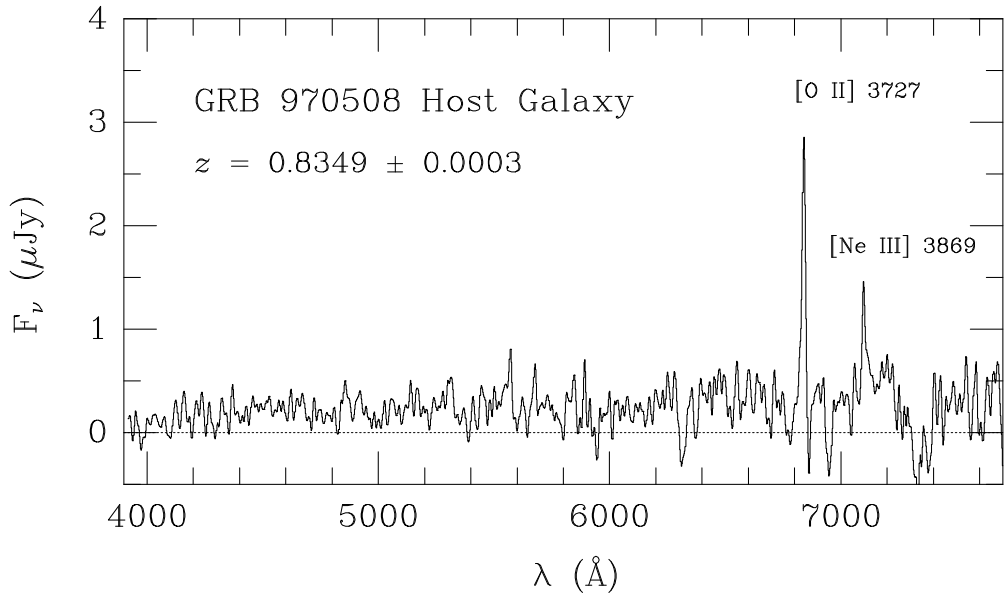,height=3.5in}}
\caption[The weighted average spectrum of the host galaxy of GRB
970508]{The weighted average spectrum of the host galaxy of GRB
970508, obtained at the Keck telescope.  The spectrum was smoothed
with a Gaussian with a $\sigma = 5$ \AA, roughly corresponding to the
instrumental resolution.  Prominent emission lines are labeled. }
\label{fig:grb970508thespect}
\end{figure*}

The averaged spectrum of the OT + host shows a very blue continuum, a
prominent emission line at \hbox{$\lambda\mathstrut{_{{\rm obs}}}$ = 6839.7
\AA}, and a somewhat weaker line at \hbox{$\lambda\mathstrut{_{{\rm obs}}}$
= 7097.7 \AA} (\S \ref{fig:grb970508thespect}). We interpret the
emission features as [O II] $\lambda\lambda$ 3727 and [Ne III]
$\lambda$ 3869 at the weighted mean redshift of $z$ = 0.8349 $\pm$
0.0003. Our inferred redshift for the host is consistent (within
errors) with that of the absorbing system discovered by
\citet{mccb97}.

The spectrum of the nearby galaxy g1 shows a relatively featureless,
blue continuum. We are unable to determine its redshift at this stage.

Our spectroscopic measurements give a magnitude $R \approx$ 25.05 mag
(OT + host) at the mean epoch ($\approx$163 days after the GRB) of our
observations, in excellent agreement with the magnitude inferred from
the fit to direct imaging data (see figure~\ref{fig:grb970508ltcurve}).
 
\section{Discussion}

After an initial brightening, the light curve of the optical transient
did not deviate significantly from a power law over the first 100 days
after the burst  (\eg\ \citeauthor{ggv+98a} \citeyear{ggv+98a};
\citeauthor{gcm+98} \citeyear{gcm+98}; although, see 
\citeauthor{pjg+98} \citeyear{pjg+98}).  Assuming the blast wave
producing the afterglow expanded relativistically (bulk Lorentz factor
$\Gamma$ greater than a few) during the beginning of the light-curve
decline, the observed flux was produced from within an angle
$\omega\mathstrut{_{{\Gamma}}} \simeq 1/\Gamma$ of the emitting
surface. As the blast wave expands, $\omega_\Gamma$ increases with
time. As long as the angle through which the blast wave is collimated
is greater than $\omega_\Gamma$, there would be no obvious break in
the light curve \citepeg{spn98}. One might expect, in addition, the
blast wave to eventually become sub-relativistic, resulting not only in
a larger observed surface area but perhaps in a change in surface
emissivity. Curiously, an apparent break expected in either scenario
did not materialize.

The spatial coincidence of the transient and the underlying galaxy may
simply be a chance projection of the transient, which lies beyond $z$
= 0.835, and the galaxy at $z$ = 0.835. The surface density of
galaxies down to $R$ = 25.7 mag is 48.3 arcmin$^{-2}$. It is important
to note that we know a priori that the host must lie in the redshift
range 0.835 $< z <$ 2.1 \citep{mdk+97}. The fraction of galaxies within
this range is $\sim$50\% of the total at the magnitude level
\citep{rsmf96}. The a posteriori Poisson probability of
finding such a galaxy within 0\arcsec.37 from the OT is $3 \times
10\mathstrut{^{-3}}$ . Keeping in mind the limitations of a posteriori
statistics, this small probability and the trend that GRB transients
appear to be nearly spatially coincident with galaxies
\citepeg{odk+98} lead us to suggest that this galaxy is the host of
GRB 970508.

Assuming a standard Friedman model cosmology with $H\mathstrut{_{0}}$
= 65 km s$^{-1}$ Mpc${^-1}$ and $\Omega\mathstrut{_{0}}$ = 0.2, we
derive a luminosity distance of 1.60 $\times$ 10$\mathstrut{^{28}}$ cm
to the host galaxy. The observed equivalent width in the [O II] line
is (115 $\pm$ 5) \AA, or about 63 \AA\ in the galaxy's rest
frame. However, this also includes the continuum light from the OT at
this epoch. Correcting for the OT contribution would then double these
values of the equivalent width. This is at the high end of the
distribution for the typical field galaxies at comparable magnitudes
and redshifts \citep{hcbp98}.  The implied [O II] line luminosity,
corrected for Galactic extinction, is $L\mathstrut{_{3727}}$ = (9.6
$\pm$ 0.7) $\times$ 10$\mathstrut{^{40}}$ ergs s$^{-1}$. Using the
relation from \citet{ken98}, we estimate the star formation rate (SFR)
$\approx$ 1.4 M$_\odot$ yr$^{-1}$.

An alternative estimate of the SFR can be obtained from the continuum
luminosity at $\lambda\mathstrut{_{{\rm rest}}}$ = 2800 \AA\
\citep{mpd98}. The observed, interpolated continuum flux from the host
itself (i.e., not including the OT light) at the corresponding
$\lambda\mathstrut{_{{\rm obs}}} \approx 5130$ \AA\ is
$F\mathstrut{_{{\nu}}} \approx 0.11 \mu$Jy, corrected for the
estimated Galactic extinction \citep[$A\mathstrut{_{{V}}} \approx
0.08$ mag;][]{dmk+97}. For our assumed cosmology, the rest-frame
continuum luminosity is then $L\mathstrut{_{2800}} \approx 1.93 \times
10\mathstrut{^{27}}$ ergs s$^{-1}$ Hz$^{-1}$, corresponding to SFR
$\approx$ 0.25 \sfr. This is notably lower than the SFR inferred from
the [O II] line. We note, however, that neither is known to be a very
reliable SFR indicator. Both are also subject to the unknown
extinction corrections from the galaxy's own interstellar medium (the
continuum estimate being more sensitive). We thus conclude that the
{\it lower limit} to the SFR in this galaxy is probably about 0.5--1 \sfr.

The observed flux in the [Ne III] $\lambda$3869 line is
$F\mathstrut{_{3869}}$ = (1.25 $\pm$ 0.1) $\times
10\mathstrut{^{-17}}$ \flux, not corrected for the extinction. The
flux ratio of the two emission lines is $F\mathstrut{_{3869}}/
F\mathstrut{_{3727}}$ = 0.44 $\pm$ 0.05. This ratio is about 10 times
higher than the typical values for H II regions. Nonetheless, it is in
the range of photo-ionization models for H II regions by \citet{stan90}
for different combinations of model parameters but generally for
effective temperatures $T\mathstrut{_{{\rm eff}}} \ge 40,000$ K.

The inferred host luminosity is in agreement with the upper limit from
earlier \hst\ observations \citep{pfb+98a}. Further, our derived $B$- and
$R$-magnitudes for the galaxy correspond to a continuum with a power law
$F\mathstrut{_{{\nu}}} \sim \nu\mathstrut{^{-1.56}}$. Extrapolating
from the observed $R$-band flux to the wavelength corresponding to the
rest-frame $B$-band (about 8060 \AA), we derive the observed flux $F
\mathstrut{_{{\nu}}}$ ($\lambda$ = 8060 \AA) $\approx 0.22 \mu$Jy. 
For our assumed cosmology, the implied rest-frame absolute magnitude
is then $M\mathstrut{_{{B}}} \approx -18.55$. Thus, the rest-frame
$B$-band luminosity of the host galaxy is about 0.12 $L_*$ today.

This galaxy is roughly 2 mag fainter than the knee of the observed
luminosity function of all galaxies between redshift $z$= 0.77 and 1.0
\citep[Canada-France redshift survey;][]{lth+95} and 1 mag fainter than
late-time, star-forming galaxies in the 2dF survey \citep{frp+99}. The
specific SFR per unit luminosity is high. This object can thus be
characterized as an actively star-forming dwarf galaxy. Objects of
this type are fairly common at comparable redshifts.

\section{Conclusions}

The high effective temperature implied by the relative line strengths
of [Ne III] and [O II] suggests the presence of a substantial
population of massive stars and thus active and recent star
formation. This, in turn, gives additional support to the ideas that
the origin of GRBs is related to massive stars
\citepeg{wbbn98,tot97,dkb+98b}. An alternative possibility for the
origin of the [Ne III] $\lambda$3869 line is photo-ionization by a
low-luminosity active galactic nucleus (AGN). While we cannot exclude
this possibility, we note that there is no other evidence in favor of
this hypothesis, and moreover we see no other emission lines, e.g., Mg
II $\lambda$2799, that would be expected with comparable strengths in
an AGN-powered object.

What may be surprising, in the neutron star binary (NS--NS) model of
GRB progenitors \citepeg{npp92,pac86}, is that GRB 970508 appears so
close (less than 2.7 kpc) to a dwarf galaxy ($L \approx 0.1
L\mathstrut{_{*}}$).  \citet{bsp99} (chapter \ref{chap:nsns}) recently
found that less than 15\% of NS-NS binaries will merge within 3 kpc of
a comparable under-massive galaxy. If GRBs are consistently found very
near (less than a few kpc) their purported host, then progenitor
models such as microquasars
\citep{pac98b}, "failed" Type Ib supernovae \citep{woo93}, or 
black hole--neutron star binaries \citep{moc93,mr97b}, all of which are
expected to produce GRBs more tightly bound to their hosts, would be
favored.

\acknowledgements

It is a pleasure to thank S.~Odewahn, M.~van Kerkwijk, R.~Gal, and
A.~Ramaprakrash for assistance during observing runs at Keck, P.~Groot
for comments, and R.~Sari for helpful discussions concerning
inferences from the light-curve.  SRK's research is supported by the
National Science Foundation and NASA.  SGD acknowledges a partial
support from the Bressler Foundation.

\begin{deluxetable}{lccccccl}
\singlespace
\rotate
\tablecolumns{8} 
\tablewidth{0pc} 
\tabletypesize{\small}
\tablecaption{Late-time GRB 970508 Imaging and Spectroscopic Observations\label{tab:grb970508obs}}
\tablehead{
\multicolumn{1}{c}{Date}& \colhead{} & 
\multicolumn{1}{c}{Integration} &  
\multicolumn{1}{c}{Seeing} & \colhead{$\Delta t$} &
\multicolumn{2}{c}{Object Magnitude} & \colhead{} \\
 & & time \\
\colhead{(UT)} & \colhead{Band/Grating} & \colhead{(sec)} & \colhead{(arcsec)} & \colhead{(days)} &
\multicolumn{1}{c}{Observed\tablenotemark{a}} & \multicolumn{1}{c}{Pure Power Law\tablenotemark{b}} & \colhead{Observers}} 
\startdata
\cutinhead{Imaging}
Nov.~28, 1997 & $R$ & 5400 & 1.2 & 203.8 & 25.09$\pm
            0.14$\tablenotemark{c} & 25.63$\pm 0.2$ & Kulkarni, van
            Kerkwijk, and Bloom \\

Nov.~29, 1997 & $R$ & 600 & 1.2 & 204.8  & & & Kulkarni, van Kerkwijk, and Bloom \\
             
Nov.~29, 1997 & $R$ & 600 & 1.1 & 204.8  & & & Kulkarni, van Kerkwijk, and Bloom \\
             
Nov.~29, 1997 & $B$ & 2400 & 1.1 & 204.8 & 26.32$\pm 0.26$ & 
26.65$\pm 0.25$ &  Kulkarni, van Kerkwijk, and Bloom \\

Feb.~22, 1998 & $R$ & 2400 & 1.1 & 290.5 & 25.29$\pm 0.16$ & 
26.07$\pm 0.20$ & Kulkarni, Ramaprakash, and van Kerkwijk \\
             
Feb.~23, 1998 & $B$ & 2400 & 1.2 & 291.5 & 26.27$\pm 0.28$ & 
27.11$\pm 0.25$ &  \\
\cutinhead{Spectroscopy}
Oct.~3, 1997 & 300 & 5400 & 0.8 & 147.8 & & & Djorgovski and Odewahn \\
Nov.~2, 1997 & 300 & 5400 & $< 1.0$ & 177.8   &  & & Djorgovski and Gal

\enddata 
\tablenotetext{a}{Magnitudes are derived from V.~Sokolov's tertiary
reference stars.  The (conservative) 1 $\sigma$ errors include
statistical uncertainties in the reference transformation and the OT +
host itself.  All nights (imaging observations) were photometric.}
\tablenotetext{b}{Predicted magnitudes are derived from a pure power-law
decline using light curve data from $\Delta t \ale 100$ days
\citep{skz+98}.  Errors are estimated using the uncertainties in both
the magnitude scaling and power-law index.}
\tablenotetext{c}{The $R$-band magnitude quoted is derived from the sum
of $R$-band images over two nights.}

\end{deluxetable}

\chapter[The Host Galaxy of GRB 990123]{The Host Galaxy of GRB 990123$^\dag$}
\label{chap:grb990123}

\secfootnote{\secfootdag}{A version of this chapter was first published 
in {\it The Astrophysical Journal Letters}, 507, p.~L25--L28 (1998).}

\vspace{-1cm}

\secauthor{J. S. Bloom$^1$, S. C. Odewahn$^1$, S. G. Djorgovski$^1$,
 S. R. Kulkarni$^1$, F. A. Harrison$^1$, C. Koresko$^1$,
 G.~Neugebauer$^1$, L.~Armus$^2$, D. A. Frail$^3$, R. R. Gal$^1$,
 R. Sari$^1$, G. Squires$^1$, G. Illingworth$^4$, D. Kelson$^5$,
 F. Chaffee$^6$, R. Goodrich$^6$, M. Feroci$^7$, E. Costa$^7$,
 L. Piro$^7$, F. Frontera$^{8,9}$, S. Mao$^{10}$, C. Akerlof$^{11}$,
 T. A. McKay$^{11}$}

\medskip
\bigskip

\secaffils{$^1$ California Institute of Technology, Palomar Observatory,
105-24, Pasadena, CA 91125}
\secaffils{$^2$ Infrared Processing and Analysis Center, Caltech, MS
   100-22, 770 S.~Wilson Avenue,}
\secaffils{Pasadena, CA 91125.}
\secaffils{$^3$ National Radio Astronomy Observatory, P.O.~Box O, 1003
   Lopezville Road, Socorro, NM}
\secaffils{87801.}
\secaffils{$^4$ University of California Observatories/Lick Observatory,
   University of California }
\secaffils{at Santa Cruz, Santa Cruz, CA 95064.}
\secaffils{$^5$  Department of Terrestrial Magnetism, Carnegie Institute of
   Washington,}
\secaffils{5241 Broad Branch Road, NW, Washington, DC 20015-1305.}
\secaffils{$^6$ W. M. Keck Observatory, 65-0120 Mamalahoa Highway, Kamuela,
   HI 96743.}
\secaffils{$^7$ Instituto di Astrofisica Spaziale CNR, via Fosso del
   Cavaliere, Roma, I-00133, Italy.}
\secaffils{$^8$ ITESRE-CNR, Via Gobetti 101, Bologna, I-40129, Italy.}
\secaffils{$^9$ Physics Department, University of Ferrara, Via Paradiso
   12, Ferrara, I-44100, Italy.}
\secaffils{$^{10}$  Max-Planck-Institut fur Astrophysik,
   Karl-Schwarzschild-Strasse 1, Postfach 1523,}
\secaffils{Garching, 85740, Germany.}
\secaffils{$^{11}$  Department of Physics, University of Michigan at Ann
   Arbor, 2071 Randall,}
\secaffils{Ann Arbor, MI 48109-1120.}

\begin{abstract}
We present deep images of the field of GRB 990123 obtained in a
broadband UV/visible bandpass with the {\it Hubble Space Telescope}
(\hst) and deep near-infrared images obtained with the Keck I 10 m
telescope.  The \hst\ image reveals that the optical transient (OT) is
offset by 0\arcsec.67 (5.8 kpc in projection) from an extended,
apparently interacting galaxy. This galaxy, which we conclude is the
host galaxy of GRB 990123, is the most likely source of the absorption
lines of metals at a redshift of $z = 1.6$ seen in the spectrum of the
OT. With magnitudes of Gunn-$r$ = 24.5 $\pm$ 0.2 mag and $K = 22.1 \pm
0.3$ mag, this corresponds to an L $\sim 0.5 L\mathstrut{_{*}}$ galaxy,
assuming that it is located at $z = 1.6$. The estimated unobscured
star formation rate is $\approx 4$ \sfr, which is typical for normal
galaxies at comparable redshifts. There is no evidence for strong
gravitational lensing magnification of this burst, and some
alternative explanation for its remarkable energetics (such as
beaming) may therefore be required. The observed offset of the OT from
the nominal host center, the absence of broad absorption lines in the
afterglow spectrum, and the relatively blue continuum of the host do
not support the notion that gamma-ray bursts (GRBs) originate from
active galactic nuclei or massive black holes. Rather, the data are
consistent with models of GRBs that involve the death and/or merger of
massive stars. 
\end{abstract}

\section{Introduction}

Following the detection of GRB 990123 by BeppoSAX \citep{piro+99}, we
discovered an optical transient (OT) \citep{obk99} and subsequently a
coincident radio transient \citep{fk99} within the error circles of
the gamma-ray burst (GRB) and the associated fading X-ray source
\citep{piro+99b}.  Examination of the ROTSE (Robotic Optical Transient
Search Experiment) images taken during the GRB itself revealed a
hitherto unseen bright ($m \simeq 8.9$ mag at peak) phase of the
optical afterglow \citep{abb+99}.

  In \citet{kdo+99} we present a comprehensive study of the optical
and infrared observations of the transient afterglow and report a
measurement of an absorption redshift of $z\mathstrut{_{{\rm abs}}}$ =
1.6. Combining the redshift with the observed fluence \citep{fpf+99}
results in an inferred energy release of $3 \times
10\mathstrut{^{54}}$ ergs (if the emission was isotropic), which
clearly poses a problem to most conventional models of GRBs. However,
noting a break in the optical afterglow decay, \citet{kdo+99} argue
that the emission geometry may have been jet-like; this would then
decrease the energy constraint.

  Both currently favored models of GRB progenitors, the death of a
massive star \citep{woo93,pac98b} and the coalescence of a neutron
star (NS) or NS--black hole (BH) binary \citep{pac86,goo86,npp92},
predict that GRB rates should correlate strongly with the cosmic star
formation rate (SFR), and so most GRBs should occur during epochs of
the highest SFR (\ie a redshift range of $z$ = 1--3). The former model
predicts a tight spatial correlation between GRBs and star-forming
regions in galactic disks. The latter, however, allows the coalescence
site of a merging binary component to be quite distant (beyond a few
kiloparsecs) from the stellar birth site (e.g., \citealt{bsp99};
chapter \ref{chap:nsns}). GRBs could also be associated with nuclear
black holes (\ie active galactic nuclei) \citepeg{car92}. In this
scenario, unlike either model described above, the GRBs will
preferentially occur in the center of the host.  The exquisite angular
resolution of the {\it Hubble Space Telescope} (\hst) is well suited
to address this issue of the locations of GRBs relative to the host
galaxies. In this Letter, we report on HST observations of the host
galaxy of GRB 990123 taken about 16 days after the burst as well as
Keck imaging in the near-infrared.

\section{Observations and Data Reduction}

  The ground-based near-IR images of the field of GRB 990123 were
obtained using the near-infrared camera \citet{ms94} on the Keck I 10
m telescope. A log of the observations and a detailed description of
the data and the reduction procedures are given by \citet{kdo+99}. The
observations were obtained in the $K$ or $K\mathstrut{_{s}}$ bands and
were calibrated to the standard $K$ band (effective wavelength = 2.195
$\mu$m). The Galactic extinction correction is negligible in the $K$
band (see below).
\begin{figure}[t]
\centerline{\psfig{file=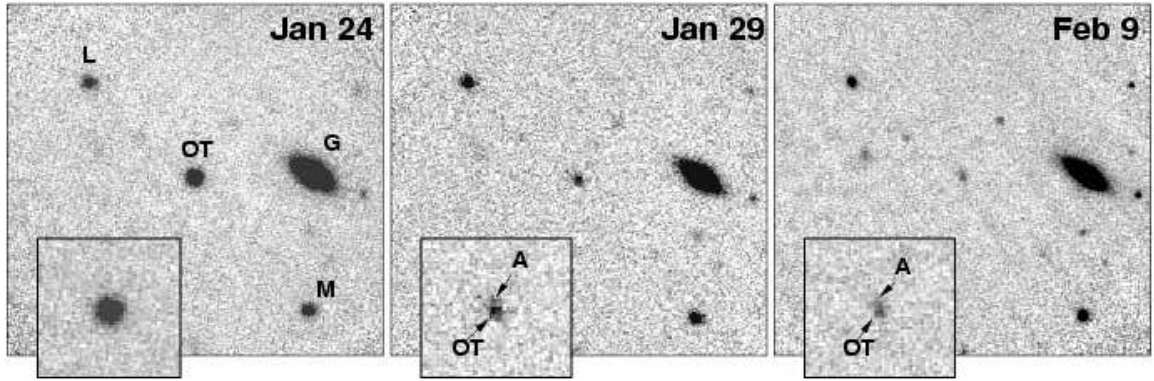,angle=-90,width=6.0in}}
\caption[Three epochs of Keck I $K$-band imaging of the field of GRB
990123]{Three epochs of Keck I $K$-band imaging of the field of GRB
990123 (1999 January 24.6, January 29.7, and February 9.6 UT). The
field shown is 32\arcsec\ $\times$ 32\arcsec, corresponding to about
270 $\times$ 270 physical kpc$^2$ in projection at $z$ = 1.6004 (for
$H\mathstrut{_{0}}$ = 65 km s$\mathstrut{^{-1}}$
Mpc$\mathstrut{^{-1}}$ and $\Omega\mathstrut{_{m}}$ = 0.2,
$\Omega\mathstrut{_{\Lambda}}$ = 0.2). The images have been rotated to
the standard orientation, so that the east is to the left and north is
up. The magnitude of the host galaxy is $K\mathstrut{_{{\rm host}}}$ =
22.1 $\pm$ 0.3 mag. In the January 24 image, the OT dominates the host
galaxy flux \citep[$K\mathstrut{_{{\rm OT}}}$ = 18.29 $\pm$
0.04;][]{kdo+99}, but by January 29 the galaxy is resolved (see inset)
from the OT.}
\label{fig:grb990123kband}
\end{figure}

The first evidence of the underlying galaxy, approximately 0\arcsec.6
from the OT, was seen in our Keck $K$-band images taken on 1999 January
27 UT. The galaxy, the putative host (which we designate as ``A"), was
then clearly detected in the images obtained on 1999 January 29 UT
\citep{dkb+99} and later, under excellent seeing conditions, on
February 9 and 10 (see figure~\ref{fig:grb990123kband}). We find the OT
and host fluxes as follows. Sets of pixels dominated by the OT or by
the galaxy were masked, and total fluxes with such censored data were
evaluated in photometric apertures of varying radii. Total fluxes of
the OT + galaxy were also measured in the same apertures using the
uncensored data. We also varied the aperture radii, and the position
and the size of the sky measurement annulus. On February 9 (February
10) UT, we found that the OT contributes 65\% (57\%) ($\pm$10\%) of
the total OT + galaxy light. The estimated errors of the fractional
contributions of the OT to the total light reflect the scatter
obtained from variations in the parameters of these image
decompositions. In both epochs, the fractional contribution of the
host implies that a flux of the host galaxy is \hbox{$0.9 \pm 0.3$
$\mu$Jy} (\hbox{$K\mathstrut{_{{\rm host}}}$ = 22.1 $\pm$ 0.3 mag}).
We assume 636 Jy for the flux zero point of the $K$ band for $K = 0$ mag
\citep{bb88}.

  The HST observations of the GRB 990123 field were obtained in 1999
February 8--9 UT in response to the director's discretionary time
proposal GO-8394, with the immediate data release to the general
community \citep{bec99a}. The CCD camera of the Space Telescope
Imaging Spectrograph (STIS) \citep{kwb+98} in CLEAR aperture (50CCD)
mode was used. Over the course of three orbits, the field was imaged
in six positions dithered in a spiral pattern for a total integration
time of 7200$\,$ s. Each position was imaged twice to facilitate
cosmic-ray removal (a total of 12 integrations).

  Initial data processing followed the STScI pipeline procedures,
including bias and dark current subtraction. The six
cosmic-ray-removed images were then combined by registering the images
and median-stacking to produce a master science-grade image. We also
produced a higher resolution image using the ``drizzle" technique
\citep{fhbm97}. Photometry and astrometry were performed
on both final image products. We find a negligible difference between
the two images compared with other uncertainties (\eg sky
determination and counting noise) in photometry and astrometry. The
mean epoch of the final images is 1999 February 9.052 UT.
\begin{figure}[tp]
\centerline{\psfig{file=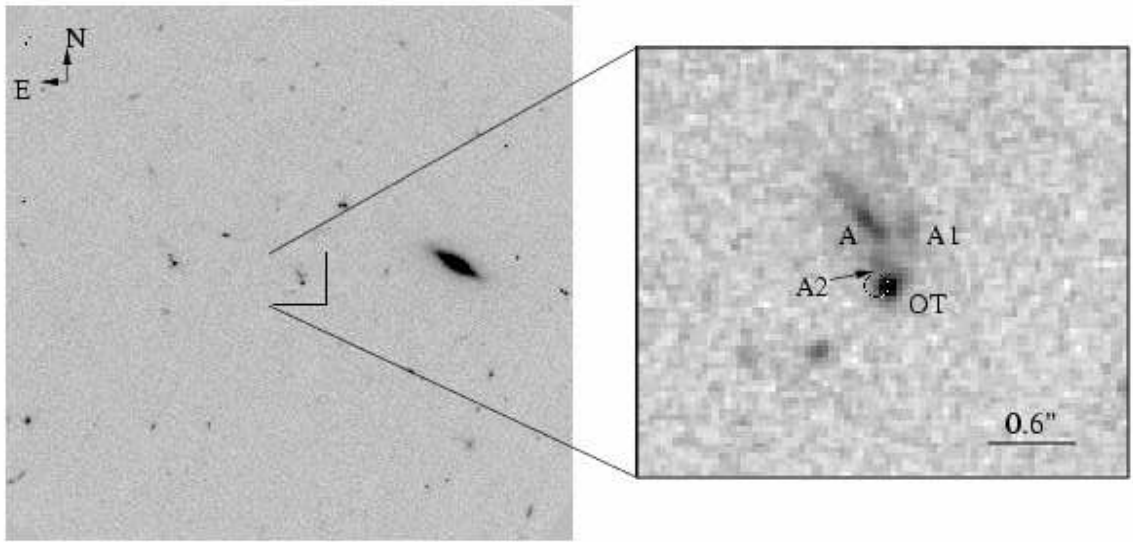,angle=-90,width=6.0in}}
\caption[The {\it HST}/STIS drizzle image of the field of 
GRB 990123]{The {\it HST}/STIS ``drizzle'' image (mean epoch 1999
February 9.052 UT) of the field of GRB 990123 rotated to the normal
orientation. (Right) The field shown is 32 $\times$ 32 arcsec$^2$,
corresponding to about 270 $\times$ 270 proper sq.~kpc (710 $\times$
710 co-moving sq.~kpc) in projection at $z$ = 1.6 (for
$H\mathstrut{_{0}}$ = 65 km s$\mathstrut{^{-1}}$
Mpc$\mathstrut{^{-1}}$ and $\Omega\mathstrut{_{m}}$ = 0.2). The
effective exposure time is 7200 s, and the pixel scale is 0\arcsec
.0254 pixel$\mathstrut{^{-1}}$ . (Left) The OT, galaxy A (the
putative host), and the bright knots (A1 and B) associated with galaxy
A are denoted.  The cross that overlays the OT point source depicts
the 1 s uncertainty in each axis for the position of the OT as
measured in ground-based imaging (see text). The positional
consistency definitively establishes that the point source is indeed
the OT. The nebulosity (A2) just to the north of the OT may be a
star-forming region \citep[see also][]{hh99}.}
\label{fig:grb990123hst}
\end{figure}

  Figure \ref{fig:grb990123hst} shows a portion of the STIS image of
the GRB 990123 field. We find (see below) the OT clearly detected as a
point source 0\arcsec.67 $\pm$ 0\arcsec.02 to the southwest of the
central region of galaxy A. Galaxy A has an elongated and clumpy
appearance, possibly indicative of star formation regions in a
late-type galaxy. A morphological classification as an irregular
galaxy or interacting galaxy system is likely most apt. Such
morphologies are typical for many galaxies at comparable flux levels,
as observed with the HST. Its extension to the south clearly overlaps
with the OT, and it is thus very likely that this galaxy is
responsible for the absorption-line system at $z\mathstrut{_{{\rm
abs}}}$ = 1.6004
\citep{kdo+99,hap+99}. The knot "B" to the east may be a satellite of
the host galaxy or a star-forming region along the interface zone of a
galaxy interacting with the host.

  We measured the centroid of the optical transient in our
discovery image from January 23 at the Palomar 60 inch telescope
(P60). The OT was bright (r = 18.65 mag) at this early epoch, and its
position is well determined with respect to other objects in the
field. Next we computed the astrometric mapping of the P60 coordinate
system to a deep Keck $R$-band image from 1999 February 9.6 UT
\citep{kdo+99} using 75 well-detected objects common to
the two images. Similarly, we tied the Keck II coordinates to the STIS
image using 19 common tie objects. We found the ground-based position
of the OT to be consistent with the STIS point source, with a
negligible offset of 0\arcsec.09 $\pm$ 0\arcsec.18. [Note: after this
chapter was published, this source faded in {\it HST} imaging,
confirming the astrometry.  See \S \ref{subsec:grb990123}.]

  The coordinates of the OT as measured in the \hst\ image are
\hbox{$\alpha =$ 15\fh 25\fm 30\fs .3026}, \hbox{$\delta =$ +44\fd
45\arcmin 59\arcsec .048} (J2000); the \hst\ plate solution is based
on the revised guide star catalog and is accurate (in an absolute
sense) to $\sim$ 0\arcsec .3. We note the excellent agreement between
this \hst\ measurement and the absolute astrometric measurement from
ground-based imaging as reported in \citet{kdo+99}. The brightest
central region of galaxy A (itself extended north-south) is located
0\arcsec .17 east, 0\arcsec .64 north of the transient. The bright,
possibly star-forming, regions ``A1'' and ``B'' are located 0\arcsec .25
west, 0\arcsec .46 north and 0\arcsec .66 east, 0\arcsec .51 south,
respectively. The uncertainties in the relative positions are $\sim$20
mas.

  No other galaxies brighter than $V \sim$ 27 mag are detected in the
STIS image closer to the OT than galaxy A, and we see no evidence for
a distant cluster (or even a sizable group) in this field. This
effectively removes the possibility \citep{dkb+99b} that the burst was
significantly magnified by gravitational lensing.

  We will assume for the Galactic reddening in this direction
E$\mathstrut{_{{B}-{V}}}$ = 0.016 mag \citep{sfd98}, and we will use
the standard Galactic extinction curve with R$\mathstrut{_{V}}$ =
A$\mathstrut{_{V}}$/E$\mathstrut{_{{B}-{V}}}$ = 3.1 to estimate
extinction corrections at other wavelengths. We assume the photometric
flux zero points as tabulated by \citet{fsi95}.

  Most objects detected in the STIS imaging are also detected in the
deep Keck II image. After photometric transformation of the R-band
Keck II image to the Gunn-$r$ system, we used objects common to both
images in order to find a zero point for the STIS image. The r.m.s.\
uncertainty of the zero point is 0.1 mag, which is mostly due to
varied color terms. This scatter implies a transformation of the
broadband STIS magnitudes to the Gunn-$r$ system that is rather
robust.  Taking the system as a whole, $r\mathstrut{_{{\rm OT}+{\rm
host}}}$ = 24.1 $\pm$ 0.1 mag, which is in excellent agreement with
the Keck II imaging \citep{kdo+99} taken 12 hr before the \hst\
imaging.  Aperture photometry on the individual components yields
$r\mathstrut{_{{\rm OT}}}$ = 25.3 $\pm$ 0.2 mag and
$r\mathstrut{_{{\rm host}}}$ = 24.5 $\pm$ 0.2 mag
($F\mathstrut{_{{\nu},\,r,\,{\rm host}}}$ = 0.5 $\pm$ 0.1 $\mu$Jy).
The errors are dominated by uncertainties in the color term of the
object and the sky value for aperture photometry. Since the host
galaxy must contribute some flux in the OT aperture, the magnitude
presented above should be considered an upper (lower) limit for the
host (OT).

  We also determined the OT and host magnitudes by converting observed
counts to flux given the instrumental response. Since the bandpass of
the STIS CLEAR is so broad, the conversion depends on the assumed
spectrum of the object. For the OT, we assumed a spectral index of
$\beta \approx -0.8$ (with $F\mathstrut{_{{\nu}}} \propto
\nu\mathstrut{^{{\beta}}}$), as inferred from theory and other OTs. For
the host, we take $\beta = -0.5$ as a good approximation for
star-forming galaxies in the redshift range of the host and also
consistent with the $K$-band measurement. In both cases, we explored a
range of plausible spectral indices. Using the STIS exposure simulator
available from STScI to convert the observed counts to flux, we find
$V\mathstrut{_{{\rm OT}}}$ = 25.4 and $V\mathstrut{_{{\rm host}}}$ =
24.6 mag. In a recent paper, \citet{ftm+99}, from analysis of the same HST
data, find $V\mathstrut{_{{\rm OT}}}$ = 25.45 $\pm$ 0.15 and
$V\mathstrut{_{{\rm host}}}$ = 24.20 $\pm$ 0.15 mag. The determination of
the host magnitude is subject to two systematic uncertainties: the
assumed galaxy spectrum and the aperture employed in photometry (see
above).  Within these errors, these measurements are in
agreement. Given the assumed spectral shapes, we also find $r
\mathstrut{_{{\rm OT}}}$ = 25.1 mag and $r\mathstrut{_{{\rm host}}}$
= 24.3 mag, in agreement with our direct photometric tie to ground-based
imaging.

  We note that the simple power-law approximation to the broadband
spectrum of the galaxy, as defined by our STIS and $K$-band
measurements, is $\beta\mathstrut{_{{\rm host}}}$ $\approx
-0.5$. This relatively blue color is suggestive of active star
formation, but it cannot be used to estimate the SFR directly.

\section{Discussion}

  In what follows, we assume a standard Friedman model cosmology with
$H\mathstrut{_{0}}$ = 65 km s$\mathstrut{^{-1}}$
Mpc$\mathstrut{^{-1}}$, $\Omega\mathstrut{_{m}}$ = 0.2, and
$\Omega\mathstrut{_{\Lambda}}$ = 0. For $z$= 1.6004, the luminosity
distance is 3.7 $\times 10\mathstrut{^{28}}$ cm, and 1\arcsec\
corresponds to 8.64 proper kpc or 22.45 co-moving kpc in projection.
These values are not appreciably different for other reasonable
cosmological world models.  For example, with $\Omega\mathstrut{_{m}}$
= 0.3 and $\Omega\mathstrut{_{\Lambda}}$ = 0.7 the luminosity distance
is 4.1 $\times 10\mathstrut{^{28}}$ cm, and 1\arcsec\ corresponds to
9.12 proper kpc.

We consider it likely that the absorption system at
$z\mathstrut{_{{\rm abs}}}$ = 1.6004 originates from galaxy A since no
other viable candidate is seen in the \hst\ images. The proximity of
the center of galaxy A to the OT line of sight (0\arcsec .67 $\pm$
0\arcsec .02), corresponding to 5.8 proper kpc at this redshift,
strongly suggests that the two are physically related. We thus propose
that galaxy A is the host galaxy of the GRB. Visual inspection of
figure \ref{fig:grb990123hst} suggests that a probability of chance
superposition at this magnitude level is very small (see \S
\ref{tab:offnorm}).

In order to estimate the rest-frame luminosity of galaxy A, we
interpolate between the observed STIS and $K$-band data points using a
power law, to estimate the observed flux at $\lambda\mathstrut{_{{\rm
obs}}}$ $\approx$ 11570 \AA, corresponding approximately to the
effective wavelength of the rest-frame $B$ band. We obtain
$F\mathstrut{_{{\nu},\,{B},\,{\rm rest}}}$ $\approx$ 0.7 $\mu$Jy,
corresponding to the absolute magnitude $M_{B} = -20.0$. Locally,
an $L\mathstrut{_{*}}$ galaxy has $M\mathstrut{_{B}}$ $\approx
-20.75$ mag. We thus conclude that this object has the rest-frame
luminosity that is $L\mathstrut{_{{\rm host}}}$ $\approx$ 0.5
$L\mathstrut{_{*,\,{\rm local}}}$. Given the uncertainty of the
possible evolutionary histories, it may evolve to become either a
normal spiral galaxy or a borderline dwarf galaxy.

  We can make a rough estimate of the SFR from the continuum
luminosity at $\lambda\mathstrut{_{{\rm rest}}}$ = 2800 \AA, following
\citet{mpd98}. Using the $F\mathstrut{_{{\nu},\,2800}}$ estimates given above, the corresponding
monochromatic rest-frame power is $P\mathstrut{_{{\nu},\,2800}} = 2.9
\times 10\mathstrut{^{28}}$ ergs s$\mathstrut{^{-1}}$
Hz$\mathstrut{^{-1}}$ (for $\beta$ = 0, since it may be appropriate in
the UV continuum itself) or $P\mathstrut{_{{\nu},\,2800}}$ = 3.6
$\times 10\mathstrut{^{28}}$ ergs s$\mathstrut{^{-1}}$
Hz$\mathstrut{^{-1}}$ (for $\beta = -0.8$). The corresponding
estimated unobscured SFRs are $\approx 3.6$ and $\approx 4.6$ \sfr,
probably accurate to within 50\% or better. This modest value is
typical for normal galaxies at such redshifts. It is of course a lower
limit, since it does not include any extinction corrections in the
galaxy itself or any fully obscured star formation.

  Further insight into the physical properties of this galaxy comes
from its absorption spectrum, presented in \citet{kdo+99}.  The lines
are unusually strong, placing this absorber in the top 10\% of all Mg
II absorbers detected in complete surveys
\citepeg{ss92}. Unfortunately, without a direct measurement of the
hydrogen column density, it is impossible to estimate the metallicity
of the gas. We note that strong metal-line absorbers are frequently
associated with high hydrogen column density systems, such as damped
Ly $\alpha$ absorbers. The small scatter of redshift in the individual
lines implies a very small velocity dispersion, less than 60 km
s$\mathstrut{^{-1}}$ in the galaxy's rest frame. This implies that the
absorber is associated with either a dwarf galaxy or a dynamically
cold disk of a more massive system.

  The OT is well offset (5.8 kpc) from the central region of the host;
this clearly casts doubt on an active galactic nucleus origin of
GRBs. However, if the host is indeed an interacting galaxy system,
then it is plausible that a massive black hole could be created
off-center \citep[as recently suggested by][]{ftm+99}, and thus the
position of the GRB 990123 could still permit massive black holes as
the progenitors of GRBs. Yet, around such massive BHs, the expectation
is that the high-velocity---enshrouding material would give rise to
absorption in the GRB afterglow. The clear absence of broad absorption
lines in our optical afterglow spectrum, then, does not bode well for
the massive black hole hypothesis for the origin of GRBs.

 The spatially resolved imaging using HST provides the clearest
picture of the relation of GRBs to their hosts. The transient of GRB
970228 is displaced from its host center \citep{slp+97,fpt+99} but
still lies within the half-light radius. GRB 970508, on the other
hand, is coincident with the nucleus of its host galaxy to 0\arcsec
.01 \citep{fp98,bdkf98}. As shown in this Letter, GRB 990123 is
separated from the central region of the host and appears to be
spatially coincident with a bright star-forming region (``A2''; see
figure~\ref{fig:grb990123hst}). Indeed, \citet{hh99} have recently
corroborated this claim by noting that the size and luminosity of A2
befit the properties of a generic star-forming region. [Note: later
{\it HST} imaging did not show the purported region A2.  It was
probably an artifact of an imperfect PSF determination in the
\citet{hh99} work.]  The close connection of GRBs to their hosts can
be extended to results from ground-based astrometry by using the host
galaxy magnitude as an objective measure \citep{owdk96} of the host
size. Using the total magnitude of all known host galaxies to date, we
note that the optical transients (except for GRB 990123) lie well
within the effective half-light radius.

Copious star formation always appears to be spatially concentrated:
along spiral arms, in bright compact H II regions in dwarf irregular
galaxies, and in the interface zone of interacting galaxies. Given the
morphology of the host (figure~\ref{fig:grb990123hst}), we suggest that
GRB 990123 arose from a star-forming complex in the interface zone of
what appears to be a pair of interacting galaxies. This is the first
clear case of a GRB associated with an interaction region.  Another
possible case is the host of GRB 980613. Until now, the GRB star
formation connection has been primarily through gross star formation
rates obtained from spectroscopic indicators. It is possible that with
increasingly larger samples of host galaxies, in analogy to
supernovae, the relationship of GRBs to the morphology of the hosts
may provide complementary insight into the progenitors of GRBs.

\acknowledgements

We are grateful to S.~Beckwith of STScI for the allocation of the
director's discretionary time for this project and to the entire
BeppoSAX team and the staff of W.~M.~Keck Observatory for their
efforts. We also thank L.~Ferrarese for aiding us with HST observing
and the anonymous referee for helpful and clarifying comments. This
work was supported in part by a grant from STScI, grants from the NSF
and NASA, and the Bressler Foundation.

\chapter[The Redshift and the Ordinary Host Galaxy of GRB
970228]{The redshift and the ordinary host galaxy of GRB
970228$^\dag$}
\label{chap:grb970228}

\secfootnote{\secfootdag}{A version of this chapter was first published 
in {\it The Astrophysical Journal}, 554, p.~678--683 (2001).}

\secauthor{J. S. Bloom, S. G. Djorgovski, S. R. Kulkarni}

\secaffils{Palomar Observatory 105--24, California Institute of Technology,
            Pasadena, CA 91125, USA; {\tt jsb,george,srk@astro.caltech.edu}}

\begin{abstract}
The gamma-ray burst of 1997 February 28 (GRB 970228) ushered in the
discovery of the afterglow phenomenon.  Despite intense study of the
nearby galaxy, however, the nature of this galaxy and the distance to
the burst eluded the community.  Here we present the measurement of
the redshift of the galaxy, the putative host galaxy of GRB 970228,
and, based on its spectroscopic and photometric properties, identify
the galaxy as a sub-luminous, but otherwise normal galaxy at redshift
\hbox{$z=0.695$} undergoing a modest level of star formation.  At this
redshift, the GRB released an isotropic-equivalent energy of
\hbox{$(1.4 \pm 0.3) \times 10^{52}$ erg} \hbox{(20--2000 keV
rest frame)}.  We find no evidence that the host is significantly
bluer or is forming stars more vigorously than the general field
population. In fact, by all accounts in our analysis
(color--magnitude, magnitude--radius, star-formation rate,
Balmer-break amplitude) the host properties appear typical for faint
blue field galaxies at comparable redshifts.
\end{abstract}

\section{Introduction}

The gamma-ray burst of 1997 February 28 (hereafter GRB 970228) was a
watershed event, especially at optical wavelengths. The afterglow
phenomenon, long-lived multi-wavelength emission, was discovered
following GRB 970228 at X-ray \citep{cfh+97} and optical
\citep{vgg+97} wavelengths.  Despite intense observations, no radio
transient of GRB 970228 was found \citep{fksw98}; the first radio
afterglow \citep{fkn+97} had to await the next {\it BeppoSAX}
localization of GRB 970508. The basic predictions of the synchrotron
shock model for GRB afterglow appeared confirmed by GRB 970228
\citepeg{wrm97}.

Despite intense efforts, early spectroscopy of the afterglow of GRB
970228 \citepeg{thcm97,kdc+97} failed to reveal the redshift.
Spectroscopy of the afterglow of GRB 970508 proved more successful
\citep{mdk+97}, revealing through absorption lines that the GRB
originated from a redshift $z \ge 0.835$.  Later spectroscopy and
imaging revealed a faint galaxy with $z=0.835$ at the same location of
the afterglow \citep{bdkf98}. Through the preponderance of subsequent
redshift determinations (currently 19 in total) and the association of
GRBs with faint galaxies, it is now widely believed that the majority
of all of long duration ($T \age 1$ s) gamma-ray bursts originate from
cosmological distances and are individually associated with faint
galaxies.

Even without a redshift, observations of the afterglow of GRB 970228
in relation to its immediate environment began to shed light on the
nature of the progenitors of gamma-ray bursts.  Ground-based
observations of the afterglow revealed a near coincidence of the GRB
with the optical light of a faint galaxy \citep{mkd+97,vgg+97}.
Later, deeper ground-based images \citep{dkg+97} and high resolution
images from the {\it Hubble Space Telescope} ({\it HST}) showed the
light from the fading transient clearly embedded in a faint galaxy
\citep{slp+97} with a discernible offset between the host galaxy
centroid and the afterglow. Given the low (albeit {\it a posteriori})
probability of chance superposition of the afterglow with a random
field galaxy (a few percent), this galaxy is presumably the host of
the GRB
\citep{vgg+97}.  This connection is supported by noting that almost
all well-localized GRBs, as a class, appear to be statistically
connected to a nearby galaxy \citep{bkd02}.  Throughout this paper we
will assume that the galaxy is indeed the host galaxy of GRB
970228. Though by no means definitive, the apparent offset of GRB
970228 from its host rendered an active galactic nucleus origin
unlikely \citep{slp+97}.

The two most popular progenitor scenarios---coalescence of binary
compact stellar remnants and the explosion of a massive star
(``collapsar'')---imply that the gamma-ray burst rate should closely
follow the massive star formation rate in the universe.  In both
formation scenarios a black hole is created as a by-product; however,
the scenarios differ in two important respects. First, only very
massive progenitors ($M_{\rm ZAMS} \age 40\, M_\odot$; Fryer, Woosley,
\& Hartmann 1999\nocite{fwh99}) will produce GRBs in the single star
model whereas the progenitors of neutron star--neutron star binaries
need only originate with $M_{\rm ZAMS} \age 8\, M_\odot$.  Second, the
scenarios predict a distinct distribution of physical offsets
\citep{pac98b,bsp99} in that the coalescence site of merging remnants
could occur far from the binary birthplace (owning to substantial
systemic velocities acquired during neutron star formation through
supernovae), whereas exploding massive stars will naturally occur in
star-forming regions.  In relation to the predicted offset of GRB
970228, \citet{bsp99} further noted the importance of redshift to
determine the luminosity (and infer mass) of the host galaxy: massive
galaxies more readily retain binary remnant progenitors.  Thus the
relationship of GRBs to their hosts is most effectively exploited with
redshift by setting the physical scale of any observed angular offset
and critically constraining the mass (as proxied by host luminosity).

The redshift of GRB 970228 also plays a critical role in the emerging
supernova--GRB link.  Following the report of an apparent supernova
(SN) component in the afterglow of GRB 980326 \citep{bkd+99}, the
afterglow light curves of GRB 970228 were reanalyzed, and both
\citet{rei99} and \citet{gtv+00} found evidence for a SN component by
way of a red ``bump'' in the light curve about 1 month after the
burst.  This interpretation, however, relies critically on the
knowledge of the redshift to GRB 970228 to set the rest-frame
wavelength of the apparent broadband turnover of the SN component in
the observed $I$ band.  A lower redshift would, for instance, help
make the dust-reradiation model of \citet{wd00} more viable since the
peak of thermal dust emission could be no bluer than 1 $\mu$m. On the
other hand, a red bump from a high-redshift ($z \age 1$) GRB is
difficult to observe from a SN component since line-blanketing of the
UV portion of SNe spectra essentially suppresses the (observer frame)
optical flux.  Instead, the late-time bumps in high-$z$ GRBs may be
more readily explained as dust echoes from the afterglow \citet{eb00}.

Finally, knowledge of the redshift is essential in order to derive the
physical parameters of the GRB itself, primarily the energy scale.  We
now know, for instance, that the typical GRB releases about $10^{52}$
ergs in gamma rays. The distribution of observed isotropic-equivalent
GRB energies is, however, very broad \citepcf{kbb+00}.

Recognizing these needs, we implemented an aggressive spectroscopy
campaign on the presumed host of GRB 970228, as detailed in \S
\ref{sec:obs}.  The redshift determination of the host (and, by
assumption, the GRB itself) was first reported by \citet{dkbf99} and
is described in more detail in \S \ref{sec:0228redshift}. We then use
this redshift and the spectrum of the host galaxy in \S
\ref{sec:0228imp} to set the physical scale of the observables:
energetics, star formation rates, and offsets.  Based on this and
photometric imaging from HST, we demonstrate in \S \ref{sec:host} that
the host is a sub-luminous, but otherwise normal galaxy.

\section{Observations and Reductions}
\label{sec:obs}

A finding chart and the coordinate location of the host galaxy of GRB
970228 are given in \citet{vgg+97}. Spectra of the host galaxy were
obtained on the W.~M.~Keck Observatory 10 m telescope (Keck II) atop
Mauna Kea, Hawaii. Observations were conducted over the course of
several observing runs: UT 1997 August 13, UT 1997 September 14, UT
1997 November 1 and 28--30, and UT 1998 February 21--24. The observing
conditions were variable, from marginal (patchy/thin cirrus or
mediocre seeing) to excellent, and on some nights no significant
detection of the host was made; such data were excluded from the
subsequent analysis.  On most nights, multiple exposures (two to five)
of 1800 s were obtained, with the object dithered on the spectrograph
slit by several arcseconds between the exposures.  The net total
useful on-target exposure was approximately 11 hr from all of the runs
combined.

All data were obtained using the Low-Resolution Imaging Spectrometer
(LRIS; Oke et al.~1995\nocite{occ+95}) with 300 lines mm$^{-1}$
grating and a 1.0 arcsec wide long slit, giving an effective
instrumental resolution FWHM $\approx 12$ \AA.  Slit position angle
was always set to $87^\circ$, with star S1 \citep{vgg+97} always
placed on the slit, and used to determine the spectrum trace along the
chip; galaxy spectra were then extracted at a position 2.8 arcsec east
of star S1.  Efforts were made to observe the target at hour
angles so as to make this slit position angle as close to parallactic
as possible.  Exposures of an internal flat-field lamp and arc lamps
were obtained at comparable telescope pointings immediately following
the target observations.  Exposures of standard stars from
\citet{og83} and \citet{msb88} were obtained and used to measure the
instrument response curve, although on some nights the flux zero
points were unreliable owing to non-photometric conditions.

Wavelength solutions were obtained from arc lamps in the standard
manner, and then a second-order correction was determined from the
wavelengths of isolated strong night-sky lines, and applied to the
wavelength solutions.  This procedure largely eliminates systematic
errors owing to the instrument flexure, and is necessary in order to
combine the data obtained during separate nights.  The final
wavelength calibrations have an r.m.s.~$\sim 0.2 - 0.5$~\AA, as
determined from the scatter of the night-sky line centers.  All
spectra were then re-binned to a common wavelength scale with a
sampling of 2.5~\AA\ (the original pixel scale is $\sim 2.45$ \AA),
using a Gaussian with a $\sigma = 2.5$ \AA\ as the
interpolating/weighting function.  This is effectively a very
conservative smoothing of the spectrum, since the actual instrumental
resolution corresponds to $\sigma \approx 5$ \AA.

Individual spectra were extracted and combined using a statistical
weighting based on the signal-to-noise ratio determined from the data
themselves (rather than by the exposure time).  Since some of the
spectra were obtained in non-photometric conditions, the final
spectrum flux zero-point calibration is also unreliable, but the
spectrum shape should be unaffected.  We use direct photometry of the
galaxy to correct this zero-point error (see below).

Our uncorrected spectrum gives a spectroscopic magnitude $V \approx
26.3$ mag for the galaxy.  Direct photometry from the {\it HST }
imaging data indicates $V = 25.75 \pm 0.3$ \citep{gtv+00} for the
host. Given that some of our spectra were obtained through thin
cirrus, this discrepancy is not surprising.  Thus, in order to bring
our measurements to a consistent system, we multiply our flux values
by a constant factor of 1.66, but we thus also inherit the systematic
zero-point error of $\sim 30$\% from the {\it HST} photometry.

There is some uncertainty regarding the value of the foreground
extinction in this direction (see discussion in \S
\ref{sec:0228host}).  We apply a Galactic-extinction correction by
assuming $E_{B-V} = 0.234$ mag from \citet{sfd98}. We assume $R_V =
A_V/E_{B-V} = 3.2$, and the Galactic extinction curve from
\citet{ccm98} to correct the spectrum.  All fluxes and luminosities
quoted below incorporate both the flux zero-point and
Galactic-extinction corrections.

\section{The Redshift of GRB 970228}
\label{sec:0228redshift}
\begin{figure*}[tp]
\centerline{\psfig{file=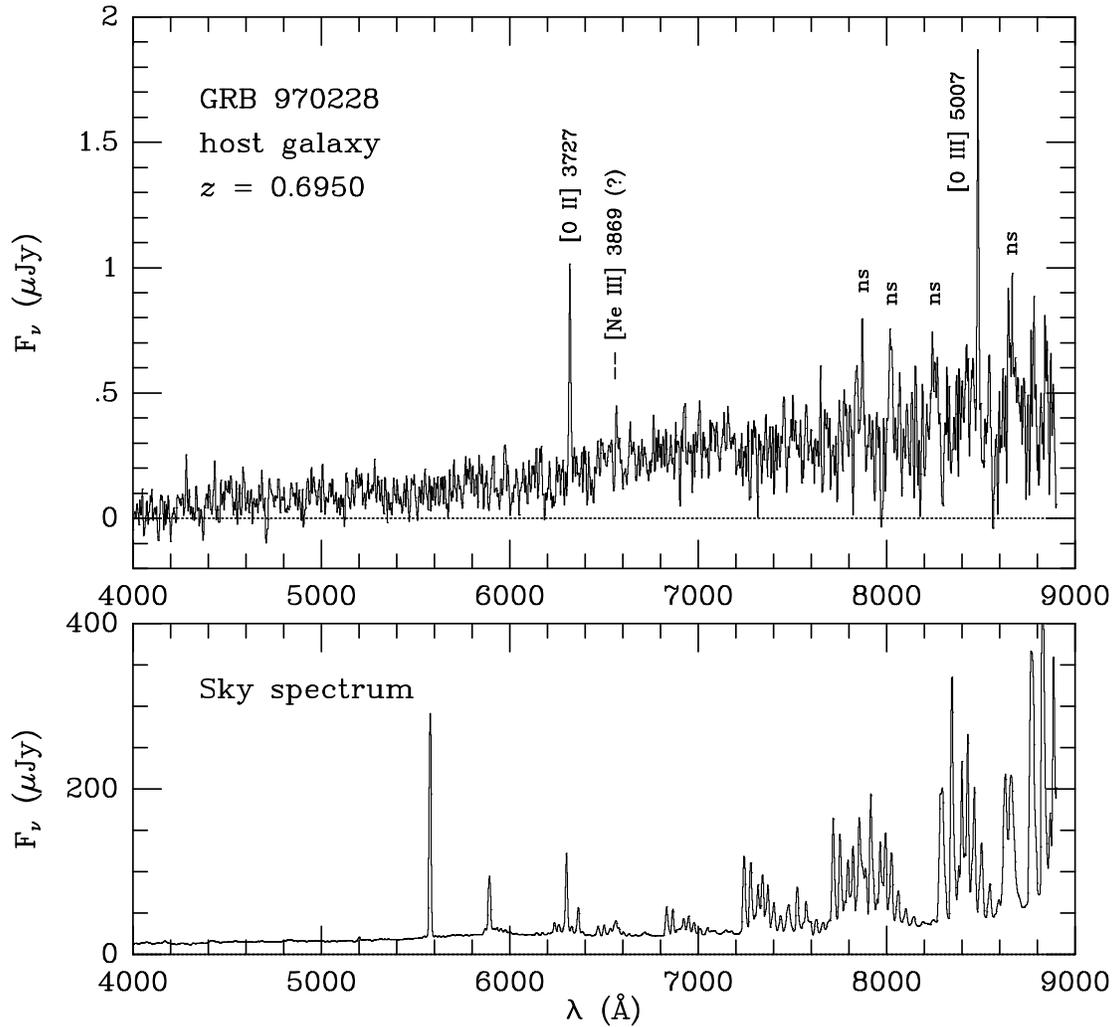,width=6.2in}}
\caption[The weighted average spectrum of the host galaxy of
GRB 970228, obtained at the Keck II telescope]{{\it Top}: The weighted
average spectrum of the host galaxy of GRB 970228, obtained at the
Keck II telescope.  Prominent emission lines [O II] 3727 and [O III]
5007 and possibly [Ne III] 3869 are labeled assuming the lines
originate from the host at redshift $z = 0.695$. The notation ``ns''
refers to noise spikes from strong night-sky lines. {\it Bottom}: The
average night-sky spectrum observed during the GRB 970228 host
observations, extracted and averaged in exactly the same way as the
host galaxy spectrum. }
\label{fig:0228spec}
\end{figure*}

The final combined spectrum of the galaxy is shown in figure
\ref{fig:0228spec}.  Two strong emission lines are seen, \hbox{[O II] 3727}
and \hbox{[O III] 5007}, thus confirming the initial redshift
interpretation based on the [O II] 3727 line alone \cite{dkbf99}.
Unfortunately, the instrumental resolution was too coarse to resolve
the [O II] 3727 doublet. The weighted mean redshift is $z = 0.6950 \pm
0.0003$.  A possible weak emission line of [Ne III] 3869 is also seen.
Unfortunately, the strong night-sky OH lines preclude the measurements
of the H$\beta$ 4861 and [O III] 4959 lines, as well as the higher
Balmer lines.

The corrected [O II] 3727 line flux is $(2.2 \pm 0.1) \times 10^{-17}$
erg cm$^{-2}$ s$^{-1}$ Hz$^{-1}$, and its observed equivalent width is
$W_\lambda = 51 \pm 4$ \AA, i.e., $30 \pm 2.4$ \AA\ in the restframe.
This is not unusual for field galaxies in this redshift range
\citep{hcbp98}.  The [Ne III] 3869 line, if real, has a flux of at
most 10\% of the [O II] 3727 line, which is reasonable for an actively
star forming galaxy. The corrected [O III] 5007 line flux is $(1.55
\pm 0.12) \times 10^{-17}$ erg cm$^{-2}$ s$^{-1}$ Hz$^{-1}$, and its
observed equivalent width is $W_\lambda = 30 \pm 2$ \AA, i.e., $17.7
\pm 1.2$ \AA\ in the restframe. For the H$\beta$ line, we derive an
upper limit of less than $3.4 \times 10^{-18}$ erg cm$^{-2}$ s$^{-1}$
Hz$^{-1}$ ($\sim 1~\sigma$), and $W_\lambda < 7$ \AA\ for its observed
equivalent width.  We note, however, that this measurement may be
severely affected by the poor night-sky subtraction.
 
The continuum flux at $\lambda_{\rm obs} = 4746\,$\AA, corresponding to
$\lambda_{\rm rest} = 2800\,$\AA, is $F_\nu = 0.29 ~\mu$Jy, with a
statistical measurement uncertainty of $\sim 10$\% and a systematic
uncertainty of 30\% inherited from the overall flux zero-point
uncertainty.  The continuum flux at $\lambda_{\rm obs} \sim 7525\,$\AA,
corresponding to the restframe $B$ band, is $F_\nu = 0.77 ~\mu$Jy,
with a statistical measurement uncertainty of $\sim 7$\% (plus 30\%
systematic).

\section{Implications of the Redshift}
\label{sec:0228imp}

For the following discussion, we will assume a flat cosmology as
suggested by recent results \citepeg{dab+00} with $H_0 = 65$ km
s$^{-1}$ Mpc$^{-1}$, $\Omega_M = 0.3$, and $\Lambda_0 = 0.7$.  For $z
= 0.695$, the luminosity distance is $1.40 \times 10^{28}$ cm, and 1
arcsec corresponds to 7.65 proper kpc or 13.0 co-moving kpc in
projection.  By virtue of the close spatial connection of GRB 970228
with the putative host galaxy (see \S 1), we assume that GRB itself
occurred at a redshift $z = 0.695$.

\subsection{Burst energetics}

The gamma-ray fluence (integrated flux over time) is converted from
count rates under the assumption of a GRB spectrum, the spectral
evolution, and the true duration of the GRB.  These quantities are
estimated from the GRB data itself but can lead to large uncertainties
(a factor of few) in the fluence determination.  In \citet{bfs01} we
developed a methodology to account for these uncertainties as well as
``$k$-correct'' each fluence measurement to a standard co-moving
bandpass. Utilizing the redshift reported herein we found that the
isotropic-equivalent energy release in GRB 970228 was \hbox{$(1.4 \pm
0.3) \times 10^{52}$ ergs} in the co-moving bandpass 20--2000 keV and a
bolometric energy release of \hbox{$(2.7 \pm 1.0) \times 10^{52}$ ergs}
\citep{bfs01}.

Since at least some GRBs are now believed to be jetted
\citepeg{fks+01}, the true energy release may have been significantly
less than that implied if the energy release was isotropic.  The
degree of ``jettedness'' in GRBs is most readily determined by the
observation of an achromatic break in the afterglow light curves. As
such, the slow decline and absence of a strong break in the optical
light curve of GRB 970228 \citepeg{ggv+97} suggest that the GRB
emission was nearly isotropic \citepcf{sph99} and so the knowledge of
the total energy release is primarily limited by the accuracy of the
fluence measurement.

\subsection{The offset of the gamma-ray burst and the host morphology}
\label{sec:off}

\begin{figure*}[tbp]
\centerline{\psfig{file=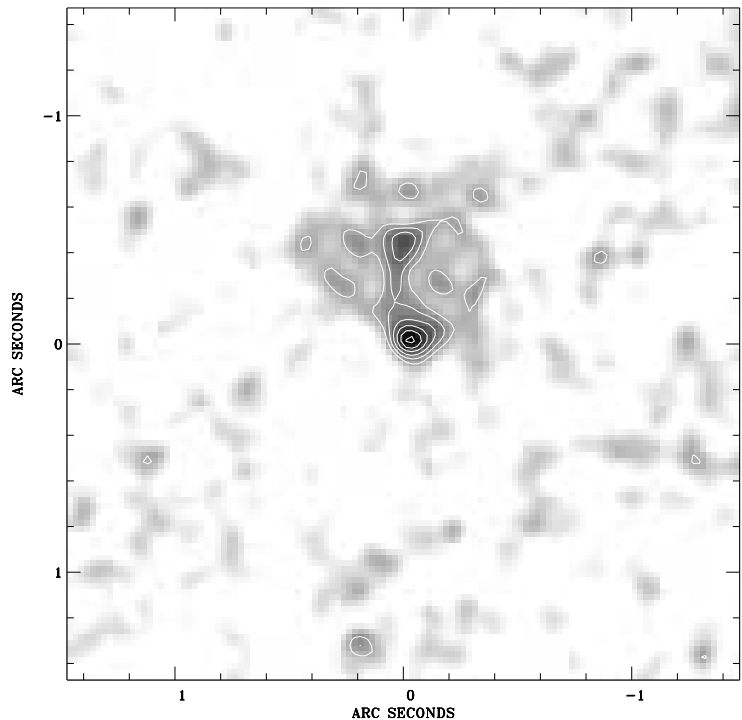,width=5.5in}}
\caption[A 3\arcsec $\times$ 3\arcsec\ region of the {\it HST}/STIS image
of the host galaxy of GRB 970228 ]{A 3\arcsec $\times$ 3\arcsec\ (23
$\times$ 23 kpc$^2$ in projection) region of the {\it HST}/STIS image
(1997 September 4.7 UT) of the host galaxy of GRB 970228.  The image
has been smoothed (see text) and is centered on the optical transient.
North is up and east is left.  Contours in units of 3,4,5,6,7, and 8
background $\sigma$ ($\sigma = 2.41$ DN) are overlaid.  The transient
is found on the outskirts of detectable emission from a faint, low
surface brightness galaxy.  The morphology is clearly not that of a
classical Hubble type, though there appears to be a nucleus and an
extended structure to the north of the transient.}
\label{fig:host}
\end{figure*}

For the purpose of determining the position of the GRB within its
host, we examined the {\it HST}/STIS observations taken on 1997 Sept
4.7 UT \citep{fpt+99}.  The observation consisted of eight 575$\,$s
STIS clear (CCD50) exposures paired into four 1150 s images to
facilitate removal of cosmic rays.  We processed these images using
the drizzle technique of \citet{fh97} to create a final image with a
plate scale of 0.0254 arcsec pixel$^{-1}$.  To enhance the low surface
brightness host galaxy, we smoothed this image with a Gaussian of
$\sigma = 0.043$ arcsec. The optical transient is well detected in
figure \ref{fig:host} (point source toward the south) and is clearly
offset from the bulk of the detectable emission of the host.

Two morphological features of the host stand out: a bright knot
manifested as a sharp 6 $\sigma$ peak near the centroid of the host
pointing north of the transient and an extension from this knot toward
the transient. Although, as we demonstrate below, this host is a
sub-luminous galaxy (i.e.,~not a classic late-type $L_*$ spiral
galaxy) and we attribute these features to a nucleus and a spiral-arm,
respectively. It is not unusual for dwarf galaxies to exhibit these
canonical Hubble-diagram structures (S.~Odewahn, private
communication).

Centroiding the transient and the nucleus components within a 3 pixel
aperture radius about their respective peak, we find an angular offset
of $436 \pm 14$ milliarcsec between the nucleus and the optical
transient.  With our assumed cosmology, this amounts to a projected
physical distance of $3.34\, \pm\, 0.11\, h_{65}^{-1}$ kpc.

\subsection{Physical parameters of the presumed host galaxy}
\label{sec:0228host}

We found the half-light radius of the host galaxy using our final
drizzled {\it HST}/STIS image: we mask a 3 $\times$ 3 pixel region
around the position of the optical transient and inspect the curve of
growth centered on the central bright knot, the supposed nucleus, and
we estimate the half-light radius to be 0.31 arcsec or
\hbox{2.4 $h_{65}^{-1}$ kpc} (physical) at a redshift of
$z=0.695$. The half-light radii of the host galaxy in the {\it
HST}/WFPC F814W ($I$-band) and F606W ($V$-band) filters estimated by
eye from the curve--of--growth plots of \citet{cl99c} are comparable
to the STIS-derived half-light radius.

Although there is some debate (at the 0.3 mag level in $A_V$) as to
the proper level of Galactic extinction toward GRB 970228
\citep{cl99d,gfd99,fpt+99}, we have chosen to adopt the value $E(B-V) =
0.234$ found from the dust maps of \citet{sfd98} and a Galactic
reddening curve $R_V = A_V/E(B-V) = 3.2$.  Using extensive reanalysis
of the {\it HST} imaging data by \citet{gtv+00}, the
extinction-corrected broadband colors of the host galaxy are $V = 25.0
\pm 0.2$ mag, $R_c = 24.6 \pm 0.2$ mag, $I_c = 24.2 \pm 0.2$ mag. 
These measurements, consistent with those of \citet{cl99c} and
\citet{fpt+99}, are derived from the WFPC2 colors and broadband STIS
flux. The errors reflect both the statistical error and the
uncertainty in the spectral energy distribution of the host galaxy.
We have not included a contribution from the uncertainty in the
Galactic extinction. Using the NICMOS measurement from \citet{fpt+99},
the extinction-corrected infrared magnitude is $H_{\rm AB} = 24.6 \pm
0.1$.  Using the zero-points from
\citet{fig+96}, the extinction-corrected AB magnitudes of the host
galaxy are $V_{\rm AB} = 25.0$, $R_{\rm AB} = 24.8$, $I_{\rm AB} =
24.7$.

To facilitate comparison with moderate-redshift galaxy surveys (\S
\ref{sec:host}), we compute the rest-frame $B$-band magnitude of the
host galaxy.  From the observed continuum in the rest-frame $B$ band,
we derive the absolute magnitude $M_B = -18.4 \pm 0.4$ mag [or $M_{\rm
{AB}}(B) =$ \hbox{$-18.6 \pm 0.4$ mag}], i.e., only slightly brighter
than the Large Magellanic Cloud (LMC) now.  For our chosen value of
$H_0$, an $L_*$ galaxy at $z \sim 0$ has $M_B \approx -20.9$ mag, and
thus the host at the observed epoch has $L \sim 0.1~L_*$ today.

\subsection{Star formation in the host}

From the [O II] 3727 line flux, we derive the line luminosity
$L_{3727} = 5.44 \times 10^{40}$ erg s$^{-1}$ ($\pm 5$\% random; $\pm
30$\% systematic).  Using the star formation rate (SFR) estimator from
\citet{ken98}, we derive that SFR $\approx 0.76 ~M_\odot$ yr$^{-1}$.
Using a 3 $\sigma$ limit on the H$\beta$ flux, we estimate $L_{H\beta}
< 2.5 \times 10^{40}$ erg s$^{-1}$.  Assuming the H$\alpha$/H$\beta$
ratio of $2.85 \pm 0.2$ for the Case B recombination and a range of
excitation temperatures, we can derive a pseudo H$\alpha$-based
estimate of the star formation rate \citep[see][]{ken98}, SFR $<
0.6~M_\odot$ yr$^{-1}$, but we consider this to be less reliable than
the [O II] 3727 measurement.  From the UV-continuum luminosity at
$\lambda_{\rm rest} = 2800$~\AA, following \citet{mpd98}, we derive SFR
$\approx 0.54 ~M_\odot$ yr$^{-1}$.
 
We note that the net uncertainties for each of these independent SFR
estimates are at least 50\%, and the overall agreement is encouraging.
While we do not know the effective extinction corrections in the host
galaxy itself, these are likely to be modest given its blue colors
(see~\S \ref{sec:host}) and are unlikely to change our results by more
than a factor of two.  (We hasten to point out that we are completely
insensitive to any fully obscured star formation component, if any is
present.)  On the whole, the galaxy appears to have a rather modest
(unobscured) star formation rate, $\sim 0.5 - 1~M_\odot$ yr$^{-1}$.
Given the relatively normal equivalent width of the [O II] 3727 line,
even the star formation per unit mass does not seem to be
extraordinarily high.
 
\section{The Nature of the Host Galaxy}
\label{sec:host}

At $M_{\rm AB}(B) = -18.6$ mag, the presumed host galaxy of GRB 970228 is
a sub-luminous galaxy, roughly 2.7 mag below $L_*$ at comparable
redshifts \citep{lth+95}. Based on the redshift of GRB 970228,
\citet{gtv+00} recently found that an Sc galaxy spectral energy
distribution reasonably fits the optical-IR photometric fluxes of the
host galaxy.  This differs from the analysis of \citet{cl99c} who
modeled the predicted galaxy colors as a function of the (then
unknown) redshift and morphological classification.  Now, given the
redshift of $z = 0.695$, the
\citet{cl99c} analysis would tend to favor classification as an
``Irregular'' galaxy, having undergone a burst of star formation over
the past few hundred Myr.  Clearly it is difficult to precisely
determine the galaxy type without more precise photometry and
knowledge of the true Galactic extinction, but our identification of a
nucleus and possible arm structure (\S \ref{sec:off}) supports the
idea that the host is a late-type dwarf. Indeed, the host has similar
characteristics to that of the LMC.  We further note that compared to
the results of \citet{skf+99}, the magnitude-size relation of the host
galaxy is consistent with that observed for late-type and
dwarf-irregular galaxies.

\begin{figure*}[tbp]
\centerline{\psfig{file=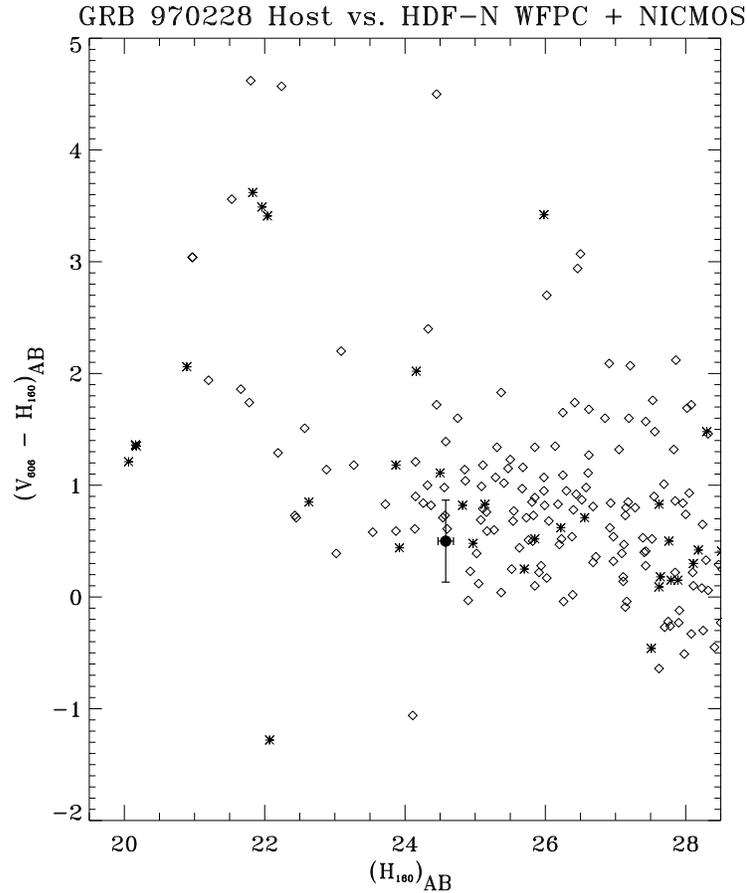,width=4.2in}}
\caption[Comparison of the color-magnitude of the host galaxy of GRB
970228 with the Hubble Deep Field-North]{Comparison of the
color-magnitude of the host galaxy of GRB 970228 with the Hubble Deep
Field-North (HDF-N) showing that the host galaxy of GRB 970228 is not
especially blue in color.  No systematic difference, assuming a
Galactic extinction toward GRB 970228 of $A_V = 0.75$ mag, is found
between field galaxies at comparable magnitudes and the host ({\it
filled circle with error bars}).  NICMOS and WFPC photometry are taken
from \citet{tsw+99} and \citet{wbd+96}, respectively.  Host galaxy
magnitudes and magnitudes of the HDF comparison field objects are
aperture-corrected.  Diamonds ($\diamond$) represent extended objects
(with the ratio of semi-major to semi-minor axes less than 0.9) and
the asterisks ($\ast$) compact galaxies and stars. The error bars on
the HDF-N data have been suppressed. See text for an explanation of
the selection criteria.}
\label{fig:hdf}
\end{figure*}

The flat continuum suggests a blue continuum of the host galaxy and
little rest-frame extinction.  How does the host galaxy color compare
to other galaxies at comparable magnitudes?  We found the $(H -
V)_{\rm AB}$ color of galaxies in the Hubble Deep Field-North (HDF-N)
using the published photometry from \citet{tsw+99} (NICMOS F160W
filter) and
\citet{wbd+96} (WFPC2 F606W filter).  All WFPC object identifications
within 0.3 arcsec of a NICMOS identification are plotted in figure
\ref{fig:hdf} along with the dereddened color of the host galaxy of
GRB 970228.  This selection from the HDF chooses only those faint
objects with detectable IR emission; galaxies with detected visual
emission in the WFPC filter but no IR emission in the NICMOS filter do
not make it in to our comparison sample.  To be clear, this selection
essentially biases the color--magnitude relation toward more {\it
red} objects and would serve to accentuate the locus of the comparison
field with a blue galaxy.  Even with this bias, there is no indication
in figure \ref{fig:hdf} that the host galaxy of GRB 970228 is
substantially more blue than other field galaxies at comparable
magnitudes.

This conclusion---that the host galaxy of GRB 970228 is not
exceptionally blue---is at odds with that of \citet{fpt+99} who have
claimed that the host galaxy is unusually blue as compared with
typical field galaxies.  The difference may be due to the fact that
the \citet{fpt+99} analysis compared the host colors with a
significantly more shallow infrared survey than the NICMOS HDF,
essentially masking the trend for faint galaxies to appear more blue.
\begin{figure*}[thp]
\centerline{\psfig{file=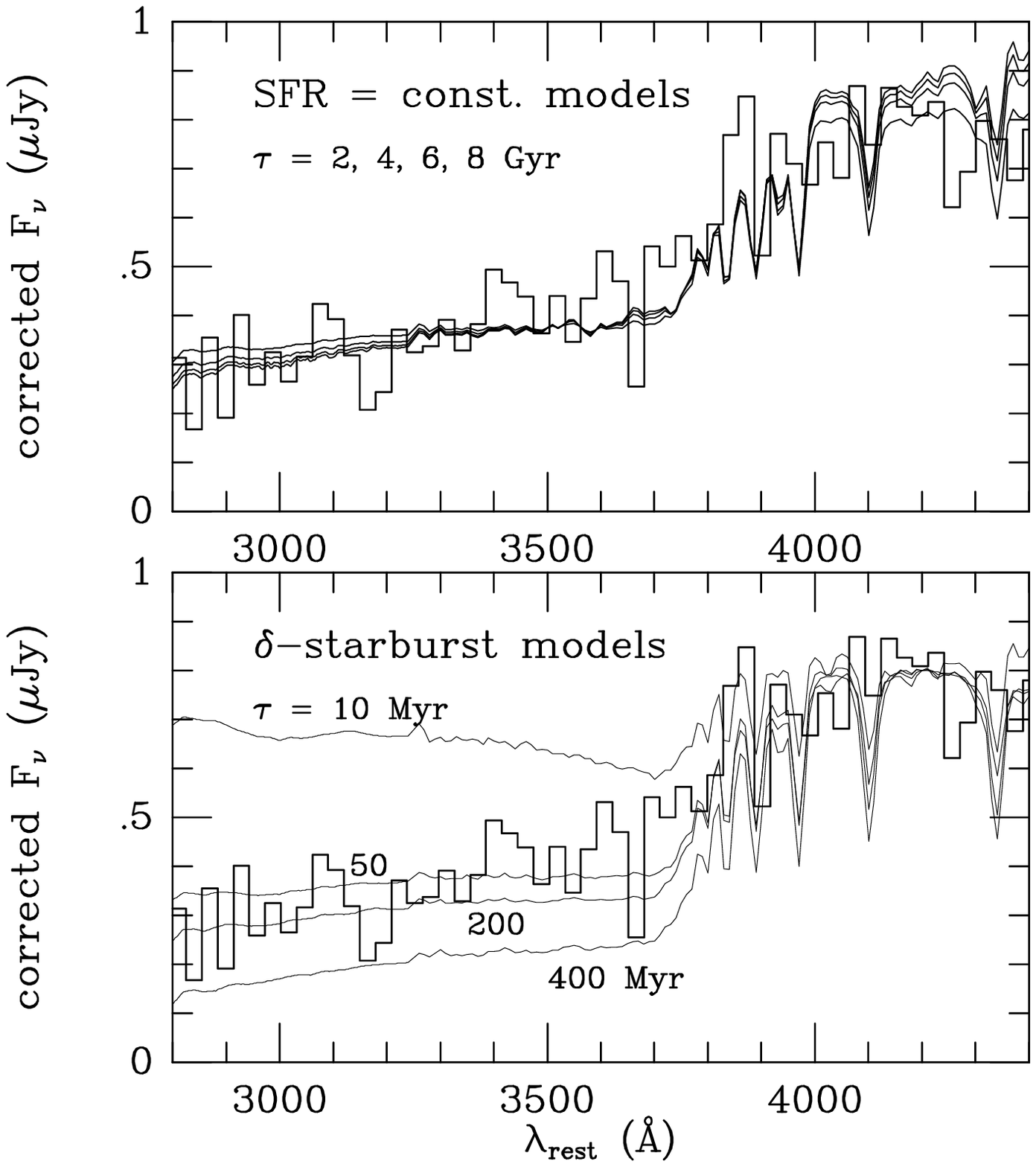,width=4.2in}}
\caption[Median-binned portion of the host spectrum near the Balmer
decrement]{Median-binned portion of the host spectrum near the Balmer
decrement. {\it Top panel}: Overlaid are Bruzual \& Charlot
(1993)\nocite{bc93} galaxy synthesis models assuming a varying time of
constant star formation. {\it Bottom panel}: Overlaid are \citet{bc93}
galaxy synthesis models assuming an instantaneous burst of
star-formation occurred $\tau$ years since observation. Clearly the
host continuum could not be dominated by a young population of stars
($\tau = 10$ Myr).  See text for a discussion.  }
\label{fig:0228balmer}
\end{figure*}

Figure \ref{fig:0228balmer} shows a section of the median-binned
spectrum of the host galaxy.  The Balmer break is clearly detected,
with an amplitude $\Delta m \approx 0.55$ mag, which is typical for
the Balmer-break-selected population of field galaxies at $z \sim 1$
(K.~Adelberger, private communication).  For reference we also plot
several population-synthesis model spectra \citep{bc93}. The top panel
shows model spectra for a galaxy with a uniform star formation rate,
which may be a reasonable time-averaged approximation for a normal
late-type galaxy.  The correspondence is reasonably good and does not
depend on the model age. The bottom panel shows models with an
instantaneous burst of star formation.  In order to match the data, we
require fine-tuning of the postburst age to be $\sim 10^{8} \times
2^{\pm 1}$ yr. No attempt was made to optimize the fit or to seek best
model parameters and the purpose of this comparison is simply
illustrative.  Clearly, if there was an ongoing or very recent burst
of star formation, the spectrum would be much flatter and have a
weaker Balmer break.
 
\section{Discussion and Conclusion}

We have determined the redshift of the presumed host galaxy of GRB
970228 to be $z=0.695$ based on [O II] 3727 and [O III] 5007 line
emission.  The implied energy release, \hbox{$(1.4 \pm 0.3) \times
10^{52}$ erg} [20---2000 keV restframe], is on the smaller end of, but
still comparable to, the other bursts with energy determinations
\citep{bfs01}.  The absence of a detectable break in the afterglow
light curve implies that any collimation of emission (i.e.,~jetting)
will not significantly reduce the estimate of total energy release in
GRB 970228 (although \citeauthor*{fwk00}
\citeyear{fwk00}, using late-time radio data, have found that even
without an optical break, GRB 970508 may have been collimated).

Most GRB transients appear spatially coincident with faint galaxies,
disfavoring the merging neutron star hypothesis
\citepeg{pac98b,bkd02}.  The coincidence of GRB 970508 with its host
\citep{fp98,bdkf98} is particularly constraining given the excellent
spatial coincidence of the GRB with the center of a dwarf galaxy
\citepcf{bdkf98}.  The transient of GRB 970228 lies $3.34 \pm 0.11\,
h_{65}^{-1}$ kpc from the center of the galaxy, about 1 kpc in
projection outside the half-light radius of the galaxy.  From the
above analysis we have shown that, like GRB 970508, the host is
sub-luminous ($L \approx 0.05 L_*$) and, by assumption, sub-massive
relative to $L_*$ galaxies at comparable redshifts.  According to
\citet{bsp99}, about 50\% of merging neutron binaries should occur
beyond 3.5 kpc in projection of such dwarf galaxies.  Thus, by itself,
the offset of GRB 970228 from its host does not particularly favor a
progenitor model.

In $R$-band magnitude the host is near the median of GRB hosts
observed to date, but in absolute $B$-band magnitude, the host at the
faint end of the distribution. Of the host galaxies detected thus far
only GRB 970508 is as comparably faint to the host of GRB
970228. Except in angular extent, the host galaxies of GRB 970228 and
GRB 970508 \citep{bdkf98} bear a striking resemblance.  Both appear to
be sub-luminous ($L \ale 0.1 L_*$), compact, and blue.  Spectroscopy of
both reveal the presence of the [Ne III] 3869 line, indicative of
recent very massive star formation. However, such properties are not
shared by all of the GRB hosts studied to date. We note too the rather
curious trend that the two GRBs themselves appear to have similar
properties in that they decay slowly, they are the two least luminous
in term of GRB energetics, and do not exhibit evidence of a strong
break in the light curve.

\acknowledgments

The authors thank the generous support of the staff of the W.~M.~Keck
Foundation.  This paper has benefited from stimulating conversations
with P.~van Dokkum and K.~Adelberger. We thank M.~van Kerkwijk for
help during observing and C.~Clemens for his use of dark-time
observing nights.  This paper has been greatly improved by the
thoughtful comments of the anonymous referee. JSB gratefully
acknowledges the fellowship from the Fannie and John Hertz
Foundation. SGD acknowledges partial funding from the Bressler
Foundation. This work was supported in part by grants from the NSF and
NASA to SRK.

\chapter[The Observed Offset Distribution of Gamma-Ray Bursts from Their
       Host Galaxies: A Robust Clue to the Nature of the
       Progenitors]{The Observed Offset Distribution of Gamma-Ray Bursts from Their
       Host Galaxies: A Robust Clue to the Nature of the
       Progenitors$^\dag$}
\label{chap:offset}
\setlength{\tabcolsep}{0.04in}
\secfootnote{\secfootdag}{A version of this chapter was first published 
in {\it The Astronomical Journal}, 123, p.~1111--11148 (2002).  Some
of the text in the published version of the introduction was used in
chapter \ref{chap:intro}; the introduction here has thus been
abbreviated.}

\secauthor{J.~S.~Bloom, S.~R.~Kulkarni, S.~G.~Djorgovski}

\secaffils{Palomar Observatory 105--24, California Institute of
       Technology, Pasadena, CA 91125, USA}

\begin{abstract}

We present a comprehensive study to measure the locations of
$\gamma$-ray bursts (GRBs) relative to their host galaxies.  In total,
we find the offsets of 20 long-duration GRBs from their apparent host
galaxy centers utilizing ground-based images from Palomar and Keck and
space-based images from the Hubble Space Telescope (HST).  We discuss
in detail how a host galaxy is assigned to an individual GRB and the
robustness of the assignment process. The median projected angular
(physical) offset is 0.17 arcsec (1.3 kpc).  The median offset
normalized by the individual host half-light radii is 0.98 suggesting
a strong connection of GRB locations with the UV light of their hosts.
This provides strong observational evidence for the connection of GRBs
to star-formation.

We further compare the observed offset distribution with the predicted
burst locations of leading stellar-mass progenitor models. In
particular, we compare the observed offset distribution with an
exponential disk, a model for the location of collapsars and promptly
bursting binaries (e.g.,~helium star--black hole binaries). The
statistical comparison shows good agreement given the simplicity of
the model, with the Kolmogorov-Smirnov probability that the observed
offsets derive from the model distribution of $P_{\rm KS} = 0.45$.  We
also compare the observed GRB offsets with the expected offset
distribution of delayed merging remnant progenitors (black
hole--neutron star and neutron star--neutron star binaries). We find
that delayed merging remnant progenitors, insofar as the predicted
offset distributions from population synthesis studies are
representative, can be ruled out at the $2 \times 10^{-3}$ level. This
is arguably the strongest observational constraint yet against delayed
merging remnants as the progenitors of long-duration GRBs.  In the
course of this study, we have also discovered the putative host
galaxies of GRB 990510 and GRB 990308 in archival HST data.
\end{abstract}

\section{Introduction}

Direct associations with other known astrophysical entities is a
possible means toward distinguishing between GRB progenitor models.
For massive stars, the energy release from the collapse of the core of
the star, just as in supernovae, is sufficient to explode the star
itself.  This may result in a supernova-like explosion at essentially
the same time as a GRB.  The first apparent evidence of such a
supernova associated with a cosmological GRB came with the discovery
of a delayed bright red bump in the afterglow light curve of GRB
980326 \citep{bkd+99}.  The authors interpreted the phenomena as due
to the light-curve peak of a supernova at redshift $z \sim 1$. Later,
\citet{rei99} and \citet{gtv+00} found similar such red bump in the 
afterglow of GRB 970228.  Merging remnant progenitors models (e.g.,
BH--NS, NS--NS systems) have difficulty producing these features in a
light curve on such long timescales and so the supernova
interpretation, if true, would be one of the strongest direct clues
that GRBs come from massive star explosions. However, the supernova
story is by no means complete.  For instance, in only one other GRB
(000911) has marginal ($\sim$2 $\sigma$) evidence of a SN signature
been found \citep{lcg+01}; further, many GRBs do not appear to show
any evidence of SNe signatures \citepeg{hhc+00}. Even the
``supernova'' observations themselves find plausible alternative
explanations (such as dust echoes) that do not strictly require a
massive star explosion \citep{eb00,rei01,wd00}. We note, however, that
all other plausible explanations of the observed late-time bumps
require high-density environments found most readily in star forming
regions.

\citet{cl99b} emphasize that if a GRB comes from a massive star, then
the explosion does not take place in a constant density medium, but in
a medium enriched by constant mass loss from the stellar winds.  One
would expect to see signatures of this wind-stratified medium in the
afterglow \citep[e.g., bright sub-millimeter emission at early times,
increasing ``cooling frequency'' with time; see][]{pk00b,kbb+00}.
However, afterglow observations have been inconclusive \citep{kbb+00}
with no unambiguous inference of GRB in such a medium to date.

We emphasize that even the connection of GRBs to stellar-mass
progenitors has yet to be established.  The most compelling arguments
we have outlined (e.g.,~temporal variability) rely on theoretical
interpretations of the GRB phenomena.  Further, direct observational
results (SNe signatures and transient Fe-line emission) are not yet
conclusive.

In this paper we examine the observed locations of GRBs with respect
to galaxies.  We find an unambiguous correlation of GRB locations with
the UV light of their hosts, providing strong indirect evidence for
the connection of GRBs to stellar-mass progenitors. Beyond this
finding, we aim to use the location of GRBs to distinguish between
stellar-mass progenitor models. In \S \ref{sec:location} we review the
expectations of GRB locations from each progenitor model.  Then in
\S\S \ref{sec:offdata}--\ref{sec:astlevels} we discuss the
instruments, techniques, and expected uncertainties involved in
constructing a sample of GRB locations about their host galaxies.  In
\S \ref{sec:indivoff} we comment on the data reductions specific to
each GRB in our sample.  The observed distribution is shown and
discussed in \S \ref{sec:offdist} and then statistically compared with
the expected offset distribution of leading progenitor models (\S
\ref{sec:compare}). Last, in \S \ref{sec:offsum} we summarize and
discuss our findings.

\section{Location of GRBs as a Clue to Their Origin}
\label{sec:location}

How have locations of GRBs within (or outside) galaxies impacted our
understanding of the progenitors of GRBs thus far?  The first accurate
localization \citep{vgg+97} of a GRB by way of an optical transient
afterglow revealed GRB 970228 to be spatially coincident with a faint
galaxy \citep{slp+97,fpt+99,bdk01}.  Though the nearby galaxy was
faint, \citet{vgg+97} estimated the {\it a posteriori} probability of
a random location on the sky falling so close to a galaxy by chance to
be low.  As such, the galaxy was identified as the host of GRB 970228.
\citet{slp+97} further noted that the OT appeared offset from the
center of the galaxy thereby calling into question an AGN origin.
Soon thereafter \citet{bdkf98} found, and then \citet{fp98} confirmed,
that GRB 970508 was localized very near the center of a dwarf
galaxy. Given that under-luminous dwarf galaxies have a weaker
gravitational potential with which to bind merging remnant binaries,
both \citet{pac98b} and \citet{bdkf98} noted that the excellent
spatial coincidence of the GRB with its putative host found an easier
explanation with a massive star progenitor rather than NS--NS
binaries.

Once the afterglow fades, one could study in detail its environment
(analogous to low-redshift supernovae). Unfortunately, however, the
current instrumentation available for GRB observations cannot pinpoint
or resolve individual GRB environments on the scale of tens of parsecs
unless the GRB occurs at a low redshift ($z \ale 0.2$) and the
transient afterglow is well-localized.  At higher redshifts (as all
GRBs localized to-date), only the very largest scales of galactic
structure can be resolved (e.g.,~spiral arms) even by HST.  Therefore,
the locations of most individual GRBs do not yield much insight into
the nature of the progenitors. Instead, the observed {\it
distribution} of GRBs in and around galaxies must be studied as a
whole and then compared with the expectations of the various
progenitor models.  This is the aim of the present study. As we will
demonstrate, while not all GRBs are well-localized, the overall
distribution of GRB offsets proves to be a robust clue to the nature
of the progenitors.

In this paper we present a sample of GRB offset measurements that
represents the most comprehensive and uniform set compiled
to-date. Every GRB location and host galaxy image has been re-analyzed
using the most uniform data available.  The compilation is complete
with well-studied GRBs until May 2000. Throughout this paper we assume
a flat $\Lambda$--cosmology \citepeg{dab+00} with $H_0 = 65$ km
s$^{-1}$ Mpc$^{-1}$, $\Omega_M = 0.3$, and $\Lambda_0 = 0.7$.

\section{The Data: Selection and Reduction}
\label{sec:offdata}

The primary goal of this paper is to measure the offsets of GRBs from
their hosts where the necessary data are available.  Ideally this
could be accomplished using a dataset of early-time afterglow and
late-time host imaging observed using the same instrument under
similar observing conditions.  The natural instrument of choice is HST
given its exquisite angular resolution and astrometric stability.
Though while most hosts have been observed with HST at late-times,
there are only a handful of early-time HST detections of GRB
afterglow.  On the other hand, early ground-based images of GRB
afterglows are copious but late-time seeing-limited images of the
hosts give an incomplete view of the host as compared to an HST image
of the same field.  Moreover, ground-based imaging is inherently
heterogeneous, taken with different instruments, at different
signal-to-noise levels, and through a variety observing  of conditions;
this generally leads to poorer astrometric accuracy.  Bearing these
imperfections in mind we have compiled a dataset of images that we
believe are best suited to find offsets of GRBs from their hosts.

\def\me{a}
\def\vgg{1}
\def\fpt{2}
\def\fkn{3}
\def\pfb{4}
\def\hp{5}
\def\ggv{6}
\def\db{7}
\def\kdr{8}
\def\odk{9}
\def\ggvc{10}
\def\bkde{11}
\def\fvn{12}
\def\lgk{13}
\def\hth{14}
\def\gvv{15}
\def\hft{16}
\def\jha{17}
\def\bdg{18}
\def\bkdg{19}
\def\hfta{20}
\def\hap{21}
\def\hthf{22}
\def\bfk{23}
\def\bk{24}
\def\fkb{25}
\def\hta{26}
\def\bod{27}
\def\ssh{28}
\def\hftb{29}
\def\tbfk{30}
\def\htab{31}
\def\vgr{32}
\def\ffp{33}
\def\bkul{34}
\def\mpp{35}
\def\hah{36}
\def\sahu{37}
\def\fsg{38}
\def\fvh{39}
\def\fra{40}
\def\bdk{41}
\def\fvsc{42}
\def\umh{43}
\def\vff{44}
\def\fjh{45}
\def\fsgb{46}
\def\fv{47}
\def\mhw{48}
\def\bdgb{49}
\def\mfm{50}

\begin{deluxetable}{llrrrrrrll}
\rotate
\tabletypesize{\footnotesize}
\tablewidth{0pt}
\tablecaption{GRB Host and Astrometry Observing Log\label{tab:offset-log}}
\tablehead{
\colhead{Name} &
\colhead{Teles./Instr./Filter} &
\colhead{Date} &
\colhead{$\alpha$ (J2000)} &
\colhead{$\delta$ (J2000)} &
\colhead{Exp.} &
\colhead{$\Delta t$} &
\colhead{Level} &
\colhead{Refs.} \\
\colhead{(1)} &
\colhead{(2)} &
\colhead{(3)} &
\multicolumn{2}{c}{(4)} &
\colhead{(5)} &
\colhead{(6)} &
\colhead{(7)} &
\colhead{(8)}}
\startdata

GRB 970228\ldots &HST/STIS/{\tt O49001040} & 4.75 Sep 1997 & 05 01 46.7 & +11 46 54  & 4600  & 189 & self-HST              & \vgg, \fpt \\

GRB 970508\ldots &HST/STIS/{\tt O41C01DIM} & 2.64 Jun 1997 & 06 53 49.5 & +79 16 20  & 5000  & 25  & HST$\rightarrow$HST   & \fkn, \pfb \\*
                 &HST/STIS/{\tt O4XB01I9Q} & 6.01 Aug 1997 &             &           & 11568 & 89  &                       & \hp \\

GRB 970828\ldots & Keck/LRIS/$R$-band    & 19.4 Jul 1998     & 18 08 34.2 & +59 18 52  & 600   & 325 & RADIO$\rightarrow$OPT  & \ggv, \db \\

GRB 971214\ldots & Keck/LRIS/$I$-band    & 16.52  Dec 1997   & 11 56 26.0 & +65 12 00  & 1080  & 1.5 & GB$\rightarrow$HST  & \kdr \\*
                 & HST/STIS/{\tt O4T301040}& 13.27 Apr 1998&            &            & 11862 & 119 &                       & \odk \\

GRB 980326\ldots & Keck/LRIS/$R$-band    & 28.25 Mar 1998    & 08 36 34.3 & $-$18 51 24& 240   & 1.4 & GB$\rightarrow$GB &\ggvc,\bkde \\*
                 & Keck/LRIS/$R$-band    & 18.50 Dec 1998    &            &            & 2400  & 267 &                       & \bkde \\
                 & HST/STIS/{\tt O59251ZWQ}& 31.80 Dec 2000 &            &            & 7080 & 1010 &                       & \fvn \\

GRB 980329\ldots & Keck/NIRC/$K$-band    & 2.31 Apr 1998    & 07 02  38.0 &  +38 50 44 & 2520  & 4.15 & GB$\rightarrow$GB  & \lgk,  \me \\
                 & Keck/ESI/$R$-band     & 1.41 Jan 2001    &  &  & 6600  & 1009 &   & \me \\
                 & HST/STIS/{\tt O65K22YXQ}& 27.03 Aug 2000 &            &            & 8012 & 884 &                       & \hth \\

SN 1998bw  \ldots &NTT/EMMI/$I$-band       & 4.41 May 1998   & 19 35 03.3  &$-$52 50 45& 120   & 8.5 & GB$\rightarrow$HST   & \gvv \\*
\phantom{~~~}(GRB 980425?)     &HST/STIS/{\tt O65K30B1Q} & 11.98 Jun 2000  &          &           & 1185  & 778 &                       & \hft \\

GRB 980519\ldots & P200/COSMIC/$R$-band   & 20.48 May 1998   & 23 22 21.5  &  +77 15 43& 480   & 1.0 &GB$\rha$GB$\rha$HST & \jha, \bdg \\*
                 & Keck/LRIS/$R$-band     & 24.50 Aug 1998   &             &           & 2100  & 97  &                    & \bkdg \\*
                 & HST/STIS/{\tt O65K41IEQ} & 7.24 Jun 2000&             &           & 8924  & 750 &                       & \hfta \\*
GRB 980613\ldots  & Keck/LRIS/$R$-band     & 16.29 Jun 1998   & 10 17 57.6  & +71 27 26 & 600   & 3.1 &GB$\rightarrow$GB$\rha$HST  & \hap, \me  \\*
                  & Keck/LRIS/$R$-band     & 29.62 Nov 1998  &             &           & 900   & 169 &                       & \me  \\*
& HST/STIS/{\tt O65K51ZZQ} & 20.31 Aug 2000 &             &           & 5851  & 799 &                       & \hthf \\*
GRB 980703\ldots  & Keck/LRIS/$R$-band     & 6.61 Jul 1998    & 23 59 06.7  & +08 35 07 & 600   & 3.4 &GB$\rightarrow$HST   & \bfk \\*
            & HST/STIS/{\tt O65K61XTQ} & 18.81 Jun 2000  &               &           & 5118  & 717 &                       & \bk \\*
GRB 981226\ldots  & Keck/LRIS/$R$-band     & 21.57 Jun 1999   & 23 29 37.2 & $-$23 55 54 & 3360 & 177 &RADIO$\rha$OPT(GB)$\rha$HST  & \fkb  \\*
            & HST/STIS/{\tt O65K71AXQ} & 3.56  Jul 2000  &              &             & 8265 & 555 &                      & \hta  \\*
GRB 990123\ldots  & HST/STIS/{\tt O59601060} & 9.12 Feb 1999  & 15 25 30.3 & +44 45 59 & 7200 & 16.7 & self-HST & \bod \\*
GRB 990308\ldots  & Keck/LRIS/$R$-band       & 19.26 Jun 1999 & 12 23 11.4 & +06 44 05& 1000& 103 &  GB$\rha$GB$\rha$HST & \ssh  \\*
                  & HST/STIS/{\tt O65K91E6Q} & 19.67 Jun 2000  &            &          & 7782& 470 &                      & \hftb \\
GRB 990506\ldots  & Keck/LRIS/$R$-band       & 11.25 Jun 1999  & 11 54 50.1 & $-$26 40 35& 1560& 36 & RADIO$\rha$OPT(GB)$\rha$HST  & \tbfk \\*
                  & HST/STIS/{\tt O65KA1UYQ} &24.55 Jun 2000 &            &            & 7856& 415  &                       & \htab \\

GRB 990510\ldots  & HST/STIS/{\tt O59273LCQ} & 17.95 Jun 1999 & 13 38 07.7& $-$80 29 49& 7440& 39   & HST$\rightarrow$HST  & \vgr, \ffp \\* 
                  & HST/STIS/{\tt O59276C7Q} & 29.45 Apr 2000 &           &            & 5840& 355  &                      & \bkul \\

GRB 990705\ldots  & NTT/SOFI/$H$-band          & 5.90  Jul  1999 & 05 09 54.5& $-$72 07 53& 1200& 0.23 & GB$\rha$GB$\rha$HST   & \mpp \\*
                  & VLT/FORS1/$V$-band     & 10.40 Jul  1999 &           &            & 1800& 4.7  &                            & \mpp \\*
                  & HST/STIS/{\tt O65KB1G2Q} & 26.06 Jul  2000 &           &            & 8792& 386  &                       & \hah \\

GRB 990712\ldots  & HST/STIS/{\tt O59262VEQ} & 29.50 Aug 1999  & 22 31 53.1& $-$73 24 28& 8160 & 48 & HST$\rha$HST & \sahu, \fsg \\*
                   & HST/STIS/{\tt O59274BNQ} & 24.21 Apr 2000  &           &            & 3720 & 287 &            & \fvh \\

GRB 991208\ldots  & Keck/NIRSPEC/$K$-band  & 16.68 Dec 1999 & 16 33 53.5  & +46 27 21 & 1560   & 8.5 & GB$\rha$GB$\rha$HST & \fra, \bdk \\*
                  & Keck/ESI/$R$-band      & 4.54 Apr 2000  &             &             & 1260 & 118 &                & \me \\* 
         & HST/STIS/{\tt O59266ODQ} & 3.58 Aug 2000  &           &            & 5120 & 239 &            & \fvsc \\

GRB 991216\ldots  & Keck/ESI/$R$-band        & 29.41 Dec 1999  & 05 09 31.2  & +11 17 07   & 600 &13&  GB$\rha$GB$\rha$HST  & \umh, \me \\*
	          & Keck/ESI/$R$-band         & 4.23 Apr 2000  &             &             & 2600 & 110 &                    & \me \\*
	          & HST/STIS/{\tt O59272GIQ} & 17.71 Apr 2000&             &             & 9440 & 123 &                    & \vff \\

GRB 000301C\ldots & HST/STIS/{\tt O59277P9Q} & 6.22 Mar 2000 & 16 20 18.6 & +29 26 36 & 1440 & 4.8 & HST$\rha$HST & \fjh, \fsgb \\
                  & HST/STIS/{\tt O59265XYQ} & 25.86 Feb 2001 &            &
        & 7361 &  361  &  & \fv \\

GRB 000418\ldots  & Keck/ESI/$R$-band & 28.41 Apr 2000 & 12 25 19.3 & +20 06 11   &  300 & 10 & GB$\rha$HST  & \mhw, \bdgb \\*
                  & HST/STIS/{\tt O59264Y6Q}           & 4.23  Jun 2000  &            &    & 2500 & 47 &                 & \mfm  

\enddata \tablecomments{(2) Telescopes: HST = {\it Hubble Space
Telescope} Keck = W.~M.~Keck 10 m Telescope II, Mauna Kea, Hawaii,
P200 = Hale 200-inch Telescope at Palomar Observatory, Palomar
Mountain, California, NTT = European Space Agency 3.5 m New Technology
Telescope, Chile, VLT = Very Large Telescope UT-1 (``Antu'');
Instruments: STIS \citep{kwb+98}, ESI \citep{em98}, LRIS
\citep{occ+95}, COSMIC \citep{kds+98}, NIRSPEC \citep{mbb+98}, SOFI
\citep{fbm+98}, FORS1 \citep{nsb+97}; Filter: all ground-based
observations are listed in standard bandpass filters while the
HST/STIS images (used for astrometry) are all in Clear Mode.  The last
dataset of the HST visit is listed. (3) Observation dates in Universal
Time (UT) corresponding to the start time of the last observation in the
dataset. (4) Position ($\alpha$: hours, minutes, seconds and $\delta$:
degrees, arcminutes, and arcseconds) of the GRB. (5) Total exposure
time in seconds. (6) Time in days since the trigger time of the
GRB. (7) The comment denotes the astrometric level as in \S
\ref{sec:astlevels}. (8) Reference to the first presentation of the
given dataset. If two references appear on a given line then the first
is a reference to the position of the GRB.}

\tablerefs{\me.~This paper; \vgg.~\citet{vgg+97};
\fpt.~\citet{fpt+99}; \fkn.~\citet{fkn+97}; \pfb.~\citet{pfb+98a};
\hp.~\citet{fp98}; \ggv.~\citet{ggv+98d}; \db.~\citet{dfk+01};
\kdr.~\citet{kdr+98}; \odk.~\citet{odk+98}; \ggvc.~\citet{ggv+98c};
\bkde.~\citet{bkd+99}; \fvn.~\citet{fvn01}; \lgk.~\citet{lgk+98};
\hth.~\citet{hth+00b}; \gvv.~\citet{gvv+98}; \hft.~\citet{hft+00};
\jha.~\citet{jha+98}; \bdg.~\citet{bdg+98}; \bkdg.~\citet{bkdg+98};
\hfta.~\citet{hfta+00}; \hap.~\citet{hap+98}; \hthf.~\citet{hth+00};
\bfk.~\citet{bfk+98}; \bk.~\citet{bk+00}; \fkb.~\citet{fkb+99};
\hta.~\citet{hta+00}; \bod.~\citet{bod+99}; \ssh.~\citet{ssh+99};
\hftb.~\citet{hft+00b}; \tbfk.~\citet{tbf+00}; \htab.~\citet{htab+00};
\vgr.~\citet{vgr+99}; \ffp.~\citet{ffp+99}; \bkul.~\citet{blo00};
\mpp.~\citet{mpp+00}; \hah.~\citet{hah+00}; \sahu.~\citet{svb+00};
\fsg.~\citet{fsg+00}; \fvh.~\citet{fvh+00}; \fra.~\citet{frail+99};
\bdk.~\citet{bdk+00}; 
\fvsc.~\citet{fvsc00}; \umh.~\citet{umh+99}; \vff.~\citet{vff+00};
\fjh.~\citet{fjh+00}; \fsgb.~\citet{fsg+00b}; \fv.~\citet{fv+01};
\mhw.~\citet{mhw+00}; \bdgb.~\citet{bdg+00}; \mfm.~\citet{mfm+00}}

\end{deluxetable}

A listing of the dataset compilation is given in table
\ref{tab:offsets}.  We include every GRB (up to and including GRB
000418) with an accurate radio or optical location and a deep
late-time optical image. There is a hierarchy of preference of imaging
conditions and instruments which yield the most accurate offsets; we
describe the specifics and expected accuracies of the astrometric
technique in \S \ref{sec:astlevels}.

\subsection{Dataset selection based on expected astrometric accuracy}
\label{sec:dataselect}

We group the datasets into five different levels ordered by decreasing
astrometric accuracy.  Levels 1--4 each utilize differential
astrometry and level 5 utilizes absolute astrometry relative to the
International Coordinate Reference System (ICRS). Specifics of the
individual offset measurements are given in \S
\ref{sec:astlevels}. The ideal dataset for offset determination is a
single HST image where both the transient and the host are
well-localized (hereafter ``self-HST''); so far, only GRB 970228, GRB
990123 and possibly GRB 991216 fall in this category. The next most
accurate offset is obtained where both the early- and late-time images
are from HST taken at comparable depth with the same filter (hereafter
``HST$\rightarrow$HST''). In addition to the centering errors of the
OT and host, such a set inherits the uncertainty in registering the
two epochs (e.g., GRB 970508).  Next, an early deep image from
ground-based (GB) Keck, Palomar 200-inch (P200), or the Very Large
Telescope (VLT) in which the OT dominates is paired with a late-time
image from HST (e.g.,~GRB 971214, GRB 980703, GRB 991216, GRB 000418;
``GB$\rightarrow$HST'').  Though in the majority of these cases most
of the objects detected in the HST image are also detected in the Keck
image (affording great redundancy in the astrometric mapping
solution), object centering of ground-based data is hampered by
atmospheric seeing.  The next most accurate localizations use
ground-based to ground-based imaging to compute offsets
(``GB$\rightarrow$GB'').  Last, radio localizations compared with
optical imaging (``RADIO$\rightarrow$OPT'') provide the least accurate
offset determinations. This is due primarily to the current difficulty
of mapping an optical image onto an absolute coordinate system (see \S
\ref{sec:vlahst}).

\subsection{Imaging reductions}

\subsubsection{Reductions of HST Imaging}

Most of the HST images of GRB afterglows and hosts were acquired using
the {\it Space Telescope Imaging Spectrograph}
\citep[STIS;][]{kwb+98}.  STIS imaging under-samples the angular
diffraction limit of the telescope and therefore individual HST images
essentially do not contain the full astrometric information
possible. To produce a final image that is closer to the diffraction
limit, inter-pixel dithering between multiple exposures is often
employed. The image reconstruction technique, which also facilitates
removal of cosmic-rays and corrects for the known optical field
distortion, is called ``drizzling'' and is described in detail in
\cite{fh97}.  We use this technique, as implemented using the
IRAF\footnotemark\footnotetext{IRAF is distributed by the National
Optical Astronomy Observatories, which are operated by the Association
of Universities for Research in Astronomy, Inc., under cooperative
agreement with the National Science Foundation.}\negafterfoot\ package
DITHER and DITHERII, to produce our final HST images.

We retrieved and reduced every public STIS dataset of GRB imaging from
the HST archive\footnotemark\footnotetext{{\tt
http://archive.stsci.edu}}\negafterfoot\ and processed the so-called
``on--the--fly calibration'' images to produce a final drizzled image.
These images are reduced through the standard HST pipeline for bias
subtraction, flat-fielding, and illumination corrections using the
best calibration data available at the time of archive retrieval.  The
archive name of the last image and the start time of each HST epoch
are given in columns 2 and 3 of table \ref{tab:offset-log}.

\begin{deluxetable}{lccccccccc}
\tabletypesize{\scriptsize}
\rotate
\tablewidth{0pt}
\tablecaption{Measured Angular Offsets and Physical Projections\label{tab:offsets}}
\tablecolumns{10}
\tablehead{
\colhead{Name} & \colhead{$X_0$ East} & \colhead{$Y_0$ North} & \colhead{$R_0$} & \colhead{$R_0
/\sigma_{R_{0}}$} & \colhead{$z$} & \colhead{$D_\theta$} & \colhead{$X_0$ (proj)} & \colhead{$Y
_0$ (proj)} & \colhead{$R_0$ (proj)} \\
\colhead{} & \colhead{\arcsec} & \colhead{\arcsec} & \colhead{\arcsec} & \colhead{} & \colhead{
} & \colhead{kpc/\arcsec} & \colhead{kpc} & \colhead{kpc} & \colhead{kpc} }
 \startdata
GRB 970228
 & $-$0.033$\pm$0.034 & $-$0.424$\pm$0.034 & 0.426$\pm$0.034 &  12.59 & 0.695
 & 7.673 & $-$0.251$\pm$0.259 & $-$3.256$\pm$0.259 & 3.266$\pm$0.259 \\
 
GRB 970508
 & 0.011$\pm$0.011 & 0.001$\pm$0.012 & 0.011$\pm$0.011 & 1.003 & 0.835
 & 8.201 & 0.090$\pm$0.090 & 0.008$\pm$0.098 & 0.091$\pm$0.090 \\
 
GRB 970828
 & 0.440$\pm$0.516 & 0.177$\pm$0.447 & 0.474$\pm$0.507 & 0.936 & 0.958
 & 8.534 & 3.755$\pm$4.403 & 1.510$\pm$3.815 & 4.047$\pm$4.326 \\
 
GRB 971214
 & 0.120$\pm$0.070 & $-$0.070$\pm$0.070 & 0.139$\pm$0.070 & 1.985 & 3.418
 & 7.952 & 0.954$\pm$0.557 & $-$0.557$\pm$0.557 & 1.105$\pm$0.557 \\
 
GRB 980326
 & 0.125$\pm$0.068 & $-$0.037$\pm$0.062 & 0.130$\pm$0.068 & 1.930 & $\sim 1$
 & \ldots & \ldots & \ldots & \ldots \\
 
GRB 980329
 & $-$0.037$\pm$0.049 & $-$0.003$\pm$0.061 & 0.037$\pm$0.049 & 0.756 & $\ale 3.5$
 & \ldots & \ldots & \ldots & \ldots \\
 
GRB 980425
 & $-$10.55$\pm$0.052 & $-$6.798$\pm$0.052 &  12.55$\pm$0.052 &  241.4 & 0.008
 & 0.186 & $-$1.964$\pm$0.010 & $-$1.265$\pm$0.010 & 2.337$\pm$0.010 \\
 
GRB 980519
 & $-$0.050$\pm$0.130 & 1.100$\pm$0.100 & 1.101$\pm$0.100 &  11.00 & \ldots
 & \ldots & \ldots & \ldots & \ldots \\
 
GRB 980613
 & 0.039$\pm$0.052 & 0.080$\pm$0.080 & 0.089$\pm$0.076 & 1.174 & 1.096
 & 8.796 & 0.344$\pm$0.454 & 0.703$\pm$0.707 & 0.782$\pm$0.666 \\
 
GRB 980703$^a$
 & $-$0.054$\pm$0.055 & 0.098$\pm$0.065 & 0.112$\pm$0.063 & 1.788 & 0.966
 & 8.553 & $-$0.460$\pm$0.469 & 0.842$\pm$0.555 & 0.959$\pm$0.536 \\
 
GRB 981226
 & 0.616$\pm$0.361 & 0.426$\pm$0.246 & 0.749$\pm$0.328 & 2.282 & \ldots
 & \ldots & \ldots & \ldots & \ldots \\
 
GRB 990123
 & $-$0.192$\pm$0.003 & $-$0.641$\pm$0.003 & 0.669$\pm$0.003 &  223.0 & 1.600
 & 9.124 & $-$1.752$\pm$0.027 & $-$5.849$\pm$0.027 & 6.105$\pm$0.027 \\
 
GRB 990308
 & $-$0.328$\pm$0.357 & $-$0.989$\pm$0.357 & 1.042$\pm$0.357 & 2.919 & \ldots
 & \ldots & \ldots & \ldots & \ldots \\
 
GRB 990506
 & $-$0.246$\pm$0.432 & 0.166$\pm$0.513 & 0.297$\pm$0.459 & 0.647 & 1.310
 & 9.030 & $-$2.221$\pm$3.901 & 1.499$\pm$4.632 & 2.680$\pm$4.144 \\
 
GRB 990510
 & $-$0.064$\pm$0.009 & 0.015$\pm$0.012 & 0.066$\pm$0.009 & 7.160 & 1.619
 & 9.124 & $-$0.584$\pm$0.082 & 0.137$\pm$0.109 & 0.600$\pm$0.084 \\
 
GRB 990705
 & $-$0.865$\pm$0.046 & 0.109$\pm$0.049 & 0.872$\pm$0.046 &  18.86 & 0.840
 & 8.217 & $-$7.109$\pm$0.380 & 0.896$\pm$0.399 & 7.165$\pm$0.380 \\
 
GRB 990712
 & $-$0.035$\pm$0.080 & 0.035$\pm$0.080 & 0.049$\pm$0.080 & 0.619 & 0.434
 & 6.072 & $-$0.213$\pm$0.486 & 0.213$\pm$0.486 & 0.301$\pm$0.486 \\
 
GRB 991208
 & 0.071$\pm$0.102 & 0.183$\pm$0.096 & 0.196$\pm$0.097 & 2.016 & 0.706
 & 7.720 & 0.548$\pm$0.789 & 1.410$\pm$0.744 & 1.513$\pm$0.750 \\
 
GRB 991216
 & 0.211$\pm$0.029 & 0.290$\pm$0.034 & 0.359$\pm$0.032 &  11.08 & 1.020
 & 8.664 & 1.828$\pm$0.251 & 2.513$\pm$0.295 & 3.107$\pm$0.280 \\
 
GRB 000301C
 & $-$0.025$\pm$0.014 & $-$0.065$\pm$0.005 & 0.069$\pm$0.007 & 9.821 & 2.030
 & 9.000 & $-$0.222$\pm$0.130 & $-$0.581$\pm$0.046 & 0.622$\pm$0.063 \\
 
GRB 000418
 & $-$0.019$\pm$0.066 & 0.012$\pm$0.058 & 0.023$\pm$0.064 & 0.358 & 1.118
 & 8.829 & $-$0.170$\pm$0.584 & 0.109$\pm$0.514 & 0.202$\pm$0.564\enddata
 
\tablenotetext{a}{Using radio detections of the host and afterglow,
\citet{bkf01} find a more accurate offset of $X_0 = -0.032 \pm 0.015$
and $Y_0 = 0.025 \pm 0.015$ ($R_0 = 0.040 \pm 0.015$), consistent with
our optical results; see \S \ref{subsec:980703}.}

\tablecomments{The observed offsets ($X_0, Y_0$) and associated
Gaussian uncertainties include all statistical errors from the
astrometric mapping and OT+host centroid measurements.  The observed
offset, $R_0 = \sqrt{X_0^2 + Y_0^2}$, and $\sigma_{R_0}$ (constructed
analogously to equation~\ref{eq:sigmar}) are given in col.~4. Note
that $R_0 - \sigma_{R_0} \le R \le R_0 + \sigma_{R_0}$ is not
necessarily the 68\% percent confidence region of the true offset
since the probability distribution is not Gaussian (see
equation~\ref{eq:pr2}). The term $R_0/\sigma_{R_0}$ in col.~5
indicates how well the offset from the host center is determined.  In
general, we consider the GRB to be significantly displaced from the
host center if $R_0/\sigma_{R_0} \age 5$. In col.~6, $z$ is the
measured redshift of the host galaxy and/or absorption redshift of the
GRB complied from the literature
\citepcf{kbb+00}. In col.~7, $D_\theta$ is the conversion of
angular displacement in arcseconds to projected physical distance.  }
\end{deluxetable}

Some HST GRB imaging has been taken using the STIS/Longpass filter
(F28x50LP) which, based on its red effective wavelength (central
wavelength $\lambda_c \approx 7100$\AA), would make for a good
comparison with ground-based $R$-band imaging.  However, the Longpass
filter truncates the full STIS field of view to about 40\% and
therefore systematically contains fewer objects to tie astrometrically
to ground-based images.  Therefore, all of the HST imaging reported
herein were taken in (unfiltered) STIS/Clear (CCD50) mode.  Unlike the
Longpass filter, the spectral response of the Clear mode is rather
broad (2000--10000 \AA).  We use the known optical distortion
coefficients appropriate to the wavelength of peak sensitivity
$\lambda \approx 5850$ \AA\ of this observing mode to produce final images
which are essentially linear in angular displacement versus
instrumental pixel location.

The original plate scale of most STIS imaging is 0\arcsec.05077 $\pm$
0.00007 pixel$^{-1}$ \citep{mb97}, though there is a possibility that
thermal expansion of the instrument could change this scale by a small
amount (see Appendix \ref{sec:errors}). The pixel scale of all our
final reduced HST images is half the original scale, i.e.,
0\arcsec.02539 pixel$^{-1}$.

\subsubsection{Reductions of Ground-based Imaging}

Ground-based images are all reduced using the standard practices for bias
subtraction, flat-fielding, and in the case of $I$-band imaging, fringe
correction.  In constructing a final image we compute the instrumental
shift of dithered exposures relative to a fiducial exposure and co-add
the exposures after applying the appropriate shift to align each
image.  All images are visually inspected for cosmic-ray contamination
of the transient, host or astrometric tie stars.  Pixels contaminated
by cosmic rays are masked and not used in the production of the final
image.

\section[Astrometric Reductions \& Dataset Levels]{Astrometric Reductions and Issues Related to Dataset Levels}
\label{sec:astlevels}

Here we provide a description of the astrometric reduction techniques
for both our ground-based and the HST images, and issues related to
the five levels of astrometry summarized in \S \ref{sec:dataselect}. A
discussion of the imaging reductions and astrometry for the individual
cases is given in section \S \ref{sec:indivoff}.

\subsection{Level 1: self-HST (differential)}

An ideal image is one where the optical transient and the host galaxy
are visible in the same imaging epoch with HST.  This typically
implies that the host galaxy is large enough in extent to be
well-resolved despite the brilliance of the nearby OT. Of course, a
later image of the host is always helpful to confirm that the putative
afterglow point source does indeed fade. In this case (as with GRB
970228 and GRB 990123) the accuracy of offset determination is limited
mostly by the centroiding errors of the host ``center'' and optical
afterglow.  Uncertainties in the optical distortion corrections and
the resulting plate scale are typically sub-milliarcsecond in size
(see Appendix \ref{sec:errors}).

In principle we expect centering techniques to result in centroiding
errors ($\sigma_c$) on a point source with a signal-to-noise, SN, of
$\sigma_c \approx \phi/SN$ \citepcf{sto89}, where $\phi$ is the
instrumental full-width half-maximum (FWHM) seeing of the final
image. Since $\phi$ is typically $\sim 75$ milliarcsecond (mas) we
expect $\sim$milliarcsecond offset accuracies with self-HST images.

\subsection{Level 2: HST$\rightarrow$HST (differential)}
\label{sec:hsthst}

Here, two separate HST epochs are used for the offset determination.
The first epoch is taken when the afterglow dominates the light and
the second when the host dominates.  In addition to the centroiding
errors, the astrometric accuracy of this level is limited by
uncertainty in the registration between the two images.

In general when two images are involved (here and all subsequent
levels), we register the two images such that an instrumental position
in one image is mapped to the instrumental (or absolute position) in
the other image. The registration process is as follows.  We determine
the noise characteristics of both of the initial and final images
empirically, using an iterative sigma-clipping algorithm.  This noise
along with the gain and effective read noise of the CCD are used as
input to the IRAF/CENTER algorithm.  In addition we measure the radial
profile of several apparent compact sources in the image and use the
derived seeing FWHM ($\phi$) as further input to the optimal filtering
algorithm technique for centering \citep[OFILTER; see][]{dav87}.  For
faint stars we use the more stable GAUSSIAN algorithm. Both techniques
assume a Gaussian form of the point-spread function which, while not
strictly matched to the outer wings of the Keck or HST point-spread
functions (PSFs), appears to reasonably approximate the PSF out to the
FWHM of the images.

\begin{figure*}[tbp]
\centerline{\psfig{file=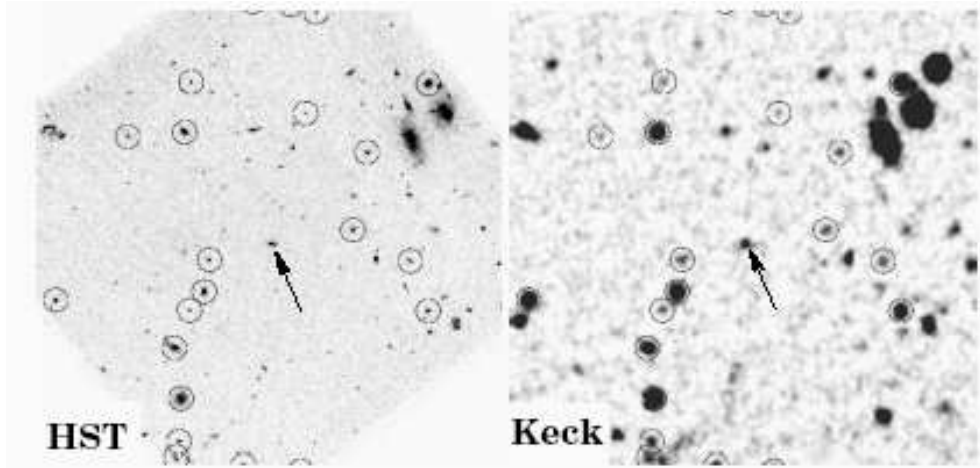,width=5.1in,angle=0}}
\caption[Example Keck and {\it HST}/STIS images of the field
of GRB 981226]{Example Keck $R$-band and HST/STIS Clear images of the
field of GRB 981226.  Twenty of the 25 astrometric tie objects are
circled in both images.  As with other Keck images used for astrometry
in the present study, most of the faint object detected in the HST
images are also detected (albeit with poorer resolution).  The optical
transient in the Keck image and the host galaxy in the HST image are
in center. The field is approximately 50\arcsec\ $\times$ 50\arcsec\
with North up and East to the left.}
\label{fig:examtie}
\end{figure*}

When computing the differential astrometric mappings between two
images (such as HST and Keck or Keck and the USNO-A2.0 catalog), we
use a list of objects common from both epochs, ``tie objects,'' and
compute the astrometric mapping using the routine IRAF/GEOMAP.  The
polynomial order of the differential fitting we use depends on the
number of tie objects.  A minimum of three tie objects are required to
find the relative rotation, shift and scale of two images, which
leaves only one degree of freedom.  The situation is never this bad;
in fact, when comparing HST images and an earlier HST image (or deep
Keck image), we typically find 20--30 reasonable tie objects, and
therefore we can solve for higher-order distortion terms. Figure
\ref{fig:examtie} shows an example Keck and HST field of GRB 981226
and the tie objects we use for the mapping. We always reject tie
objects that deviate by more than $3 \sigma$ from the initial
mapping. A full third-order two-dimensional polynomial with
cross-terms requires 18 parameters which leaves, typically, $N \approx
30$ degrees of freedom.  Assuming such a mapping adequately
characterizes the relative distortion, and it is reasonable to expect
that mapping errors will have an r.m.s.~error $\sigma \approx
30^{-1/2} \phi/\langle {SN}
\rangle$, where $\langle {SN} \rangle$ is the average
signal--to--noise of the tie objects.  For example, in drizzled HST
images $\phi \approx 75$ mas and $\langle {SN} \rangle \approx 20$ so
that we can expect differential mapping uncertainties at the 1 mas
level for HST$\rightarrow$HST mapping. Cross-correlation techniques,
such as IRAF/CROSSDRIZZLE, can in principle result in even better
mapping uncertainties, but, in light of recent work by \citet{ak99},
we are not confident that the HST CCD distortions can be reliably removed
at the sub-mas level.

\subsection{Level 3: GB$\rightarrow$HST (differential)}

This level of astrometry accounts for the majority of our dataset.  In
addition to inheriting the uncertainties of centroiding errors and
astrometric mapping errors described above, we must also consider the
effects of differential chromatic refraction (DCR) and optical image
distortion in ground based images.  In Appendix \ref{sec:errors} we
demonstrate that these effects should not dominate the offset
uncertainties. Following the argument above (\S \ref{sec:hsthst}) the
astrometric mapping uncertainties scale linearly with the seeing of
ground-based images which is typically a factor of 10--20 larger than
the effective seeing of the HST images.

An independent test of the accuracy of the transference of
differential astrometry from ground-based images to space-based
imaging is illustrated by the case of GRB 990123. In \citet{bod+99} we
registered a Palomar 60-inch (P60) image to a Keck image and thence to
an HST image. The overall statistical uncertainty introduced by this
process (see Appendix \ref{sec:osh-derive} for a derivation) is
$\sigma_r = 107$ mas (note that in the original paper we mistakenly
overstated this error as 180 mas uncertainty).  The position we
inferred was 90 mas from a bright point source in the HST image.  This
source was later seen to fade in subsequent HST imaging and so our
identification of the source as the afterglow from
P60$\rightarrow$Keck$\rightarrow$HST astrometry was vindicated. Since
the P60$\rightarrow$Keck differential mapping accounted for
approximately half of the error (due to optical field distortion and
unfavorable seeing in P60), we consider $\sim 100$ mas uncertainty in
Keck$\rightarrow$HST mapping as a reasonable upper limit to the
expected uncertainty from other cases.  In practice, we achieved
r.m.s.~accuracies of 40--70 mas (see table \ref{tab:offsets}).

\subsection{Level 4: GB$\rightarrow$GB (differential)}

This level contains the same error contributions as in
GB$\rightarrow$HST level, but in general, the uncertainties are larger
since the centroiding uncertainties are large in both epochs.  The
offsets are computed in term of pixels in late-time image.  Just as
with the previous levels with HST, we assume an average plate scale to
convert the offset to units of arcseconds.  For the Low Resolution
Imaging Spectrometer \citep[LRIS;][]{occ+95} and the Echellete
Spectrograph Imager \citep[ESI;][]{em98} imaging, we assume a plate
scale of 0\arcsec.212 pixel$^{-1}$ and 0\arcsec.153 pixel$^{-1}$,
respectively.  We have found that these plate scales are stable over
time to better than a few percent; consequently, the errors introduced
by any deviations from these assumed plate scales are negligible.

\subsection{Level 5: RADIO$\rightarrow$OPT (absolute)}
\label{sec:vlahst}

Unfortunately, the accuracy of absolute offset determination is
(currently) hampered by systematics in astrometrically mapping deep
optical/infrared imaging to the ICRS.  Only bright stars ($V \ale 9$
mag) have absolute localizations measured on the milliarcsecond level
thanks to astrometric satellite missions such as Hipparcos.  The
density of Hipparcos stars is a few per square degree so the
probability of having at least two such stars on a typical CCD frame
is low.  Instead, optical astrometric mapping to the ICRS currently
utilizes ICRS positions of stars from the USNO-A2.0 Catalog,
determined from scanned photographic plates \citep{mon98}.  Even if
all statistical errors of positions are suppressed, an astrometric
plate solution can do no better than inherit a systematic 1-$\sigma$
uncertainty of 250 mas in the absolute position of any object on the
sky \citep[$\sigma_\alpha = 0\arcsec.18$ and $\sigma_\delta =
0\arcsec.17$;][]{deu99}. By contrast, very-long baseline array (VLBA)
positions of GRB radio afterglow have achieved sub-milliarcsecond
absolute positional uncertainties relative to the ICRS \citep{tbf+99}.
So, until optical systematics are beaten down and/or sensitivities at
radio wavelengths are greatly improved (so as to directly detect the
host galaxy at radio wavelengths), the absolute offset astrometry can
achieve 1-$\sigma$ accuracies no better than $\sim 300$ mas ($\approx
2.5$ kpc at $z=1$).  In fact, one GRB host (GRB 980703) has been
detected at radio wavelengths \citep{bkf01} with the subsequent offset
measurement accuracy improving by a factor of $\sim$3 over the optical
measurement determined herein (\S \ref{subsec:980703}).

There are three GRBs in our sample where absolute astrometry (level 5)
is employed. In computing the location of the optical transient
relative to the ICRS we typically use 20--40 \hbox{USNO A2.0}
astrometric tie stars in common with Keck or Palomar images.  We then
use IRAF/CCMAP to compute the mapping of instrumental position ($x$,
$y$) to the world coordinate system ($\alpha$, $\delta$).

\section{Individual Offsets and Hosts}
\label{sec:indivoff}

Below we highlight the specific reductions for each offset, the
results of which are summarized in table \ref{tab:offsets}. In total
there are 21 bursts until May 2000 that have been reliably localized
at the arcsecond level and 1 burst with an uncertain association with
the nearby SN 1998bw (GRB 980425). In our analysis we do not include
\hbox{GRB 000210} \citep{scm+00} due to lack of late-time imaging
data.  Thereby, the present study includes 20 ``cosmological'' GRBs
plus the nearby SN 1998bw/GRB 980425.  Offset measurements should be
possible for the recent bursts \hbox{GRB 000630} \citep{hur+00},
\hbox{GRB 000911} \citep{hcm+00b}, \hbox{GRB 000926} \citep{hmg+00}
and GRB 010222 \citep{pir+01}.

To look for the hosts, we generally image each GRB field roughly a few
months to a year after the burst with Keck.  Typically, these
observations reach a limiting magnitude of $R \approx 24$--26 mag
depending on the specifics of the observing conditions. If an object
is detected within $\sim 1$ arcsec from the afterglow position and has
a brightness significantly above the extrapolated afterglow flux at
the time of observation, this source is deemed the host (most GRB
hosts are readily identified in such imaging).  If no object is
detected, we endeavor to obtain significantly deeper images of the
field. Typically these faint host searches require 1--3 hours of Keck
(or VLT) imaging to reach limiting magnitudes at the $R \approx 27$
mag level.  If no object is detected at the location of the afterglow,
HST imaging is required and the host search is extended to limiting
magnitudes of $R \approx 28$--29 mag.  Only 3 hosts in our sample (GRB
990510, GRB 000301C, and GRB 980326) were first found using HST after
an exhaustive search from the ground.

Note that the assignment of a certain observed galaxy as the host of a
GRB is, to some extent, a subjective process.  We address the
question of whether our assignments are ``correct'' in \S
\ref{sec:angoffs} where we demonstrate on statistical grounds that at
most only a few assignments in the sample of 20 could be spurious. In
\S \ref{sec:angoffs} we also discuss how absorption/emission redshifts
help strengthen the physical connection of GRBs to their assigned
hosts.

Irrespective of whether individual assignments of hosts are correct,
we uniformly assign the nearest (in angular distance) detected galaxy
as the host.  In practice this means that the nearest object
(i.e.,~galaxy) brighter than $R \simeq 25$--26 mag detected in Keck
imaging is assigned as the host.  In almost all cases, there is a
detected galaxy within $\sim$1 arcsecond of the transient position.
For the few cases where there is no object within $\sim$1 arcsecond,
deeper HST imaging {\it always} reveals a faint galaxy within $\sim$1
arcsecond. In most cases, the estimated probability that we have
assigned the ``wrong'' host is small (see \S \ref{sec:angoffs}).
After assigning the host, the center of host is then determined,
except in a few cases, as the centroid near the brightest component of
the host system. In a few cases where there is evidence for
significant low-surface brightness emission (e.g.,~980519) or the host
center is ambiguous, we assign the approximate geometric center as the
host center.

A summary of our offset results is presented in table
\ref{tab:offsets}.  Since all our final images of the host galaxies
are rotated to the cardinal orientation before starting the
astrometric mapping process, these uncertainties are also directly
proportional to the uncertainties in $\alpha$ and $\delta$. It is
important to note, however, that the projected radial offset is a
positive-definite number and the probability distribution is not
Gaussian.  Thereby, the associated error ($\sigma_r$) in offset
measurements does not necessarily yield a 68\% confidence region for
the offset (see Appendix \ref{sec:osh-derive}) but is, clearly,
indicative of the precision of the offset measurement.

Once the offsets are determined from the final images, we then measure
the half-light radii of the host galaxies.  For extended hosts, the
value of the half-light radius may be obtained directly from aperture
curve-of-growth analysis.  However, for compact hosts, the
instrumental resolution systematically spreads the host flux over a
larger area and biases the measurement of the half-light radius to
larger values.  We attempt to correct for this effect (for all hosts,
not just compact hosts) by deconvolving the images with IRAF/SCLEAN
using an average STIS/Clear PSF derived from 10 stars in the final HST
image of the GRB 990705 field (which were obtained through low
Galactic latitude). We then fit curve-of-growth photometry about the
host centers and determine the radius at which half the detected light
was within such radius. These values, along with associated errors are
presented in table \ref{tab:offnorm}.  We tested that the PSFs derived
at differing roll angles and epochs had little impact upon the
determined value of the half-light radius.
\begin{deluxetable}{lcccccccc}
\tablewidth{0pt}
\tablecaption{Host Detection Probabilities and Host Normalized Offsets\label{tab:offnorm}}
\tablecolumns{8}
\tablehead{
\colhead{Name} & \colhead{R$_{c, {\rm host}}$} & \colhead{$A_{R_c}$} & \colhead{$P_{\rm chance}$} & \colhead{$R_{\rm half}$ (obs)} & \colhead{$R_{\rm half}$ (calc)} & \colhead{$r_e$} & \colhead{$r_0$} \\
\colhead{} & \colhead{mag} & \colhead{mag} & \colhead{} & \colhead{\arcsec} & \colhead{\arcsec} & \colhead{kpc} & \colhead{} }
 \startdata

GRB 970228
 &  24.60$\pm$0.20 & 0.630 & 0.00935 &  0.345$\pm$0.030 & 0.316$\pm$0.095
 &  1.6 & 1.233$\pm$0.146 \\
 
GRB 970508
 &  24.99$\pm$0.17 & 0.130 & 0.00090 &  0.089$\pm$0.026 & 0.300$\pm$0.090
 &  0.4 & 0.124$\pm$0.129 \\
 
GRB 970828
 &  25.10$\pm$0.30 & 0.100 & 0.07037 & \ldots & 0.296$\pm$0.089
 &  1.5 & 1.603$\pm$1.780 \\
 
GRB 971214
 &  25.65$\pm$0.30 & 0.040 & 0.01119 &  0.226$\pm$0.031 & 0.273$\pm$0.082
 &  1.1 & 0.615$\pm$0.321 \\
 
GRB 980326
 &  28.70$\pm$0.30 & 0.210 & 0.01878 &  0.043$\pm$0.028 & 0.116$\pm$0.035
 &  0.2 & 3.023$\pm$2.532 \\
 
GRB 980329
 &  27.80$\pm$0.30 & 0.190 & 0.05493 &  0.245$\pm$0.033 & 0.168$\pm$0.050
 &  1.3 & 0.152$\pm$0.202 \\
 
GRB 980425
 &  14.11$\pm$0.05 & 0.170 & 0.00988 & 18.700$\pm$0.025 & \ldots
 &  2.1 & 0.671$\pm$0.003 \\
 
GRB 980519
 &  25.50$\pm$0.30 & 0.690 & 0.05213 &  0.434$\pm$0.041 & 0.279$\pm$0.084
 &  2.2 & 2.540$\pm$0.332 \\
 
GRB 980613
 &  23.58$\pm$0.10 & 0.230 & 0.00189 &  0.227$\pm$0.031 & 0.352$\pm$0.106
 &  1.2 & 0.392$\pm$0.338 \\
 
GRB 980703
 &  22.30$\pm$0.08 & 0.150 & 0.00045 &  0.169$\pm$0.026 & 0.392$\pm$0.117
 &  0.9 & 0.663$\pm$0.385 \\
 
GRB 981226
 &  24.30$\pm$0.01 & 0.060 & 0.01766 &  0.336$\pm$0.030 & 0.327$\pm$0.098
 &  1.7 & 2.227$\pm$0.996 \\
 
GRB 990123
 &  23.90$\pm$0.10 & 0.040 & 0.01418 &  0.400$\pm$0.028 & 0.341$\pm$0.102
 &  2.2 & 1.673$\pm$0.117 \\
 
GRB 990308
 &  28.00$\pm$0.50 & 0.070 & 0.31659 &  0.213$\pm$0.028 & 0.156$\pm$0.047
 &  1.1 & 4.887$\pm$1.776 \\
 
GRB 990506
 &  24.80$\pm$0.30 & 0.180 & 0.04365 &  0.090$\pm$0.027 & 0.308$\pm$0.092
 &  0.5 & 3.297$\pm$5.196 \\
 
GRB 990510
 &  27.10$\pm$0.30 & 0.530 & 0.01218 &  0.167$\pm$0.041 & 0.205$\pm$0.061
 &  0.9 & 0.393$\pm$0.111 \\
 
GRB 990705
 &  22.00$\pm$0.10 & 0.334$^{\rm a}$ & 0.01460 &  1.151$\pm$0.030 & 0.400$\pm$0.120
 &  5.7 & 0.758$\pm$0.045 \\
 
GRB 990712
 &  21.90$\pm$0.15 & 0.080 & 0.00088 &  0.282$\pm$0.026 & 0.403$\pm$0.121
 &  1.0 & 0.175$\pm$0.284 \\
 
GRB 991208
 &  24.20$\pm$0.20 & 0.040 & 0.00140 &  0.048$\pm$0.026 & 0.330$\pm$0.099
 &  0.2 & 4.083$\pm$2.994 \\
 
GRB 991216
 &  25.30$\pm$0.20 & 1.640 & 0.00860 &  0.400$\pm$0.043 & 0.288$\pm$0.086
 &  2.1 & 0.898$\pm$0.127 \\
 
GRB 000301C
 &  28.00$\pm$0.30 & 0.130 & 0.00629 &  0.066$\pm$0.028 & 0.156$\pm$0.047
 &  0.4 & 1.054$\pm$0.462 \\
 
GRB 000418
 &  23.80$\pm$0.20 & 0.080 & 0.00044 &  0.096$\pm$0.027 & 0.345$\pm$0.103
 &  0.5 & 0.239$\pm$0.670\enddata

\tablenotetext{a}{Since the GRB position pierces through the Large
Magellanic Cloud \citep{dkh+99}, we have added 0.13 mag extinction to
the Galactic extinction quoted in \citet{pia01}.  This assumes an
average extinction through the LMC of $E(B-V)$ = 0.05 \citep{dbc+01}.}

\tablecomments{Column 2 gives the de-reddened host magnitude as
referenced in \citet{pia01}, \citet{dfk+01a} and
\citet{sfc+01}. Column 3 gives the estimated extinction in the
direction of the GRB host galaxy \citet{pia01}. Column 4 gives the
estimated probability that the assigned host is a chance superposition
and not physically related to the GRB (following \S
\ref{sec:angoffs}).  The half-light radii $R_{\rm half}$ are observed
from HST imaging (col.~5) or calculated using the magnitude-radius
empirical relationship (col.~6; see text).  For HST imaging, the
uncertainty is taken as the sum of the statistical error and estimated
systematics error (0\arcsec.025 which is approximately the size of one
de-convolved pixel). Otherwise, the uncertainty is taken as 30\% of the
calculated radius (col.~6). Column 7 gives the estimated host disk
scale length. The host-normalized offset $r_0 = R_0/R_{\rm half}$
given in col.~8 is derived from (if possible) the observed half-light
radius or the calculated half-light radius (otherwise).  The error on
$r_0$ is $\sigma_r$ from equation~\ref{eq:sigmar}. Note that $r_0 -
\sigma_r \le r \le r_0 + \sigma_r$ is not necessarily the 68\% percent
confidence region of the true offset since the probability
distribution is not Gaussian.}

\end{deluxetable}

\subsection{GRB 970228}

\begin{figure*}[tp]
\centerline{\psfig{file=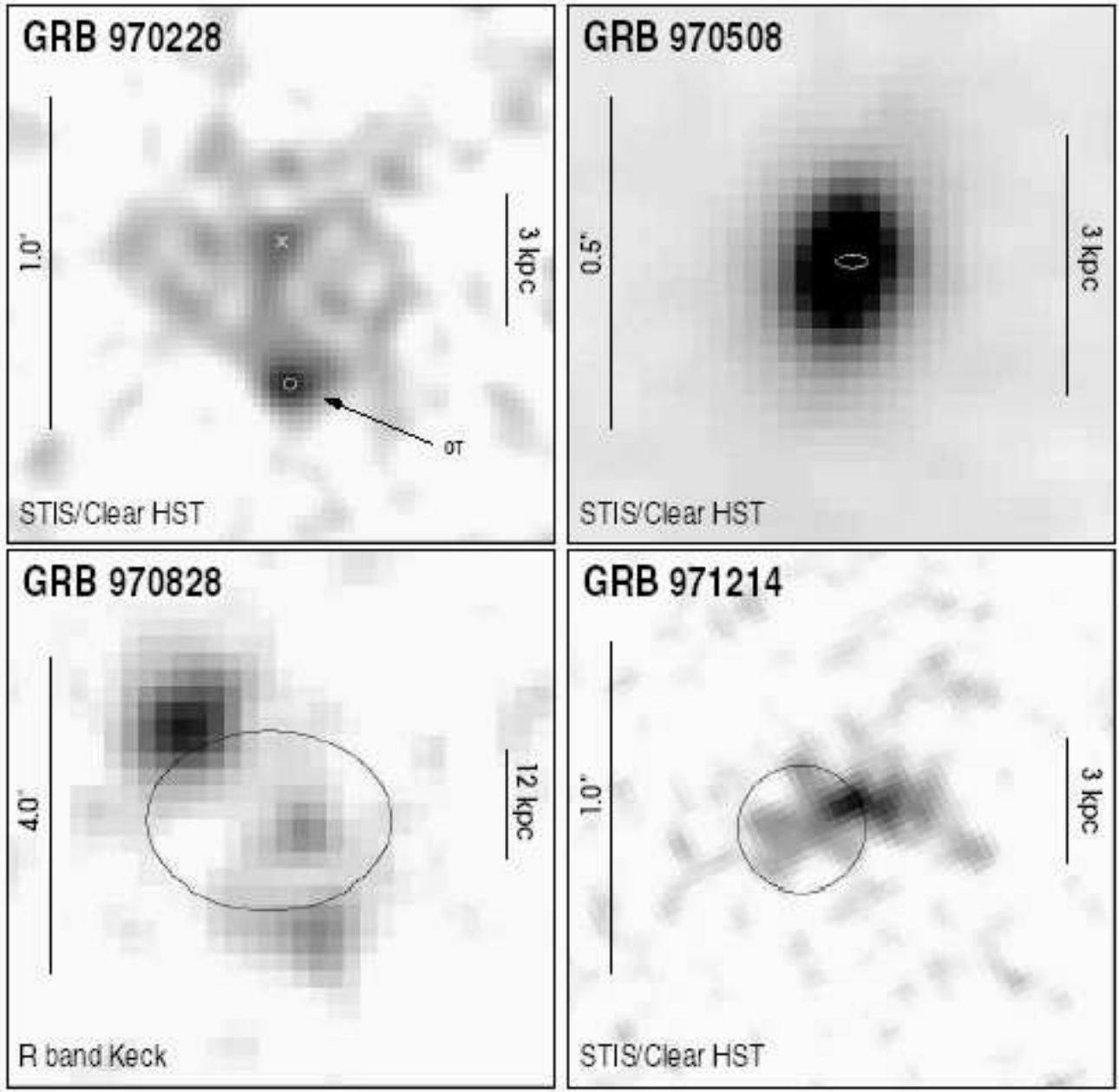,width=6.8in,angle=0}}
\caption[The location of individual GRBs about their host
galaxies]{The location of individual GRBs about their host
galaxies. The ellipse in each frame represents the 3-$\sigma$ error
contour for the location of the GRB as found in \S \ref{sec:indivoff}
and in table \ref{tab:offsets}.  The angular scale of each image is
different and noted on the left-hand side.  The scale and stretch was
chosen to best show both the detailed morphology of the host galaxy
and the spatial relationship of the GRB and the host.  The GRB
afterglow is still visible is some of the images (GRB 970228, GRB
991216).  In GRB 980425, the location of the associated supernova is
noted with an arrow. In all cases where a redshift is available for
the host or GRB afterglow, we also provide a physical scale of the
region on the right-hand side of each image.  For clarity, the host
centers are marked with ``$\times$'' when the centers are not obvious.
For all images, North is up and East is to the left.}
\label{fig:offset1}
\end{figure*}

\begin{figure*}[tbp]
\figurenum{\ref{fig:offset1}}
\centerline{\psfig{file=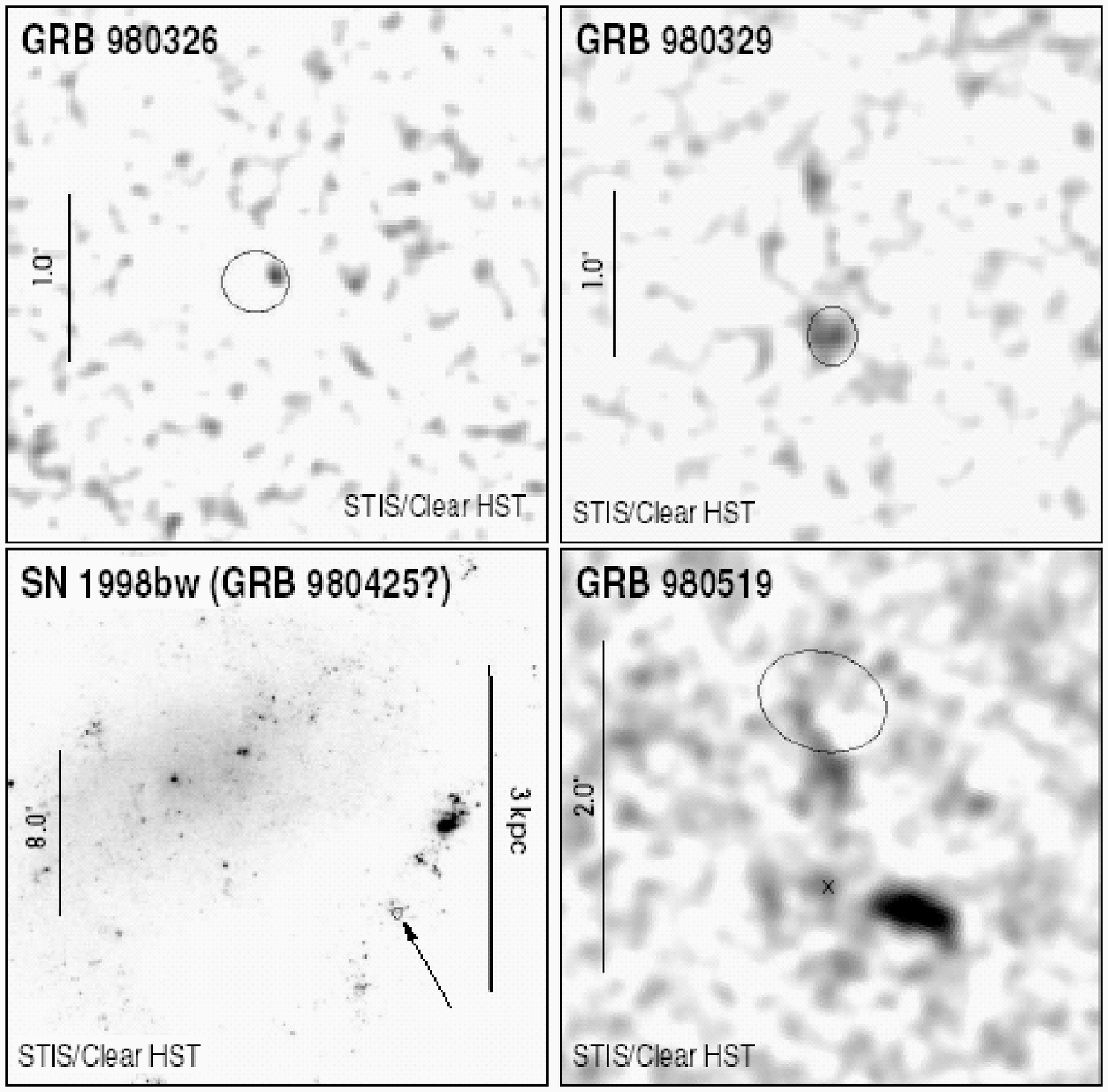,width=6.8in,angle=0}}
\caption[The location of individual GRBs about their host
galaxies (cont.)]{(cont.) The location of individual GRBs about their
host galaxies.}
\label{fig:offset2}
\end{figure*}

\begin{figure*}[tbp]
\figurenum{\ref{fig:offset1}}
\centerline{\psfig{file=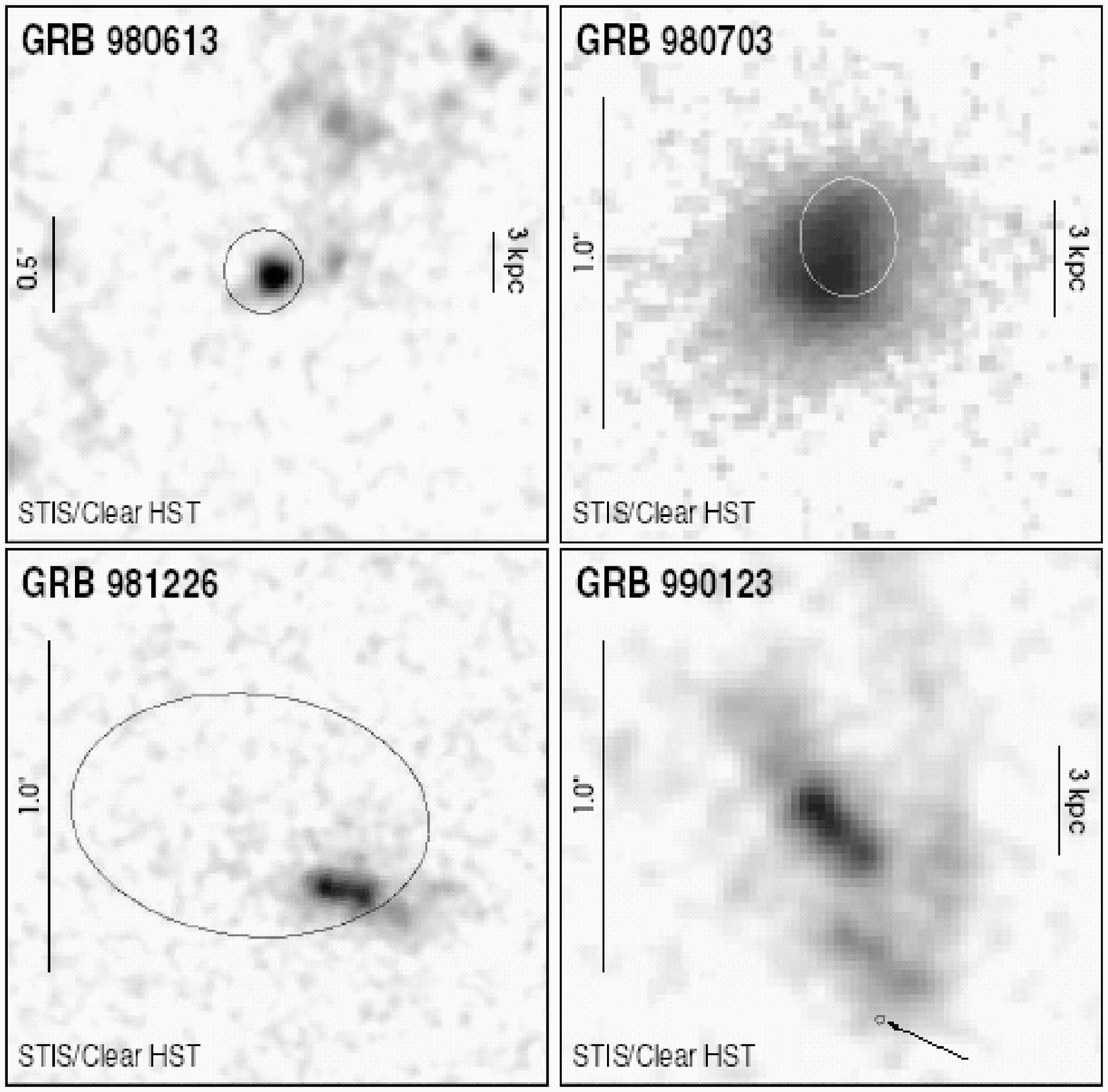,width=6.8in,angle=0}}
\caption[The location of individual GRBs about their host
galaxies (cont.)]{(cont.) The location of individual GRBs about their host
galaxies.}
\label{fig:offset3}
\end{figure*}

\begin{figure*}[tbp]
\figurenum{\ref{fig:offset1}}
\centerline{\psfig{file=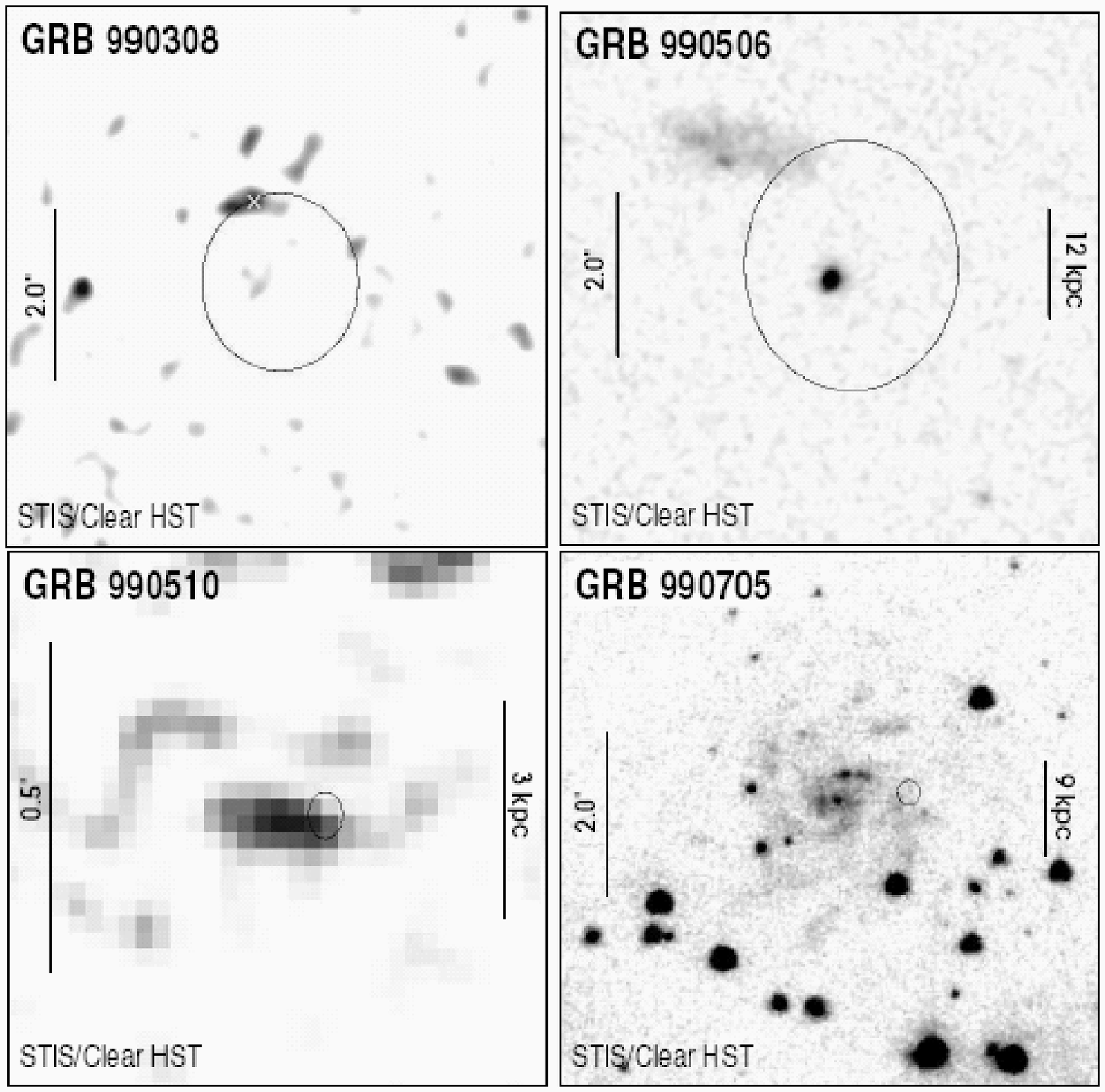,width=6.8in,angle=0}}
\caption[The location of individual GRBs about their host
galaxies (cont.)]{(cont.) The location of individual GRBs about their host
galaxies.}
\label{fig:offset4}
\end{figure*}

\begin{figure*}[tbp]
\figurenum{\ref{fig:offset1}}
\centerline{\psfig{file=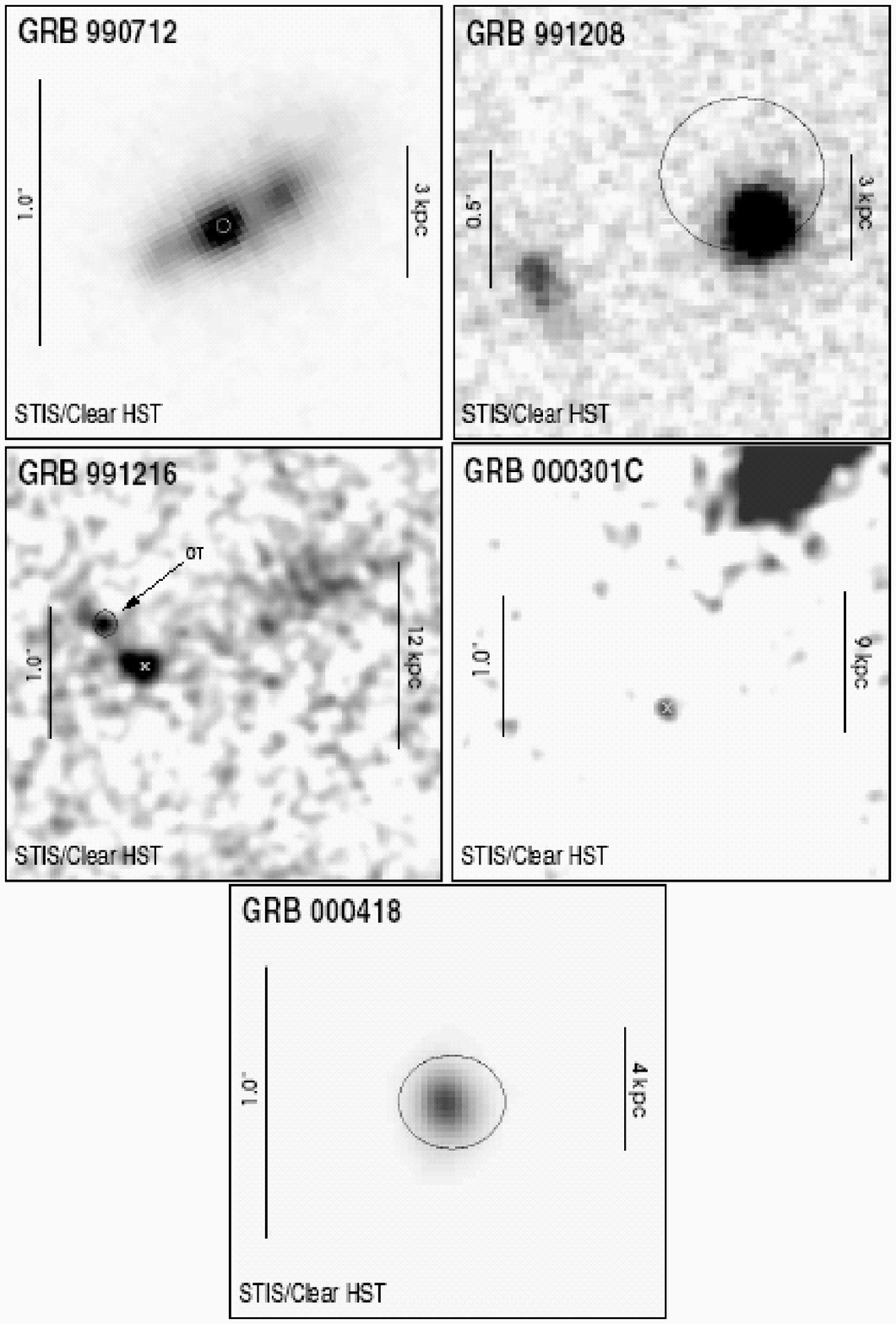,width=6.0in,angle=0}}
\caption[The location of individual GRBs about their host
galaxies (cont.)]{(cont.) The location of individual GRBs about their host
galaxies.}
\label{fig:offset5}
\end{figure*}

The morphology and offset derivation have been discussed extensively
in \citet{bdk01} and we briefly summarize the results.  In the
HST/STIS image (figure~\ref{fig:offset1}), the host appears to be
essentially a face-on late-type blue dwarf galaxy.  At the center is
an apparent nucleus manifested as a 6-$\sigma$ peak north of the
transient.  There is also an indication of arm-like structure
extending toward the transient.

This image represents the ideal for astrometric purposes (level 1):
both the transient and the host ``center'' are well-localizable in the
same high-resolution image. The transient appears outside the
half-light radius of the galaxy.

\subsection{GRB 970508}

The host is a compact, elongated and blue galaxy \citep{bdkf98} and is
likely undergoing a starburst phase.  The optical transient was
well-detected in the early time HST image \citep{pfb+98a} and the host
was well-detected (figure~\ref{fig:offset1}) in the late-time image
\citep{fp98}.  We masked out a 2\arcsec\ $\times$ 2\arcsec\ region
around the OT/host and cross-correlated the two final images using the
IRAF/CROSSDRIZZLE routine.  We used the IRAF/SHIFTFIND routine on the
correlation image to find the systematic shift between the two epochs.
The resulting uncertainty in the shift was quite small, $\sigma =
0.013$, $0.011$ pix ($x$, $y$ direction).  We also found 37 compact objects
in common to both images and performed an astrometric mapping in the
usual manner. We find $\sigma = 0.344, 0.354$ pix in the ($x$,$y$
directions). We centered the OT and the host in the normal manner
using the IRAF/ELLIPSE task.

The resulting offset is given in table \ref{tab:offsets} where we use
the more conservative astrometric mapping uncertainties from using the
tie objects, rather than the CROSSDRIZZLE routine.  As first noted in
\citet{bdkf98} (Keck imaging) and then in \citet{fp98} (HST imaging),
the OT was remarkably close to the apparent center of the host galaxy.
The P200$\rightarrow$Keck astrometry from \citet{bdkf98} produced an
r.m.s.~astrometric uncertainty of 121 mas, compared to an
r.m.s.~uncertainty of 11 mas from HST$\rightarrow$HST astrometry.  The
largest source of uncertainty from the HST$\rightarrow$HST is the
centroid position of the host galaxy.

\subsection{GRB 970828}

The host is identified as the middle galaxy in an apparent
three-component system.  We discuss the host properties and the
astrometry (RADIO$\rightarrow$OPT) in more detail in
\citet{dfk+01}. The total uncertainty in the radio to Keck tie is 506
mas ($\alpha$) and 376 mas ($\delta$).

\subsection{GRB 971214}

By all accounts, the host appears to be a typical $L_*$ galaxy at
redshift $z=3.42$.  The Keck$\rightarrow$HST astrometry is discussed
in detail in \citet{odk+98}.  The offset
uncertainty found was $\sigma_r$ = 70 mas. The GRB appears located to
the east of the host galaxy center, but consistent with the east-west
extension of the host (see figure \ref{fig:offset1}).

\subsection{GRB 980326}
\label{subsec:980326}

No spectroscopic redshift for this burst was found. However, based on
the light-curve and the SN hypothesis, the presumed redshift is
$z\sim$1 \citep[see][]{bkd+99}. \citet{bkd+99} reported that no galaxy
was found at the position of the optical transient down to a
3-$\sigma$ limiting magnitude of $R \approx 27.3$ mag.  Given the
close spatial connection of other GRBs with galaxies, we posited that
a deeper integration would reveal a nearby host.  Indeed,
\citet{fvn01} recently reported the detection with HST/STIS imaging of
a very faint ($V = 29.25 \pm 0.25$ mag) galaxy within 25 mas of the OT
position.

For astrometry we used an $R$-band image, from 27 April 1998 when the
OT was bright and found the position of the OT on our deep $R$-band
image from 18 December 1998.  In this deep $R$-band image we found 34
objects in common with the HST/STIS drizzled image.  We confirm the
presence of this faint and compact source near the OT position though
our astrometry places the OT at a distance of 130 mas ($\sigma_r = 68$
mas).  

The galaxy and OT position are shown in figure \ref{fig:offset1}.  The
low-level flux to the Southeast corner of the image is a remnant from
a diffraction spike of a nearby bright star. \citet{fvn01} find that
the putative host galaxy is detected at the 4.5-$\sigma$ level.
Adding to the notion that the source is not some chance superposition,
we note that the galaxy is the brightest object within 3 $\times$ 3
arcsec$^2$ of the GRB position.  There is also a possible detection of
a low-surface brightness galaxy $\sim$0.5 arcsec to the East of the
galaxy.

\subsection{GRB 980329}

The afterglow of GRB 980329 was first detected at radio wavelengths
\citep{tfk+98}.  Our best early time position was obtained using Keck
K-band image of the field observed by J.~Larkin and collaborators
\citep{lgk+98}.  We recently obtained deep R- imaging of the field
with Keck/ESI and detected the host galaxy at $R = 26.53 \pm 0.22$
mag.  We found the location of the afterglow relative to the host
using 13 stars in common to the early K-band and late R-band image.

As shown in figure~\ref{fig:offset1}, the GRB is coincident with a
slightly extended faint galaxy.  Our determined angular offset (see
table \ref{tab:offsets}) of the GRB from this galaxy is significantly
closer to the putative host than the offset determined by
\citet{hth+00} in late-time HST imaging (our astrometric uncertainties
are also a factor of $\sim$ 9 smaller).  The difference is possibly
explained by noting that the \citet{hth+00} analysis used the VLA
radio position and just three USNO-A2.0 stars to tie the GRB position
to the HST image.

\subsection{GRB 980425}

The SN 1998bw was well-localized at radio wavelengths \citep{kfw+98}
with an astrometric position relative to the ICRS of 100 mas in each
coordinate.  Ideally, we could calibrate the HST/STIS image to ICRS to
ascertain where the radio source lies. However, without
Hipparcos/Tycho astrometric sources or radio point sources in the STIS
field, such absolute astrometric positioning is difficult.

Instead, we registered an early ground-based image to the STIS field
to determine the differential astrometry of the optical SN with
respect to its host.  Unfortunately, most early images were relatively
shallow exposures to avoid saturation of the bright SN and so
many of the point sources in the STIS field are undetected. The best
seeing and deepest exposure from ground-based imaging is from the
EMMI/ESO NTT 3.5 meter Telescope on 4.41 May 1998 \citep{gal99} where
the seeing was 0.9 arcsec FWHM.  We found 6 point sources which were
detected in both the STIS/CLEAR and the ESO NTT $I$-band image. The use
of $I$-band positions for image registration is justified since all 6
point sources are red in appearance and therefore unlikely to
introduce a systematic error in the relative positioning.  Since the
number of astrometric tie sources is low, we did not fit for
high-order distortions in the ESO image and instead we fit for the
relative scale in both the $x$ and $y$ directions, rotation, and shift (5
parameters for 12 data points).  We compute an r.m.s.~uncertainty of
40 mas and 32 mas in the $x$ and $y$ positions of the astrometric tie
sources. These transformation uncertainties dominate the error in the
positional uncertainty of the SN in the ESO NTT image and so we take
the transformation uncertainties as the uncertainty in the true
position of the supernova with respect to the STIS host image.

The astrometric mapping places the optical position of SN 1998bw
within an apparent star-forming region in the outer spiral arm of the
host 2.4 kpc in projection at $z=0.0088$ to the south-west of the
galactic nucleus.  Within the uncertainties of the astrometry the SN
is positionally coincident with a bright, blue knot within this
region, probably an HII region. This is consistent with the
independent astrometric solutions reported by \citet{fha+00}.

\subsection{GRB 980519}

The GRB afterglow was well-detected in our early-time image from the
Palomar 200-inch.  We found 150 objects in common to this image and
our intermediate-time Keck image. An astrometric registration between
the two epochs was performed using IRAF/GEOMAP.  Based on this
astrometry, \citet{bkdg+98} reported the OT to be astrometrically
consistent with a faint galaxy, the putative host. This is the second
faintest host galaxy (after GRB 990510; see below) observed to date
with $R = 26.1 \pm 0.3$.

We found 25 objects in common with the intermediate-time Keck image
and the HST/STIS image. These tie objects were used to further
propagate the OT position onto the HST frame.  Inspection of our final
HST image near the optical transient location reveals the presence of
low surface-brightness emission connecting the two bright elongated
structures.  Morphologically, the ``host'' appears to be tidally
interacting galaxies, although this interpretation is subjective.  The
GRB location is coincident with the dimmer elongated structure to the
north.  Using the approximate geometric center of the host, we estimate
the center, albeit somewhat arbitrarily, as the faint knot south of
the GRB location and $\sim$0\arcsec .3 to the east of the brighter
elongated structure. The half-light radius of the system was also
measured from this point.  From this ``center'' we find the offset of
the GRB given in table \ref{tab:offsets}.

\subsection{GRB 980613}

The morphology of the system surrounding the GRB is complex and
discussed in detail in \citet{dbk00}.  There we found the OT to be
within $\sim$3 arcsec of five apparent galaxies or galaxy fragments,
two of which are very red ($R - K > 5$). In more recent HST imaging,
the OT appears nearly coincident with a compact high-surface
brightness feature, which we now identify as the host center.  Given
the complex morphology, we chose to isolate the feature in the
determination of the half-light radius by truncating the
curve-of-growth analysis at 0.5 arcsec from the determined center.

\subsection{GRB 980703}
\label{subsec:980703}

The optical transient was well-detected in our early time image and,
based on the light curve and the late-time image, the light was not
contaminated by light from the host galaxy.
\citet{bkf01} recently found that the radio transient was very near
the center of the radio emission from the host. 

We found 23 objects in common to the Keck image and our final reduced
HST/STIS image and computed the geometric transformation.  The
r.m.s.~uncertainty of the OT position on the HST image was quite
small: 49 mas and 60 mas in the instrumental $x$ and $y$ coordinates,
respectively.  We determined the center of the host using
IRAF/ELLIPSE and IRAF/CENTER which gave consistent answers to 2
mas in each coordinate.  

Recently, \citet{bkf01} compared the VLBA position of the afterglow
with the position of the persistent radio emission from the
host. Since both measurements were referenced directly to the ICRS,
the offset determined was a factor of $\sim$3 times more accurate than
that found using the optical afterglow; the two offset measurements
are consistent within the errors.  In the interest of uniformity, we
use the optical offset measurement in the following analysis.

\subsection{GRB 981226}

Unfortunately, no optical transient was found for this burst though a
radio transient was identified \citep{fkb+99}.  We rely on the
transformation between the USNO-A2.0 and the Keck image to place the
host galaxy position on the ICRS \citep[see][for further
details]{fkb+99}.  We then determine the location of the radio
transient in the HST frame using 25 compact sources common to both the
HST and Keck image. In figure \ref{fig:examtie} we show as example the
tie objects in both the Keck and HST image.  The tie between the two
images is excellent with an r.m.s. uncertainty of 33 mas and 47 mas in
the instrumental $x$ and $y$ positions.  Clearly, the uncertainty in
the radio position on the Keck image dominates the overall location of
the GRB on the HST image.

The host appears to have a double nucleated morphology, perhaps
indicative of a merger or interacting system.  \citet{hta+00} noted,
by inspecting both the STIS Longpass and the STIS clear image, that
the north-eastern part of the galaxy appeared significantly bluer then
the south-western part. As expected from these colors, the center of the
host, as measured in our late-time $R$-band Keck, lies near ($\sim 50$
mas) the centroid of the red (south-western) portion of the host.  We
assign the $R$-band centroid in Keck image as the center of the host.

\subsection{GRB 990123}
\label{subsec:grb990123}

This GRB had an extremely bright prompt optical afterglow emission
which was found in archived images from a robotic telescope, the
Robotic Optical Transient Search Experiment \citep{abb+99}.
We reported on the astrometric comparison of ground-based data with
HST imaging and found that the bright point source on the southern
edge of a complex morphological system was the afterglow
\citep{bod+99}.  Later HST imaging revealed that indeed this source
did fade \citepeg{ftp+99} as expected of GRB afterglow.

As seen in figure \ref{fig:offset2}, the host galaxy is fairly
complex, with two bright elongated regions spaced by $\sim 0\arcsec.5$
which run approximately parallel to each other.  The appearance of
spatially curved emission to the west may be a tidal tail from the
merger of two separate systems or a pronounced spiral arm of the
brighter elongated region to the north.  We choose, again somewhat
subjectively, the peak of this brighter region as the center of the
system and find the astrometric position of the GRB directly from the
first HST epoch.

\subsection{GRB 990308}

An optical transient associated with GRB 990308 was found by
\citet{ssh+99}.  Though the transient was detected at only one epoch
\citep[3.3 hours after the GRB;][]{ssh+99}, it was observed in three
band-passes, twice in $R$-band.  Later-time Keck imaging revealed no
obvious source at the location of the transient to $R$ = 25.7 mag,
suggesting that the source had faded by at least $\sim$7.5 mag in
$R$-band.

A deep HST exposure of the field was obtained by \citet{hft+00b} who
reported that the \citeauthor{ssh+99} position derived from the
USNO-A2.0 was consistent with two faint galaxies.

We found the offset by two means. First, we found an absolute
astrometric solution using 12 USNO-A2.0 stars in common with the
later-time Keck image.  The HST/STIS and the Keck $R$-band image were
then registered using 27 objects in common. Second, we found a
differential position by using early ground-based images kindly
provided by B.~Schaefer to tie the optical afterglow position directly
to the Keck (then to HST) image.  Both methods give consistent results
though the differential method is, as expected, more accurate.

Our astrometry places the OT position further East from the two faint
galaxies than the position derived by \citet{hft+00b}. At a distance
of 0.73 arcsec to the North of our OT position, there appears to be a
low-surface brightness galaxy near the detection limit of the STIS
image (see figure~\ref{fig:offset2}), similar to the host of GRB
980519 (there is also, possibly, a very faint source 0.23 arcsec
southwest of the OT position, but the reality of its detection is
questionable).  Due to the faintness and morphological nature of the
source, a detection confidence limit is difficult to quantify, but we
are reasonably convinced that the source is real.  At $V \sim 27$ mag,
the non-detection of this galaxy in previous imaging is consistent
with the current STIS detection. Since the angular extent of the
galaxy spans $\sim$25 drizzled STIS pixels ($\sim$0.63 arcsec), more
high-resolution HST imaging is not particularly useful for confirming
the detection of the galaxy. Instead, deeper ground-based imaging with
a large aperture telescope would be more useful.

\subsection{GRB 990506}

The Keck astrometric comparison to the radio position was given in
\citet{tbf+00}, with a statistical error of 250 mas. We transferred
this astrometric tie to the HST/STIS image using 8 compact sources
common to both the Keck and HST images of the field near GRB 990506.
The resulting uncertainty is negligible compared to the uncertainties
in the radio position on the Keck image.  As first reported in
\citet{tbf+00}, the GRB location appears consistent with a faint
compact galaxy.  \citet{htab+00} later reported that the galaxy
appears compact even in the STIS imaging.

\subsection{GRB 990510}
\label{subsec:990510}

This GRB is well-known for having exhibited the first clear evidence
of a jet manifested as an achromatic break in the light curve
\citepeg{hbf+99,sgk+99b}.  Recently, we discovered the host galaxy in
late-time HST/STIS imaging \citep{blo00} with $V = 28.5 \pm 0.5$.
Registration of the early epoch where the OT was bright reveals the OT
occurred 64 $\pm$ 9 mas west and 15 $\pm$ 12 mas north of the center
of the host galaxy.  This amounts to a significant displacement of 66
$\pm$ 9 mas or ~600 pc at a distance of $z = 1.62$ \citep{gvr+99}.
The galaxy is extended with a position angle PA = 80.5 $\pm$ 1.5
degree (east of north) and an ellipticity of about $\sim$0.5.

In retrospect, the host does appear to be marginally detected in the
July 1999 imaging as well as the later April 2000 image although, at
the time, no galaxy was believed to have been detected
\citep{ffp+99}.

\subsection{GRB 990705}

\citet{mpp+00} discovered the infrared afterglow of GRB 990705
projected on the outskirts of the Large Magellanic Cloud.  At the
position of the afterglow, \citet{mpp+00} noted an extended galaxy
seen in ground-based $V$-band imaging; they identified this galaxy as
the host.  \citet{hah+00} reported on HST imaging of the field and
noted, thanks to the large size ($\sim 2$ arcsec) of the galaxy and
resolution afforded by HST, an apparent face-on spiral at the location
of the transient. We retrieved the public HST data and compared the
early images provided by N.~Masetti with our final reduced HST image.
Consistent with the position derived by \citet{hah+00}, we find that
the transient was situated on the outskirts of a spiral arm to the
west of the galaxy nucleus and just north of an apparent star-forming
region.

\subsection{GRB 990712}

This GRB is the lowest measured redshift of a ``cosmological'' GRB
with $z=0.4337$ \citep{hhc+00}.  Unfortunately, the astrometric
location of the GRB appears to be controversial, though there is no
question that the GRB occurred within the bright galaxy pictured in
figure~\ref{fig:offset2}. \citet{hhc+00} found that the only source
consistent with a point source in the earlier HST image was the faint
region to the Northwest side of the galaxy and concluded that the
source was the optical transient.  However, \citet{fvh+00} found that
this source did not fade significantly.  Instead the Fruchter et
al.~analysis showed, by subtraction of two HST epochs, that a source
did fade near the bright region to the southeast.  While the fading
could be due to AGN activity instead of the presence of a GRB
afterglow, we adopt the conclusion of Fruchter et al.~for astrometry
and place conservative uncertainties on the location relative to the
center as 75 mas (3 pixels) in both $\alpha$ and $\delta$ for
1-$\sigma$ errors.  We did not conduct an independent analysis to
determine this GRB offset.

\subsection{GRB 991208}

In our early $K$-band image of the field, we detect the afterglow as
well as 7 suitable tie stars to our late ESI image.  The host galaxy
is visible in the ESI image and the subsequent offset was reported by
\citet{dbg+00}. An HST image was later obtained and reported by
\citet{fvsc00} confirming the presences of the host galaxy.

We reduced the public HST/STIS data on this burst and found the offset
in the usual manner by tying the OT position from Keck to the HST
frame.  The GRB afterglow position falls near a small, compact galaxy.
A fainter galaxy, to the Southeast, may also be related to the
GRB/host galaxy system (see figure ~\ref{fig:offset2}).

\subsection{GRB 991216}

We used nine compact objects in common to our early Keck image (seeing
FWHM = 0\arcsec.66) and the late-time HST/STIS imaging to locate the
transient.  As noted first by \citet{vff+00}, the OT is spatially
coincident with a faint, apparent point source in the HST/STIS image.
Our astrometric accuracy of $\sigma_r = 32$ mas of the OT position is
about four times better than that of \citet{vff+00}. Thanks to this we
can confidently state that the OT coincides with a point source on the
HST/STIS image. We believe this point source, as first suggested by
Vreeswijk et al., is the OT itself.

The ``location'' of the host galaxy is difficult to determine.  The OT
does appear to reside to the southwest of faint extended emission
(object ``N'' from Vreeswijk et al.~2000) but it is also located to
the northeast of a brighter extended component (object ``S'' from
Vreeswijk et al.~2000).  There appears to be a faint bridge of
emission connecting the two regions as well as the much larger region
to the west of the OT (see figure~\ref{fig:offset2}).  In fact, these
three regions may together comprise a large, low-surface brightness
system.  Again, somewhat arbitrarily, we take the center of the
``host'' to be the peak of object ``S.''

\subsection{GRB 000301C}

\citet{fmp+00}, in intermediate-time (April 2000) imaging of the field
of GRB 000301C, detected a faint unresolved source coincident with the
location of the GRB afterglow; the authors reckoned the source to be
the faded afterglow itself. In the most recent imaging on February
2001 the same group detected a somewhat fainter, compact object very
near the position of the transient.  Given the time ($\sim$ one year)
since the GRB, the authors suggested that the afterglow should have
faded below the detection level and that therefore this object is the
host of GRB 000301C \citep{fv+01}.

We confirm the detection of this source and measure the offset using
earlier time imaging from HST. Though no emission line redshift of
this source has been obtained given its proximity to the GRB, it
likely resides at $z = 2.03$, inferred from absorption spectroscopy
\citep{sfg+01,cdd+00} of the OT.

In figure \ref{fig:offset3} we present the late-time image from
HST/STIS.  A galaxy 2\arcsec.13 from the transient to the northwest is
detected at $R = 24.25 \pm 0.08$ mag and may be involved in possible
microlensing of the GRB afterglow \citepeg{gls00}.

\subsection{GRB 000418}

We reported the detection of an optically bright component and an
infrared bright component at the location of GRB 000418
\citep{bdg+00}. \citet{mfm+00} later reported that HST/STIS imaging of
the field revealed that the OT location was 0\arcsec.08 $\pm$
0\arcsec.15 east of the center of the optically bright component, a
compact galaxy.

For our astrometry we used an early Keck $R$-band image and the late
HST/STIS image.  The astrometric uncertainty is improved over the
\citet{mfm+00} analysis by a factor of 2.4. Within errors, the OT is
consistent with the center of the compact host.

\section{The Observed Offset Distribution}
\label{sec:offdist}

\subsection{Angular offset}
\label{sec:angoffs}

\begin{figure*}
\centerline{
\psfig{file=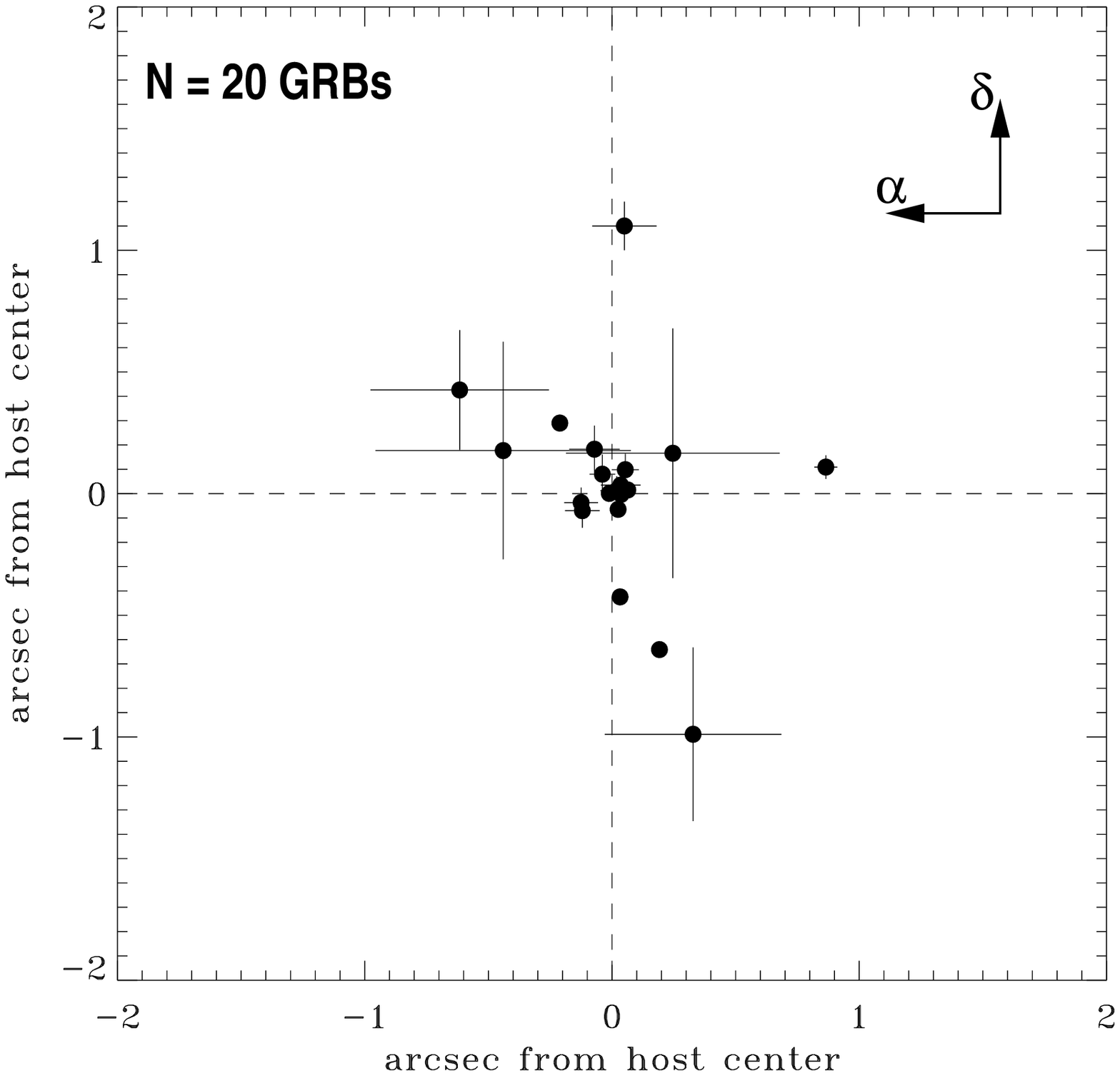,width=3.0in,angle=270}
\psfig{file=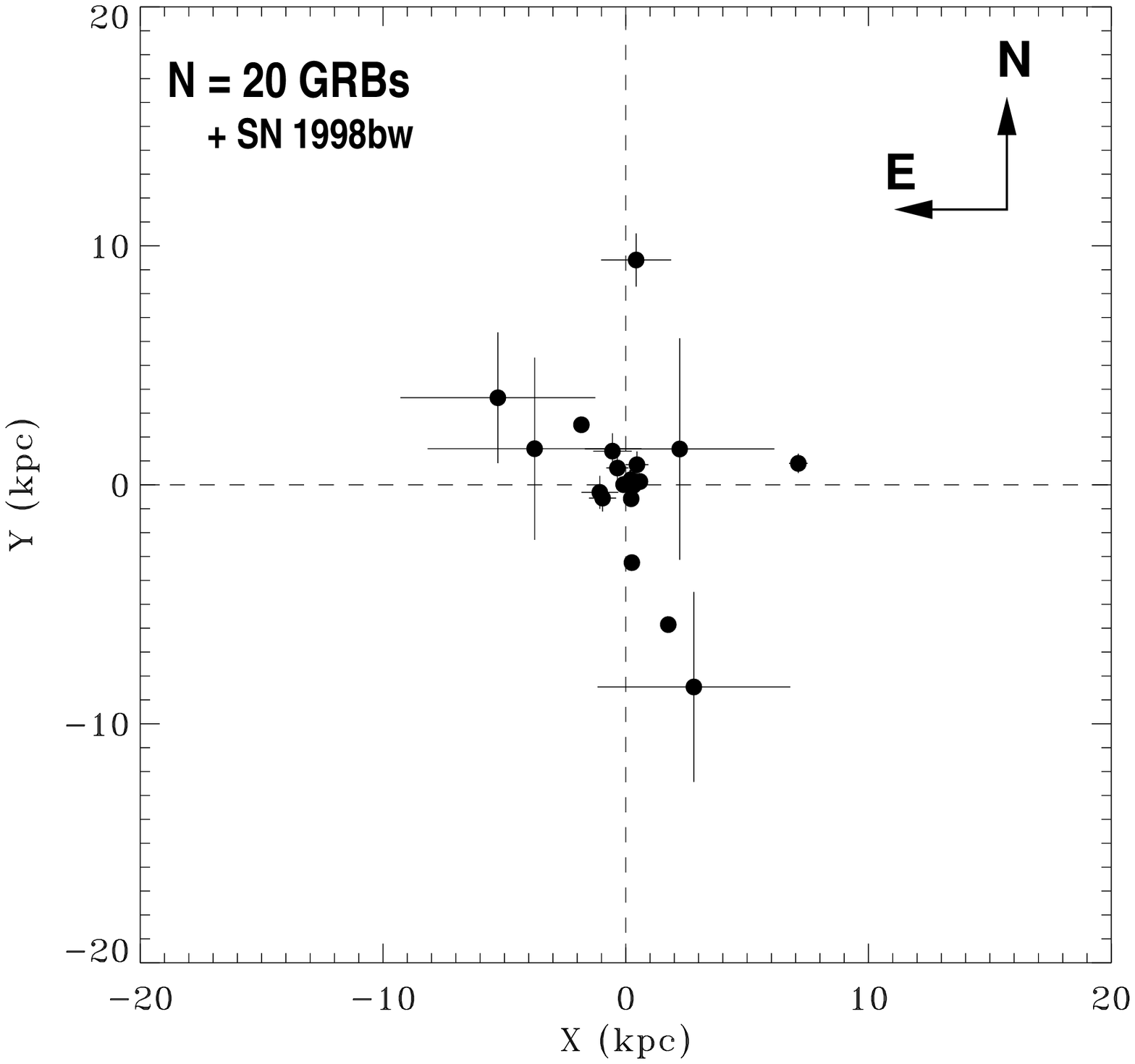,width=3.0in,angle=270}}
\caption[The angular distribution of 20 gamma-ray bursts
about their presumed host galaxy]{(left) The angular distribution of
20 gamma-ray bursts about their presumed host galaxy. The error bars
are 1 $\sigma$ and reflect the total uncertainty in the relative
location of the GRB and the apparent host center.  The offset of GRB
980425 from its host is suppressed for clarity since the redshift,
relative to all the others, GRB was so small. (right) The projected
physical offset distribution of 20 $\gamma$-ray bursts (now including
SN1998bw/GRB 980425) about their presumed host galaxies .  The
physical offset is assigned assuming $H_0 = 65$ km/s Mpc$^{-1}$,
$\Lambda = 0.7$, and $\Omega_m = 0.3$ and assuming the GRB and the
presumed host are at the same redshift. Where no redshift has been
directly measured a redshift is assigned equal to the median redshift
($z = 0.966$) of all GRBs with measured redshifts (see text).}
\label{fig:skyoff}
\end{figure*}

As seen in table \ref{tab:offsets} and \S \ref{sec:indivoff}, there
are 20 GRBs for which we have a reliable offset measurement from
self-HST, HST$\rightarrow$HST, HST$\rightarrow$GB, GB$\rightarrow$GB,
or RADIO$\rightarrow$OPT astrometric ties.  There are several
representations of this data worth exploring.  In figure
\ref{fig:skyoff} we plot the angular distribution of GRBs about their
presumed host galaxy.  In this figure and in the subsequent analysis
we exclude GRB 980425 because the association of this GRB with SN
1998bw is still controversial.  And more importantly (for the purposes
of this paper) the relation of GRB 980425 with the classical
``cosmological'' GRBs is unclear \citep{scm99} given that, if the
association proved true, the burst would have been under-luminous by a
factor of $\sim 10^{5}$ \citep{gvv+98,bkf+98}.

\begin{figure*}[tbp]
\centerline{\psfig{file=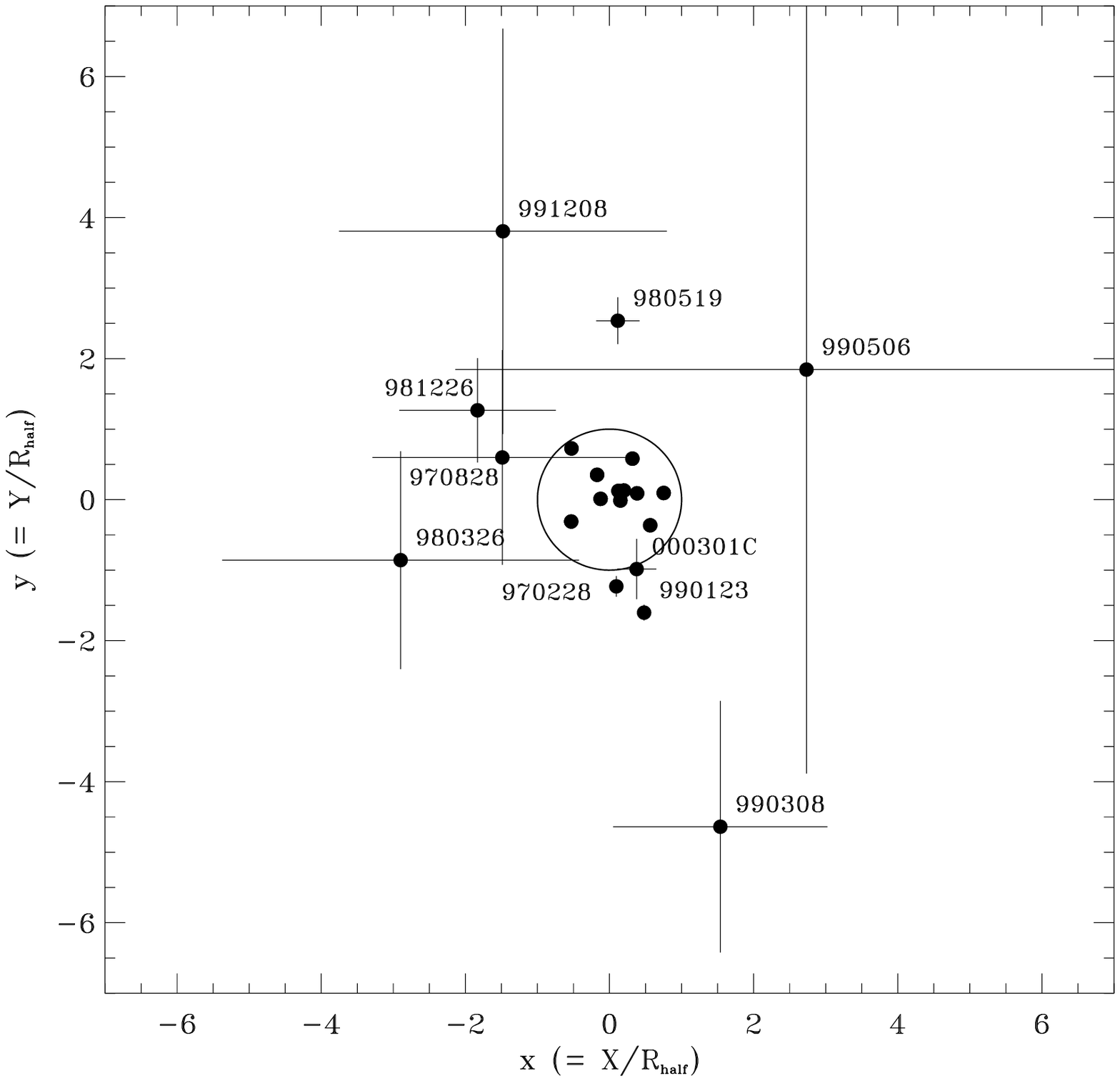,height=2.9in,angle=0}
\psfig{file=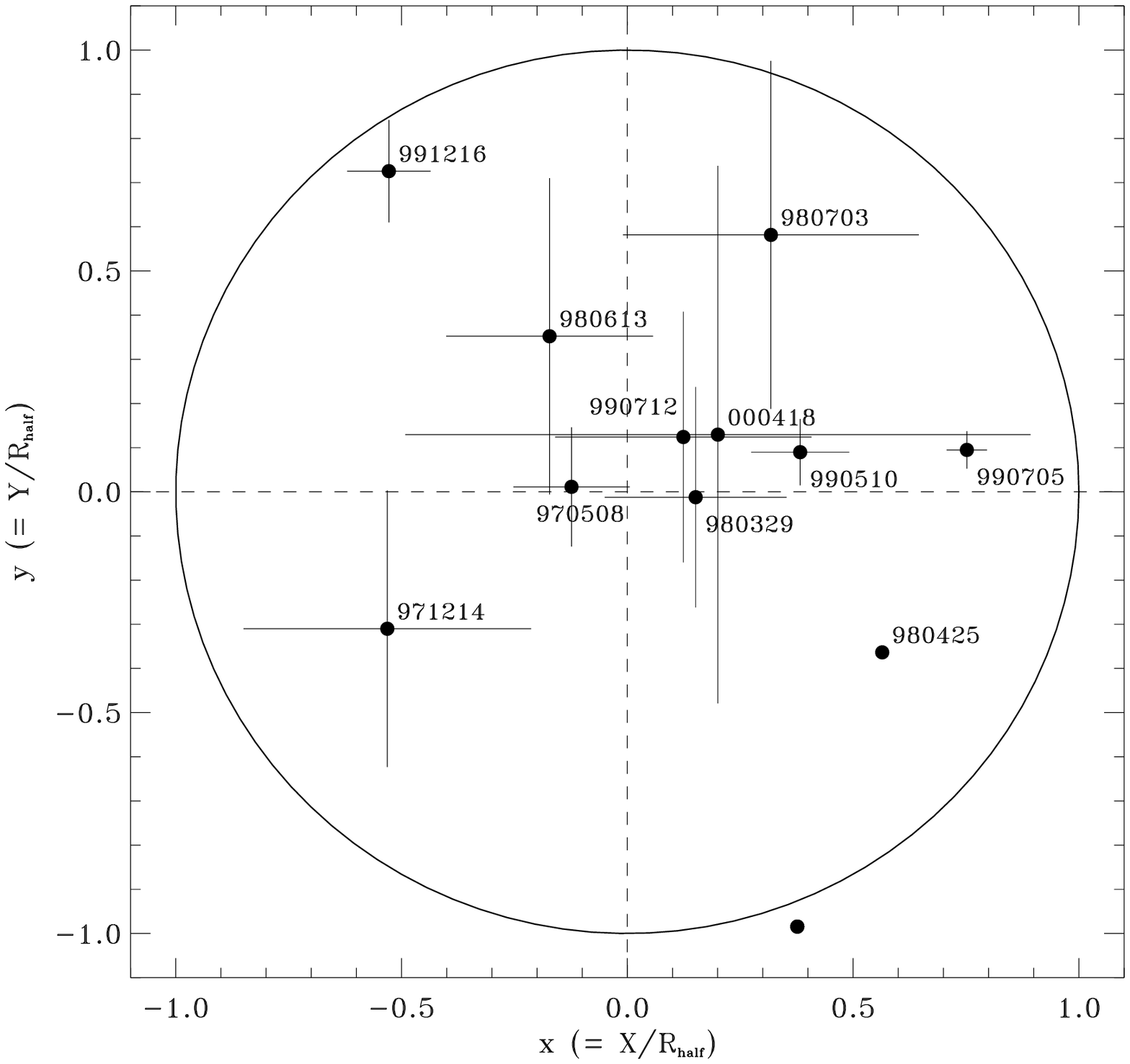,height=2.9in,angle=0}}
\caption[Host-normalized offset distribution]{Host-normalized offset
 distribution.  The dimensionless offsets are the observed offsets
($X_0, Y_0$) normalized by the host half-light radius ($R_{\rm half}$)
of the presumed host galaxy. See text for an explanation of how the
half-light radius is found.  The 1-$\sigma$ error bars reflect the
uncertainties in the offset measurement and in the half-light radius.
As expected if GRBs occur where stars are formed, there are 10 GRBs
(plus 1998bw/GRB 980425) inside and 10 GRBs outside the half-light
radius of their host.  (left) All GRBs outside of one half-light
radius (small circle) are labeled. (right) All GRBs observed to be
internal to one half-light radius are labeled.}
\label{fig:hostnorm}
\end{figure*}

As can be seen from table \ref{tab:offsets} and figure
\ref{fig:skyoff}, well-localized GRBs appear on the sky close to
galaxies. The median projected offset of the 20 GRBs from their
putative host galaxies is 0.17 arcsecond---sufficiently small that
almost all of the identified galaxies must be genuine hosts (see
below). In detail, three of the bursts show no measurable offset from
the centroid of their compact hosts (970508, 980703, 000418) whereas
five bursts appear well displaced ($\age 0\arcsec.3$) from the center
of their host at a high level of significance.  Three additional
bursts detected via radio afterglows (GRB 970828, GRB 981226, GRB
990506) and GRB 990308 (poor astrometry of the discovery image due to
large pixels and shallow depth) suffer from larger uncertainties
(r.m.s.~$\approx 0.3$ arcsecond) but have plausible host galaxies.
\begin{figure*}[tbp]
\centerline{\psfig{file=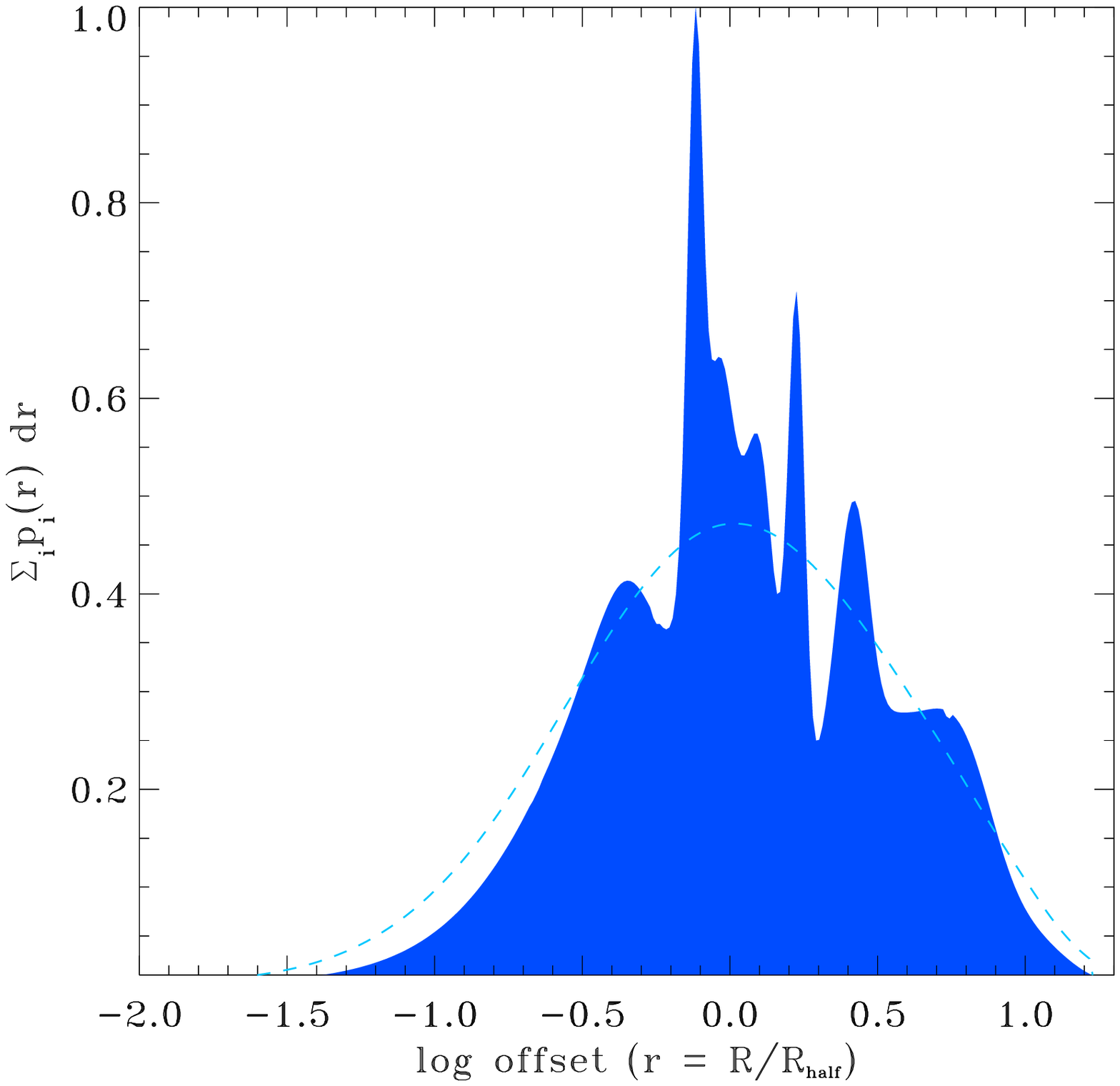,height=5.3in}}
\caption[The GRB offset distribution as a function of
normalized galactocentric radius]{The GRB offset distribution as a
function of normalized galactocentric radius.  The normalized offset
is $r = R/R_{\rm half}$, where $R$ is the projected galactocentric
offset of the GRB from the host and $R_{\rm half}$ is the half-light
radius of the host. This distribution is essentially a smooth
histogram of the data, but one which takes into account the
uncertainties in the measurements: the sharper peaks are due to
individual offsets where the significance ($r_0/\sigma_{r_0}$) of the
offset is high. That is, if a GRB offset is well-determined, its
contribution to the distribution will appear as a $\delta$-function
centered at $r = r_0$.  The dashed curve is the distribution under the
blue (dark) curve but smoothed with a Gaussian of FWHM = 0.7 dex in
$r$.  Strikingly, the peak of the probability is near one half-light
radius, a qualitative argument for the association of GRBs with
massive star formation.  We compare in detail this distribution with
predicted progenitor distributions in \S \ref{sec:compare}.}
\label{fig:offset-log}
\end{figure*}

As discussed in Appendix \ref{sec:errors}, GB$\rightarrow$GB or
GB$\rightarrow$HST astrometry could systematically suffer from the
effects of differential chromatic refraction, albeit on the
5--10 mas level. The HST$\rightarrow$HST measured offsets of GRB
970228, GRB 970508, GRB 990123, GRB 990510, GRB 990712, and GRB 000301C
are immune from DCR effects.  Since optical transients are, in
general, red in appearance and their hosts blue, DCR will
systematically appear to pull OTs away from their hosts in the
parallactic direction toward the horizon.  Comparing the observed
offsets directions parallactic at the time of each OT observation in
table \ref{tab:offsets}, we find no systematic correlation thus
confirming that DCR does not appear to play a dominant role in
determining the differential offsets of OTs from their hosts.

On what basis can we be confident that the host assignment is correct
for a particular GRB? Stated more clearly in the negative is the
following question: "What is the probability of finding an unrelated
galaxy (or galaxies) within the localization error circle of the
afterglow (3-sigma) or, in the case where the localization error
circle is very small, whether a galaxy found close to a GRB
localization is an unrelated galaxy seen in projection?"  This
probability, assuming that the surface distribution of galaxies is
uniform and thus follows a Poisson distribution (i.e.,~we ignore
clustering of galaxies) is
\begin{equation}
        P_{i,{\rm chance}} = 1 - \exp (-\eta_i).
\label{eq:thep}
\end{equation}
Here 
\begin{equation}
        \eta_i = \pi r_i^2 \sigma (\leq m_i)
\nonumber
\end{equation}
is the expected number of galaxies in a circle with an effective
radius, $r_i$, and
\begin{eqnarray}
  \sigma(\leq m_i) &=& \frac{1}{3600^2 \times 0.334 \log_e
10} \times \nonumber\\
    && 10^{0.334\,\left(m_i - 22.963\right) + 4.320} {\rm ~~galaxy~arcsec}^{-2}
\nonumber
\end{eqnarray}
is the mean surface density of galaxies brighter than R-band magnitude
of $m_i$, found using the results from \citet{hpm+97}. Since GRB are
observed through some Galactic extinction, the surface density of
galaxies at a given limiting flux is reduced; therefore, we use the
reddened host galaxy magnitude for $m_i$ (= col.~2 $-$ col.~3 of table
\ref{tab:offnorm}). 

There are few possible scenarios for determining $r_i$ at a given
magnitude limit. If the GRB is very well localized inside the
detectable light of a galaxy, then $r_i \approx 2\, R_{\rm half}$ is a
reasonable estimate to the effective radius.  If the localization is
poor and there is a galaxy inside the uncertainty position, then $r_i
\approx 3\sigma_{R_0}$. If the localization is good, but the position
is outside the light of the nearest galaxy, then $r_i \approx
\sqrt{R_0^2 + 4\, R_{\rm half}^2}$. Therefore, we take $r_i = \max
[2\, R_{\rm half}, 3\sigma_{R_0}, \sqrt{R_0^2 + 4\, R_{\rm half}^2}]$
as a conservative estimate to the effective radius.  Here, the
quantity $R_0$ is the radial separation between the GRB and the
presumed host galaxy, $R_{\rm half}$ is the half-light radius, and
$\sigma_{R_0}$ is the associated r.m.s.~error (see table
\ref{tab:offsets}).

If no ``obvious'' host is found (i.e.,~$P_{\rm chance} \age 0.1$) then
we often seek deeper imaging observations, which will, in general,
decrease the estimated $r_i$ as more and more galaxies are
detected. However, the estimate for $\eta_i$ should remain reasonable
since the surface density of background galaxies continues to grow
larger with increasing depth. This is to say that there is little
penalty to pay in statistically relating sky positions to galaxies by
observing to fainter depths.

The values for $P_{i,{\rm chance}}$ are computed and presented in
table \ref{tab:offnorm}. As expected, GRBs which fall very close to a
galaxy (e.g.,~GRB 970508, GRB 980703, GRB 990712) are likely to be
related to that galaxy. Similarly, GRB localizations with poor
astrometric accuracy (e.g.,~GRB 990308, 970828) yield larger
probabilities that the assigned galaxy is unrelated. 

In the past, most authors (including ourselves) did not endeavor to
produce a probability of chance association, instead opting to assume
that these assigned galaxies are indeed the hosts.  Nevertheless, we
believe these estimates are conservative; for instance, \citet{vgg+97}
estimated that $P_{\rm chance}$ (970228) = 0.0016 which is a value 5.8
times smaller than our estimate.  Again, we emphasize that the
estimated probabilities are constructed {\it a posteriori} so there is
no exact formula to the determine the true $P_{\rm chance}$.

The probability that {\bf all} supposed host galaxies in our sample
are random background galaxies is
$$
P(n_{\rm chance} = m = {\rm all}) = \prod_{k=1}^m P_k,
$$
with $m = 20$ and $P_k$ found from equation \ref{eq:thep} for each GRB
$k$.  Not surprisingly, this number is extremely small, $P({\rm all})
= 2 \times 10^{-60}$, insuring that at least some host assignments must be
correct.  

The probability that {\bf all} galaxies are physically
associated (i.e.,~that none are chance super-position of a random field
galaxies) is
$$
P(n_{\rm chance} = 0) = \prod_{k=1}^m (1 - P_k) = 0.483.
$$
In general, the chance that $n$ assignments will be spurious out of a
sample of $m \ge n$ assignments is
\begin{eqnarray}
P(n_{\rm chance}) &=& \frac{1}{n_{\rm chance}{\rm !}} \times \\ 
	&&\overbrace{\sum_i^m \sum_{j \ne i}^m \cdots\mathstrut}^{n_{\rm chance}}
	\left[\overbrace{P_i \times P_j \times \cdots\mathstrut}^{n_{\rm chance}}
	\prod_{k \not= i \not= j \not= \cdots}^m (1 - P_k)\right]. \nonumber
\end{eqnarray}
$P(n_{\rm chance})$ reflects the probability that we have generated a
number $n_{\rm chance}$ of spurious host galaxy identifications.  For
our sample, we find that $P(1) = 0.395$, $P(2) = 0.106$, and $P(3) =
0.015$ and so the number of spurious identifications is likely to be
small, $\sim 1$--2. Indeed, if the two GRBs with the largest $P_{\rm
chance}$ are excluded (GRB 970828, GRB 990308), then $P(n_{\rm
chance}=0$) jumps to 0.76.  Thus we are confident that almost all of
our identifications are quite secure.

The certainty of our host assignments of the nearest galaxy to a GRB
finds added strength by using redshift information.  In {\it all}
cases where an absorption redshift is found in a GRB afterglow (GRB
970508, GRB 980613, GRB 990123, GRB 980703, GRB 990712, GRB 991216),
the highest redshift absorption system is observed to be at nearly
the same emission redshift of the nearest galaxy.  Therefore, with
these bursts, clearly the nearest galaxy cannot reside at a higher
redshift than the GRB.  The galaxy may simply be a foreground object
which gives rise both to nebular line emission and the absorption of
the afterglow originating from a higher redshift.  However, using the
observed number density evolution of absorbing systems, \citet{bsw+97}
calculated that statistically in $\age 80$\% of such absorption cases,
the GRB could reside no further than 1.25 times the absorption
redshift.  For example, if an emission/absorption system is found at
$z = 1.0$, then there is only a $\ale 20$\% chance that the GRB could
have occurred beyond redshift $z = 1.25$ without another absorption
system intervening.  Though this argument cannot prove that a given
GRB progenitor originated from the assigned host, the effect of
absorption/emission redshifts is to confine the possible GRB redshifts
to a shell in redshift-space, reducing the number of galaxies that
could possibly host the GRB, and increasing the chance that the host
assignment is correct. Therefore, given this argument and the
statistical formulation above, we proceed with the hypothesis that, as
a group, GRBs are indeed physically associated with galaxies assigned
as hosts.

\subsection{Physical projection}
\label{sec:physproj}

Of the 20 GRBs with angular offsets, five have no confirmed redshift,
and the angular offset is thus without a physical scale.  These bursts
have hosts fainter than $R \approx 25$ mag and, given the distribution
of other GRB redshifts with these host magnitudes, it is reasonable to
suppose that the five bursts originated somewhere in the redshift
range $z =$ 0.5--5. It is interesting to note (with our assumed
cosmology) that despite a luminosity distance ratio of 37 between
these two redshifts, the angular scales are about the same:
$D_\theta(z=0.5)/D_\theta(z=5) \approx 1$.  In fact, over this entire
redshift range, \hbox{6.6 kpc arcsec$^{-1}$ $< D_\theta(z) < 9.1$ kpc
arcsec$^{-1}$} which renders the conversion of angular displacement to
physical projection relatively insensitive to redshift.  For these
five bursts, then, we assign the median $D_\theta$ of the other bursts
with known redshifts so that \hbox{$D_\theta = 8.552$ kpc
arcsec$^{-1}$} (corresponding to a redshift of $z = 0.966$) and scale
the observed offset uncertainty by an additional 30\%.  Here, we use
the GRB redshifts (and, below, host magnitudes) compiled in the review
by \citet{kbb+00}. The resulting physical projected distribution is
depicted in figure \ref{fig:skyoff} and given in table
\ref{tab:offsets}. The median projected physical offset of the 20 GRBs
in the sample is 1.31 kpc or 1.10 kpc including only those 15 GRBs for
which a redshift was measured.  The minimum offset found is just 91
$\pm$ 90 pc from the host center (GRB 970508).

\subsection{Host-normalized projected offset}
\label{sec:hostnorm}

If GRBs were to arise from massive stars, we would then expect that
the distribution of GRB offsets would follow the distribution of the
light of their hosts.  As can be seen in figure \ref{fig:offset1},
qualitatively this appears to be the case since almost all
localizations fall on or near the detectable light of a galaxy.

The next step in the analysis is to study the offsets but normalized
by the half-light radius of the host.  This step then allows us to
consider all the offsets in a uniform manner.  The half-light radius,
$R_{\rm half}$, is estimated directly from STIS images with
sufficiently high signal-to-noise ratio and in the remaining cases we
use the empirical half-light radius--magnitude relation of
\citet{owd+96}; we use the de-reddened R-band magnitudes found in the
GRB host summaries from \citet{dfk+01a} and \citet{sfc+01}. Table
\ref{tab:offnorm} shows the angular offsets and the effective radius
used for scaling.  Where the empirical half-light radius--magnitude
relation is used, we assign an uncertainty of 30\% to $R_{\rm half}$.

The median of the distribution of normalized offsets is 0.976 (table
\ref{tab:offnorm}).  That this number is close to unity suggests a
strong correlation of GRB locations with the light of the host
galaxies.  The same strong correlation can be graphically seen in
figure \ref{fig:hostnorm} where we find that half of the galaxies lie
inside the half-light radius and the remaining, outside the half-light
radius.  We remark that the effective wavelength of the STIS band-pass
and the ground-based $R$ band correspond to rest-frame UV and thus GRBs
appear to be traced quite faithfully by the UV light which mainly
arises from the youngest and thus massive stars. We will examine the
distribution in the context of massive star progenitors more closely
in \S \ref{sec:compare-collapsars}.

\subsection{Accounting for the uncertainties in the offset measurements}

A simple way to compare the normalized offsets to the expectations of
various progenitor models (see \S \ref{sec:compare}) is through the
histogram of the offsets. However, due to the small number of offsets,
the usual binned histogram is not very informative. In addition, the binned
histogram implicitly assumes that the observables can be represented
by $\delta$-functions and this is not appropriate for our case, in which
several offsets are comparable to the measurement uncertainty.

To this end we have developed a method to construct a probability
histogram (PH) that takes into account the errors on the
measurements. Simply put, we treat each measurement as a probability
distribution of offset (rather than a $\delta$-function) and create a
smooth histogram by summing over all GRB probability distributions.
Specifically, for each offset $i$ we create an individual PH
distribution function, $p_i(r)\,$d$r$, representing the probability of
observing a host-normalized offset $r$ for that burst. The integral of
$p_i(r)\,{\rm d}r$ is normalized to unity. The total PH is then
constructed as $p(r)\,$d$r$ = $\sum_i \, p_i(r)\,{\rm d}r$ and plotted
as a shaded region curve in figure \ref{fig:offset-log}; see Appendix
\ref{sec:osh-derive} for further details.

The total cumulative probability histogram, $\int_0^{r} p(r)\,$d$r$,
is depicted as the solid smooth curve in figure \ref{fig:offset-cum}.
There is, as expected, a qualitative similarity between the cumulative
total PH distribution and the usual cumulative histogram distribution.
 
\section{Testing Progenitor Model Predictions}
\label{sec:compare}

Given the observed offset distribution, we are now in the position to
pose the question: which progenitor models are favored by the data?
Clearly, GRBs as a class do not appear to reside at the centers of
galaxies, and so we can essentially rule out the possibility that {\it
all} GRBs localized to-date arise from nuclear activity.

\subsection{Delayed merging remnants binaries (BH--NS and NS--NS)}

In general, the expected distributions of merging remnant binaries are
found using population synthesis models for high-mass binary evolution
to generate synthetic remnant binaries.  The production rate of such
binaries from other channels (such as three-body encounters in dense
stellar clusters) are assumed to be small relative to isolated binary
evolution. Due to gravitational energy loss, the binary members
eventually coalesce but may travel far from their birth-site before
doing so.  The locations of coalescence are determined by integrating
the synthetic binary orbits in galactic potential models.

\citet{bsp99}, \citet{fwh99} and \citet{bbz99} have simulated the
expected radial distribution of GRBs in this manner.  All three
studies essentially agree on the NS--NS differential offset
distributions as a function of host galaxy
mass\footnotemark\footnotetext{It appears that figure 22 of
\citet{fwh99}, the cumulative distribution of merger sites, is
mislabeled (showing NS--NS merger sites to be at radii a factor of
$\sim$10 times larger than suggested in their differential
distributions and tables).  Accepting the radii given in figure 21 and
table 10 of \citet{fwh99}, all three of the aforementioned studies are
in approximate agreement on the NS--NS merger sites.} The NS--NS
distributions of \citet{bsp99} are slightly more concentrated toward
smaller offset radii than those from \citet{fwh99} and \citet{bbz99};
To account for this, \citet{fwh99} suggested that the
\citet{bsp99} synthesis may have incorrectly predicted an
over-abundance of compact binaries with small merger ages, because the
population synthesis did not include a non-zero helium star radius;
this is not the case, although an arithmetic error in our code may
account for the discrepancy (Sigurdsson, priv.~communication). We
emphasize that, within a factor of a $\sim$two, population synthesis
studies that use the classical channels for NS--NS production, are in
agreement with NS--NS merger sites with respect to host galaxies.

The formation scenarios of BH--NS binaries are less certain than that of
NS--NS binaries.  Both \citet{fwh99} and \citet{bbz00} suggest that
so-called ``hypercritical accretion'' \citep{bb98} dominates the
birthrate of BH--NS binaries.  Briefly, hypercritical accretion occurs
when the primary star evolves off the main sequence and explodes as a
supernova, leaving behind a neutron star.  Mass is rapidly accreted
from the secondary star (in red giant phase) during common envelope
evolution, causing the primary neutron star to collapse to a black
hole.  The secondary then undergoes a supernova explosion leaving
behind a NS. As in NS--NS binary formation, only some BH--NS systems
will remain bound after having received systemic velocity kicks from
two supernovae explosions.  

One important difference is that BH--NS binaries are in general more
massive (total system mass $M_{\rm tot} \approx 5 M_\odot$) than
NS--NS binaries ($M_{\rm tot} \approx 3 M_\odot$).  Furthermore, the
coalescence timescale after the second supernova is shorter than in
NS--NS binaries because of the BH mass.  Therefore, despite similar
evolutionary tracks, BH--NS binaries could be retained more tightly to
host galaxies than NS--NS binaries \citep{bsp99,bbz00}. \citet{bbz00}
quantified this expected trend, showing that on average, BH--NS
binaries merge $\sim$few times closer to galaxies than NS--NS
binaries. Surprisingly, \citet{fwh99} found that BH--NS binaries
merged {\it further} from galaxies than NS--NS binaries, but this
result was not explained by \citet{fwh99}.  Nevertheless, just as with
NS--NS binaries, a substantial fraction of BH--NS binaries will escape
the potential well of the host galaxy and merge well-outside of the
host. For example, even in massive galaxies such as the Milky Way,
these studies show that roughly 25\% of mergers occur $> 100$ kpc from
the center of a host galaxy.

Before comparing in detail the predicted and observed distributions,
it is illustrative to note that the observed distribution appears
qualitatively inconsistent with the delayed merging remnant binaries.
All the population synthesis studies mentioned thus far find that at
approximately 50\% of merging remnants will occur outside of $\approx
10$ kpc when the mass of the host is less than or comparable to the
mass of the Milky Way. Comparing this expectation with
figure~\ref{fig:skyoff}, where no bursts lie beyond 10 kpc from their
host, the simplistic Poisson probability that the observed
distribution is the same as the predicted distribution is no larger
than 2 $\times 10^{-3}$.

To provide a more quantitative comparison of the observed distribution
with the merging remnant expectation, we require a model of the
location probability of GRB mergers about their hosts.  These models,
which should in principle vary from host to host, have a complex
dependence on the population synthesis inputs, the location of star
formation within the galaxies and the dark-matter halo mass.

No dynamical or photometric mass of a GRB host has been reported
to-date. However, since many GRB hosts are blue starbursts
\citepeg{dfk+01a}, it is not unreasonable to suspect that their masses
will lie in the range of 0.001 -- 0.1 $\times$ 10$^{11} M_\odot$
\citepeg{oab+01}. The most obvious exceptions to this are the hosts of
GRB 971214 and GRB 990705 which are likely to be near $L_*$.  The
observed median effective disk scale length of GRB hosts is $r_e =
1.1$ kpc though GRB hosts clearly show a diversity of sizes (table
\ref{tab:offnorm}, col.~5).  This value of $r_e$ is also close to the
median effective scale radii found in the \citet{oab+01} study of
nearby compact blue galaxies.

To compare the observed and predicted distributions, we use galactic
models a--e from \citet{bsp99} corresponding to hosts ranging in mass
from 0.009 -- 0.62 $\times 10^{11} M_\odot$ and disk scale radii
($r_e$) of 1 and 3 kpc.  Following the discussion above, we also
construct a new model ($a^*$) which we consider the most representative
of GRB hosts galaxies with $v_{\rm circ} = 100$ km s$^{-1}$, $r_{\rm
break} = 1$ kpc, and $r_e = 1.5$ kpc ($M_{\rm gal} = 9.2 \times 10^{9}
M_\odot$).

We project these predicted {\it radial} distribution models by
dividing each offset by a factor of $1.15$ since the projection of a
merger site on to the plane of the sky results in a smaller observed
distance to the host center than the radial distance. We determined
the projection factor of 1.15 by a Monte Carlo simulation projecting a
three-dimensional (3-D) distribution of offsets onto the sky. The median
projected offset is thus 87\% of the 3-D radial offset.

\begin{deluxetable}{lcccccccc}
\rotate
\tablewidth{0pt}
\tablecaption{Comparison of Observed Offset Distributions to Various Progenitor model
Predictions\label{tab:nsks}} 
\tablecolumns{9}
\tablehead{
\colhead{Progenitor} & \multicolumn{4}{c}{Comparison Model} & \multicolumn{3}{c}{$P_{KS}$} & \colhead{Fraction of} \\
\colhead{} & \colhead{Name} & \colhead{$r_e$} & \colhead{$v_{\rm circ}$} &
\colhead{Mass ($M_\odot$)} & \colhead{observed} & \colhead{synth} & \colhead{synth (replaced)} & \colhead{ $P_{KS} \ge 0.05$} }
\startdata

Collapsar\ldots& expon. & $R_{\rm half}/1.67$ & & & 0.454 & 0.409 & 0.401 & 0.996 \\
~~/Promptly Bursting Binary & disk  \\
NS--NS, BH--NS\ldots & a$^*$ & 1.5 kpc &  100 km s$^{-1}$ & $9.2 \times 10^9$ & 9.5 $\times 10^{-4}$  & 2.2 $\times 10^{-3}$ & 2.2 $\times 10^{-3}$  & 0.003 \\
                     & a     & 1.0 kpc & 100 km s$^{-1}$ & $9.2 \times 10^{9}$ & 6.9 $\times 10^{-3}$ & 1.7 $\times 10^{-2}$ & 1.8 $\times 10^{-2}$  & 0.139 \\
                     & b     & 1.0 kpc & 100 km s$^{-1}$ & $2.8 \times 10^{10}$ & 4.9 $\times 10^{-3}$& 2.0 $\times 10^{-2}$ &2.0 $\times 10^{-2}$  & 0.243 \\
                     & c     & 3.0 kpc & 100 km s$^{-1}$ & $2.8 \times 10^{10}$ & 3.5 $\times 10^{-5}$& 4.7 $\times 10^{-5}$ &4.7 $\times 10^{-5}$  & 0.001 \\
                     & d     & 1.0 kpc & 150 km s$^{-1}$ & $6.3 \times 10^{10}$ & 2.5 $\times 10^{-2}$& 6.0 $\times 10^{-2}$ &6.3 $\times 10^{-2}$  & 0.553 \\
                     & e     & 3.0 kpc & 150 km s$^{-1}$ & $6.3 \times 10^{10}$ & 5.7 $\times 10^{-5}$& 1.3 $\times 10^{-4}$ &1.3 $\times 10^{-4}$  & 0.001  \\
\enddata

\tablecomments{The names in column 2 correspond to the model
distribution compared against the data, with the letters corresponding
to population synthesis models from \citet{bsp99}.  For the NS--NS,
BH--NS comparisons, we believe that $a^*$ best represents the observed
host galaxy properties (see text).  There are three columns that give
the KS probability that the data are drawn from the model.  The first,
marked ``observed,'' is the direct comparison of the models to the
data without accounting for the uncertainties and measurements in the
data.  The second is the median KS probability derived from our Monte
Carlo modeling (\S \ref{sec:robust}). The third is the median KS
probability derived from our Monte Carlo modeling but where offsets
thrown out due to high $P_{\rm chance}$ are replaced by synthetic
offsets drawn from the model distribution. This pushes $P_{\rm KS}$ to
larger values, in general, but the resulting median of the
distribution is strongly affected (since $P_{\rm chance}$ is near zero
for most offsets).  The last column gives the fraction of synthetic
datasets with $P_{KS} \ge 0.05$.}

\end{deluxetable}

The observed distribution is compared with the predicted distributions
and shown in figure \ref{fig:rem-comp1} (later, in
fig.~\ref{fig:sfr-comp1}, we compare the observed distribution with the
massive star prediction).  We summarize the results in table
\ref{tab:nsks}. Only model $d$ ($M = 6.3 \times 10^{10} M_\odot$, $r_e
= 3$ kpc) could be consistent with the data ($P_{\rm KS} = 0.063$), but
this galactic model has a larger disk and is probably more massive
than most GRB hosts.  Instead, for the ``best bet'' model $a^*$, the
one-sided Kolmogorov-Smirnov probability that the observed sample
derives from the same predicted distribution is $P_{\rm KS} =
2.2\times 10^{-3}$, in agreement with our simplistic calculation
above; that is, the location of GRBs appears to be inconsistent with
the NS--NS and NS--BH hypothesis.

\begin{figure*}[tbp]
\centerline{\psfig{file=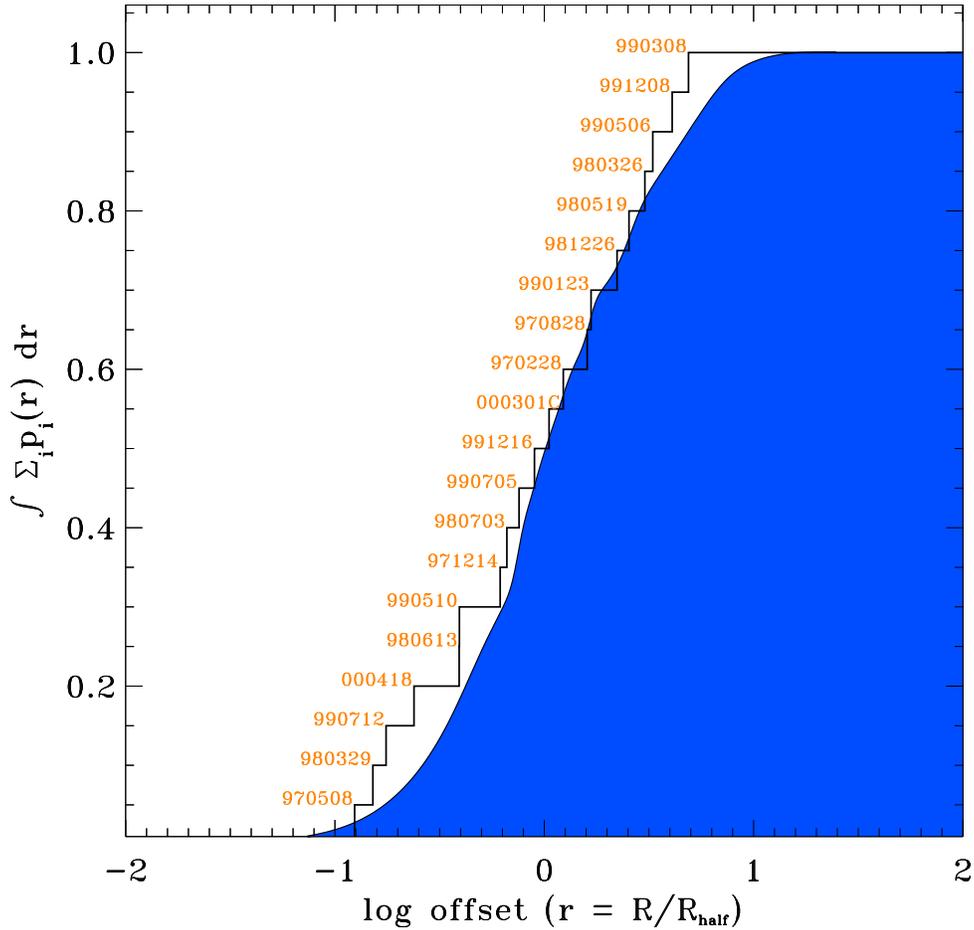,height=5.3in,angle=0}}
\caption[The cumulative GRB offset distribution as a
function of host half-light radius]{The cumulative GRB offset
distribution as a function of host half-light radius.  The solid
jagged line is the data in histogram form. The smooth curve is the
probability histogram (PH) constructed with the formalism of Appendix
\ref{sec:osh-derive} and is the integral of the curve depicted in
figure \ref{fig:offset-log}. The GRB identifications are noted
alongside the solid histogram. In this figure and in figure
\ref{fig:offset-log}, SN 1998bw/GRB 980425 has not been included.}
\label{fig:offset-cum}
\end{figure*}

If GRBs do arise from systems which travel far from their birthsite,
then there is a subtle bias in determining the offset to the host.  If
the progenitors are ejected from the host by more than half the
distance between the host and the nearest (projected) galaxy, then the
transient position will appear unrelated to any galaxies (the wrong
host will be assigned, of course) but $P_{\rm chance}$ will always
appear high no matter how deep the host search is.  We try to account
for this effect in our modeling (Appendix \ref{sec:robust}) by
synthetically replacing observed (small) offsets that are associated
with a high value of $P_{\rm chance}$ with new, generally larger,
offsets drawing from the expected distribution of offsets for a
particular galactic model.  This then biases the distribution of
$P_{KS}$ statistics toward {\it higher} values (by definition), but
the median values of $P_{KS}$ are largely unaffected (see table
\ref{tab:nsks}).

\subsection{Massive stars (collapsars) and promptly bursting
binaries (BH--He)}
\label{sec:compare-collapsars}

As discussed earlier, collapsars produce GRBs in star-forming regions,
as will BH--He binaries. The localization of GRB 990705 near a spiral
arm is, of course, tantalizing smaller-scale evidence of the
GRB--star-formation connection.  Ideally, the burst sites of
individual GRBs could be studied in detail with imaging and
spectroscopy and should, if the collapsar/promptly bursting binary
origin is correct, reveal that the burst sites are HII regions.
Unfortunately, the distances to GRBs preclude a detailed examination
of the specific burst sites on a resolution scale of tens of parsecs
(the typical size for a star-forming region) with current
instrumentation. Adaptive optics laser-guide star imaging may prove
quite useful in this regard as will IR imaging with the {\it Next
Generation Space Telescope}.

Weaker evidence for a star-formation connection exists in that no GRB
to date has been observed to be associated with an early-type galaxy
(morphologically or spectroscopically), though in practice it is often
difficult to discern galaxy type with the data at hand.  Indeed most
well-resolved hosts appear to be compact star forming blue galaxies,
spirals, or morphological irregulars.

Above we have demonstrated that GRBs follow the UV (restframe) light
of their host galaxies. However, the comparison has been primarily
mediated by a single parameter, the half-light radius and the median
normalized offset. We now take this comparison one step further. For
the GRB hosts with high signal-to-noise HST detections (e.g.,~GRB
970508, GRB 971214, GRB 980703), our analysis shows that the surface
brightness is well-approximated by an exponential disk. We use this
finding as the point of departure for a simplifying assumption about
all GRB hosts: we assume an exponential disk profile such that the
surface brightness of the host galaxy scales linearly with the
galactocentric radius in the disk. We further assume that the star
formation rate of massive stars scales with the observed optical light
of the host; this is not an unreasonable assumption given that
HST/STIS imaging probes restframe UV light, an excellent tracer of
massive stars, at GRB redshifts.

Again, clearly not all host galaxies are disk-like
(figure~\ref{fig:offset1}), so this assumption is not strictly valid
in all cases. If $r_e$ is the disk scale length, the half-light radius
of a disk galaxy is $R_{\rm half} = 1.67 \times r_e$, so that the
simplistic model of the number density of massive star-formation
regions in a galaxy is
\begin{equation}
N(r)\, dr \propto r \exp(-1.67\, r)\, dr,
\label{eq:sfr}
\end{equation}
where $r = R/R_{\rm half}$.  In reality, the distribution of massive
star formation in even normal spirals is more complex, with a strong
peak of star formation in the nuclear region and troughs between
spiral arms \citepeg{rw86,bdd89}. We make an important assumption when
comparing the observed distribution with the star-formation disk
model: that each GRB occurs in the disk of its host (see discussion
below).  Dividing the observed offset by the apparent half-light
radius host essentially performs a crude de-projection.

We find the probability that the observed distribution could be
derived from the simplistic distribution of massive star regions
(equation~\ref{eq:sfr}) is $P_{\rm KS} = 0.454$ (i.e.,~the two distributions
are consistent).  In Appendix \ref{sec:robust} we show that these
results are robust even given the measurement uncertainties.  This
broad agreement between GRB positions and the UV light of their hosts
is remarkable in the sense that the model for massive star locations
is surely too simplistic; even in classic spiral galaxies (which most
GRB hosts are not) star-formation is a complex function of
galactocentric radius, with peaks in galactic centers and spiral arms.
Furthermore, surface brightness dimming with redshift causes galaxies
to appear more centrally peaked, resulting in a systematic
underestimate of $R_{\rm half}$.

\begin{figure*}[tp]
\centerline{\psfig{file=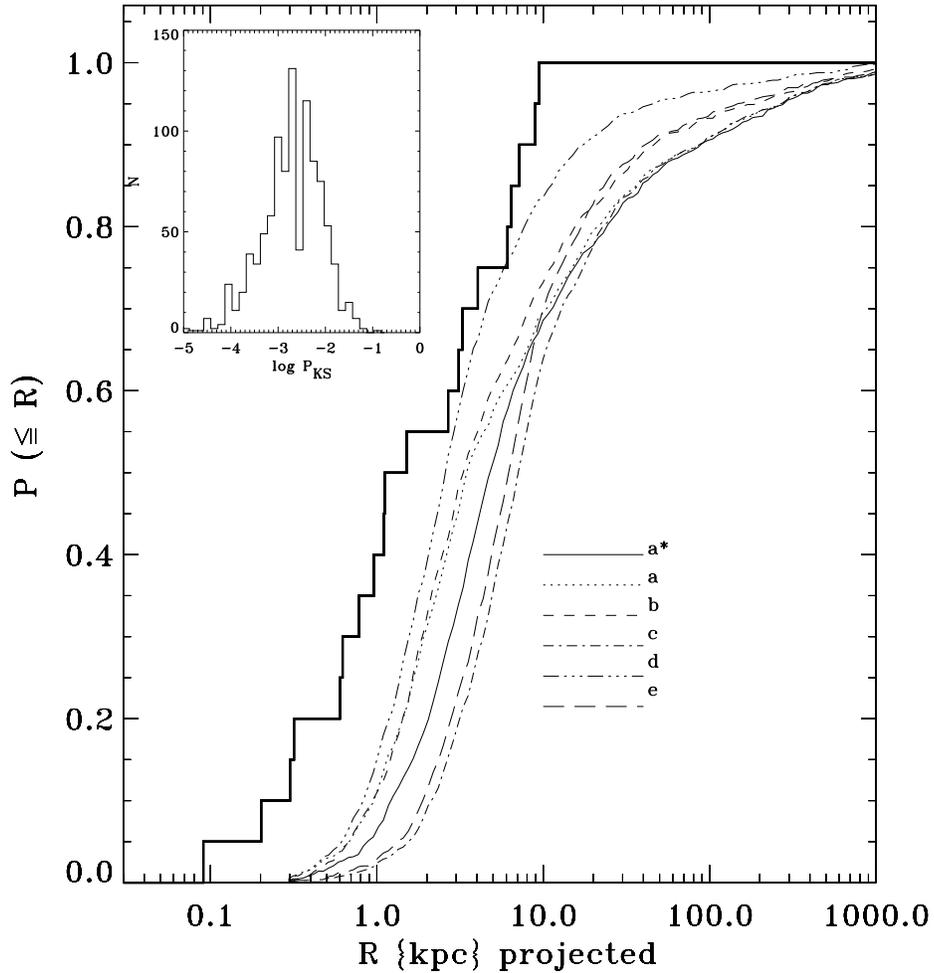,width=5.2in,angle=0}}
\caption[Offset distribution of GRBs compared with delayed
merging remnant binaries (NS--NS and BH--NS) prediction]{Offset
distribution of GRBs compared with delayed merging remnant binaries
(NS--NS and BH--NS) prediction. The models, depicted as smooth curves,
are the radial distributions in various galactic systems that have
been projected by a factor of 1.15 (see text).  The letters denote the
model distributions from table 2 of
\citet{bsp99}; $a^*$ is the galactic model which we consider as the
most representative of GRB hosts galaxies ($v_{\rm circ} = 100$ km
s$^{-1}$, $r_{\rm break} = 1$ kpc, $r_e = 1.5$ kpc, $M_{\rm gal} = 9.2
\times 10^{9} M_\odot$).  The cumulative histogram is the observed
data set.  Inset is the distribution of KS statistics (based on the
maximum deviation from the predicted and observed distribution) of
1000 synthetic data sets compared with model $a^*$.  Even with
conservative assumptions (see text) the observed GRB distribution is
inconsistent with the prediction: in only 0.3\% of synthetic datasets
is $P_{\rm KS} \ge 0.05$. Instead, the collapsar/promptly bursting
remnant progenitor model appears to be a better representation of the
data (see figure~\ref{fig:sfr-comp1}).}
\label{fig:rem-comp1}
\end{figure*}

\section{Discussion and Summary}
\label{sec:offsum}

We have determined the observed offset distribution of GRBs by
astrometrically comparing localizations of GRB afterglows with optical
images of the field surrounding each GRB. In all cases, the GRB
location appears ``obviously'' associated with a galaxy---either
because the position is superimposed atop a galaxy or very near ($\ale
1.2\arcsec$) a galaxy in an otherwise sparse field.  In fact,
irrespective of the validity of individual assignments of hosts, the
offset distribution may be considered a distribution of GRB positions
from the nearest respective galaxy at least as bright as $R \approx
28$ mag (note that in most cases the host galaxies are much brighter,
typically $R = 24$--26 mag). We find that at most a few of the 20 GRBs
could be unrelated physically to their assigned host and about a 50\%
chance that all GRBs are correctly assigned to their hosts (see \S
\ref{sec:angoffs}).

We then compare the distribution of GRB locations about their
respective hosts with the {\it predicted} radial offset distribution
of merging binary remnants. This comparison is complicated by an
unknown projection factor for each burst: if a GRB occurs near an
edge-on disk galaxy there exists no model-independent manner to
determine the true 3-D radial offset of the GRB from the center of the
host.  Indeed, in a few cases (e.g., GRB 980519, GRB 991216) even the
``center'' of the host is not well defined and we must estimate a
center visually.  In all other cases, we find the centers using a
luminosity-weighted centroid surrounding the central peak of the
putative host.

To compare the GRB offsets with those predicted by the NS--NS and
NS--BH binary models, we make a general assumption about the
projection factor and, to facilitate a comparison in physical units
(that is, offsets in kiloparsec rather than arcseconds), we assign an
angular diameter distance to the 5 hosts without a confirmed distance
(\S \ref{sec:physproj}).  We have shown that the conversion of an
angular offset to physical projection is relatively insensitive to the
actual redshift of the host. We estimate that the probability that the
observed GRB offset distribution is the same as the predicted
distribution of NS--NS and BH--NS binaries is $P \ale 2 \times
10^{-3}$.  Insofar as the observed distribution is representative (see
below) and the predicted distribution is accurate, our analysis
renders BH--NS and NS--NS progenitor scenarios unlikely for
long-duration GRBs.

Having cast doubt on the merging remnant hypothesis, we test whether
the offset distribution is consistent with the collapsar (or BH--He)
class.  Since massive stars (and promptly merging binaries) explode
where they are born, we have compared the observed GRB offset
distribution with a very simplistic model of massive star formation in
late-type galaxies: an exponential disk.  After normalizing each GRB
offset by their host half-light radius we compare the distribution
with a KS test and find good agreement: $P_{\rm KS} = 0.454$. We have
shown that these KS results, based on the assumption of
$\delta$-function offsets, are robust even after including the
uncertainties in the offset measurements.

Thus far we have neglected discussion of the observational biases that
have gone into the localizations of these 20 GRBs.  The usual problems
plaguing supernova detection, such as the brightness of the central
region of the host and dust obscuration, are not of issue for
detection of the {\it prompt} high-energy emission (i.e., X-rays
and $\gamma$-rays) of GRBs since the high-energy photons penetrate
dust.  If the intrinsic luminosity of GRBs is only a function of the
inner-workings of the central engine (that is, GRBs arise from
internal shocks and not external shocks) then the luminosity of a GRB
is independent of ambient number density.  Therefore, prompt X-ray
localizations from BeppoSAX and $\gamma$-ray locations from the IPN
should not be a function of the global properties of GRB environment;
only intrinsic GRB properties such as duration and hardness will
affect the prompt detection probability of GRBs.

The luminosity of the afterglows is, however, surmised to be a function
of the ambient number density. Specifically, the afterglow luminosity
will scale as $\sqrt{n}$ where $n$ is the number density of hydrogen
atoms in the 1--10 pc region surrounding the GRB explosion site
\citep[see][]{mrw98b}.  While $n \approx$ 0.1--10 cm$^{-3}$ in the
interstellar medium, the ambient number density is probably $n \approx
10^{-4}$--10$^{-6}$ in the intergalactic medium.  Thus GRB afterglows
in the IGM may appear $\sim 10^{-3}$ times fainter than GRB afterglows
in the ISM (and even more faint compared to GRBs that occur in
star-forming regions where the number densities are higher than in the
ISM). If only a small fraction of GRBs localized promptly in X-rays
and studied well at optical and radio wavelengths were found as
afterglow, the ambient density bias may be cause for concern. However,
this is not the case. As of June 2001, 29 of 34 bursts localized by
prompt emission were later found as X-ray, optical, and/or radio
afterglow \citepcf{fkw+00}; that is, almost all GRBs have detectable
X-ray afterglow.  Therefore, {\it no more than about $10\%$ of GRBs
localized by BeppoSAX could have occurred in significantly lower
density environments} such as in the IGM; thus, we do not believe that
our claim against the delayed merging binaries is affected by this
bias.
\begin{figure*}[tp]
\centerline{\psfig{file=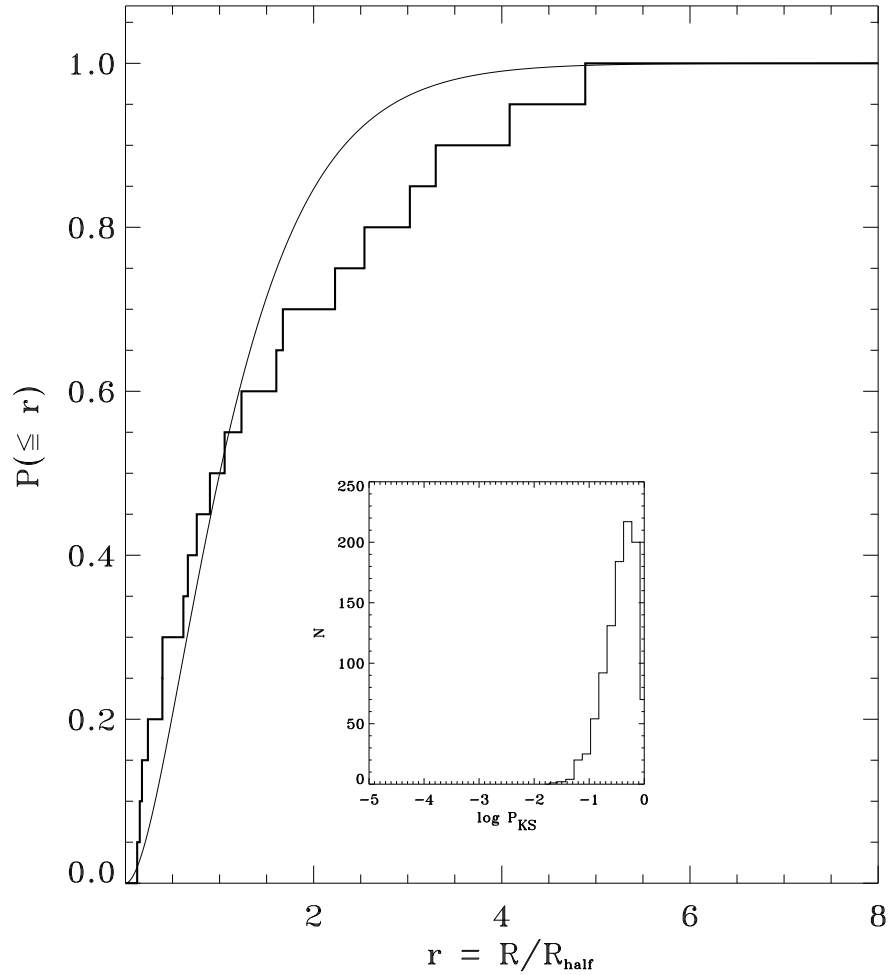,width=5.2in,angle=0}}
\caption[GRB Offset distribution compared with host
galaxy star formation model]{Offset distribution of GRBs compared with
host galaxy star formation model. The model, an exponential disk, is
shown as the smooth curve and was chosen as an approximation to the
distribution of the location of collapsars and promptly bursting
remnant binaries (BH--He). The cumulative histogram is the observed
data set. Inset is the distribution of KS statistics (based on the
maximum deviation from the predicted and observed distribution) of
1000 synthetic data sets.  Since the observed KS statistic is near the
median in both cases, we are assured that errors on the measurements
do not bias the results of the KS test, and therefore the KS test is
robust. The observed GRB distribution provides a good fit to the model
considering we make few assumptions to perform the comparison.  In
reality the location of star formation in GRB hosts will be more
complex than a simple exponential disk model.}
\label{fig:sfr-comp1}
\end{figure*}

What about the non-detection of GRB afterglow at optical/radio
wavelengths?  Roughly half of GRBs promptly localized in the gamma-ray
or X-ray bands are not detected as optical or radio afterglow
\citep{fkw+00,lcg01}.  While many of these ``dark'' GRBs must be due
to observing conditions (lunar phase, weather, time since burst, etc.)
at least some fraction may be due to intrinsic extinction local to the
GRB.  If so, then these GRBs are likely to be centrally biased since
the optical column densities are strongest in star-forming regions and
giant molecular clouds.  Therefore, {\it any optically obscured GRBs
which do not make it in to our observed offset sample will be
preferentially located in the disk}. We do not therefore believe the
ambient density bias plays any significant role in causing GRBs to be
localized preferentially closer to galaxies; in fact, the opposite may
be true.

The good agreement between our simplistic model for the location of
massive stars and the observed distribution is one of the strongest
arguments yet for a collapsar (or promptly bursting binaries) origin
of long-duration GRBs. However, the concordance of the predicted and
observed distributions are necessary to prove the connection, although
not sufficient. 

We may now begin to relate the offsets to the individual host and GRB
properties. For instance, of the GRBs which lie in close proximity to
their host centers (GRB 970508, GRB 980703, and GRB 000418), there is
a striking similarity between their hosts---all appear compact and
blue with high-central surface brightness suggesting that these hosts
are nuclear starburst galaxies (none show spectroscopic evidence for
the presence of an AGN).  

In fact, the closeness of some GRBs to their host centers signifies
that our simplistic model for star-formation may require modification.
This is not unexpected since, in the Galaxy, star formation as a
function of Galactocentric radius does not follow a pure exponential
disk, but is vigorous near the center and is strongly peaked around $R
\sim 5$ kpc \citepcf{ken89}.  As more accurate offsets are amassed,
these subtle distinctions in the GRBs offset distribution may be
addressed.

\acknowledgments

The authors thank the staff of the W.~M.~Keck Foundation and the staff
at the Palomar Observatories for assistance.  We thank the anonymous
referee for very insightful comments; in particular, the referee
pointed out (and suggested the fix for) the bias in offset assignment
if GRBs are ejected far from their host galaxies. The referee also
pointed out that we did not establish that GRBs are preferentially
aligned with the major axes of their hosts (as we had claimed in an
earlier version of the paper). We applaud E.~Berger and D.~Reichart
for close reads of various drafts of the paper.  We also thank
M.~Davies for encouraging us to compare several NS--NS models to the
data rather than just one.  We acknowledge the members of the
Caltech-NRAO-CARA GRB collaboration and P.~van Dokkum, K.~Adelberger,
and R.~Simcoe for helpful discussions. We thank N.~Masetti for
allowing us access to early ground-based data on GRB 990705 and
B.~Schaefer for kindly providing {\it QUEST} images of the afterglow
associated with GRB 990308. This work was greatly enhanced by the use
of data taken as part of the {\it A Public Survey of the Host Galaxies
of Gamma-Ray Bursts} with HST (\#8640; S.~Holland, P.I.). JSB
gratefully acknowledges the fellowship and financial support from the
Fannie and John Hertz Foundation.  SGD acknowledges partial support
from the Bressler Foundation. SRK acknowledges support from NASA and
the NSF.  The authors wish to extend special thanks to those of
Hawaiian ancestry on whose sacred mountain we are privileged to be
guests. Without their generous hospitality, many of the observations
presented herein would not have been possible.

\begin{subappendices}

\section{Potential Sources of Astrometric Error}
\label{sec:errors}

\subsection{Differential chromatic refraction}

Ground-based imaging always suffers from differential chromatic
refraction (DCR) introduced by the atmosphere.  The magnitude of this
refraction depends strongly ($\propto 1/\lambda_{\rm eff}^2$) on the
effective wavelength ($\lambda_{\rm eff}$) of each object, the airmass
of the observation, and the air temperature and pressure.  With
increasing airmass, images are dispersed by the atmosphere and
systematically stretched in the parallactic direction in the sense
that bluer objects shift toward the zenith and redder objects shift
toward the horizon.  Other sources of refraction, such as turbulent
refraction \citepeg{lin80}, are statistical in nature and will only
serve to increase the uncertainty in our astrometric solution.

Here we show that DCR, in theory, will not dominate our offset
determinations.  Since all of our early ground-based imaging were
conducted with airmass ($\sec(z)$) $\ale 1.6$, we take as an extreme
example an image with airmass $\sec (z) = 2$, where $z$ is the
observed zenith angle. It is instructive to determine the scale of
systematic offset shifts introduced when compared with either
late-time ground-based or HST imaging where refractive distortions are
negligible.  Following \citet{gt98}, the differential angular
distortion between two point sources at an apparent angular separation
along the zenith, $\Delta z$, may be broken into a color and a zenith
distance term.  Assuming nominal values for the altitude of the Keck
Telescopes on Mauna Kea, atmospheric temperature, humidity and
pressure, at an effective wavelength of the $R$-band filter,
$\lambda_{\rm eff} (R) = 6588$ \AA\ \citep{fig+96}, the zenith
distance term is 16 mas for an angular separation of 30 arcsec at an
airmass of $\sec (z) = 2$.  The zenith term is approximately linear in
angular distance and so, in practice, even this small effect will be
accounted for as a first-order perturbation to the overall rotation,
translation, and scale mapping between a Keck and HST image.  In other
words, we can safely neglect the zenith term contribution to the DCR.

We now determine the color term contribution. Optical transients of
GRBs are, in general, redder in appearance (apparent $V - R \approx
0.5$ mag) than their host galaxies (apparent $V - R \sim 0.2$ mag).
We assume the average astrometric tie object has $V-R = 0.4$ mag.  If
the OT is observed through an airmass of $\sec(z) = 2$ and then the
galaxy is observed at a later time through and airmass of, for
example, $\sec(z) = 1.2$, then DCR will induce a $\sim 30$ mas
centroid shift between the OT and the host galaxy if the two epochs
are observed in $B$-band \citep[see~figure~2 of][]{aaa+99}.  In
$R$-band, the filter used in almost all of our ground-based imaging
for the present work, the DCR strength is about 20\% smaller than in
$B$-band because of the strong dependence of refraction on
wavelength. Therefore we can reasonably assume that DCR should only
{\it systematically} affect our astrometric precision at the 5---10
mas level.  Such an effect could, in principle, be detected as a
systematic offset in the direction of the parallactic angles of the
first epochs of GRB afterglow observations. In \S \ref{sec:angoffs} we
claim that no such systematic effect is present in our data. DCR could
of course induce a larger {\it statistical} scatter in the uncertainty
of an astrometric transformation between epochs since individual tie
objects are not, in general, the same color and each will thus
experience its own DCR centroid shift.

Bearing in mind that DCR is probably negligible we can minimize the
effects of DCR by choosing small fields and similar spectral responses
of the offset datasets.  The HST fields are naturally small and there
are enough tie stars when compared with deep ground-based imaging.
However, since the spectral response of the HST/STIS CCD is so broad,
extended objects with color gradients will have different apparent
relative locations when compared with our deep ground-based $R$-band
images.  As such, in choosing astrometric tie objects, we pay
particular attention to choosing objects which appear compact
(half-light radii $\ale$ 0\arcsec.3) on the STIS image.

\subsection{Field distortion}
Optical field distortion is another source of potential error in
astrometric calibration.  Without correcting for distortion in STIS,
the maximum distortion displacement (on the field edges) is $\sim 35$
mas \citep{mb97}.  This distortion is corrected to a precision at the
sub-milliarcsec level on individual STIS exposures with IRAF/DITHER
\citep{mb97}. \citet{mb97} also found that the overall plate scale
appears to be quite stable with r.m.s.~changes at the 0.1\% level.  We
confirmed this result by comparing two epochs of imaging on GRB 990510
and GRB 970508 which span about 1 year. The relative plate scale of
the geometric mapping between final reductions was unity to within
0.03\%.

We do not correct for optical field distortion before mapping
ground-based images to HST.  While there may be considerable distortion
($\sim $ few $\times 100$ mas) across whole ground-based CCD images, these
distortions are correlated on small scales.  Therefore, when mapping a
50 $\times$ 50 arcsec$^2$ portion of a Keck image with an HST image,
the intrinsic differential distortions in the Keck image tend to be
small ($\ale 30$--50 mas).  Much of the distortion is accounted for in
the mapping by the higher-order terms of the fit, and any residual
differential distortions simply add scatter to the mapping
uncertainties.

\section{Derivation of the Probability Histogram (PH)}
\label{sec:osh-derive}

\def\xp{{\ensuremath{x}}}
\def\yp{{\ensuremath{y}}}
\def\rp{{\ensuremath{r}}}
\def\dxp{{\ensuremath{{\rm d}}\ensuremath{\xp}}}
\def\dyp{{\ensuremath{{\rm d}}\ensuremath{\yp}}}

Histogram binning is most informative when there are many more data
points than bins and the bin sizes are much larger than the errors on
the individual measurements.  Unfortunately, the set of GRB offsets is
contrary to both these requirements. We require a method to display
the data as in the traditional histogram, but where the errors on the
measurements are accounted for.  Instead of representing each
measurement as a $\delta$-function, we will represent each measurement
as a probability distribution as a function of offset.

What distribution function is suitable for offsets?  When the offset
is much larger than the error, then the probability that the burst
occurred at the measured displacement should approach a
$\delta$-function.  When the offset is much larger than zero, then the
probability distribution should appear essentially Gaussian (assuming
the error on the measurement is Gaussian).  However, when the observed
offset is small and the error on the measurement non-negligible with
respect to the observed offset, the probability distribution is
decidedly non-Gaussian since the offset is a positive quantity.  The
distribution we seek is similar to the well-known Rice distribution
\citepcf{wax54}, only more general.

We derive the probability histogram (PH) as follows.  For each GRB
offset, $i$, we construct an individual probability distribution
function $p_i(r)\,$d$r$ of the host-normalized offset ($r_i$) of the
GRB given the observed values for $X_{0,i}$, $Y_{0,i}$ and host
half-light radius $R_{i,{\rm half}}$ and the associated uncertainties.
To simplify the notation in what follows, we drop the index $i$ and
let all lower case parameters represent dimensionless numbers; for
example, the value $x_{0} = X_{0}/R_{\rm half}$, where $R_{\rm half}$
is the host half-light radius.  Without loss of generality, we can
subsume (by quadrature summation) the uncertainties in the host
center, the astrometric transformation, and the GRB center into the
error contribution in each coordinate. We assume that these
statistical coordinate errors are Gaussian distributed with
$\sigma_{\xp}$ and $\sigma_{\yp}$ with, for example,
$$
\sigma_{\xp} = \frac{X_0}{R_{\rm half}} \sqrt{ \frac{\sigma_{X_0}^2} {X_0^2} +
  \frac{\sigma_{R_{\rm half}}^2}{R_{\rm half}^2}}.
$$
Therefore, we can construct the probability $p(\xp,\yp)\,\dxp \, \dyp$
of the true offset at some distance $\xp$ and $\yp$ from the measured
offset location ($x_0, y_0$):
\begin{equation}
p(\xp,\yp)\,\dxp \, \dyp = 
	\frac{1}{2\pi\sigma_{\xp}\,\sigma_{\yp}}\,
	\exp\left[-\frac{1}{2}\left(\frac{\xp^2}{\sigma_{\xp}^2}
	+ \frac{\yp^2}{\sigma_{\yp}^2}\right)\right]\dxp \, \dyp,
\label{eq:pxy}
\end{equation}
assuming the errors in the $\xp$ and $\yp$ are uncorrelated. This is a
good approximation since, while the astrometric mappings generally
include cross-terms in $X$ and $Y$, these terms are usually small.  If
$\sigma_\xp = \sigma_\yp$, then equation~\ref{eq:pxy} reduces to the
Rayleigh distribution in distance from the observed offset, rather
than the host center.

\begin{figure*}[tbh]
\centerline{\psfig{file=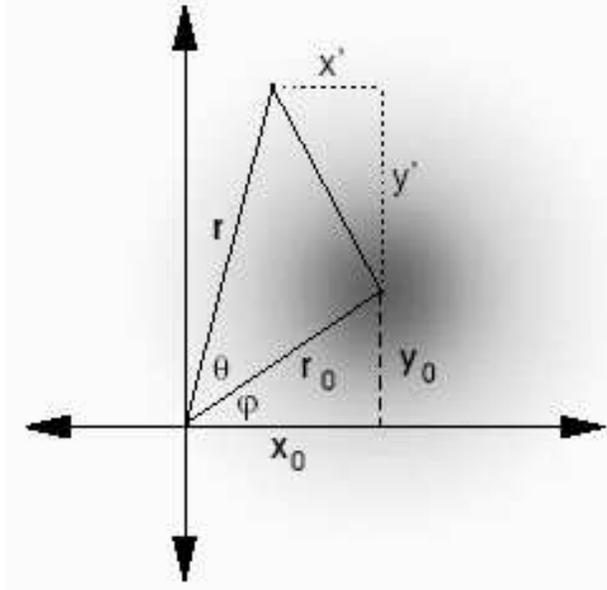,width=3.2in,angle=0}}
\caption[Geometry for the offset distribution probability calculation
in Appendix \ref{sec:osh-derive}]{Geometry for the offset distribution probability calculation
in Appendix \ref{sec:osh-derive}. }
\label{fig:proboff}
\end{figure*}

The probability distribution about the host center is found with an
appropriate substitution for $\xp$ and $\yp$ in equation~\ref{eq:pxy}.  In
figure \ref{fig:proboff} we illustrate the geometry of the
problem. The greyscale distribution shows $p(\xp,\yp)\,\dxp \, \dyp$
about the offset point $x_0$ and $y_0$.  Let $\phi = \tan^{-1}
(y_0/x_0)$ and transform the coordinates in equation~\ref{eq:pxy} using
$\psi = \phi + \theta$, $\xp = r\,\cos \psi - x_0$, and $\yp = r\,\sin
\psi - y_0$.  The distribution we seek, the probability that the true
offset lies a distance $r$ from the host center, requires a
marginalization of $\int_\psi p_i(r,\psi)\,{\rm d}r\,{\rm d} \psi$
over $\psi$,
\begin{eqnarray}
p_i(r)\,{\rm d}r &=& \int_\psi p_i(r,\psi)\,{\rm d}r\,{\rm d} \psi
       \nonumber \\
       &=&  \frac{J\, {\rm d}r}{2\pi\sigma_{\xp}\,\sigma_{\yp}}\,
	\int_{0}^{2\pi}
	\exp\left[-\frac{1}{2}\left(\frac{\xp(r,\psi)^2}{\sigma_{\xp}^2}
	+ \frac{\yp(r,\psi)^2}{\sigma_{\yp}^2}\right)\right]
       \,{\rm d} \psi\label{eq:pr1},
\end{eqnarray} 
finding $J = r$ as the Jacobian of the coordinate transformation.  In
general, equation \ref{eq:pr1} must be integrated numerically using
the observed values $x_0$, $y_0$, $\sigma_{\xp}$, and $\sigma_{\yp}$.
The solution is analytic, however, if we assume that $\sigma_\xp
\rightarrow \sigma_\rp$ and $\sigma_\yp \rightarrow \sigma_\rp$, so
that
\begin{eqnarray}
p_i(r)\,{\rm d}r &\approx& \frac{r}{\pi \sigma_{\rp}^2}\,
   \exp\left[-\frac{r^2 + r_0^2}{2\sigma_\rp^2}\right] \,
	\int_{\theta = 0}^{\pi}\exp \left[\frac{r\,r_0\,\cos
   \theta}{\sigma_\rp}\right] {\rm d} \, \theta {\rm d}r  \nonumber \\
		 &\approx& \frac{r}{ \sigma_{\rp}^2}\,
   \exp\left[-\frac{r^2 + r_0^2}{2\sigma_{\rp}^2}\right] \, 
           I_0\left(\frac{r\,r_{0}}{\sigma^2_{\rp}}\right) {\rm d} r,
	\label{eq:pr2}
\end{eqnarray}
where $I_0(x)$ is the modified Besel function of zeroth order and $r_0
= \sqrt{x_0^2 + y_0^2}$. 

The equation \ref{eq:pr2} is readily recognized as the Rice
distribution and is often used to model the noise characteristics of
visibility amplitudes in interferometry; visibility amplitudes, like
offsets, are positive-definite quantities.  Only when $\sigma_\xp =
\sigma_\yp = \sigma_r$ is the probability distribution exactly a Rice
distribution, which is usually the case for interferometric
measurements since the real and imaginary components of the fringe phasor
have the same r.m.s.

\begin{figure*}
\centerline{\psfig{file=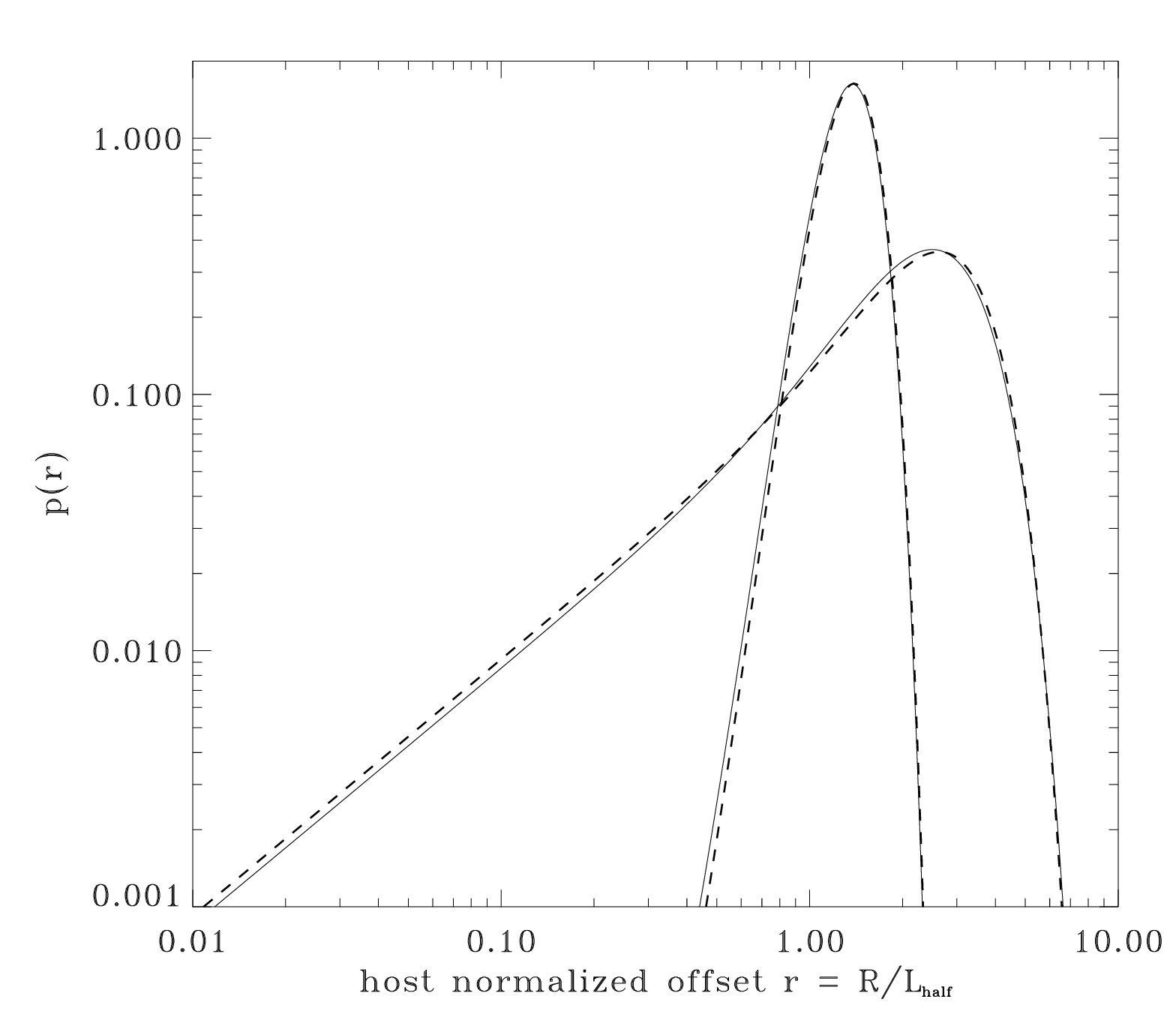,width=3.2in}}
\caption[Example offset distribution functions $p(r)$]{Example offset distribution functions $p(r)$. Depicted are
two probability distribution curves for ($X_0$, $Y_0$, $\sigma_{X_0}$,
$\sigma_{Y_0}$, $R_{\rm half}$, $\sigma_{R_{\rm half}}$) =
[0\arcsec.033, 0\arcsec.424, 0\arcsec.034, 0.\arcsec034, 0\arcsec.31,
0\arcsec.05] (GRB 970228) and [0\arcsec.616, 0\arcsec.426,
0\arcsec.361, 0\arcsec.246, 0\arcsec.314, 0\arcsec.094] (GRB 981226)
for the lower and upper peaked distributions, respectively. The solid
line is the exact solution (equation~\ref{eq:pr1}) and the dashed line is
the approximate solution (equation~\ref{eq:pr2}). Here, as in the text, the
host-normalized offset $r = R/R_{\rm half}$, where $R$ is the
galactocentric offset of the GRB from the host and $R_{\rm half}$ is
the half-light radius of the host.}
\label{fig:probexam}
\end{figure*}

Equation \ref{eq:pr1} is a generalized form of the Rice distribution
but can be approximated as a Rice distribution by finding a suitable
value for $\sigma_r$.  We find that by letting,
\begin{equation}
\sigma_\rp = \frac{1}{r_0}\sqrt{ \left(x_0\, \sigma_{\xp}\right)^2 +
	\left(y_0\, \sigma_{\yp}\right)^2},
\label{eq:sigmar}
\end{equation}
equation \ref{eq:pr2} approximates (to better than 30\%) the exact
form of the probability distribution in equation~\ref{eq:pr1} as long as
$\sigma_{\xp} \ale 2\,\sigma_{\yp}$ (or vise versa). In figure
\ref{fig:probexam} we show two example offset probability
distributions in exact and approximate form.  Note that $r_0 -
\sigma_r \le r \le r_0 + \sigma_r$ is not necessarily the 68\% percent
confidence region of the true offset since the probability
distribution is not Gaussian. The exact form is used to construct the
data representations in figures
\ref{fig:offset-cum}--\ref{fig:sfr-comp1}.

\section{Testing the Robustness of the KS Test}
\label{sec:robust}

How robust are the estimates of probabilities found comparing the
observed distribution and the predicted progenitor offset
distributions?  Since there are different uncertainties on each offset
measurement, the KS test is not strictly the appropriate statistic to
determine the likelihood that the observed distribution could be drawn
from the same underlying (predicted) distribution.  One possibility is
to construct synthetic sets of observed data from the model using the
observed uncertainties.  However, a small uncertainty (say 0.2 arcsec
in radius observed to be paired with an equally small offset) which is
randomly assigned to a large offset from a Monte Carlo distribution
has a different probability distribution then if assigned to a small
offset (since the distribution in $r$ is only physical for positive
$r$).  Instead, we approach the problem from the other direction by
using the data themselves to assess the range in KS statistics given
our data. We construct $k=1000$ synthetic cumulative physical offset
distributions using the smoothed probability offset distributions
$p_i(r)\,$d$r$ for each GRB.  As before $r$ is the offset in units of
host half-light radius.  For each simulated offset distribution $k$, we
find a set \{$r_i$\}$_k$ such that
$$
P[0,1] = \frac{\int_0^{r_i}  p_i(l)\,dl}{\int_0^{\infty}
p_i(l)\,dl},
$$
where $P[0,1]$ is a uniform random deviate over the closed interval
[0,1].  In addition, since some of the host assignments may be
spurious chance superpositions, we use the estimate of $P_{\rm
chance}$ (\S \ref{sec:angoffs}; table \ref{tab:offnorm}) to
selectively remove individual offsets from a given Monte Carlo
realization of the offset dataset. GRBs with relatively secure host
assignments remain in more realizations than those without.  So, for
instance, the offset of GRB 980703 ($P_{\rm chance}$ = 0.00045) is
used in all realizations but the offset of GRB 970828 ($P_{\rm
chance}$ = 0.07037) is retained in only 93\% of the synthetic
datasets.

We evaluate the KS statistic as above for each synthetic set and
record the result. Figure \ref{fig:sfr-comp1} depicts the cumulative
probability distribution compared with the simple exponential disk
model.  The inset of the figure shows the distribution of KS
statistics for the set of synthetic cumulative distributions
constructed as prescribed above.  In both cases, as expected, the {\it
observed} KS probability falls near the median of the synthetic
distribution.  The distribution of KS statistics is not significantly
affected by retaining all GRB offsets equally (that is, assuming
$P_{\rm chance}$ = 0.0 for every GRB offset).  In table \ref{tab:nsks}
we present the result of the Monte Carlo modeling.  Using this
distribution of KS statistics we can now assess the robustness of our
comparison result: given the data and their uncertainties, the
probability that the observed GRB offset distribution is the same as
the model distribution of star formation (exponential disk) is $P_{\rm
KS} \ge 0.05$ in 99.6\% of our synthetic datasets.

\end{subappendices}

\mypart{The GRB/Supernova Connection}{The GRB/Supernova Connection}{
\begin{singlespace}
\begin{quote}
{\small I never cared much for moonlit skies \\
I never wink back at fireflies \\
But now that the stars are in your eyes \\
I'm beginning to see the light \\
\medskip
I never went in for afterglow \\
Or candlelight on the mistletoe \\
But now when you turn the lamp down low \\
I'm beginning to see the light}
\end{quote}
\medskip\medskip
\begin{flushright}
{\small ``I'm Beginning To See The Light,'' 1944 \\
written by \\
Johnny Hodges, Don George, Harry James, \& Duke Ellington \\
performed by Ella Fitzgerald, \underline {Duke Ellington Songbook, disc 5}}
\end{flushright}
\end{singlespace} }

\chapter[The Unusual Afterglow of the Gamma-ray Burst of 26 March 1998 as Evidence for a Supernova Connection]{The Unusual Afterglow of the Gamma-ray Burst of 26 March 1998 as Evidence for a Supernova Connection$^\dag$}
\label{chap:sn-grb}

\secfootnote{\secfootdag}{A version of this chapter was first published 
in {\it Nature},  401, p.~453--456, (1999).}

\def\Palomar  {$^1$}
\def\NRAO     {$^2$}
\def\UCB      {$^3$}
\def\JHU      {$^4$}
\def\IGPP     {$^5$}
\def\CPA      {$^6$}
\def\LBL      {$^7$}
\def\ESO      {$^8$}
\def\IAS      {$^9$}

\secauthor{\vskip -1cm    
            J. S. Bloom\Palomar,             
            S. R. Kulkarni\Palomar,          
            S. G. Djorgovski\Palomar,        
            A. C. Eichelberger\Palomar,      
            P.    C\^ot\'e\Palomar,          
            J. P. Blakeslee\Palomar,         
            S. C. Odewahn\Palomar,           
	    F. A. Harrison\Palomar,          
            D. A. Frail\NRAO,	             
            A. V. Filippenko\UCB,            
            D. C. Leonard\UCB,               
            A. G. Riess\UCB,                 
            H. Spinrad\UCB,                  
            D. Stern\UCB,                    
            A. Bunker\UCB,                   
            A. Dey\JHU,                      
            B. Grossan\CPA,                  
            S. Perlmutter\LBL,               
            R. A. Knop\LBL,                  
            I. M. Hook\ESO,                  
            \&\
            M.    Feroci\IAS                
}

\secaffils{\Palomar Palomar Observatory 105-24, Caltech, Pasadena, CA 91125,
USA}

\secaffils{\NRAO National Radio Astronomy Observatory, P. O. Box O,
       Socorro, NM 87801, USA}

\secaffils{\UCB Department of Astronomy, University of California, Berkeley,
CA 94720-3411 USA}

\secaffils{\JHU National Optical Astronomy Observatories, 950 N. Cherry,
Ave.~Tucson, AZ 85719, USA}

\secaffils{\CPA Center for Particle Astrophysics, University of California,
Berkeley, CA 94720 USA}

\secaffils{\LBL Lawrence Berkeley National Laboratory, Berkeley, CA 94720,
USA}

\secaffils{\ESO European Southern Observatory, D-85748 Garching, Germany}

\secaffils{\IAS Istituto di Astrofisica Spaziale, CNR,
           via Fosso del Cavaliere, Roma I-00133, Italy}

\begin{abstract}
Cosmic gamma-ray bursts have now been firmly established as one of the
most powerful phenomena in the universe, releasing almost the
rest-mass energy of a neutron star in a few seconds
\citep{kdo+99}. The most popular models to explain gamma-ray bursts
are the coalescence of two compact objects such as neutron stars or
black holes, or the catastrophic collapse of a massive star in a very
energetic supernova-like explosion \citep{mw99,pac98b}.  An
unavoidable consequence of the latter model is that a bright
supernova should accompany the GRB.  The emission from this supernova
competes with the much brighter afterglow produced by the relativistic
shock that gives rise to the GRB itself.  Here we show that about 3
weeks after the gamma-ray burst of 26 March 1998, the transient
optical source associated with the burst brightened to about 60 times
the expected flux, based upon an extrapolation of the initial light
curve.  Moreover, the spectrum changed dramatically, with the color
becoming extremely red. We argue that the new source is an underlying
supernova. If our hypothesis is true then this provides evidence
linking cosmologically located gamma-ray bursts with deaths of massive
stars.

\end{abstract}

\section{Introduction}

The origin of GRBs remained elusive for a period of nearly three
decades after their discovery \citep{kso73}.  Beginning in 1997,
however, the prompt localization of GRBs by the Italian-Dutch
satellite BeppoSAX \citep{bbp+97} and the All Sky Monitor
\citep{lbc+96} on board the X-ray Timing Explorer led to the discovery
of the GRB afterglow phenomenon -- emission at lower energies: X-ray
\citep{cfh+97}, optical \citep{vgg+97}, and radio \citep{fkn+97}.

The persistence of the afterglow emission (days at X-ray wavelengths,
weeks to months at optical wavelengths, months to a year at radio
wavelengths) enabled astronomers to carry out detailed observations
which led to fundamental advances in our understanding of these
sources: (1) the demonstration that GRBs are at cosmological distances
\citep{mdk+97}; (2) the proof that these sources expand with
relativistic speeds \citep{fkn+97}; and (3) the realization that the
electromagnetic energy released in these objects exceeds that in
supernovae \citep{wkf98} and, in some cases, the released energy is
comparable to the rest mass energy of a neutron star
\citep{kdr+98,dkb+98b,kdo+99,mia+99}.

Despite these advances, we are still largely in the dark about the
nature of the GRB progenitors. Though there are a number of models for
their origin, the currently popular models involve the formation of
black holes resulting from either the coalescence of neutron stars
\citep{pac86,goo86,npp92} or the death of massive stars
\citep{woo93,pac98b}.  The small offsets of GRBs with respect to their
host galaxies and the association of GRBs with dusty regions and
star-formation regions favors the latter, the so-called hypernova
scenario \citep{pac98b}.  However, this evidence is indirect and also
limited by the small number of well-studied GRBs (see, however,
chapter \ref{chap:offset} which was published after this chapter).

The most direct evidence for a massive star origin would be the
observation of a supernova coincident with a GRB.  Here we present
observations of GRB 980326 and argue for the presence of such an
underlying supernova. If our conclusions are correct, then the
implication is that at least some fraction of GRBs, perhaps the entire
class of long duration GRBs, represent the endpoint of the most
massive stars. Furthermore, if the association \citep{gvv+98,kfw+98}
of GRB 980425 with a bright supernova in a nearby galaxy holds, then
the apparent $\gamma$-ray luminosity of GRBs ranges over six orders of
magnitude.

\section{The Unusual Optical Afterglow}

Following the localization of GRB 980326 by BeppoSAX \citep{ccf+98},
\citet{ggv+98c} quickly identified the optical afterglow.
Our optical follow-up program began at the Keck Observatory,
approximately 10 hr after the burst.  A log of these observations is
given in table \ref{tab:9803260bs}.

\begin{deluxetable}{lcccccccccr}
\singlespace
\tablecolumns{6} 
\tablewidth{0pc} 
\tablecaption{Keck II Optical Observations of GRB 980326\label{tab:9803260bs}}
\tablehead{
\colhead{Date$^a$} & \colhead{Band/} & \colhead{Int.~Time} & \colhead{Seeing} & \colhead{Magnitude} & \colhead{Observers} \\
\colhead{(UT)} & \colhead{Grating} & \colhead{(sec)} & \colhead{(FWHM)} }
\startdata
Mar 27.35 & ~$R$ & 240 & 0\arcsec .74 & $21.25 \pm 0.03$
	& AVF, DCL, AGR & \\
Mar 28.25   & ~$R$ & 240 & 0\arcsec .66 & $23.58 \pm 0.07$
	& HS, AD, DS, SAS \\
Mar 29.27   & 300 & 3600 &    & $24.45 \pm 0.3$  & HS, AD, DS, SAS \\
Mar 30.24   & ~$R$ & 900 & 0\arcsec .93 & $24.80 \pm 0.15$
	 & SP, BG, RK, IH \\
Apr 17.25   & ~$R$ & 900 & 0\arcsec .82 & $25.34 \pm 0.33$ 
	& PC, JB \\
Apr 23.83   & 300 & 5400 &   & $24.9 \pm 0.3^c$ & SGD, SCO \\ 
Dec 18.50   & ~$R$ & 2400 & 0\arcsec .74 & $> 27.3$ 
	 & SRK, JSB, MvK \\
Dec 18.54   & ~$I$ & 2100 & 0\arcsec .74 & $> 25.3$ 
	& SRK, JSB, MvK \\
Mar 24  & ~$I$ & 5450 & 0\arcsec .80 & $> 26.6$ 
	& SRK, JSB
\enddata

\tablecomments{We used the Keck II 10-m Telescope 2,048 $\times$
2,048 pixel CCD (charged coupled device) Low-Resolution Imaging
Spectrometer \citep[LRIS;][]{occ+95} for imaging and spectroscopy of
the GRB field. The epoch of GRB 980326 is 26.888 March 1998
\citep{ccf+98}. For details of the data reductions, see \S \ref{sec:grb980326-data}.}

\tablenotetext{a}{Mean epoch of the image.  The year is 1998 for all images
	except for that on March 24 for which it is 1999.}
\end{deluxetable}

\begin{figure*}[tp]
\centerline{\psfig{file=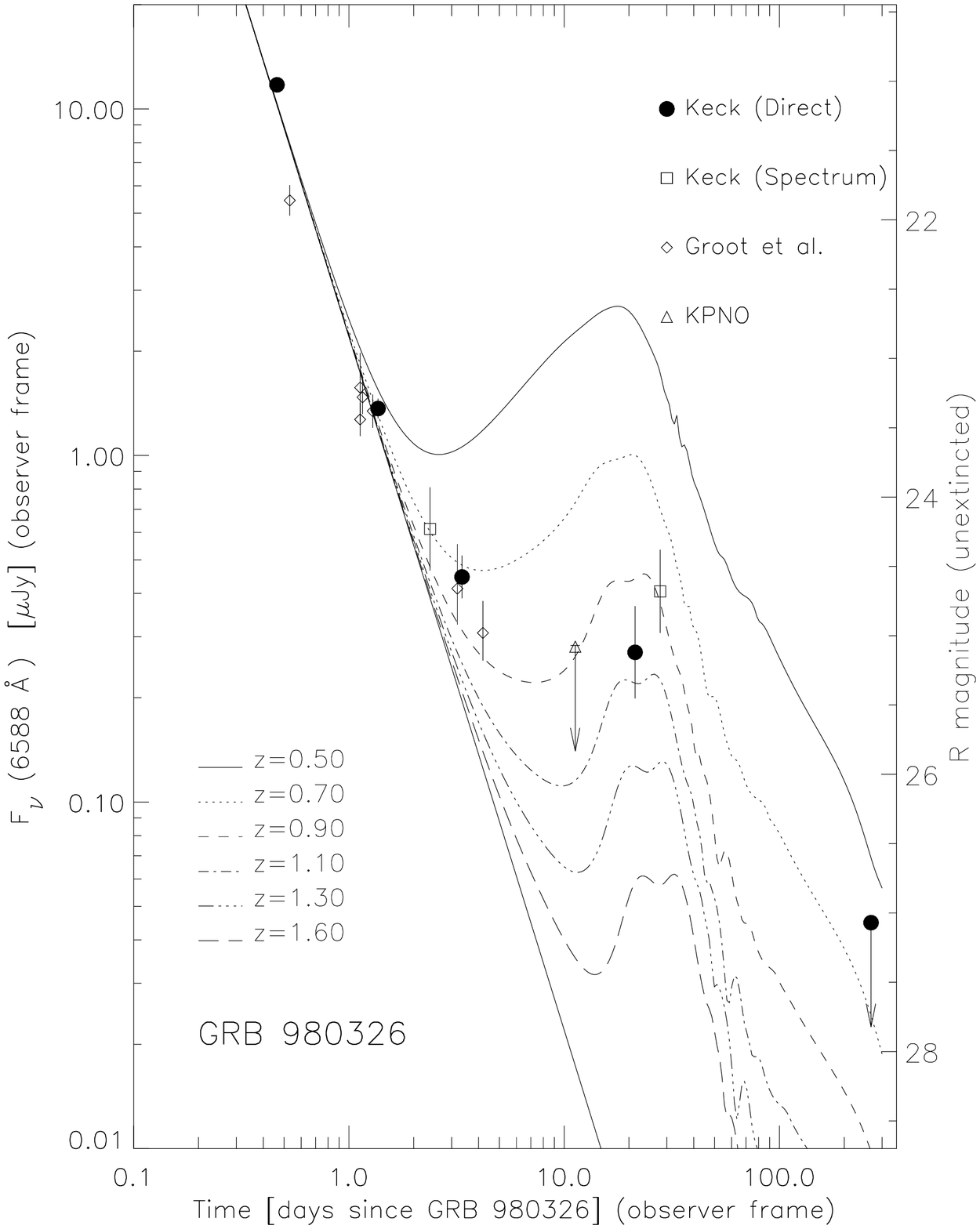,width=5in}}
\caption[The $R$-band light curve of the afterglow of GRB 980326]{
The $R$-band light curve of the afterglow of GRB 980326.  Overlaid is
a power-law afterglow decline summed with a bright supernova light
curve at different redshifts.  (Although we use as as a template the
multi-band light curve of SN 1998bw \citep{gvv+98,ms99}, the bright
supernova potentially associated with GRB 980425, we emphasize that
the exact light-curve shapes of a supernova accompanying a GRB is not
known $a~priori$.)  The GRB+SN model at redshift of about unity
provides an adequate description of the data. See \S
\ref{sec:grb980326-light} for future details.}
\label{fig:980326image}
\end{figure*}

\begin{figure*}[tp]
\centerline{\psfig{file=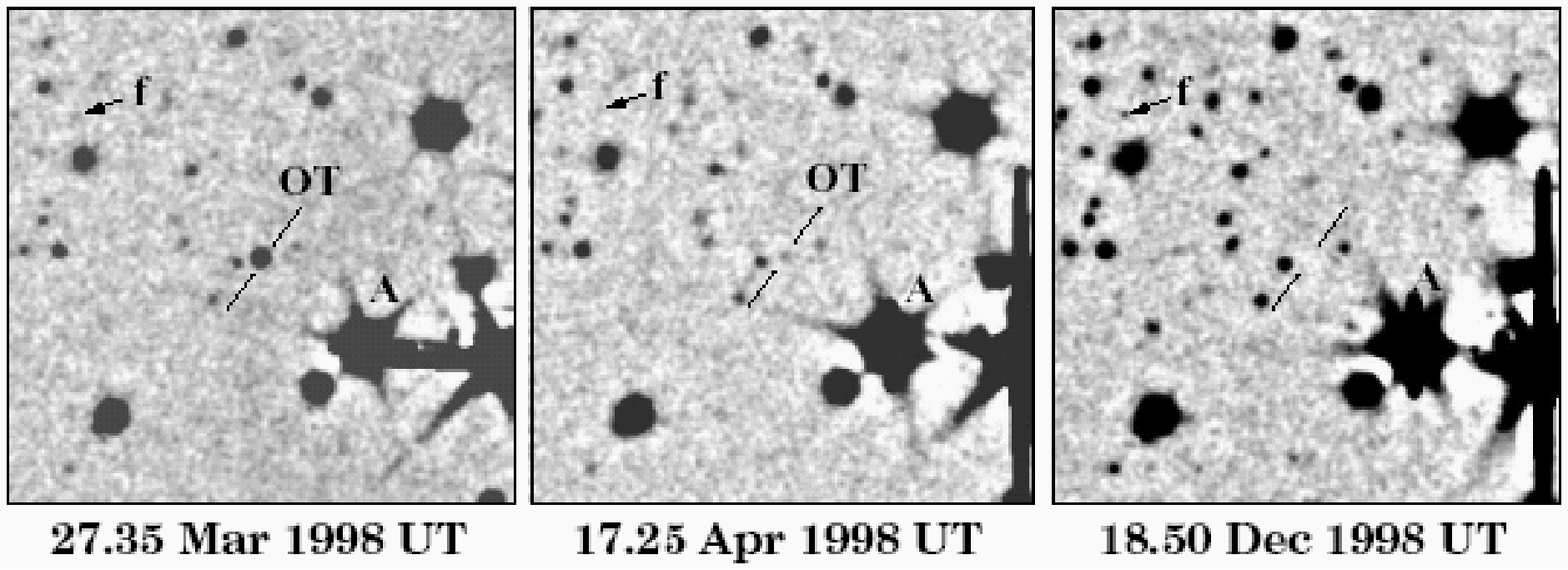,width=6.0in,angle=0}}
\caption[Images of the field of GRB 980326 at three epochs]{Images of the field of GRB 980326 at three epochs. Each images shows
a 54\arcsec $\times$ 54\arcsec\ region centered on the optical
transient (labeled ``OT'').  In all the images, the local background
has been subtracted by a median filter and the resulting image
smoothed (with a two-dimensional Gaussian with $\sigma =0\arcsec.23$).
An unrelated faint source ``f'' in the field is noted for comparison
of the relative limiting flux between the three epochs: it is
marginally detected (at the $\sim 2$-$\sigma$ level) on March 27 and
April 17 but well detected on December 18.  In contrast the OT is
brighter and better detected (at the 4.6-$\sigma$ level, see text and
\S \ref{sec:grb980326-upperlim}) on April 17 but clearly not detected
to fainter levels on December 18 ($R > 27.3$ mag; see table
\ref{tab:9803260bs}).}
\label{fig:980326light-curve}
\end{figure*}

In figure \ref{fig:980326light-curve} we present our $R$-band
photometry, along with values reported by other workers.  Considering
only reported data taken within the first month of the burst, we find
a characteristic power law decay in the flux versus time, followed by
an apparent flattening.  The usual interpretation is that the decaying
flux is the afterglow emission, while the constant flux is due to the
host galaxy.  Indeed, earlier \citep{dkc+98} we attributed the entire
observed flux on April 17th to the host galaxy.

But to our surprise, our more recent observations (first performed
nine months after the GRB event) showed no galaxy at the position of
the optical transient (OT); see figure \ref{fig:980326image}.  We
estimate a 2-$\sigma$ upper limit the $R$-band magnitude of $R > 27.3$
mag (see table \ref{tab:9803260bs}). This is almost a factor of 10
less flux than that reported from our 17 April detection.  A secure
conclusion is that the presumed host galaxy of GRB 980326, assuming
that the GRB was coincident with the host (as appears to be the case
for all other well-studied GRBs to date), is fainter than $R \approx
27$ magnitude.  This conclusion is not alarming as such faint (or
fainter) galaxies are indeed expected from studies \citep{mm98,hf99}
of the properties of cosmological GRB host galaxies.

Having established that the host galaxy of GRB 980326 is faint, we are
forced to conclude that the OT did not continue the rapid decay it
exhibited initially.  Instead, we find three phases of the light curve
(figure \ref{fig:980326light-curve}): a steeply declining initial
phase (at times since the burst of $t \ale 5$ day), a subsequent
re-brightening phase ($t \sim $ 3--4 weeks) and, finally, a phase in
which the source appears to have faded away to an undetectable level
by the time of our next observation (9 months after the burst).

In previously studied bursts, the optical afterglow emission has been
modeled by a power-law function, flux $\propto t^{\alpha}$; here
$\alpha$ is the power-law index.  In some bursts, at early times ($t$
less than a day or so), significant deviations have been seen, for
example, GRB 970508 \citep{dmk+97}.  At late times, in some bursts,
deviations manifest as steepening (that is, $\alpha$ becoming smaller)
of light curves, for example, GRB 990123 \citep{kdo+99} and GRB 990510
\citep{hbf+99,sgk+99b}.

It is against this backdrop of the observed afterglow phenomenology
that we now analyze the light curve in figure
\ref{fig:980326light-curve}.  The declining phase cannot be fitted by
a simple power law ($\chi^2=72$ for 9 degrees of freedom). From figure
\ref{fig:980326light-curve} it is clear that the flux had already started flattening
by day 3.  Restricting the analysis to the first two days, we obtain
$\alpha = -2.0 \pm 0.1$, consistent with a previous analysis
\citep{ggv+98c}.

Such power-law decays are usually interpreted as arising from
electrons shocked by the explosive debris sweeping up the ambient
medium \citep{katz94a,mr97a,viet97,wax97c}.  Assuming that the
electrons behind the shock are accelerated to a power-law differential
energy distribution with index $-p$, on general grounds \citep{spn98}
we expect that the afterglow flux, $f_\nu(t)$, is proportional to
$t^{\alpha}\nu^{\beta}$; here $f_{\nu}(t)$ is the flux at frequency
$\nu$ and time $t$.  The value of $\alpha$ and $\beta$ depend on $p$,
the geometry of the emitting surface \citep{mrw98b} (spherical versus
collimation) and the radial distribution of the medium around the
burst \citep{cl99}.

\begin{figure*}[tp]
\centerline{\psfig{file=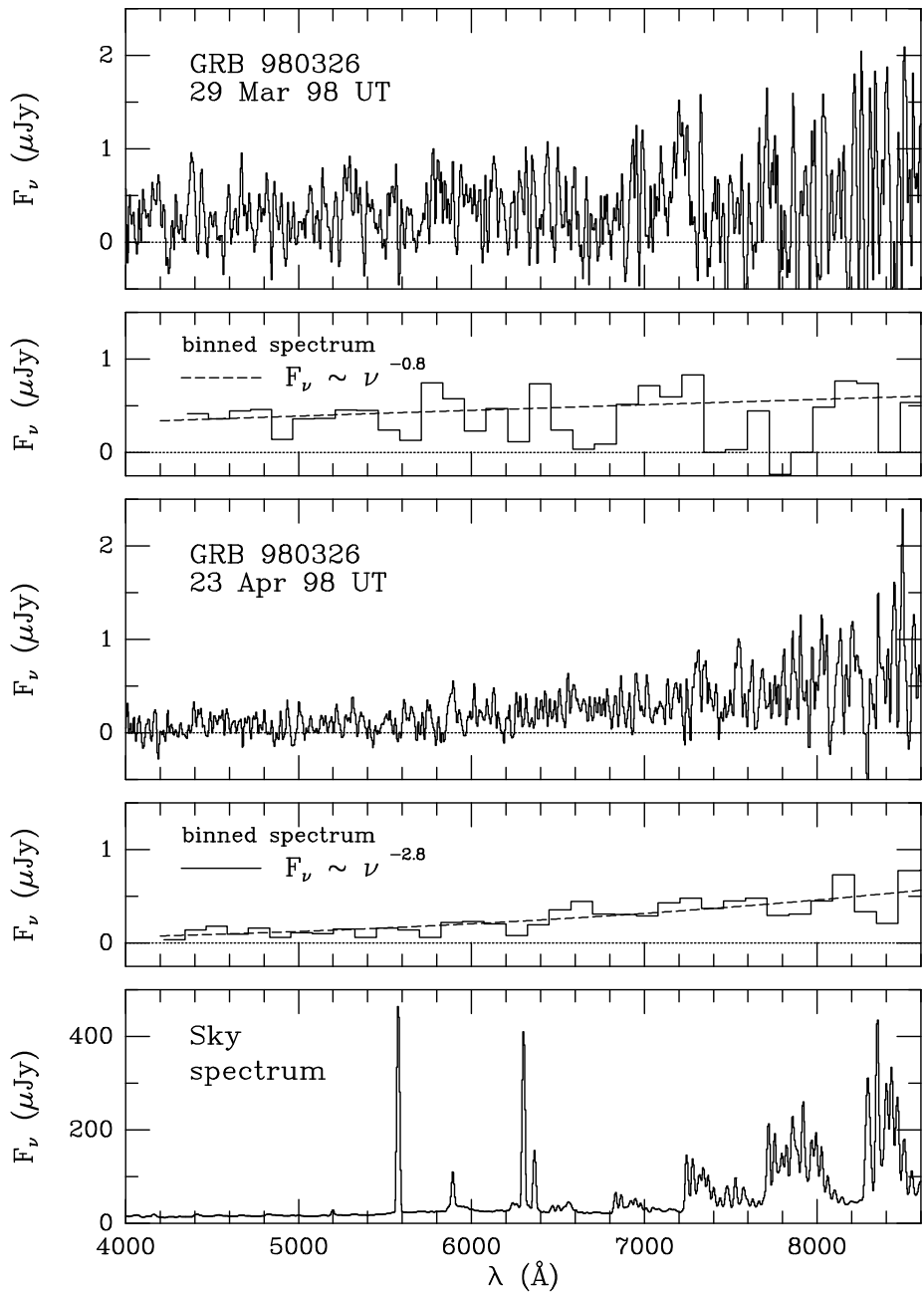,width=5in,angle=0}}
\vskip 0.6cm
\caption[The spectra of the transient on March 29.27 and April 23.83 1998 UT.]
{The spectra of the transient on March 29.27 and April 23.83 1998 UT.
The two spectra are shown at two different spectral resolutions.
Starting from the top, panels 1 and 3 show the spectra at the two
epochs at the full spectral resolution (see below for details) and
panels 2 and 4 show the same two spectra but binned in groups of 51
channels. Panel 5 is the spectrum of the sky.  The best-fit power law
models ($f_\nu\propto \nu^{\beta}$) to the binned spectra are shown by
dashed lines; the fits were restricted to the wavelength range
4500--8500~\AA. The scatter of individual channel values within each
bin was used to assign relative weights to the median fluxes in each
bin when performing the fits.  On March 29.27, we obtain $\beta=
-0.8\pm 0.4$ and $\beta=-2.8\pm 0.3$ on April 23.83.  The derived
power-law indices include the correction of Galactic extinction.  From
the absence of continuum breaks in the spectrum of March 29, we can
place an upper limit to the redshift, $z_{\rm OT} \ale 2.3$. See \S
\ref{sec:grb980326-spec} for details.}
\label{fig:980326spectrum}
\end{figure*}

From our spectroscopic observations of 29 March (figure
\ref{fig:980326spectrum}), we find $\beta = -0.8 \pm 0.4$.  This
combination of ($\alpha,\beta$) is similar to the ($\alpha=-2.05 \pm
0.04, \beta=-1.20 \pm 0.25$) seen in GRB 980519 \citep{hkp+99}, and
can be reasonably interpreted
\citep{sph99} as arising from a standard $p \sim 2.2$ shock with a
jet-like emitting surface. Alternatively, the emission could arise in
a $p \sim 3$ shock propagating in a circumburst medium \citep{cl99}
whose density falls as the inverse square of the distance from the
explosion site.

\section{A New Transient Source}
\label{sec:transient}

We now discuss the bright source seen in the re-brightening phase
(corresponding to observations of April 17 and April 23). This source
is $\sim 60$ times brighter than that extrapolated from the rapidly
declining afterglow. Given that a magnitude of excess at late times
has not been reported before, it is important to review the crucial
observation of mid-April.

First, the re-brightened source is coincident with the OT in the image
of 27 March to within the expected astrometric error ($0.04\pm 0.18$
arcsecond).  As noted in the legend to figure \ref{fig:980326image},
the source is consistently detected in three separate frames taken on
17 April.  In the summed image, the source is detected at 4.6-$\sigma$
(chance probability of $2 \times 10^{-6}$) on 17 April and all other
objects in the field at this flux level are reliably detected in our
deeper 18 December image. Next, the source is clearly detected
(spectroscopically) on 23 April (figure \ref{fig:980326spectrum}) at
the same position as that of the OT. The inferred spectrophotometric
$R$-band magnitude is plotted as open squares on the light curve
(figure
\ref{fig:980326light-curve}).  Thus we conclude that there was indeed
a source at the position of the OT which brightened three weeks after
the burst and subsequently faded to undetectable levels. We now
investigate possible explanations for this source.

The simplest picture is that afterglow re-brightened. \citet{pcf+99}
have recently suggested that the doubling of the X-ray flux of GRB
970508 three days after the GRB event arises from the relativistic
shell running into a dense gas cloud.  Such an explanation for the GRB
980326 light curve would require a large dense region, with a size
comparable to the timescale of re-brightening, about $\Gamma\, \times$
10 light days ($\sim 0.01$ pc) and located at a distance $\Gamma^2\,
c\, \times 20$ days ($\sim 0.1$ pc) from the explosion site.  Here,
$\Gamma$ is the bulk Lorentz factor of the shock and is expected to be
order unity three weeks after the burst.  \citet{pmr98} suggest that
the re-brightening of GRB 970508 may be due to a shock refreshment--
delayed energy injection by the extremely long-lived central engine
that produced the GRB.  Alternatively, the delayed energy could come
from the spin-down of a newly formed milli-second pulsar through
magnetic dipole radiation \citet{dl00}. For all these models, however,
the expected spectrum would be the typical synchrotron spectrum, flux
$f_\nu\propto \nu^{-1}$ (or flatter). The very red spectrum of 21
April (figure \ref{fig:980326spectrum}) allows us to essentially rule out a
synchrotron origin for the re-brightening phase of GRB 980326.

Alternatively, as suggested by \citet{Loeb99}, the GRB could have
occurred in a dusty region and the afterglow would re-brighten as the
dust is sublimated by the afterglow.  However, the observed spectral
evolution from a relatively blue spectrum (29 March) to red (23 April)
moves in a direction opposite to that expected in this model.  Last, a
non-relativistic, thermal expanding envelope powered by radioactive
decay could accompany the merger of a compact binary system
\citep{lp98b} that gives rise to the GRB and the afterglow: this
envelop would radiate at luminosities comparable to that of a bright
supernova and should produce a red spectrum similar to that seen on 23
April; however, the time-scale for the peak emission is more than an
order of magnitude shorter than the timescale for the re-brightening
that we see with GRB 980326.

\section{The Supernova Interpretation}

We advance the hypothesis that the new source is due to an underlying
supernova revealed only after the afterglow emission has vanished.
Woosley and collaborators (\citealt{mw99,woo93} and references
therein) have pioneered the ``collapsar'' model in which GRBs arise
from the death of massive stars---stars which produce black hole
remnants rather than neutron stars.  In this model, the iron core of a
massive star collapses to a black hole and releases up to a few
$\times 10^{52}$ erg of kinetic energy. Some fraction of this energy
is expected to emerge in the form of a jet with little entrained
matter; bursts of gamma-rays result from internal shocks in this jet.
The remaining energy is absorbed by the star, causing it to explode
and thereby produce a supernova \citep{han99}.

Thus in this model, the total light curve has two distinct
contributions: a power-law decaying afterglow component, and emission
from the underlying supernova.  In figure \ref{fig:980326light-curve}
we show the light curve expected in this model and use the light curve
of the well observed \citep{gvv+98,ms99} SN 1998bw as a template for
the supernova contribution.  We find the $R$-band and $I$-band data to
be consistent with a bright supernova at $z \approx 1$.

The very red spectrum of the source on 23 April finds a natural
explanation in the supernova hypothesis.  On theoretical \citep{mw99}
and phenomenological \citep{kfw+98} grounds, we expect GRBs to arise
from massive stars which have lost their hydrogen envelope, that is,
type Ibc supernovae. At low redshifts, all type I supernovae are
observed to exhibit a strong ultraviolet deficit relative to the
blackbody fit to their spectra.  This deficit is due to absorption by
prominent atomic resonance lines starting below $\sim 3900$ \AA. Below
$\lambda_c \sim 2900$ \AA\ we expect to see very little flux.  In the
near-ultraviolet range (3000--4000 \AA) all spectra from type I
supernovae have a red appearance. Approximating the flux by a power
law ($f_\nu \propto \nu^\beta$), the ultraviolet power-law index
(depending on the wavelength range chosen) is $-3$ or even smaller, as
found in the prototypical type Ic SN 1994I \citep{kjl+93}.  Fitting
the spectrum of figure \ref{fig:980326spectrum} to a power law, we
obtain $\beta=-2.8\pm 0.3$.  Such a red spectrum (negative $\beta$)
requires that the observed spectrum corresponds to the
ultraviolet--blue region in the restframe of the object.  A smaller
redshift would lead to a larger $\beta$. A larger redshift would
substantially suppress the light in the observed $R$ band (which
covers the wavelength range 5800--7380 \AA).  The detection in $R$
band then provides an independent constraint (figure
\ref{fig:980326light-curve}), $z \ale 1.6$.

We have used the light curve of SN 1998bw because it is a very well
studied type Ibc SN with a possible association with GRB 980425.  We
do not know {\it a priori} the precise spectrum and light curve of a
supernova accompanying GRBs.  In the collapsar model, the progenitor
stars are expected to have no significant envelopes and thus the
expected supernovae are of type Ibc.  The general shape of the spectra
of all type Ibc supernovae are expected to be the same and are
summarized above. Given the low signal-to-noise ratio of the 23 April
spectrum and the expected line broadening due to high photospheric
velocity, we do not, as seems to be the case, expect to see any
features in our spectrum.

Independently, from the absence of strong spectral breaks in our
spectrum of the OT, we can firmly place the redshift of the OT at
$\ale 2.3$. This constraint is consistent with our deduction that $z
\ale 1.6$ (see above).  Thus from a variety of accounts we find a
plausible redshift of around unity for GRB 980326. Such a redshift is
not entirely unexpected. Indeed, we note that five out of eight
spectroscopically confirmed redshifts of GRBs lie in the range $0.7 <
z < 1.1$.

\section{Implications of the Supernova Connection}

The direct evidence for an accompanying SN can be seen in the light
curve at timescales comparable to the time for SNe to peak, $\sim
20(1+z)$ days.  However, in our opinion three conditions must be
satisfied in order to see the underlying SN even when one was present.
(1) The GRB afterglow should decline rapidly, otherwise the SN will
remain overpowered by the afterglow for all epochs.  (2) Given the
strong ultraviolet absorption (discussed above), only GRBs with redshift $z
\ale 1.6$ have an observable SN component in the optical band. (3) The
host must be dimmer than the peak magnitude of the SN ($M_V \sim
-19.5$ mag). The last requirement is not needed if the GRB can be resolved
from the host (for example, by using the Hubble Space
Telescope). Finally, one caveat is worth noting: the peak magnitudes
of Type Ibc SNe are not constant (unlike those of Type Ia), and can
vary \citep{imn+98} from $-16$ mag to a maximum of $-19.5$ mag
\citep[see][]{grs+00}.  We have investigated the small sample of GRBs
with adequate long-term follow up and conclude that perhaps only GRB
980519 satisfies the first and the third observational conditions for
supernova detection; the redshift of this GRB is unfortunately
unknown.

The dynamics of the relativistic blast wave is strongly affected by
the distribution of circumstellar matter.  \citet{cl99} note that
massive stars, through their active winds, leave a circumstellar
medium with density falling as the inverse square of the distance from
the star.  One expects smaller $\alpha$ for GRBs exploding such a
circumstellar medium.  In this framework, GRB afterglows which decline
rapidly and are at modest redshifts will again be prime targets to
search for the underlying SN.

If we accept the SN interpretation for GRB 980326, a long-duration (5
sec) GRB, then it is only reasonable to posit that all other long
duration GRBs are also associated with supernovae. We suggest that
sensitive observations be made---especially at longer wavelengths, to
avoid the UV cutoff of supernovae---of GRBs satisfying the above three
conditions. If our proposed hypothesis is correct, then the light
curves and the spectra of such GRBs would exhibit the behavior shown
in figure \ref{fig:980326light-curve} and figure
\ref{fig:980326spectrum} and discussed here.  Indeed, motivated by
this work, evidence for underlying supernovae in other GRBs is now
being reported (GRB 970228;
\citealt{rei99}).

We end with a discussion of one interesting point.  The total energy
release in $\gamma$-rays of GRB 980326 was $E_\gamma = (3.42 \pm 3.74)
\times 10^{51} f_b$ erg where $f_b$ is the fractional solid angle of
the jet (if any); here we have used the measured fluence
\citep{ggv+98c} and assumed $z \sim 1$ ($H_0 = 65$ km s$^{-1}$
Mpc$^{-2}$, $\Omega_0 = 0.3$, $\Lambda_0 = 0.7$) \citep[see
also][]{bfs01}.  If this GRB was beamed, then $E_\gamma \sim 10^{49}$
erg.  Curiously enough, this rather small energy requirement places
GRB 980326 as close in energetics to GRB 980425 ($E_\gamma = 7.16
\times 10^{47}$ erg; \citealt{gvv+98} and chapter \ref{chap:sgrbs}) as to the classic gamma-ray
 bursts ($E_\gamma \age 5
\times 10^{50}$ erg; \citealt{fks+01}).

{\it Note added in proof:} \citet{gtv+00} have also recently reported
supernova-like behavior in the light curve and spectrum of 970228.



\bigskip
\bigskip

\noindent{\bf Acknowledgment.} 
We thank M.~H.~van Kerkwijk for help with the December 18 observations
at the Keck II telescope and R.~Sari for helpful discussions.  We
gratefully acknowledge the excellent support from the staff at the
Keck Observatory.  The observations reported here were obtained at the
W.~M.~Keck Observatory, made possible by the generous financial
support of the W.~M.~Keck Foundation, which is operated by the
California Association for Research in Astronomy, a scientific
partnership among California Institute of Technology, the University
of California and the National Aeronautics and Space Administration.
SRK's and AVF's research is supported by the National Science
Foundation and NASA.  SGD acknowledges partial support from the
Bressler Foundation.


\begin{subappendices}

\section[Details of the Data Reduction for Table \ref{tab:9803260bs}]{Details of the Data Reduction for Table \ref{tab:9803260bs}}
\label{sec:grb980326-data}

\subsection{Photometric calibration}

The absolute zero-point of the $R$ (effective wavelength of
$\lambda_{\rm eff}
\approx$ 6588 \AA;  \citealt{fsi95}) and $I$-bands ($\lambda_{\rm eff}
\approx 8060$ \AA ) were calibrated to the standard Cousins bandpass 
using standard-stars in the field SA98 \citep{lan92} and assuming the
standard atmospheric correction on Mauna Kea (0.1 mag and 0.06 mag per
unit airmass, respectively).  The estimated statistical error on the
absolute zero-point is 0.01 mag.  We estimate the systematic error
(due to lack of inclusion of color term) to be less than 0.1 mag.  We
propagated all photometry to the absolute zero-point derived in the
first epoch of observation using 8 ``secondary'' stars which were
detected with high signal-to-noise ratio, unsaturated, near to the
transient, and common to every epoch; the typical uncertainty in the
zero-point propagation is 0.01 mag.  Thus any systematic error in our
absolute zero-point will not affect the conclusions based on {\it
relative} flux.  The uncertainties quoted in the table
\ref{tab:9803260bs} contain all known sources of error (aperture
correction, etc.).  The calibrated magnitudes of the secondary stars
reported in \citet{ggv+98c} agree to within the measurement errors.

\subsection{Spectrophotometric measurement}

The flux in $\mu$Jy is determined at 6588 \AA, the central wavelength
of the $R_c$ band; the conversion to magnitude assumes 0 mag equal to
3020 Jy
\citep{fsi95}. The spectrophotometric magnitudes
are relative to a bright star that was on the slit (for which
we have obtained independent photometry from our images).

\subsection{Photometry of the faint source}

Since the transient was not detected to significantly fainter levels
in later epochs, it is safe to assume that the April 17 detection was
that of a point-source (and not an extended galaxy as we had earlier
believed;
\citealt{dkc+98}).  To maximize the signal-to-noise ratio, we choose to
measure the photometry in an aperture radius equal to the FWHM of the
seeing and correct for the missing flux outside the aperture by using
the radial flux profiles of bright isolated stars in the image.  The
determination of the optimum sky level (from which we subtract the
total flux in the aperture) is not well-defined.  We estimate the
systematic uncertainty introduced by the uncertainty in the sky level
as 0.25 mag.  The statistical uncertainty (weighted mean over
different background determinations) of the flux was 0.22 mag.  Thus
we quote the quadrature sum of the statistical and systematic
uncertainty of 0.33 mag.

\subsection{Upper-limits}

On 1998 Dec 18 UT and 1999 March 24 UT there was no detectable flux
above the background at the position of the optical transient.  We
centered 1000 apertures randomly in our image (approximately $1800
\times 2048$ pixels in size) and performed weighted aperture
photometry with a local determination of sky background and recorded
the counts (``DN'') above background at each location. The flux
contribution from an individual pixel, some radius $r$ from the center
of the aperture, to the total flux was weighted by a Gaussian with a
radial width FWHM equal to the seeing. A histogram of the resulting
flux was constructed. This histogram was decomposed into two
components---a Gaussian with median near zero DN and a long tail of
positive DN corresponding to actual source detections.  We fit a
Gaussian to the zero-median component, iteratively rejecting outlier
aperture fluxes.  Based on the photometric zero-point and using
isolated point sources in the image for aperture corrections, we
computed the relationship between DN within the weighted aperture and
the total magnitude.  In table \ref{tab:9803260bs} we quote an upper
limit (95\%-confidence level corresponding to 2-$\sigma$ of the
Gaussian fit) at the position of the optical transient.

\section[Notes on Figure \ref{fig:980326light-curve}]
{Notes on Figure \ref{fig:980326light-curve}}
\label{sec:grb980326-light}

\subsection{Transient light curve}

From \citet{sfd98} we estimate the Galactic extinction in the
direction of the optical transient $(l,b = 242{^\circ}.36,
13^{\circ}.04)$ to be E($B-V$) = 0.08. Thus, assuming the average
Galactic extinction curve ($R_V$ = 3.1), the extinction measure is
$A_R = 0.22$ mag, $A_I$ = 0.16 mag.  Plotted are the extinction
corrected magnitudes (see table \ref{tab:9803260bs}) of the transient
converted to the standard flux zero-point of the Cousins $R$ filter
from
\citet{fsi95}. In addition to our data, we include photometric
detections from \citet{ggv+98c} and an upper-limit from
\citet{vjr98} (KPNO).  The GRB transient flux dominates at early
times, but with a power-law decline slope $\alpha = -2$ (straight
solid line).

\subsection{Supernova light curve} 

The supernova light curve template was constructed by spline-fitting
the broadband spectrum measured by
\citet{gvv+98} of the bright supernova 1998bw at
various epochs (augmented with late-time observations of SN 1998bw by
\citealt{ms99}) and transforming back to the
restframe of SN 1998bw ($z=0.0088$).  As discussed in the text, we
expect the rest-frame UV emission (below 3900\AA) to be suppressed due
to absorption by resonance lines.  We assume that the UV flux declines
as $f_\nu\propto\nu^{-3}$.  Theoretical light curves are then
constructed by red-shifting the template to various redshifts and
determining the flux in the $R$ (observer frame) by interpolating (or
extrapolating, for $z \age 1$).  The flux normalization of the
redshifted SN 1998bw curves are independent of the Hubble constant but
are dependent upon the value of $\Omega_0$ and $\Lambda_0$ (here we
show the curves for $\Omega_0 = 0.2$ and $\Lambda_0=0$).  Beyond $z
\approx 1.3$ the observed $R$-band corresponds to restframe $\lambda
\ale 2900$ \AA. As stated in the text, the spectrum in this range has
been modeled along simple lines.  Qualitatively, the peak flux derived
from the SN model as a function of redshift and shown here agrees with
the theoretical peak flux-redshift relation for type Type
Ia \citep{ssp+98}. This suggests that our adopted model for the UV
spectrum is reasonable.

\section[Upper-limit Determination for Figure \ref{fig:980326image}]
{Upper-limit Determination for Figure \ref{fig:980326image}}
\label{sec:grb980326-upperlim}

In keeping with standard practice, our 17 April observations consisted
of three separate 300-s observations (dithered by 5 arcseconds).
Visual inspection of the three frames reveals a faint source near the
position of the optical transient.  In no frames did a diffraction
spike of the nearby bright star ``A'' overlap the OT position.  Also,
there were no apparent cosmic-ray hits at the transient position nor
were there any strong gain variations (that is, no apparent problem
with the flat-fielding) at the three positions on the CCD.

In both the sum and mode-scaled median of the three shifted images, we
detect a faint source consistent with the centroid location (angular
offset $0.04 \pm 0.18$ arcsec) of the optical transient on 1998 March
27.  Lastly, we computed the point source sensitivity in the 17 April
image by computing Gaussian-weighted photometry in 1000 random
apertures (see discussion accompanying table \ref{tab:9803260bs}).
Relative to this distribution, the flux at the location of the
transient is positive and equal to 4.6-$\sigma$; the probability that
the measured flux is due to noise is $2\times 10^{-6}$.  All objects
at the flux level of the transient are reliably detected in the deeper
image from December, thereby providing an independent validation of
our methodology.  We conclude that indeed the transient was
significantly detected on 17 April.  We discuss the photometric
calibration of the detection in table \ref{tab:9803260bs}.

\section[Reduction Details for Figure \ref{fig:980326spectrum}]
{Reduction Details for Figure \ref{fig:980326spectrum}}
\label{sec:grb980326-spec}

\subsection{Observing details} 

Spectroscopic observations of the OT were obtained on 29 March 1998
UT, using the Low Resolution Imaging Spectrometer
\citep[LRIS;][]{occ+95} at the Keck-II 10-m telescope on Mauna Kea,
Hawaii.  We used a grating with 300 lines mm$^{-1}$ blazed at
$\lambda_{\rm blaze} \approx 5000$ \AA\ and a 1.0 arcsec wide slit.
The effective wavelength coverage was $\lambda \sim 4000 - 9000$
\AA\ and the instrumental resolution was $\sim 12$ \AA.  Two exposures
of 1800 s each were obtained.  We used Feige 34 \citep{msb88} for flux
calibration.  The estimated uncertainty of the flux zero point is
about 20\%.  Additional spectra were obtained on 23 April 1998 UT, in
photometric conditions, using the same instrument, except that the
spectrograph slit was 1.5 arcsec wide. The effective spectral
resolution for this observations was $\sim$ 16 \AA.  Three exposures
of 1800 s each were obtained.
For these observations we used HD 84937 \citep{og83} for flux
calibration. The estimated zero-point uncertainty is about 10\%.  On
both epochs, exposures of arc lamps were used for primary wavelength
calibration. Night sky lines were used to correct for calibration
changes due to flexure.  In both cases, slit position angles were
close to the parallactic angles. Thus the differential slit losses
were negligible.

The spectra shown were convolved with a Gaussian with $\sigma = 5$
\AA\ (that is, less than the instrumental resolution) and re-binned to a
common 5 \AA\ sampling.  None of the apparent features in the spectra
are real, on the basis of a careful examination of two-dimensional,
sky-subtracted spectroscopic images: apparent emission of absorption
features are all due to an imperfect sky subtraction noise.  A sky
spectrum from the April 23 observation, extracted in the same
aperture, is shown for the comparison.  These spectra are shown before
the correction for the Galactic foreground extinction.

\subsection{Spectrophotometry}

In both epoch, we chose a slit position angle close to parallactic so
that the slit would cover both the transient and a relatively bright
star ($R \sim 19$ mag).  The spectroscopic $R$-band magnitudes
reported in table \ref{tab:9803260bs} were derived relative to the
calibrated $R$-band magnitude of these stars. This calibration serves
to eliminate most of the systematics and calibration errors; that is,
the spectrophotometric magnitudes were put on the direct CCD system,
and are not based on the flux calibration of the spectra (which do,
nevertheless, agree to 20 percent). This procedure bypasses most
of the systematic errors in comparing our spectroscopic magnitudes
with those from direct CCD images.

\end{subappendices}

\chapter[Detection of a supernova signature associated with
       GRB 011121]{Detection of a supernova signature associated with
       GRB 011121$^\dag$}
\label{chap:grb011121}

\secfootnote{\secfootdag}{A version of this chapter was published in
the {\it The Astrophysical Journal Letters}, vol.~572, L45--L49.}

\vskip -1cm

\def\cit{1}
\def\mso{2}
\def\vla{3}
\def\uva{4}
\def\uta{5}
\def\col{6}
\def\tap{7}
\def\car{8}
\def\god{9}
\def\nms{{10}}
\def\ucb{{11}}
\def\cnr{{12}}
\def\fer{{13}}
\def\rom{{14}}
\begin{singlespace}

\secauthor{J. S. Bloom$^\cit$,
S. R. Kulkarni$^\cit$,
P. A. Price$^{\cit,\mso}$,
D.    Reichart$^\cit$,
T. J. Galama$^\cit$,
B. P. Schmidt$^\mso$,
D. A. Frail$^{\cit,\vla}$,
E.    Berger$^\cit$,
P. J. McCarthy$^\car$,
R. A. Chevalier$^\uva$,
J. C. Wheeler$^\uta$,
J. P. Halpern$^\col$,
D. W. Fox$^\cit$,
S. G. Djorgovski$^\cit$,
F. A. Harrison$^\cit$,
R.    Sari$^\tap$,
T. S. Axelrod$^\mso$,
R. A. Kimble$^\god$,
J.    Holtzman$^\nms$,
K.    Hurley$^\ucb$,
\hbox{F.    Frontera$^{\cnr, \fer}$},
L.    Piro$^\rom$,
\&\
E.  Costa$^\rom$}

\secaffil{$\cit$ Division of Physics, Mathematics and Astronomy,
  105-24, California Institute of Technology, Pasadena, CA 91125}
\secaffil{$\mso$ Research School of Astronomy \& Astrophysics, 
Mount Stromlo Observatory, via Cotter Rd., Weston Creek 2611, Australia}
\secaffil{$\vla$ National Radio Astronomy Observatory, Socorro,
NM 87801} 
\secaffil{$\uva$ Department of Astronomy, University of Virginia, P.O. Box 3818, Charlottesville, VA 22903-0818}
\secaffil{$\uta$ Astronomy Department, University of Texas, Austin, TX 78712}
\secaffil{$\col$ Columbia Astrophysics Laboratory, Columbia University, 550 West 120th Street, New York, NY 10027}
\secaffil{$\tap$ Theoretical Astrophysics 130-33, California Institute
  of Technology, Pasadena, CA 91125}
\secaffil{$\car$ Carnegie Observatories, 813 Santa Barbara Street, Pasadena, CA 91101}
\secaffil{$\god$ Laboratory for Astronomy and Solar Physics, NASA Goddard Space Flight Center, Code 681, Greenbelt, MD 20771}
\secaffil{$\nms$ Department of Astronomy, MSC 4500, New Mexico State University, P.O.~Box 30001, Las Cruces, NM 88003}
\secaffil{$\ucb$ University of California at Berkeley, Space Sciences Laboratory, Berkeley, CA 94720-7450}
\secaffil{$\cnr$ Istituto Astrofisica Spaziale and Fisica Cosmica, C.N.R., Via Gobetti, 101, 40129 Bologna, Italy}
\secaffil{$\fer$ Physics Department, University of Ferrara, Via Paradiso, 12, 44100 Ferrara, Italy}
\secaffil{$\rom$ Istituto Astrofisica Spaziale, C.N.R., Area di Tor Vergata, Via Fosso del Cavaliere 100, 00133 Roma, Italy}
\end{singlespace}

\begin{abstract}
  Using observations from an extensive monitoring campaign with the
  {\it Hubble Space Telescope}, we present the detection of an
  intermediate-time flux excess that is redder in color relative to
  the afterglow of GRB 011121, currently distinguished as the
  gamma-ray burst with the lowest known redshift.  The red ``bump,''
  which exhibits a spectral roll-over at $\sim$7200 \AA, is well
  described by a redshifted Type Ic supernova that occurred
  approximately at the same time as the gamma-ray burst event.  The
  inferred luminosity is about half that of the bright supernova
  1998bw. These results serve as compelling evidence for a massive
  star origin of long-duration gamma-ray bursts. Models that posit a
  supernova explosion weeks to months preceding the gamma-ray burst
  event are excluded by these observations. Finally, we discuss the
  relationship between spherical core-collapse supernovae and
  gamma-ray bursts.
\end{abstract}

\section{Introduction}

Two broad classes of long-duration gamma-ray burst (GRB) progenitors
have survived scrutiny in the afterglow era: the coalescence of
compact binaries (see \citealt{fwh99} for review) and massive stars
\citep{woo93}. More exotic explanations \citepeg{pac88,car92,der96}
fail to reproduce the observed redshift distribution, detection of
transient X-ray lines, and/or the distribution of GRBs about host
galaxies.

In the latter viable scenario, the so-called ``collapsar'' model
\citep{woo93,mw99,han99}, the core of a massive star collapses to a
compact stellar object (such as a black hole or magnetar) which then
powers the GRB while the rest of the star explodes.  We expect to see
two unique signatures in this scenario: a rich circumburst medium fed
by the mass-loss wind of the progenitor \citep{cl99} and an underlying
supernova (SN). Despite extensive broadband modeling of afterglows,
unambiguous signatures for a wind-stratified circumburst media have
not been seen \citepeg{fks+00,bdf+01}.

There has, however, been been tantalizing evidence for an underlying
SN.  The first association of a cosmologically distant GRB with the
death of a massive star was found for GRB 980326, where a clear excess
of emission was observed, over and above the rapidly decaying
afterglow component. This late-time ``bump'' was interpreted as
arising from an underlying SN \citep{bkd+99} since, unlike the
afterglow, the bump was very red.  GRB 970228, also with an
intermediate-time bump and characteristic SN spectral rollover, is
another good candidate \citep{rei99,gtv+00}.

Suggestions of intermediate-time bumps in GRB light curves have since
been put forth for a number of other GRBs
\citep{lcg+01,svb+00,fvh+00,bhj+01,csg+01,sok01,ddr02}. Most of these
results are tentative or suspect with the SN inferences relying on a
few mildly deviant photometric points in the afterglow light curve. Even
if some of the bumps are real, a number of other explanations for
the physical origin of such bump have been advanced: for example, dust
echoes \citep{eb00,rei01b}, shock interaction with circumburst density
discontinuities \citepeg{rdm+01}, and thermal re-emission of the
afterglow light \citep{wd00}. To definitively distinguish between the
SN hypothesis and these alternatives, detailed spectroscopic
and multi-color light curve observations of intermediate-time bumps are
required. 

It is against this background that we initiated a program with the
{\it Hubble Space Telescope} (HST) to sample afterglow light curves at
intermediate and late-times. The principal attractions of HST
are the photometric stability and high angular resolution. These are
essential in separating the afterglow from the host galaxy and in
reconstructing afterglow colors.

On theoretical grounds, if the collapsar picture is true, then we
expect to see a Type Ib/Ic SN \citep{woo93}.  In the first month,
core-collapsed supernova spectra are essentially characterized by a
blackbody (with a spectral peak near $\sim$5000 \AA) modified by broad
metal-line absorption and a strong flux suppression blueward of
$\sim4000$\, \AA\ in the restframe.  For GRBs with low redshifts, $z
\ale 1$, the effect of this blue absorption blanketing is a source
with an apparent red spectrum at observer-frame optical wavelengths;
at higher redshifts, any supernova signature is highly suppressed. For
low redshift GRBs, intermediate-time follow-up are, then, amenable to
observations with the Wide Field Planetary Camera 2 (WFPC2). In this
{\it Letter} we report on WFPC2 multi-color photometry of GRB 011121
($z=0.36$; \citealt{igsw01}) and elsewhere we report on observations
of GRB 010921 ($z=0.451$; \citealt{psk02}).  In a companion paper
(\citealt{price02a}; hereafter Paper II), we report a multi-wavelength
(radio, optical and NIR) modeling of the afterglow.

\section{Observations and Reductions}

\subsection{Detection of GRB 011121 and the afterglow}

On 21.7828 November 2001 UT, the bright GRB 011121 was detected and
localized by {\it BeppoSAX} to a 5-arcmin radius uncertainty
\citep{pir+01a}.  Subsequent observations of the error circle refined
by the IPN and {\it BeppoSAX} (see Paper II) revealed a fading optical
transient (OT) \citep{wsg01,sgw01}.  Spectroscopic observations with
the Magellan 6.5-m telescope revealed redshifted emission lines at the
OT position ($z=0.36$), indicative of a bright, star-forming host
galaxy of GRB 011121 \citep{igsw01}.

\subsection{HST Observations and reductions}

\begin{figure*}[tbp]
\centerline{\psfig{file=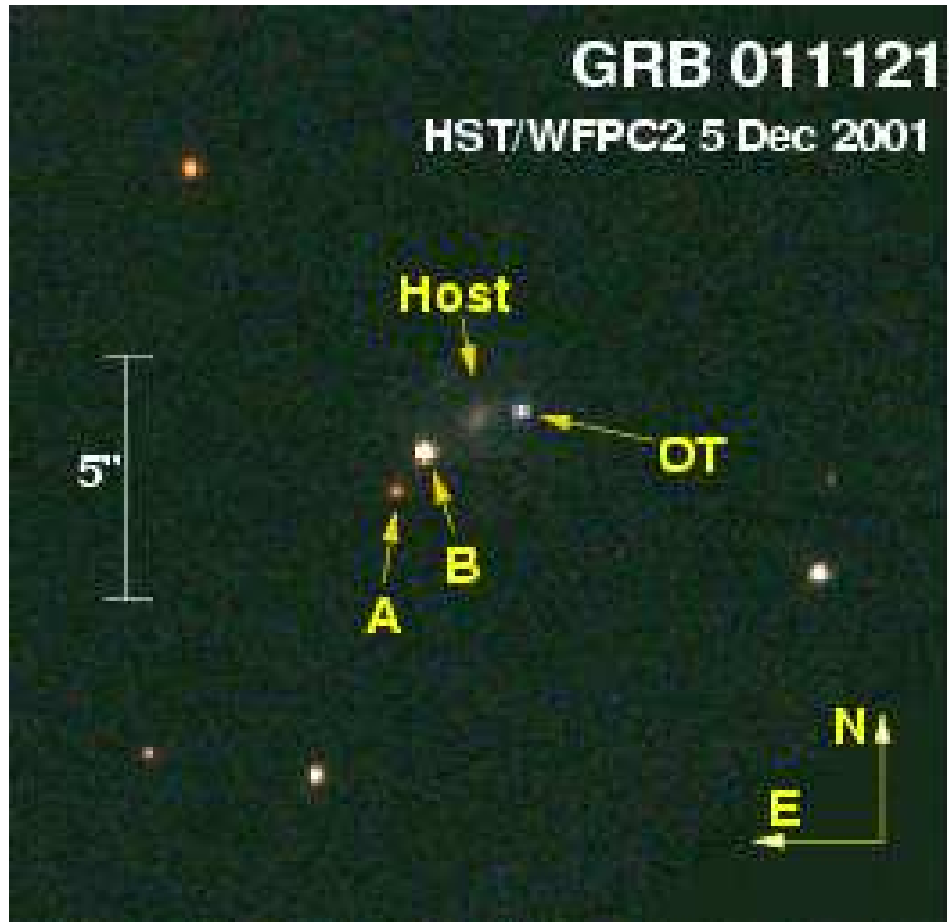,width=5in,angle=0}}
\caption[{\it Hubble Space Telescope} image of the field of GRB
011121]{{\it Hubble Space Telescope} image of the field of GRB 011121
on 4--6 December 2001 UT.  This false-color image was constructed by
registering the final drizzled images in the F555W (blue), F702W
(green) and F814W (red) filters.  The optical transient (OT) is
clearly resolved from the host galaxy and resides in the outskirts of
the morphologically smooth host galaxy. Following the astrometric
methodology outlined in \citet{bkd02}, we find that the transient is
offset from the host galaxy (883 $\pm$ 7) mas west, (86 $\pm$ 13) mas
north.  The projected offset is (4.805 $\pm$ 0.035) kpc, almost
exactly at the host half-light radius. Sources ``A'' and ``B'' are
non-variable point sources that appear more red than the OT and are
thus probably foreground stars.\vskip 0.2cm}
\label{fig:1121im}
\end{figure*}

For all the HST visits, the OT and its underlying host were placed
near the serial readout register of WF chip 3 (position {\sc WFALL})
to minimize the effect of charge transfer (in)efficiency (CTE).  The
data were pre-processed with the best bias, dark, and flat-field
calibrations available at the time of retrieval from the archive
(``on--the--fly'' calibration).  We combined all of the images in each
filter, dithered by sub-pixel offsets, using the standard IRAF/DITHER2
package to remove cosmic rays and produce a better sampled final image
in each filter.  An image of the region surrounding the transient is
shown in figure \ref{fig:1121im}. The point source was detected at
better than 20 $\sigma$ in epochs one, two and three in all filters,
and better than 5 $\sigma$ in epoch four.

\begin{deluxetable}{lccccr}
\singlespace
\tablecolumns{6} 
\tablewidth{0in}
\tablecaption{Log of HST Imaging and Photometry of the OT of GRB 011121\label{tab:hst-mag}}
\tabletypesize{\small}
\tablehead{
\colhead{Filter} & \colhead{$\Delta t$\tablenotemark{a}}  & \colhead{Integration Time} & \colhead{$\lambda_{\rm eff}$} &
\colhead{$f_\nu(\lambda_{\rm eff})$} & \colhead{Vega Magnitude\tablenotemark{b}} \\
\colhead{ } & \colhead{(days)} & \colhead{(sec)} & \colhead{(\AA)} & \colhead{($\mu$Jy)} & \colhead{(mag)}}

\startdata
\cutinhead{Epoch 1}
F450W & 13.09 & 1600 & 4678.52 & 0.551 $\pm$ 0.037 & $B$ = 24.867
$\pm$ 0.073 \\ F555W & 13.16 & 1600 & 5560.05 & 0.996 $\pm$ 0.049 &
$V$ = 23.871 $\pm$ 0.056 \\ F702W & 13.23 & 1600 & 7042.48 & 1.522
$\pm$ 0.072 & $R$ = 23.211 $\pm$ 0.054 \\ F814W & 14.02 & 1600 &
8110.44 & 1.793 $\pm$ 0.042 & $I$ = 22.772 $\pm$ 0.032 \\ F850LP &
14.15 & 1600 & 9159.21 & 1.975 $\pm$ 0.103 & \\
\cutinhead{Epoch 2}
F555W \ldots     & 23.03   & 1600  & 5630.50   & 0.647 $\pm$ 0.035    & $V$ = 24.400 $\pm$  0.061    \\  
F702W \ldots     & 23.09   & 1600  & 7002.71   & 1.271 $\pm$ 0.051    & $R$ = 23.382 $\pm$  0.048    \\  
F814W \ldots     & 24.83   & 1600  & 8105.05   & 1.495 $\pm$ 0.053    & $I$ = 22.982 $\pm$  0.043    \\  
F850LP \ldots    & 24.96   & 1600  & 9166.39   & 1.708 $\pm$ 0.100    &     \\  
\cutinhead{Epoch 3}
F555W \ldots     & 27.24   & 1600  & 5711.00   & 0.378 $\pm$ 0.027    & $V$ = 25.071 $\pm$  0.076    \\  
F702W \ldots     & 27.30   & 1600  & 7043.85   & 0.981 $\pm$ 0.036    & $R$ = 23.697 $\pm$  0.044    \\  
F814W \ldots     & 28.10   & 1600  & 8164.90   & 1.301 $\pm$ 0.070    & $I$ = 23.157 $\pm$  0.061    \\  
F850LP \ldots    & 28.16   & 1600  & 9188.39   & 1.635 $\pm$ 0.092    &     \\
\cutinhead{Epoch 4}
F555W \ldots     & 77.33   & 2100  & 5604.61   & 0.123 $\pm$ 0.014    & $V$ = 26.173 $\pm$  0.118    \\
F702W \ldots     & 76.58   & 4100  & 7042.09   & 0.224 $\pm$ 0.019    & $R$ = 25.264 $\pm$  0.092    \\ 
F814W \ldots     & 77.25   & 2000  & 8149.18   & 0.294 $\pm$ 0.020    & $I$ = 24.762 $\pm$  0.073   

\enddata

\tablecomments{In the fourth column, the effective wavelength of
the filter based upon the observed spectral flux distribution of the
transient at the given epoch. In the fifth column, the flux is given
at this effective wavelength in an 0\arcsec.5 radius. The observed
count rate, corrected for CTE effects, was converted to flux using the
{\sc IRAF/SYNPHOT} package. An input spectrum with $f_\nu =$ constant
was first assumed. Then approximate spectral indices between each
filter were computed and then used to re-compute the flux and the
effective wavelength of the filters.  This bootstrapping converged
after a few iterations. The HST photometry contains an unknown but
small contribution from the host galaxy at the OT location. We
attempted to estimate the contamination of the host at the transient
position by measuring the host flux in several apertures at
approximate isophotal levels to the OT position. We estimate the
contribution of the host galaxy to be $f_\nu(F450W) = (0.098 \pm
0.039)~\mu$Jy, $f_\nu(F555W) = (0.087 \pm 0.027)~\mu$Jy, $f_\nu(F702W)
= (0.127 \pm 0.026)~\mu$Jy, $f_\nu(F814W) = (0.209 \pm 0.059)~\mu$Jy,
and $f_\nu(F850LP) = (0.444 \pm 0.103)~\mu$Jy. To correct these
numbers to ``infinite aperture,'' multiply the fluxes by 1.096
\citep{hbc+95}. These fluxes have not been corrected for Galactic or
host extinction. } 
\tablenotetext{a}{Mean time since GRB trigger on 21.7828 Nov 2001 UT.}
\tablenotetext{b}{Tabulated brightnesses in the
Vega magnitude system ($B_{\rm Vega} = 0.02$ mag, $V_{\rm Vega} =
0.03$ mag, $R_{\rm Vega} = 0.039$ mag, $I_{\rm Vega} = 0.035$ mag;
\citealt{hbc+95}). Subtract 0.1 mag from these values to get the
infinite aperture brightness. These magnitudes have not been corrected
for Galactic or host extinction.}
\end{deluxetable}

Given the proximity of the OT to its host galaxy, the final HST images
were photometered using the {\sc IRAF/DAOPHOT} package which
implements PSF-fitting photometry on point-sources \citep{ste87}.  The
PSF local to the OT was modeled with {\sc PSTSELECT} and {\sc PSF}
using at least 15 isolated stars detected in the WF chip 3 with an
adaptive kernel to account for PSF variations across the image ({\sc
VARORDER = 1}).  The resulting photometry, reported in Table
\ref{tab:hst-mag}, was obtained by finding the flux in an 0\arcsec .5
radius using a PSF fit.  We corrected the observed countrate using the
formulation for CTE correction in \citet{dol00} with the most
up-to-date parameters\footnotemark\footnotetext{See {\tt
http://www.noao.edu/staff/dolphin/wfpc2\_calib/.}}\nadaafterfoot; such
corrections, computed for each individual exposure, were never larger
than 8\% (typically 4\%) for a final drizzled image. We estimated the
uncertainty in the CTE correction, which is dependent upon source
flux, sky background, and chip position, by computing the scatter in
the CTE corrections for each of the images that were used to produce
the final image.  The magnitudes reported in the standard bandpass
filters in Table \ref{tab:hst-mag} were found using the
\citeauthor{dol00} prescription.

\section{Results}

\begin{figure*}[tbp]
\centerline{\psfig{file=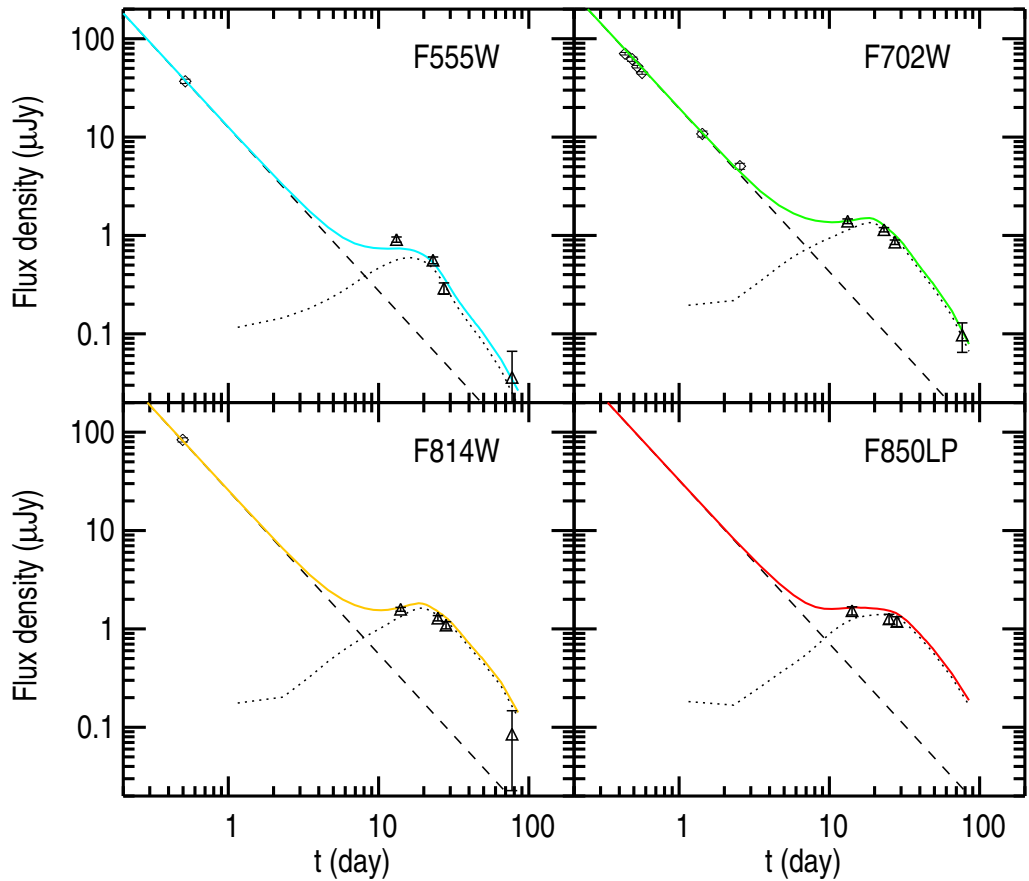,width=5.5in,angle=0}}
\caption[Light curves of the intermediate-time red bump of
GRB~011121]{Light-curves of the afterglow and the intermediate-time
red bump of GRB~011121.  The triangles are our HST photometry in the
F555W, F702W, F814W and F850LP filters (all corrected for the
estimated contribution from the host galaxy), and the diamonds are
ground-based measurements from the literature \citep{obs+01,kw01}.
The dashed line is our fit to the optical afterglow (see Paper II),
the dotted line is the expected flux from the template SN at the
redshift of GRB~011121, with foreground extinction applied and dimmed
by 55\% to approximately fit the data, and the solid line is the sum
of the afterglow and SN components. Corrections for color effects
between the ground-based filters and the HST filters were taken to be
negligible for the purpose of this exercise.}
\label{fig:lc}
\end{figure*}

In figure~\ref{fig:lc} we plot the measured fluxes from our four HST
epochs in the F555W, F702W, F814W and F850LP filters.  We also plot
measurements made at earlier times (0.5 days $< t <$ 3 days) with
ground-based telescopes and reported in the literature.  These
magnitudes were converted to fluxes using the zero-points of
\citet{fsi95} and plotted in the appropriate HST filters.

\begin{figure*}[tbp]
\centerline{\psfig{file=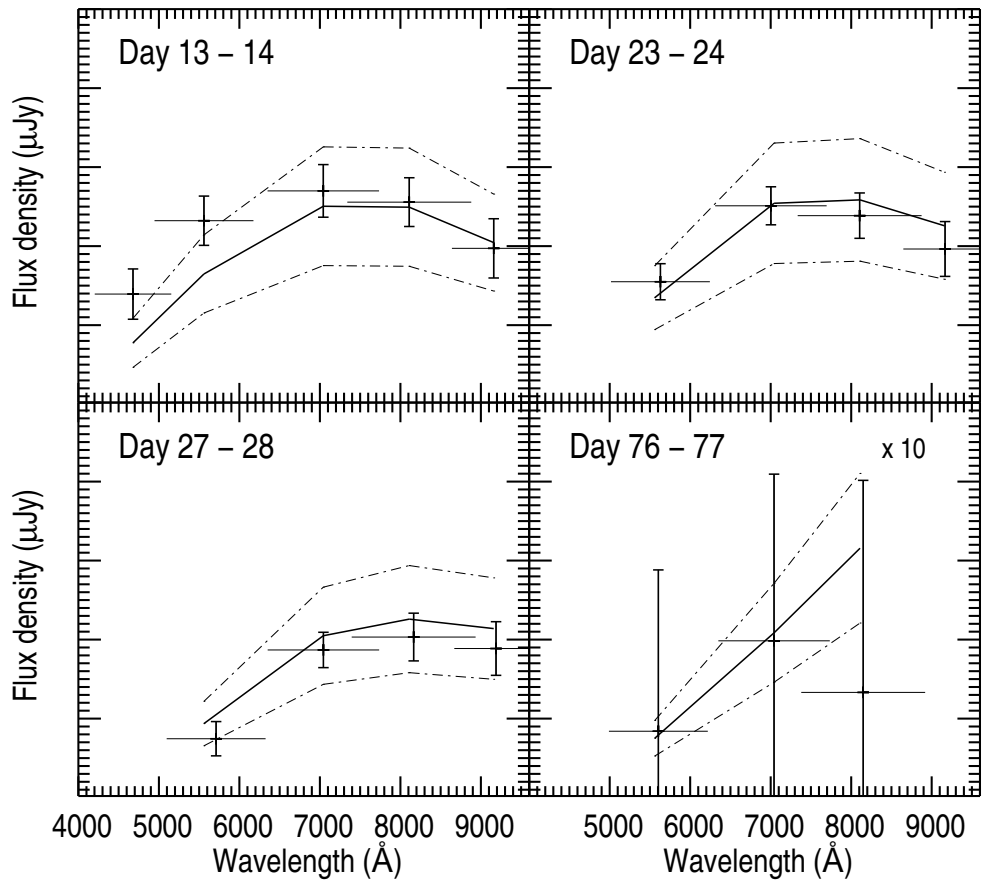,width=5.5in,angle=0}} 
\caption[The spectral flux distributions of the red bump at the time
of the four HST epochs]{The spectral flux distributions of the red
bump at the time of the four HST epochs. The fluxes are dereddened
using $A_V = 1.16$ mag.  Spectral evolution, and more important, a
turn-over in the spectra of the first three epochs, are clearly seen.
The peak of the turn-over (around $7200$ \AA) corresponds to a peak in
the red bump spectrum at $\sim$5300 \AA. For comparison, we show a
template broadband SN spectra (a dimmed version of SN 1998bw; solid
curve) as it would appear at the redshift of GRB 011121 and the
associated 2 $\sigma$ errors (see text).  The vertical error bars on
the red bump reflect the 1 $\sigma$ statistical uncertainty flux from
only the red bump.  There are large ($\sim$1 mag) systematic
uncertainties (e.g., Galactic reddening, relative distance moduli
between SN 1998bw and GRB 011121) in both the data and the model;
these are suppressed for clarity.}
\label{fig:sfd}
\end{figure*}
Corrections for color effects between the ground-based filters and HST
filters were taken to be negligible for the purpose of this exercise.

The estimated contribution from the afterglow is heavily weighted by
the available data: our ground-based data (and those reported in the
literature so far) are primarily at early times. Roughly, over the
first week, the afterglow exhibits a simple power law decay. The
afterglow contribution derived from our NIR data and optical data from
the literature (see Paper II) is shown by the dashed line in each
panel. No afterglow light curve breaks (e.g., from jetting) were assumed.

\citet{ghj+02} drew attention to an excess of flux (in $R$-band), at a
time 13~days after the GRB, with respect to that expected from the
power-law extrapolation of early-time afterglow emission; they
suggested the excess to arise from an underlying SN. As can vividly be
seen from our multi-color data, the excess is seen in all bands and
over several epochs.

We used the light curve and spectra\footnotemark\footnotetext{Spectra
were obtained through the Online Supernova Spectrum Archive (SUSPECT)
at {\tt
http://tor.nhn.ou.edu/$\sim$suspect/index.html}.}\nadaafterfoot\ of
the well-studied Type Ic supernova SN~1998bw \citep{gvv+98,ms99} to
create a comparison template broad-band light curve of a Type Ic
supernova at redshift $z = 0.36$. Specifically, the spectra of SN
1998bw were used to compute the $K$-corrections between observed
photometric bands of 1998bw and {\it HST} bandpasses (following
\citealt{kgp96} and
\citealt{ssp+98}).  A flat $\Lambda$ cosmology with $H_0 = 65$ km
s$^{-1}$ Mpc$^{-1}$ and $\Omega_M=0.3$ was assumed and we took the
Galactic foreground extinction to SN 1998bw of $A_V=0.19$ mag
\citep{gvv+98}.

Since dimmer Ic SNe tend to peak earlier and decay more quickly (see
fig.~1 of \citealt{imn+98}), much in the same way that SN Ia do, we
coupled the flux scaling of SN 1998bw with time scaling in a method
analogous to the ``stretch'' method for SN Ia distances
\citep{pgg+97}.  To do so, we fit an empirical relation between 1998bw
and 1994I to determine the flux-time scaling. We estimate that a
1998bw-like SN that is dimmed by 55\% (see below), would peak and
decay about 17\% faster than 1998bw itself. Some deviations from our
simple one-parameter template are apparent, particularly in the F555W
band and at late-times.

In figure \ref{fig:sfd}, we plot the spectral flux distributions
(SFDs) of the intermediate-time bump at the four HST epochs. A clear
turn-over in the spectra in the first 3 epochs is seen at about 7200
\AA. The solid curve is the SFD of SN~1998bw transformed as described
above with the associated 2-$\sigma$ errors. Bearing in mind that
there are large systematic uncertainties in the template (i.e., the
relative distance moduli between SN 1998bw and GRB 011121) and in the
re-construction of the red bump itself (i.e., the Galactic extinction
toward GRB~011121 and the contribution from the afterglow in the early
epochs), the consistency between the measurements and the SN is
reasonable. We consider the differences, particularly the bluer bands
in epoch one, to be relatively minor compared with the overall
agreement. This statement is made in light of the large observed
spectral diversity of Type Ib/Ic SNe (see, for example, figure 1 of
\citealt{mdm+02}).

\section{Discussion and Conclusions}
\label{sec:bw-dis}

We have presented unambiguous evidence for a red, transient excess
above the extrapolated light curve of the afterglow of GRB~011121.  We
suggest that the light curve and spectral flux distribution of this
excess appears to be well represented by a bright SN.  While we have
not yet explicitly compared the observations to the expectations of
alternative suggestions for the source of emission (dust echoes,
thermal re-emission from dust, etc.), the simplicity of the SN
interpretation---requiring only a (physically motivated) adjustment in
brightness---is a compelling (i.e., Occam's Razor) argument to accept
our hypothesis.  Given that the red bump detections in a number of
other GRBs occur on a similar timescale as in GRB 011121, any model
for these red bumps should have a natural timescale for peak of $\sim
20 (1 + z)$ day; in our opinion, the other known possibilities do not
have such a natural timescale as compared with the SN
hypothesis. Indeed, if our SN hypothesis is correct, then the flux
should decline as an exponential from epoch four onward.  The ultimate
confirmation of the supernova hypothesis is a spectrum which should
show characteristic broad metal-line absorption of the expanding
ejecta (from, e.g., Ca II, Ti II, Fe II).

We used a simplistic empirical brightness--time stretch relation to
transform 1998bw, showing good agreement between the observations and
the data. If we neglect the time-stretching and only dim the 1998bw
template, then the data also appear to match the template reasonably
well, however, the discrepancies in the bluer bands become somewhat
larger and the flux ratios between epochs are slightly more
mismatched.  The agreement improves if we shift the time of the
supernova to be about $\sim$3--5 days (restframe) before the GRB time.
Occurrence times more than about ten days (restframe) before the GRB
can be ruled out. This observation, then, excludes the original
``supranova'' idea \citep{vs98}, that posited a supernova would
precede a GRB by several years (see eq.~[1] of \citealt{vs98}).
Modified supranova scenarios that would allow for any time delay
between the GRB and the accompanying SN, albeit {\it ad hoc}, are
still consistent with the data presented
herein\footnotemark\footnotetext{The explosion date of even very
well-studied supernovae, such as 1998bw, cannot be determined via
light curves to better than about 3 days (e.g.,
\citealt{imn+98}). This implies that future photometric studies might
not be equipped to distinguish between contemporaneous SN/GRB events
and small delay scenarios.}\periodafterfoot

Regardless of the timing between the SN explosion and the GRB event
(constrained to be less than about 10 days apart), the bigger picture
we advocate is that GRB 011121 resulted from an explosive death of a
massive star.  This conclusion is independently supported by the
inference, from afterglow observations of GRB 011121 (Paper II), of a
wind-stratified circumburst medium.

The next phase of inquiry is to understand the details of the
explosion and also to pin down the progenitor population. A large
diversity in any accompanying SN component of GRBs is expected from
both a consideration of SNe themselves and the explosion mechanism.
The three main physical parameters of a Type Ib/Ic SN are the total
explosive energy, the mass of the ejecta, and the amount of Nickel
synthesized by the explosion ($M_{\rm Ni}$). The peak luminosity and
time to peak are roughly determined by the first two whereas the
exponential tail is related to $M_{\rm Ni}$.  Ordinary Ib/Ic SNe
appear to show a wide dispersion in the peak luminosity
\citep{imn+98}. There is little {\it ab initio} understanding of this
diversity (other than shifting the blame to dispersion in the three
parameters discussed above).

It is now generally accepted that GRBs are not spherical explosions
and are, as such, usually modeled as a jetted outflow. \citet{fks+01}
model the afterglow of GRBs and have presented a compilation of
opening angles, $\theta$, ranging from less than a degree to 30
degrees and a median of 4 degrees. If GRBs have such strong
collimation then it is not reasonable to assume that the explosion,
which explodes the star, will be spherical. We must be prepared to
accept that the SN explosion is extremely asymmetric and thus even a
richer diversity in the light curves. This expected diversity may
account for both the scale factor difference between the SN component
seen here and in SN 1998bw seen in figure \ref{fig:sfd}.  Indeed,
there has been a significant discussion as to the degree to which the
central engine in GRBs will affect the overall explosion of the star
\citep{woo93,kho99,mw99,hww99,mwh01}.  These models have focused
primarily on the hydrodynamics and lack the radiative modeling
necessary to compare observations to the models.

Clearly, the observational next step is to obtain spectroscopy (and
perhaps even spectropolarimetry) and to use observations to obtain a
rough measure of the three-dimensional velocity field and geometry of
the debris.  As shown by GRB 011121 the SN component is bright enough
to undertake observations with the largest ground-based telescopes.

We end by noting the following curious point.  The total energy yield
of a GRB is usually estimated from the gamma-ray fluence and an
estimate of $\theta$ \citepcf{fks+01}. Alternatively, the energy in the
afterglow is used \citepeg{pkpp01}. However, for GRB~011121, the
energy in the SN component (scaling from the well-studied SN~1998bw)
is likely to be comparable or even larger than that seen in the burst
or the afterglow. In view of this, the apparent constancy of the
$\gamma$-ray energy release is even more mysterious.

\acknowledgments

We thank S.~Woosley, who, as referee, provided helpful insights toward
the improvement of this work. A.~MacFadyen and E.~Ramirez-Ruiz are
acknowledged for their constructive comments on the paper. J.~S.~B.~is
a Fannie and John Hertz Foundation Fellow. F.~A.~H.~acknowledges
support from a Presidential Early Career award. S.~R.~K.~and
S.~G.~D.~thank the NSF for support. R.~S.~is grateful for support from
a NASA ATP grant. R.~S.~and T.~J.~G.~acknowledge support from the
Sherman Fairchild Foundation.  J.~C.~W.~acknowledges support from NASA grant
NAG59302.  KH is grateful for Ulysses support under JPL Contract
958056, and for IPN support under NASA Grants FDNAG 5-11451 and NAG
5-17100. Support for Proposal number HST-GO-09180.01-A was provided by
NASA through a grant from Space Telescope Science Institute, which is
operated by the Association of Universities for Research in Astronomy,
Incorporated, under NASA Contract NAS5-26555.

\chapter[Expected Characteristics of the Subclass of Supernova Gamma-Ray Bursts]{Expected Characteristics of the Subclass of Supernova Gamma-Ray Bursts$^\dag$}
\label{chap:sgrbs}

\secfootnote{\secfootdag}{A version of this chapter was first published 
in {\it The Astrophysical Journal Letters},  506, p.~L105--L108, (1998).}

\secauthor{J. S. Bloom$^1$, S. R. Kulkarni$^1$, F.~Harrison$^1$, T.~Prince$^1$, E.~S.~Phinney$^1$, D.~A.~Frail$^2$}

\medskip
\bigskip

\secaffils{$^1$ Palomar Observatory 105--24, California Institute of Technology,
            Pasadena, CA 91125, USA}
\secaffils{$^2$ National Radio Astronomy Observatory, P.O.~Box O, 1003
   Lopezville Road, Socorro, NM}

\begin{abstract}
The spatial and temporal coincidence of gamma-ray burst (GRB) 980425
and supernova (SN) 1998bw has prompted speculation that there exists a
subclass of GRBs produced by SNe (``S-GRBs''). A physical model
motivated by radio observations lead us to propose the following
characteristics of S-GRBs: (1) prompt radio emission and an implied
high brightness temperature close to the inverse Compton limit, (2)
high expansion velocity ($\age$50,000 km s-1) of the optical
photosphere as derived from lines widths and energy release larger
than usual, (3) no long-lived X-ray afterglow, and (4) a single-pulse
GRB profile. Radio studies of previous SNe show that only (but not
all) Type Ib and Ic SNe potentially satisfy the first condition. We
investigate the proposed associations of GRBs and SNe within the
context of these proposed criteria and suggest that $\sim$1\% of GRBs
detected by BATSE may be members of this subclass.
\end{abstract}

\section{Introduction}

With the spectroscopic observations of the optical afterglow of
gamma-ray burst (GRB) 970508 by \citet{mdk+97} came proof that at
least one GRB is at a cosmological distance. \citet{kdr+98} later
added another cosmological GRB, which, based on an association with a
high-redshift galaxy, had an implied energy release of $E_{\gamma}
\age 10^{53}$ erg. However, not all GRBs have been shown to be associated with
distant host galaxies. Only about half of all GRBs are followed by
long-lived optical afterglow, and one in four produce a longer-lived
radio afterglow at or above the 100 $\mu$Jy level. In contrast, X-ray
afterglow has been seen for almost all \emph{BeppoSAX}-localized
bursts. Until recently, the emerging picture had been that all GRBs
are located at cosmological distances and these GRBs (hereafter
cosmological GRBs, or C-GRBs) are associated with star-forming regions
and that C-GRBs are the death throes of massive stars.

The discovery of a supernova (SN 1998bw) both spatially (chance
probability of $10^{-4}$) and temporally coincident with GRB 980425
\citep{gvp+98,gvv+98} suggests the existence of another class
of GRBs. Remarkably, SN 1998bw showed very strong radio emission with
rapid turn-on; it is, in fact, the brightest radio SN to date
\citep{wfk+98}. This rarity further diminishes the probability
of chance coincidence \citep{ssbe98}. From the radio observations,
\citet{kfw+98} concluded that there exists a relativistic
shock [bulk Lorentz factor, $(\Gamma \equiv
(1-\beta^{2})\mathstrut{^{-1{/}2}} \age 2$] even 4 days after the SN
explosion. \citeauthor{kfw+98} argue that the young shock had all the
necessary ingredients (high $\Gamma$, sufficient energy) to generate
the observed burst of gamma rays.

We feel that the physical connection between GRB 980425 and SN 1998bw
is strong. Accepting this connection then implies that there is at
least one GRB that is not of distant cosmological origin but is
instead related to an SN event in the local universe ($\ale$100
Mpc). We refer to this category of GRBs as supernova-GRBs or
S-GRBs. Many questions arise: How common are S-GRBs? How can they be
distinguished from C-GRBs? What are their typical energetics?

In this paper, accepting the physical model advocated by
\citet{kfw+98}, we enumerate the defining characteristics of the class
of S-GRBs. We then apply these criteria to members of this proposed
class and conclude with a discussion of the potential number of
S-GRBs.

\section{How to Recognize S-GRBs}
\label{sec:recognize}

The expected characteristics of S-GRBs is motivated by the model
developed to explain the radio observations of SN 1998bw. Briefly,
from the radio data, \citet{kfw+98} conclude that the radio emitting
region is expanding at least at 2\emph{c} (4 days after the explosion)
and slowing down to $c$, one month after the burst. Indeed, one
expects the shock to slow down as it accretes ambient matter. Thus, it
is reasonable to expect the shock to have had a higher $\Gamma$ when
it was younger. The expectation is that this high-$\Gamma$ shock is
also responsible for the observed burst of gamma rays (synchrotron or
inverse Compton scattering). Of note, whereas in C-GRBs the primary
afterglow is optical, in S-GRBs the primary afterglow is in the radio
band. We now enumerate the four criteria of S-GRBs:

  1. \emph{Prompt radio emission and high brightness
temperature}.---An unambiguous indication of a relativistic shock in
an SN is when the inferred brightness temperature,
\emph{T}$\mathstrut{_{B}}$, exceeds \emph{T}$\mathstrut{_{{\rm
icc}}}$$\sim$4$\times$10$\mathstrut{^{11}}$ K, the so-called inverse
Compton catastrophe temperature. $T_B$ is given by
$$
      T_B = 6\times 10^{8} \Gamma^{-3}\beta^{-2} S({\rm mJy})
          (\nu/5\,{\rm GHz})^{-2} t_d^{-2}d_{\rm Mpc}^2 \,\,{\rm K};
          \eqno (1)
$$
here $t_d$ is the time in days since the burst of gamma-rays, $d_{\rm
Mpc}$ is the distance in Mpc, and $S$, the flux density at frequency
$\nu$. The energy in the particles and the
magnetic field is the smallest when
\emph{T}$\mathstrut{_{B}}$$\simeq$\emph{T}$\mathstrut{_{{\rm eq}}}$,
the so-called ``equipartition'' temperature
\citep[\hbox{\emph{T}$\mathstrut{_{{\rm eq}}}$$\sim$5$\times$10$\mathstrut{^{10}}$
K};][]{read94}. The inferred energy increases sharply with increasing
\emph{T}$\mathstrut{_{B}}$. For SN 1998bw, even with
\emph{T}$\mathstrut{_{B}}$=\emph{T}$\mathstrut{_{{\rm eq}}}$, the
inferred energy in the relativistic shock is 10$\mathstrut{^{48}}$
ergs, which is already significant. If
\emph{T}$\mathstrut{_{B}}$$>$\emph{T}$\mathstrut{_{{\rm icc}}}$, the
inferred energy goes up by a factor of 500 and thus approaches the
total energy release of a typical SN ($\sim10^{51}$ ergs). Thus, the
condition \emph{T}$\mathstrut{_{B}}$$<$\emph{T}$\mathstrut{_{{\rm
icc}}}$ is a reasonable inequality to use. This then leads to a lower
limit on $\Gamma$. We consider the shock to be relativistic when
$\Gamma$$\beta$$>$1. For SN 1998bw, \citet{kfw+98} find
$\Gamma$$\beta$$\age$2.

It is well known that prompt radio emission (by this we mean a
timescale of a few days) is seen from Type Ib/Ic SNe
\citep{ws98a,chev98}. Radio emission in Type II SNe peaks on very long
timescales (months to years). No Type Ia SN has yet been detected in
the radio. Thus, the criterion of prompt radio emission (equivalent to
high \emph{T}$\mathstrut{_{B}}$) will naturally lead to selecting only
Type Ib/Ic SNe. High brightness temperature is achieved when the radio
flux is high. Indeed, the radio luminosity of SN 1998bw was 2 orders
of magnitude larger than the five previously studied Type Ic/Ic SNe
\citep{vdsw+93}.

2. \emph{No long-lived X-ray afterglow}.---In our physical picture
   above, we do not expect any long-lived X-ray emission since the
   synchrotron lifetime of X-ray--emitting electrons is so short. The
   lack of X-ray afterglow from GRB 980425 in the direction of SN
   1998bw is consistent with this picture.

3. \emph{A simple GRB profile}.---In the model we have adopted, the
   gamma-ray and the radio emission is powered by an energetic
   relativistic shock. Is it likely that there is more than one
   relativistic shock? Our answer is no. There is no basis to believe
   or expect that the collapse of the progenitor core will result in
   multiple shocks. It is possible that the nascent pulsar or a black
   hole could be energetically important, but the envelope matter
   surely will dampen down rapid temporal variability of the
   underlying source. From this discussion we conclude that there is
   only one relativistic shock. Thus, the gamma-ray burst profile
   should be very simple: a single pulse (SP).

\begin{figure*}[p]
\centerline{\psfig{file=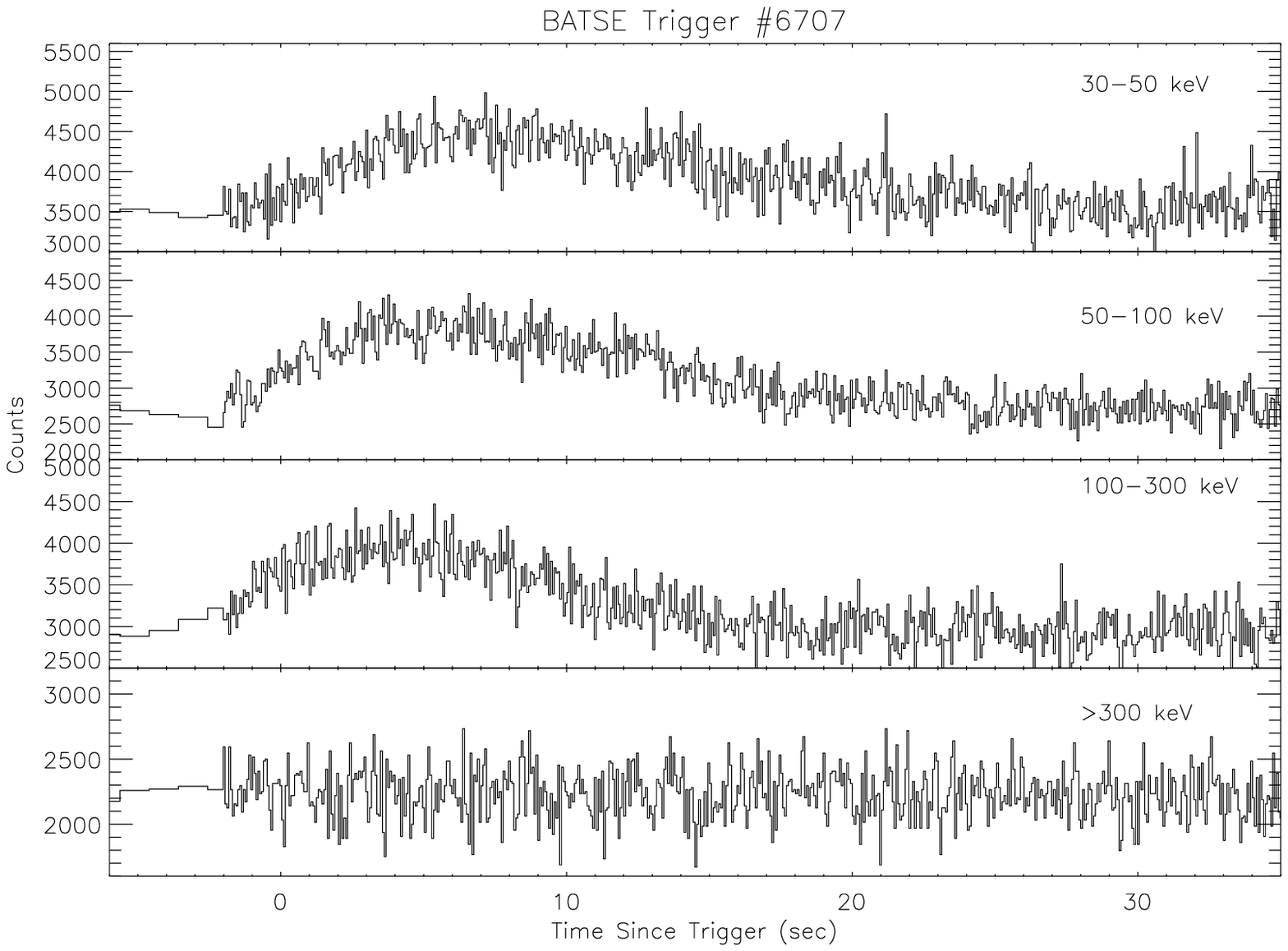,height=3in,width=5.5in}}
\centerline{\psfig{file=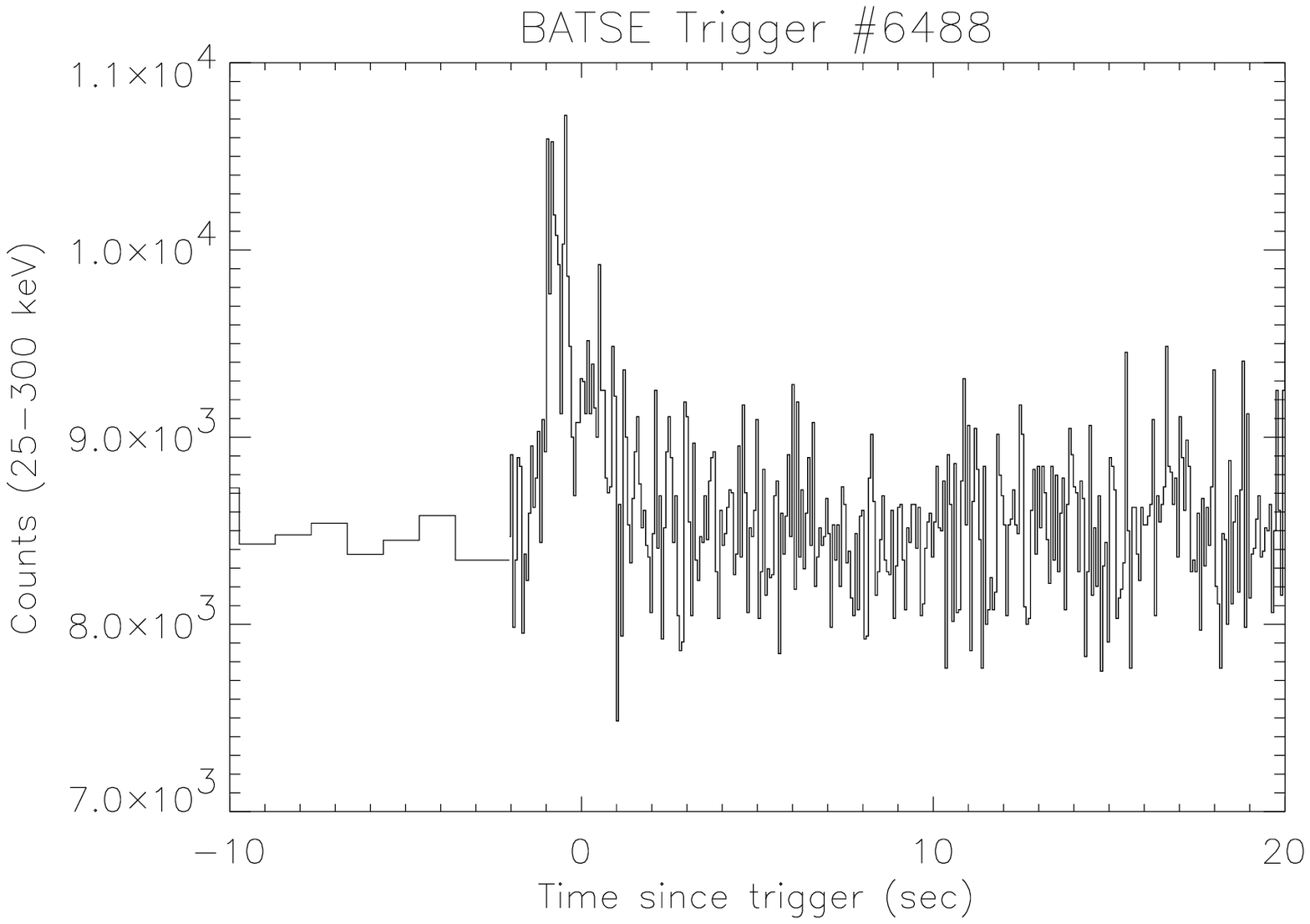,height=2in,width=2.5in}
        \psfig{file=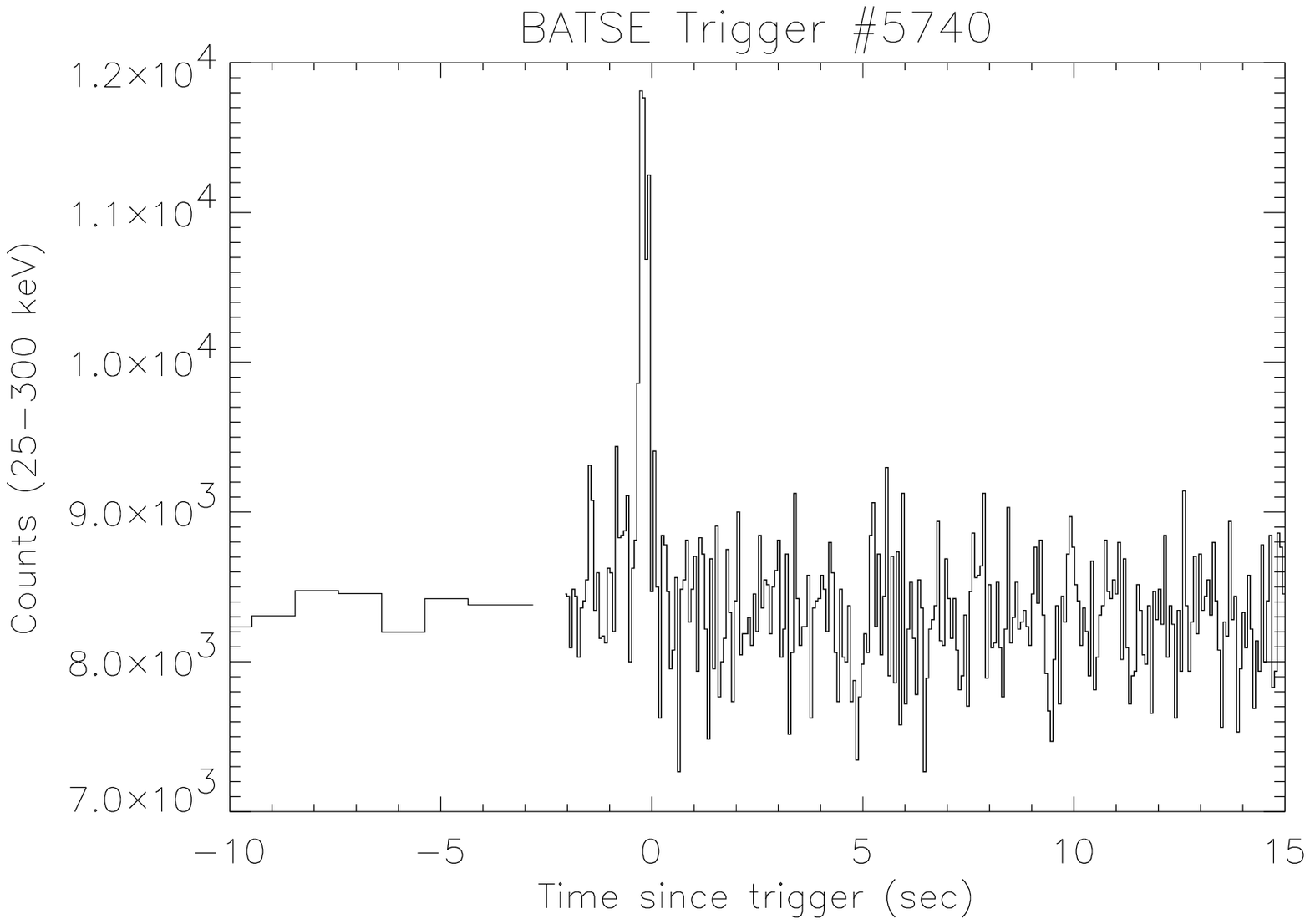,height=2in,width=2.5in}}
\centerline{\psfig{file=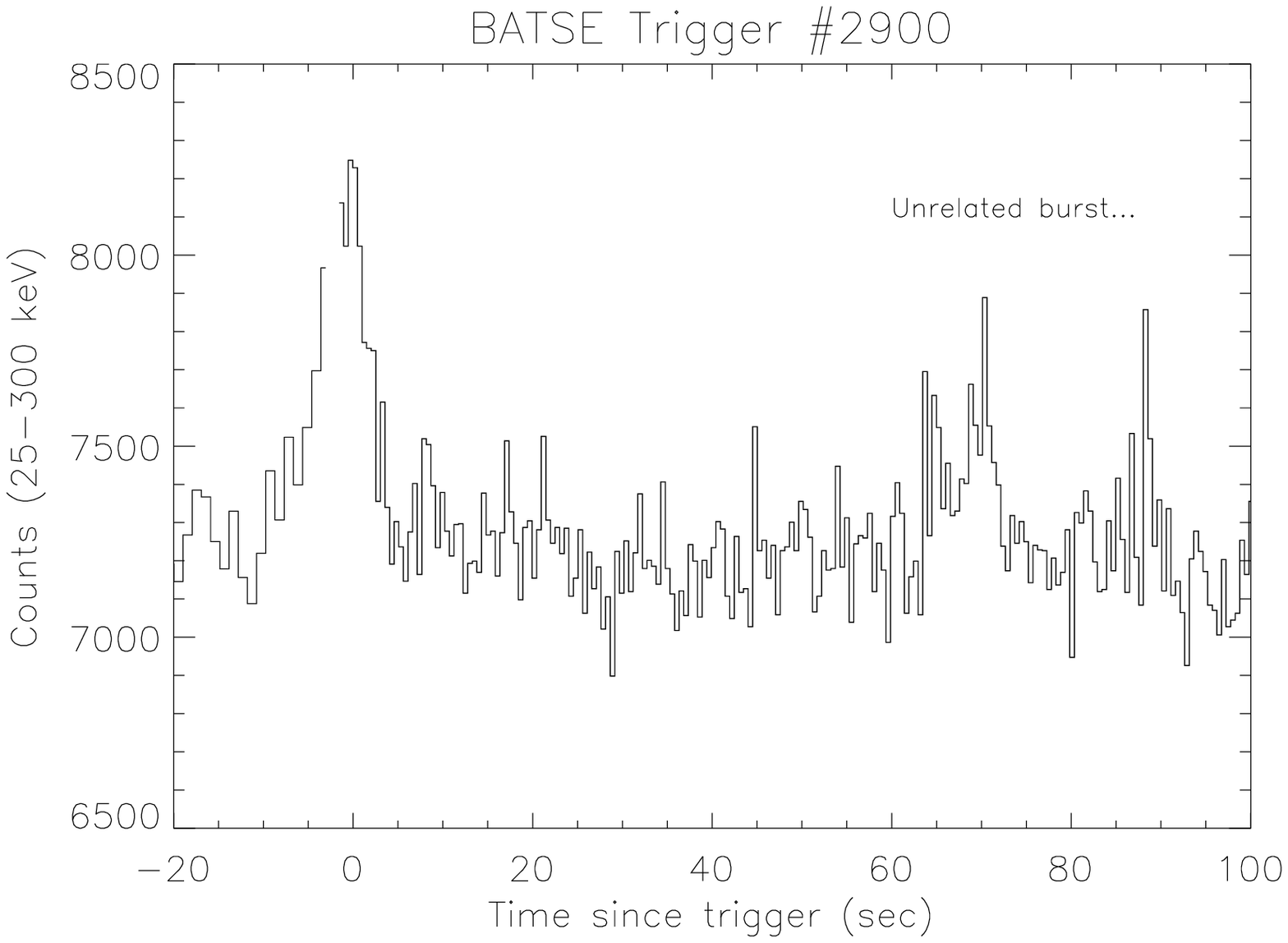,height=2in,width=2.5in}
        \psfig{file=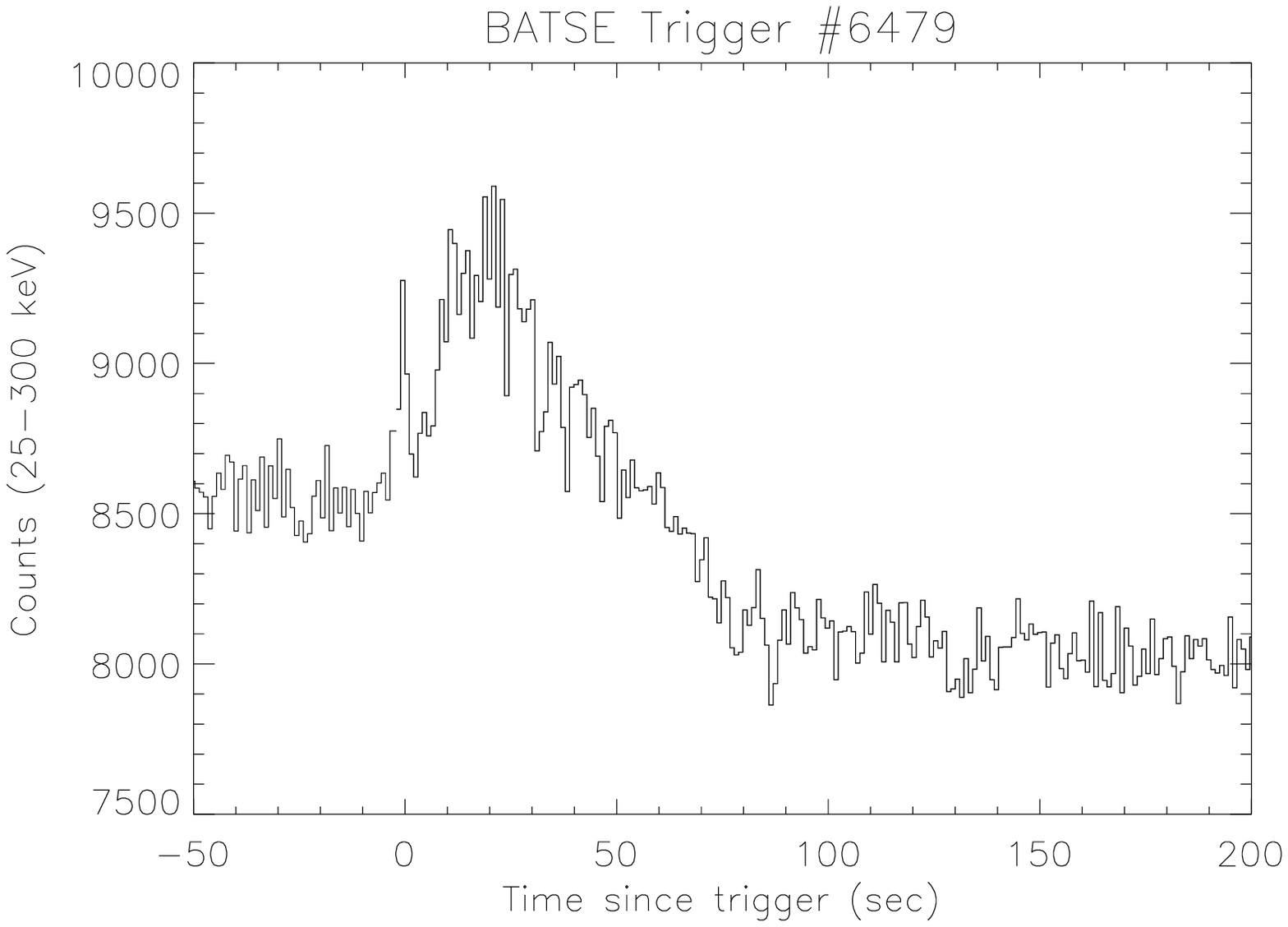,height=2in,width=2.5in}}
\caption[The $\gamma$-ray light curves of GRB 980425 and other potential 
S-GRBs]{(top) The 4-channel light curve of GRB 980425 (Trigger \#6707)
associated with SN 1998bw. The single pulse (SP) appears cusp-less
unlike most SP BATSE bursts.  The hard-to-soft evolution is clear from
the progression of the peak from channels 3 to 1 over time.  After
BATSE triggers, the light curve is sampled on 64-ms
timescales. Continuous DISCLA data is augmented to the pre-trigger
light curve; this data is basically a 16 bin (1.024 sec) averaged over
the more finely sampled 64-ms data.  In the case of longer bursts, we
average the 64-ms bins over 16-sec intervals to reduce noise.  (bottom
panel; clockwise from top left) Light curve of GRB 97112 (Trigger
\#6488)/SN 1997ef or SN 1997ei; GRB 970103 (Trigger
\#5740)/SN 1997X; GRB 971115 (Trigger \#6479)/SN 1997ef; GRB 940331
(Trigger \#2900)/SN 1994I.  According to the BATSE archive, the pulse
beginning $t \simeq 65$ sec is an unrelated (i.e.,~not spatially
coincident) GRB.}
\label{fig:sn-6707}
\end{figure*}

The light curve of GRB 980425 (fig.~\ref{fig:sn-6707}) is a simple
single pulse (SP) with a $\sim$5 s rise (HWHM) and a $\sim$8 s
decay. Like most GRBs \citepeg{cls+97,band97}, the harder emission
precedes the softer emission with channel 3 (100--300 keV) peaking
$\sim$1 s before channel 1 (25--50 keV). Unlike most GRB light curves,
the profile of GRB 980425 has a rounded maximum instead of a cusp.

4. \emph{Broad line emission and bright optical
luminosity}.---\citet{kfw+98} noted that the minimum energy in the
relativistic shock, \emph{E}$\mathstrut{_{{\rm min}}}$, is 10$^{48}$
ergs and that the true energy content could be as high as 10$^{52}$
ergs. Even the lower value is a significant fraction of energy of the
total supernova release of ordinary SNe (\emph{E}$\mathstrut{_{{\rm
tot}}}$$\sim$10$\mathstrut{^{51}}$ ergs). Clearly, a larger energy
release in the supernova would favor a more energetic shock and,
hence, increase the chance such a shock could produce a burst of gamma
rays. Indeed, there are indications from the modeling of the light
curve and the spectra that the energy release in SN 1998bw was
3$\times$10$\mathstrut{^{52}}$ ergs \citep{wes99,imn+98}, a factor of
$\sim$30 larger than the canonical SN. This then leads us to propose
the final criterion: indications of a more-than-normal release of
energy. Observationally, this release is manifested by large expansion
speed, which leads to the criterion of broad emission lines and bright
optical luminosity.

\citet{naka98} suggests that S-GRBs derive their energy from the
formation of a strongly magnetized pulsar rotating at millisecond
period. Furthermore, he advocates that S-GRBs must possess
``non-high-energy'' \citep[NHE; see ][]{ppb+97} profiles (i.e., little
flux above 300 keV). However, Nakamura's model does not address the
most outstanding feature of SN 1998bw---its extremely unusual radio
emission. Our model is silent on whether the bursts should be NHE or
HE since that would depend on the details of the emission mechanism
and the importance of subsequent scattering.

There is an implicit assumption on the part of several authors
\citepeg{naka98,wes99,ww98} that S-GRBs are intimately connected with
Type Ic SNe. Within the framework of our model, the key issue is
whether there exists a relativistic shock that can power the gamma
rays. Clearly this relativistic shock is distinctly different from the
low-velocity shock that powers the optical emission. Thus, the
connection between GRB emission and the optical properties of the SN
is bound to be indirect (e.g., our fourth criterion).

\section{Application of Criteria to Proposed Associations}

We now apply the above four criteria motivated by a specific physical
model to proposed S-GRBs \citep{ww98,wes99}. We searched for more
potential GRB-SN associations by cross-correlating the earlier WATCH
and Interplanetary Network (IPN) localizations
\citep{abh+87b,lund95,hhk+97} with an archive catalog of
supernovae\footnotemark\footnotetext{The updated Asiago Supernova
Catalog of \citet{bct89} (maintained by E.~Capellaro) is available at
{\tt http://athena.pd.astro.it/$\sim$supern/.}}\periodafterfoot We
found no convincing associations in archival GRB/SN data before the
launch of the Burst and Source Transient Experiment (BATSE). Thus, our
total list remains at nine, seven from \citet{ww98} and two from
\citet{wes99}.

We reject the following proposed associations: (1) SN 1996N/GRB 960221
\citep{ww98}. The IPN data rule out this association on
spatial grounds alone. This lack of association was independently
recognized by \citet{kbk+98}. (2) SN 1992ar/GRB 920616
\citep{wes99}. The associated GRB appears not to exist in the BATSE 4B
Catalog\footnotemark\footnotetext{The BATSE Gamma-Ray Burst Catalog
(maintained by C. A. Meegan et al.), including the BATSE 4B Catalog,
is available at {\tt
http://www.batse.msfc.nasa.gov/data/grb/catalog/}.}\commaafterfoot
$\!\!$and furthermore, there are no other GRBs within a month that are
spatially coincident with the SN. (3) SN 1998T/GRB 980218
\citep{ww98}. This is ruled out on spatial grounds from the IPN data
\citep{kbk+98}.

\begin{deluxetable}{lcccccccccr}
\singlespace
\tabletypesize{\scriptsize}
\tablecolumns{8} 
\tablewidth{0pc} 
\tablecaption{GRB/Supernovae Associations: Which are Truly S-GRBs? \label{tab:sgrb-obs}}
\tablehead{
\multicolumn{1}{l}{Association} & SN & {\small Prompt} & $\delta
\theta^\spadesuit$ & \multicolumn{1}{c}{$v_{\rm max}$} & GRB Type &
D$^{\clubsuit}$ & {\small S$^\heartsuit$ ($\times 10^{-7}$)} &
\multicolumn{1}{c}{$E^\diamondsuit$}\\
 & {\small Type} & Radio? & $(N\sigma)$ & (km s$^{-1}$) & & {\small (Mpc)} &
(erg cm$^{-2}$) & (erg)}

\startdata

1998bw/6707 & Ic & Y & 0.0 & $60,000^a$ & SP/NHE & 39.1$^b$ & 44$^c$
 & 8.1 $\times 10^{47}$\\

1997ei/6488 & Ic & NA & 2.4 & 13,000$^d$ & SP/NHE & 48.9$^e$ &
7.69 & 2.2 $\times 10^{47}$\\

1997X/5740 & Ic & NA & 3.1 & 16,000$^f$ & SP/HE & 17.0$^e$ & 5.88
& 2.0 $\times 10^{46}$\\

1994I/2900 & Ib/c & Y & 4.4 & 14,000$^g$ & SP/NHE & 7.10$^e$ &
32.6 & 2.0 $\times 10^{46}$ \\

1997ef/6488 & Ib/c? & NA & NA & 15,000$^h$ & SP/NHE & 53.8$^e$ &
7.69 & 2.7 $\times 10^{47}$\\

1997ef/6479 & Ib/c? & NA & NA & 15,000$^h$ & MP/HE & 53.8$^e$ &
99.5 & 3.4 $\times 10^{48}$\\

1992ad/1641 & Ib & NC$\dagger$ & 2.0 & NA & NA & 19.504$^e$ & NA & NA\\

1997cy/6230 & IIPec & NA & NA & 5000$^i$ & SP/HE & 295$^i$ & 2.22 & 2
$\times 10^{48}$\\

1993J/2265 & IIt & N & NA & 13,000$^j$ & SP/NHE & 3.63$^k$ &
1.53$^c$ & 2.4 $\times 10^{44} $\\

\enddata

\tablecomments{GRB and SN properties of the suggested pairs by
\citet{ww98} and \citet{wes99} are compared against the expected criteria of S-GRBs (see \S \ref{sec:recognize}).  The
associations are listed in order of decreasing likelihood that the
SN/GRB falls into the S-GRB subclass. Those with the least amount of
information are placed at the bottom of the list.  The list last two
entries are SNe which are of type II and thus listed separately.  NA =
not available. $^\spadesuit$ Distance of SN from BATSE position in
units of number of BATSE sigma from \citet{kbk+98}. In the case of GRB
980425, the SN 1998bw lies near the center of the small ($\sim 8$
arcmin) BeppoSAX error circle.
\noindent $^\clubsuit$ Assuming $H_0$ = 65 km s$^{-1}$ Mpc$^{-1}$ with
$D \simeq cz/H_0$ with $z$, the heliocentric redshift, from noted
reference.
\noindent$^\heartsuit$ Fluence in BATSE channels 1 -- 4 (24--1820 keV).
 From \citet{mpp+98} (BATSE Database) unless noted.
\noindent$^\diamondsuit$ Required isotropic energy ($> 25$ keV).
\noindent $^\dagger$ The prompt radio criterion is not constrained (NC)
by the late radio detections.  }

\tablerefs{$^a$ R.~A.~Stathakis communication in
\citet{kfw+98}; $^b$ \citet{tsc+98}; $^c$ \citet{gvp+98,gvv+98};
$^d$ Based on a spectrum provided in \citet{whw98}; $^e$
\citet{dvdvc91};  $^f$ \citet{btp+97}; $^g$ \citet{whc+94}; $^h$
Based on \citet{fili98}; $^i$ \citet{bpw97}; $^j$ \citet{fm93}; $^k$
\citet{fhm+94}}

\end{deluxetable}
In table \ref{tab:sgrb-obs} we summarize the proposed
associations. They are ranked according to the viability of the
association based on the four criteria discussed in the previous
section. The pulse profile for each GRB is characterized as either
simple/single pulse (SP) or multipulse (MP). The SN type was drawn
from the literature, as was the distance to the host galaxy. The
isotropic gamma-ray energy release is computed from the publicly
available fluence (BATSE 4B Catalog) and the assumed distance.

It is unfortunate that crucial information---the early radio emission
observations---are missing for all but one SN (1994I). SN 1994I does
have early radio emission \citep{rsvd+94}. However, according to
\citet{kbk+98}, the associated candidate GRB 940331 is more than 4
$\sigma$ away from the location of SN 1994I. Thus, either the GRB
associated with this event is not observed by BATSE, or this event is
not an S-GRB.

\section{Discussion}

From the observations (primarily radio) and analysis of SN 1998bw we
have enumerated four criteria to identify S-GRBs. We have attempted to
see how well the proposed associations of S-GRBs fare against these
criteria. Unfortunately, we find the existing data are so sparse that
we are unable to really judge if the proposed criteria are supported
by the observations.

Independent of our four criteria, the expected rate of S-GRBs is
constrained by the fact that this subclass is expected, with the
assumption of a standard candle energy release, to have a homogeneous
Euclidean ($\langle$\emph{V} /\emph{V}$\mathstrut{_{{\rm max}}}
\rangle$=0.5) brightness distribution. Since there is a significant
deviation from Euclidean in the BATSE catalog \citepeg{feh+93}, S-GRBs
cannot comprise a majority fraction of the BATSE catalog. Indeed, as
studies show \citepeg{ppb+97}, $\approx 25$\% of the BATSE GRB
population can derive from homogeneous population.

From the \emph{BeppoSAX} observations we know that at least 90\% of
\emph{BeppoSAX} -identified GRBs have an X-ray afterglow. Thus, at
least in the \emph{BeppoSAX} sample, the population of S-GRBs is
further constrained to be no more than 10\% using the criterion of no
X-ray afterglow. However, it is well known that \emph{BeppoSAX} does
not trigger on short bursts---duration $\ale$ a few seconds---and thus
this statement applies only to the longer bursts.

The small number of candidate associations prohibits us from drawing
any firm conclusions based on common characteristics. Nonetheless, it
is of some interest to note that four of our top five candidate S-GRBs
(the exception is \# 6479) are single-pulsed (SP) bursts. We clarify
that the ordering in table \ref{tab:sgrb-obs} did not use the
morphology of the pulse profile in arriving at the rank. We remind the
reader that roughly half of all BATSE bursts are SP, and these mostly
are sharp spikes ($<$1 s) or exhibit a fast rise followed by an
exponential decay---the so-called FREDs. Thus, only a subclass of SP
bursts could be S-GRBs.

What could be the special characteristics of this sub-class of SPs? In
search of this special subclass, we note that the profile of GRB
980425 (fig.~\ref{fig:sn-6707}) exhibits a rounded maximum and is
quite distinctive. A visual inspection of the BATSE 4B catalog shows
that there are only 15 bursts with similar profiles; we note that such
bursts constitute 1\% of the BATSE bursts. Interestingly, most of
these bursts appear to have the same duration as GRB 980425, although
this may be due to bias in our selection. It is heartening to note
that an independent detailed analysis of GRB light curves by
\citet{nbw99} confirmed the small fraction (1\%--2\%) of GRB light
curves that meet our proposed criteria.

We end with some thoughts and speculation on the population of
S-GRBs. Assuming the fluence of the GRB 980425 is indicative of the
subclass, we find a canonical gamma-ray energy of $E \simeq 8 \times
10^{47}$ \emph{h}$\mathstrut{^{-2}_{65}}$ ergs. Although BATSE
triggers on flux (rather than fluence), 80\% of the bursts with
fluence \emph{S} $\age 8 \times$10$\mathstrut{^{-7}}$ ergs
cm$\mathstrut{^{-2}}$ will be detected \citep{bfi96}. Thus, BATSE can
potentially probe the class of S-GRBs out to $\sim$100
\emph{h}$\mathstrut{^{-1}_{65}}$ Mpc. \citet{vdbt91}
concluded that the rate of Ib/Ic SNe is roughly half that of Type II
SNe. Thus, the expected rate of Type Ib/Ic SNe is 0.3 per day out to a
distance of 100 \emph{h}$\mathstrut{^{-1}_{65}}$ Mpc. This can be
compared with the daily rate of $\sim$3 GRBs per day at the BATSE flux
limit. Thus, if all Type Ib/Ic SNe produced an S-GRB, then the
fraction of S-GRBs is 10\%, consistent with the upper limit on the
fraction due to the X-ray afterglow criterion found above. But, since
most known SNe do not fit our criteria 1 and 4, the fraction
constrained by the Type Ib/Ic rates is likely much smaller.

The sky distributions of SNe and GRBs that fit our four criteria but
are not necessarily correlated (as in SN 1998bw/GRB 980425) can be
used as an indirect test of the S-GRB hypothesis. \citet{nbw99}
have shown that the anisotropy of the 21 Type Ib/Ic SNe are marginally
inconsistent with the isotropy of the 32 SP GRBs. We note, however,
that most current search strategies are optimized to discover SNe in
regions of large galaxy overdensity (presumably biased toward the
supergalactic plane), which may cause the observed SNe anisotropy to
be larger than it truly is. Further, most SNe Ib/Ic do not fit our
criteria 1 and 4, and thus it is unwarranted to simply correlate all
Type Ib/Ic SNe to SP GRBs.

We conclude with two suggestions for observations that directly test
the S-GRB hypothesis. Even with the poor localization of BATSE, a
Schmidt telescope equipped with large plates can be employed to search
for SNe out to a few hundred megaparsecs. This is the best way to
constrain the S-GRB population frequency. Second, S-GRBs will dominate
the GRB number counts at the faint end of the flux distribution. From
this perspective, future missions should be designed to have the
highest sensitivity with adequate localization.

\acknowledgements

We thank J.~Sievers, R.~Simcoe, E.~Waxman, B.~Kirshner, A.~Filippenko,
S.~Sigurdsson, and E.~E.~Fenimore for helpful discussions and
direction at various stages of this work. The National Radio Astronomy
Observatory is a facility of the National Science Foundation operated
under cooperative agreement by Associated Universities, Inc. SRK and
JSB are supported by the NSF and NASA.

\part{An Instrument to Study the Small-scale Environments of GRBs}

\chapter[JCAM: A Dual-Band Optical Imager for the Hale 200-inch
Telescope at Palomar Observatory]{JCAM: A Dual-Band Optical Imager for the Hale 200-inch
Telescope at Palomar Observatory$^\dag$}
\label{chap:jcam}

\secfootnote{\secfootdag}{This chapter was submitted to the {\it Publications of the Astronomical Society of the Pacific} on 7 March 2002.}

\vspace{-1cm}

\secauthor{J.~S.~Bloom$^{a}$, S.~R.~Kulkarni$^a$,
J.~C.~Clemens$^{a,b}$, A.~Diercks$^{a,c}$, R.~A.~Simcoe$^{a}$, \&
B.~B.~Behr$^{a,d}$}

\secaffil{\vskip 0.5cm$^a$ California Institute of Technology, MS~105-24, Pasadena, CA 91125 USA}
\secaffil{$^b$ Department of Physics and Astronomy, University of North Carolina, Chapel Hill, NC 27599-3255, USA}
\secaffil{$^c$ The Institute for Systems Biology, 1441 North 34th Street,
Seattle, WA 98103-8904 USA}
\secaffil{$^d$ RLM 15.308, C1400, UT-Austin, Austin, TX 78712 USA}

\begin{abstract}

We describe the design and construction of the {\it Jacobs Camera}
(JCAM), a dual-CCD optical imaging instrument now permanently mounted
at the East Arm f/16 focus of the Hale 200 inch Telescope at Palomar
Observatory.  JCAM was designed to provide quick ($\ale 30$ min) and
ready access to high-quality photometry simultaneously in two optical
bandpasses, albeit over a small field--of--view (3.2 arcmin diameter).
The prime motivating science is as a follow-up imager to gamma-ray
burst afterglows in the first hour to days after a burst. However,
given the quick frame readout of each CCD (9.5 sec), JCAM may also be
useful for time-resolved dual color photometry of other faint
variables and as an effective tool for targeted surveys.  JCAM, built
for under \$65 k, is the first instrument at Palomar to be fully
operated remotely over the Internet.

\end{abstract}

\section{Introduction}

The golden age of optical transient astronomy began in earnest over
the past few years, fueled in part by the scientific promises of
early gamma-ray burst (GRB) afterglow observations but also in part by
the relative ease of constructing sophisticated instruments at low
cost and mounting on inexpensive, dedicated telescopes. Robotic $\ale
0.5$ meter-class instruments \citepeg{abb+00,ppw+00,bab+01} were
designed to follow-up on GRB positions rapidly ($t \ale 1$ min from
GRB trigger) and proved a great success upon the discovery of a 9th
magnitude optical transient following GRB 990123 \citep{abb+99}.

Owing to the large fields of view (typically larger than 1 deg in
diameter), robotic telescopes are ideally suited to the rapid, but
crude burst localizations from BATSE (5--10 deg diameter uncertainty
radii). By virtue of aperture size and plate scale, the robotic
telescopes are, however, limited in sensitivity and cannot provide
detailed light curve information once an afterglow is older than 1--3
hours ($V \age 18$ mag). For more detailed, longer-term observations
of GRBs afterglows, a readily available imager on a large-aperture
telescope was clearly warranted.

Recognizing this need and potential, we built and commissioned a
dual-band optical imager for the Hale 200 inch Telescope at Palomar
Observatory.  The instrument, named the {\it Jacobs Camera} (JCAM)
after the private donor who provided the funds to build the
instrument, has a small field--of--view (1.6 arcmin radius) and was
built for under \$65 k. It is now permanently mounted (until replaced
by some future instrument) at the East Arm f/16 focus and was dedicated as
a rapid ($t \ale 30$ min from trigger) follow-up photometer of
gamma-ray burst (GRB) afterglows.  The advantages of JCAM were to be
to uniquely provide accurate simultaneous color information while
sampling the light curve on short (seconds to minutes) to long (hours
to days) timescales.  The instrument was designed to complement
accurate ($\ale$3 arcmin diameter) and rapid locations of GRBs by {\it
HETE-II} \citep{ricker01} and later, {\it Swift} \citep{geh00}.

\section{Scientific Motivation}
\label{sec:motivation}

Beginning with the first optical detection of a GRB afterglow in 1997
\citep{vgg+97}, our group at Caltech has been involved in a campaign
to locate and photometrically monitor optical afterglows associated
with GRBs.  This optical afterglow radiation is thought to result from
synchrotron radiation generated by a relativistic blastwave as it
interacts with the surrounding medium \citepeg{pir99}.  Because of the
effort required to alert the community and localize optical
transients, our knowledge of the early-time behavior of GRB
afterglows is still sparse \citep[cf.][]{abb+99}.

Theoretical models of the afterglow emission are principally
constrained by well-char\-act\-erized, long time series of data taken
over days to months where the temporal and spectral evolution of the
afterglow is relatively mild.  However, the same models predict very
strong evolution in the total flux and spectral slope of the afterglow
in the first 1--2 hours after the explosion when emission from the
reverse shock decays and emission from the forward shock brightens.
\begin{figure}[tp]
\centerline{\psfig{figure=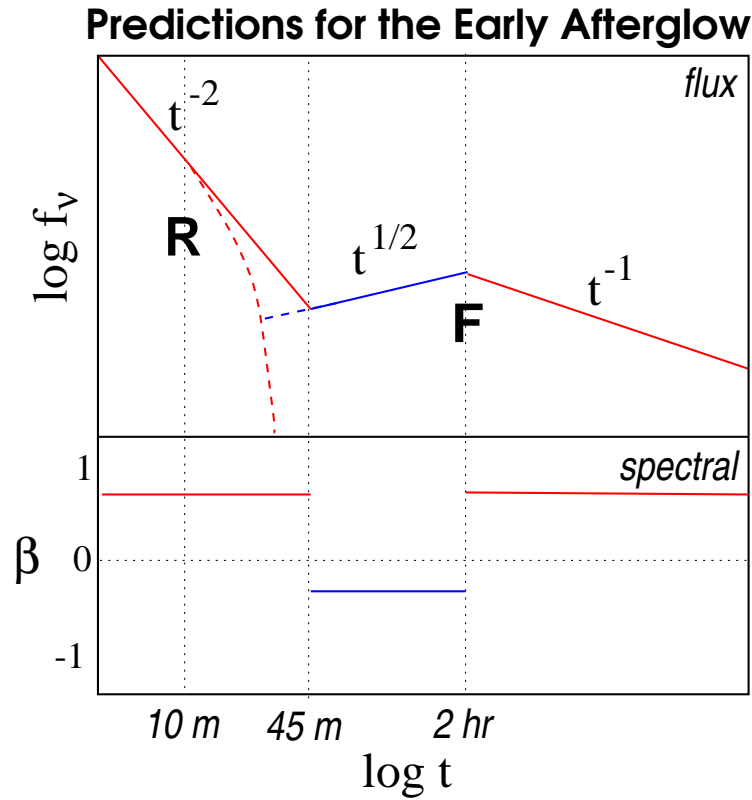,width=4in,angle=0}}
\caption[Theoretical evolution of the reverse-forward shock transition
in the early afterglow]{Theoretical evolution of the reverse-forward
shock transition in the early afterglow.  The dramatic change in
spectral slope $\beta$ ($f_\nu \propto \nu^{-\beta}$) during the onset
of the forward shock as well as the temporal decays should be
detectable by JCAM. Here ``R'' refers to emission from the reverse
shock and ``F'' refers to emission dominated by the forward
shock. After the reverse shock sweeps through the ejecta completely,
the optical flux density will decay roughly as $t^{-2}$.  A transition
then occurs whereby emission from the forward shock dominates and the
classical GRB afterglow begins. On a 45 minute timescale after the
burst, the optical flux begins to be dominated by the forward shock
emission and will appear blue ($\beta = -1/3$).  At this time, the
afterglow is expected to be in the $R \sim 15$ -- 18 mag range. Later
the afterglow declines as $\sim t^{-1}$ with a red spectrum ($\beta
\approx 0.7$).  See \citet{sp99c} for details. These time dependencies
and spectral evolution are distinct signatures of the synchrotron
model. }
\label{fig:revshock}
\end{figure}

If the synchrotron hypothesis for the origin of GRB afterglows is
correct, then the early-time behavior of the afterglows should show two
as yet unobserved transitions on timescales of tens of minutes to
hours. Figure \ref{fig:revshock} depicts schematically the temporal
and spectral prediction of the reverse-forward shock transition.
Observations of the transitions carry an important diagnostic of the
initial parameters of the GRB itself since the reverse shock liberates
kinetic energy before the blastwave begins to decelerate
self-similarly.  Prompt observations are the only way to directly
measure the initial Lorentz factor, $\Gamma_0$ \citep{sp99c}. The
timescales of these transitions have yet to be explored
observationally.

After \citet{abb+99} discovered the bright prompt optical emission of
GRB 990123 attributed to a reverse shock, \citet{kfs+99} also observed
a bright radio flare about one day after the burst that authors
attributed to the same reverse shock.  Though no other reverse shocks
have been observed optically \citepeg{abb+00}, it is now believed that
at least 25\% of radio afterglows show evidence for the presence of a
reverse shock.  Optical reverse shock signatures should be just as
pervasive if observed to fainter levels than the robotic telescopes
have allowed.

The degree of inhomogeneity surrounding the burst will also leave an
imprint on the afterglow in the form of small deviations from
power-law behavior. \citet{wl00} have predicted the degree and
character of temporal variability induced by inhomogeneities in the
immediate environment of the GRB.  Specifically, they calculate that
the r.m.s.~variability at optical wavelengths should be observed at
the 0.1 -- 10\% level about 1 hour after the GRB.  The fluctuation
amplitude and timescale are directly related to the length scale of
the density perturbations and is essentially undetectable after a few
days. Thus is it possible to probe the immediate environs of the GRB
by continuously and rapidly sampling the light curve at early times.

\section{Instrumentation}

We chose to use the East Arm f/16 focus of the Hale 200 inch since the
port was available and we could be insured at least several years of
permanent mounting, a timescale to coincide with the lifetime of the
{\it HETE-II} satellite and the beginning of the {\it Swift} mission.
Given the science goals we then formulated several practical
objectives and limitations for the instrument.  We required JCAM:

\begin{itemize}
\item{} To be capable of background-limited imaging in $UBVRI$ bands,
with unfiltered throughput $\geq$ 65\% in $B$, $V$, and $R$, and $>$
40\% in $U$ and $I$, inclusive of detector Q.E., exclusive of
telescope throughput (that is,~a 4-mirror system).

\item{} To have a large enough image field to encompass at least one
comparison star $\leq$ 17th mag in the frame with the GRB afterglow
source at any Galactic latitude.
\smallskip

\item{} To have image quality equivalent to r.m.s.~spot diameters of $<$
0\arcsec.5 at field corners.
\smallskip

\item{} To be capable of integration times from 1 s to 30 minutes with rapid response time and low latency.
\smallskip

\item{} To cost $<$ \$65,000 in total using mostly ``off-the-shelf'' parts.

\item{} To be operated remotely (and efficiently) over the Internet
with minimal on-site assistance.

\end{itemize}
\begin{figure*}[tp]
\centerline{\psfig{file=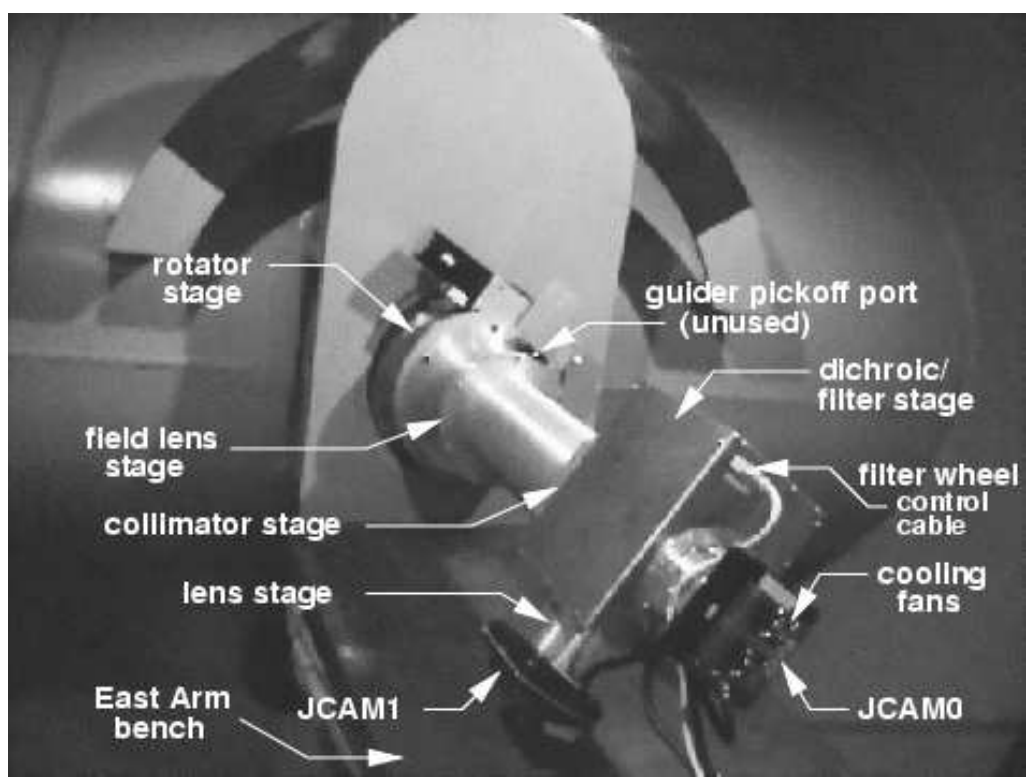,height=4.0in}}
\caption[Picture of JCAM mounted at the East Arm]{Picture of JCAM mounted at the East Arm with labeling of
JCAM components; looking north-easterly from the stairs inside the
East Arm. The f/16 image plane rests at the field lens stage. The
light is collimated at the collimator stage, then split by the
dichroic and passed through the filter wheel for each camera (JCAM0 =
red side; JCAM 1 = blue side).  The black and light gray cables from
the two Apogee cameras and the filter wheels are connected to the JCAM
computer and peripherals (not shown) lower down on the East Arm bench
(see fig.~\ref{fig:jcam-soft}).}
\label{fig:jcampict}
\end{figure*}

Here we describe the construction of JCAM and discuss the rationale
behind the design choices in the context of the science and practical
limitations described in the previous sections.  The instrument, shown
in figure \ref{fig:jcampict}, was first mounted in November 2000 and
finished full commissioning in February 2002.

As JCAM is now permanently mounted on the 200 inch Telescope, the
observations such as those described in \S \ref{sec:motivation} are
now feasible even in lunar conditions when optical imagers are not
normally mounted on the telescopes of major observatories
(i.e.,~bright time). Practically, this implies that optical imaging of
the observations any afterglow localized northward of $-35^\circ$
declination can be followed up to depths comparable with other optical
imagers available less than half of the nights.

\subsection{Optical design}

A detailed listing of the optical elements of the 200 inch Telescope
and JCAM are given in table \ref{tab:optical-elements}.  The total
un-vignetted field size available at the f/16 focus of the East Arm is
about 4 arcmin diameter, large enough to cover some prompt {\it
HETE-II} (and all {\it Swift}) localizations and adequate to find
suitable photometric and astrometric tie stars for imaging of known
(i.e., precisely localized) transients.

\begin{deluxetable}{llrrllll}
\rotate
\tabletypesize{\small} 
\tablewidth{0pt}
\tablecolumns{7}
\tablecaption{Properties of Optical Elements in
JCAM\label{tab:optical-elements}} 
\tablehead{ \colhead{Surface} &
\colhead{Radius of} & \colhead{Thickness} & \colhead{Material} &
\colhead{Diameter} & \colhead{Manufacturer} & \colhead{Comments} \\
\colhead{ } & \colhead{Curvature} \\
\colhead{ } & \colhead{(mm)} & \colhead{(mm)} & \colhead{ } &
\colhead{(mm)} & \colhead{ }} 

\startdata

\cutinhead{Hale 200 inch Telescope Elements}

primary mirror & 33926.78 & -13713.46$^d$ & Pyrex; Al coated 
               & 10210.8 & Corning Glass & paraboloid ($k=-1$)\\
secondary mirror& 8089.39 & -8379.46$^d$ & Al coated & 1041.4 
                &        & convex hyperboloid  \\
                &        &              &           & 
                &        & ($k=-1.53$) \\
tertiary flat  & $\infty$ & 5511.673 & Al coated & $^e$ \\
~~~(``Coud\'e'' mirror) \\
quaternary mirror & $\infty$ & 1266.952 & Al coated & 292.1 \\
\cutinhead{JCAM Elements}
field lens & 1000.0 & 6.4 & fused silica, MgF coating & 101.6 & JML Optical & plano-convex; FPX12080/100 \\
collimator & 1000.0 & 7.7 & fused silica, MgF coating & 50.8 & JML Optical & FPX12060/100 \\
Dichroic & $\infty$ & 3.25 & Float glass substrate & $^{\rm b}$ &  Custom Scientific & 45 deg angle of incidence$^{\rm c}$ \\
\cutinhead{Filters}
\\
Bessel $U$ & $\infty$ & 5.0 & UG, S-8612, WG305 & 50.0 &  Omega Optical & $^{\rm a}$ \\
Bessel $B$ & $\infty$ & 5.0 & GG385, S-8612, BG12  & 50.0 & Omega Optical & \\
Bessel $V$ & $\infty$ & 5.0 & GG495, S-8612 &  50.0&  Omega Optical  & \\
Bessel $R$ & $\infty$ & 5.0 & OG570, KG3  & 50.0 & Omega Optical\\
Bessel $I$ & $\infty$ & 5.0 & WG305, RG9  & 50.0 & Omega Optical\\
Sloan  $g^\prime$ & $\infty$ & 5.0 & OG400, S-8612  & 50.0 & Custom Scientific\\
Sloan  $r^\prime$ & $\infty$ & 5.0 & OG550  & 50.0 & Custom Scientific \\
Re-imaging Lens ($\times$2) & 170.0 & proprietary &  proprietary & 80.5 &
	Nikon (Nikkor)  & F/1.4; 7 element lens in 5 groups\\
\enddata 

\tablecomments{ $^{\rm a}$ All filters are specified with a maximum 
transmitted wavefront distortion $\lambda$/4 per inch, minimum
flatness of $\lambda$/4 (or better), and are coated with dialectic
anti-reflective coatings on both sides to increase transmission and
reduce the effects of ghosting. $^{\rm b}$ 4 in $\times$ 3 in
rectangular.  $^{\rm c}$ Typical flatness of 4$\times\lambda$ per in,
parallelism of 8 arcmin, and anti-reflectance coated. $^{\rm d}$
Negative values are for mirrors. The absolute value of this number is
the distance to the next optical element. $^{\rm e}$ 53 in $\times$ 36
in elliptical mirror.}

\end{deluxetable}

Obtaining an acceptable field size of the instrument was a trade-off
between CCD format and the amount of focal reduction.  Based on our
cost analysis, we opted for a rather large reduction ratio of 6:1
rather than the much more expensive large format CCD camera.  This
provides a near optimum scale (with respect to the typical seeing at
Palomar of 1--1.5 arcsec FWHM) of 15.5 arcsec mm$^{-1}$, or
$\sim 0$\arcsec.37 pixel$^{-1}$ for $24 \mu$m pixels.  To reduce the
size of the optics which follow, the initial element is a field lens,
followed by a collimator which both reduces the power required in the
camera optics, and provides a collimated beam for the dichroic and
filters; the field lens and the collimator have the same focal length
(500 mm) and are set apart at that distance.  These lenses are coated
with a MgF coating, offering the best transmission performance over
the broad range of wavelengths. The cameras are equipped with f/1.4
85mm Nikon Nikkor lenses which likely exceed a standard telephoto lens
in image quality; however, one unknown in the design process was the
throughput of the lenses at ultraviolet wavelengths.

We recognized that a 6:1 focal reduction would be difficult to manage
over the entire specified wavelength range with a single set of camera
optics and coatings.  Given this, and the science objectives, we opted
to separate the blue and red light with a dichroic. After the light is
split by the dichroic, each beam passes through its own filter wheel,
camera lens, and CCD.  However, since the field-lens and collimator
are singlets, there is a chromatic dependence on the optimum focus for
each filter.  We found, however, that with an appropriate internal
focus setting on each camera such that the Sloan $r^\prime$ and Sloan
$g^\prime$ are parfocal (i.e.,~optimally small point-spread function
r.m.s.~and good image quality in both filters for a given position of the
secondary mirror), as are Bessel $U$ and Bessel $I$. The best setting of
the focus value for secondary mirror of the telescope is consistently
0.35--0.40 mm larger for Bessel $U$/$I$ than for Sloan
$r^\prime$/$g^\prime$.

Our design target for image quality of no greater than an r.m.s.~of
0.5 arcsec on the field edges appears to have been accomplished in
practice as we have not measured any substantial image degradation
across on-sky images; this is not unexpected given we have only had
$\sim$5 hr of imaging below 1 arcsec seeing.

\subsubsection{Filters and Dichroic}
\begin{figure*}[p]
\centerline{\psfig{file=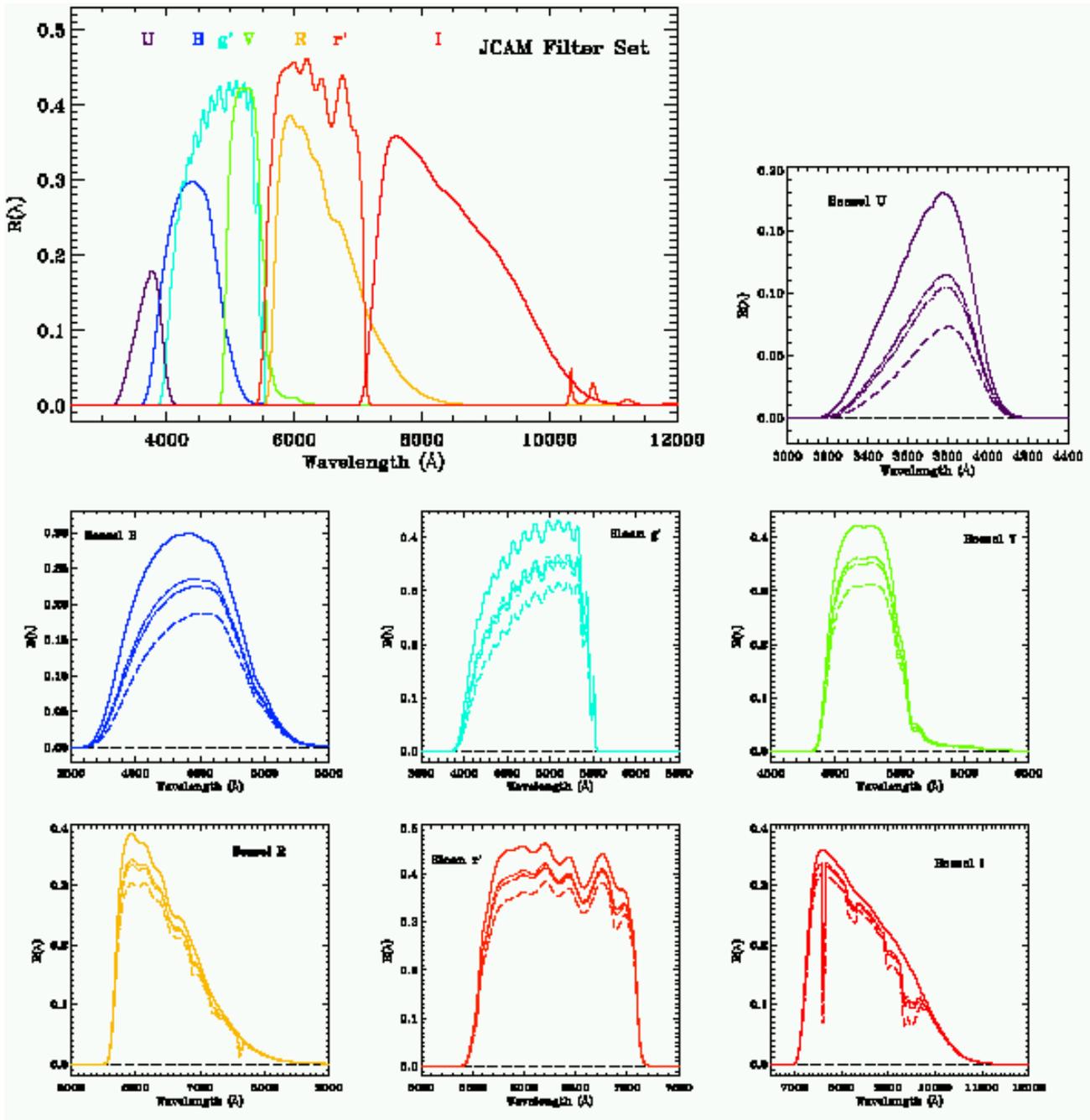,height=7.0in,angle=0}}
\caption[The calculated filter response curves of JCAM]{The calculated
filter response curves of JCAM. Shown at top left are the effective
response curves through the 7 filters of JCAM ($U$, $B$, $g^\prime$, $V$, $R$,
$r^\prime$, $I$) found using the laboratory transmission curves of the
filters themselves, the laboratory transmission and reflection curves
of the dichroic, the nominal quantum efficiency of the SITe backside
illuminated CCD, and the nominal reflectivity of aluminum (4 mirror
elements).  The effect of the remainder of the optical
elements---field lens, collimator, 7-element Nikon lens (all of which
are anti-reflective coded)---are each modeled as a 98.5\% transmissive
element.  The individual response curves are also shown with the solid
curve representing the instrument response alone.  The other curves in
each plot correspond to the calculated response at an airmass of 1.0,
1.2, and 2.0 (dash-dot, dash-dot-dot-dot, and dashed, respectively)
using an average atmospheric transmission curve for Palomar. The small
``red leak'' at $\sim$1.05 micron is from the Sloan $r^\prime$
filter.}
\label{fig:filter-curves}
\end{figure*}

The motorized filter wheels (Oriel part \#77384), one for each camera,
can carry up to five, 2 in diameter filters.  The movement of the
filters is controlled via TTL pulses generated from the rotator
control system which, in turn, is initiated through a serial-port
interface with the host computer (see \S \ref{sec:software}).  We
purchased the pre-fabricated Bessel filters ($U$, $B$, $V$, $R$, $I$)
from Omega Optical,
Inc.$\!$\footnotemark\footnotetext{http://www.omegafilters.com/}\negafterfoot\
based in Brattleboro, VT. In addition, given the higher throughput of
Sloan filters, we also decided to purchase custom designed Sloan
$g^\prime$ and Sloan $r^\prime$ from Custom Scientific,
Inc.$\!$\footnotemark\footnotetext{http://www.CustomScientific.com/}\negafterfoot\
based in Phoenix, Arizona. The $U$, $B$, $V$ and Sloan $g^\prime$
filters are installed on the blue-side camera and the remainder are
installed on the red-side camera.  A summary of the relevant physical
specifications for the filters is provided in table
\ref{tab:optical-elements}.

Recently (December 2001) we replaced our initial dichroic, which was
slightly undersized but purchased off--the--shelf for only
$\sim$\$300, with a more expensive (\$1600), higher-throughput custom
built dichroic from Custom Scientific. The dichroic was fabricated and
designed to accommodate the full 2 in diameter incoming collimated
beam. At a 45 degree tilt angle, the dichroic, reflective at blue
wavelengths and transmissive at red wavelengths, presents a 2.82 in
$\times$ 3 in (height) to the incident beam.  The central cut-off
wavelength, nominally 5577 \AA\ at half-power, was designed to
coincide with the half-power cut-off (and turn-on) wavelengths of the
Sloan $g^\prime$ and Sloan $r^\prime$ filters.  Given the large size
of the dichroic, and the fact that the optical bread-board can be
turned upside-down at certain rotator angles, we designed and built a
custom holder to secure the dichroic.

The effective filter transmission curves are shown in figure
\ref{fig:filter-curves}, created by the convolution of the quantum
efficiency curve of the CCD and the transmission of the filters and
dichroic.  We provide the tabular version of these curves as well as
estimates of the transmission curves through the entire telescope +
JCAM system in table \ref{tab:filter-curves1}.  In table
\ref{tab:filter-curves2} we provide a summary of the basic wavelength
properties of the curves, such as FWHM and peak response efficiency,
as well as a spectrum-dependent tabulation of the effective wavelength
of the filters as a function of airmass. Table \ref{tab:synth-mags}
shows the synthetic JCAM AB magnitudes of two primary stars in the
Sloan filter system tabulated using equation~7 of \citet{fig+96}.

\begin{deluxetable}{lcccccccccccc}
\tabletypesize{\scriptsize}
\rotate
\tablewidth{0in}
\tablecaption{Summary of JCAM Filter Properties\label{tab:filter-curves2}}
\tablecolumns{14}
\tablehead{
\colhead{Filter} & \colhead{Side$^{\rm a}$} & \colhead{$\lambda_{\rm peak}^{\rm b}$} & \colhead{$R(\lambda_{\rm peak})^{\rm b}$} & \colhead{Airmass} & \colhead{FWHM} &  \multicolumn{7}{c}{$\lambda_{\rm eff}$ (\AA)$^{\rm c}$} \\
\colhead{} & \colhead{} & \colhead{\AA} & \colhead{} & \colhead{} & \colhead{\AA} & \colhead{$\beta = -3.0$} & \colhead{$-2.5$} & \colhead{$-2.0$} & \colhead{$-1.5$} & \colhead{$-1.0$} & \colhead{$-0.5$} & \colhead{$0.0$} }
 \startdata
Bessel $U$
 & 1 &  3787.5 & 0.179 & 0.0 &  453.2 & 3712.3 & 3705.9 & 3700.3 & 3695.1 & 3690.2 & 3685.4 & 3680.7 \\ 
 &  &  &  & 
1.2 &  422.3 & 3735.5 & 3728.3 & 3722.2 & 3716.8 & 3711.8 & 3707.1 & 3702.5 \\ 
 &  &  &  & 
2.0 &  406.1 & 3751.0 & 3742.9 & 3736.2 & 3730.5 & 3725.3 & 3720.6 & 3716.1 \\ 
 
Bessel $B$
 & 1 &  4411.5 & 0.299 & 0.0 &  945.1 & 4428.7 & 4416.1 & 4403.6 & 4391.3 & 4379.1 & 4366.9 & 4354.8 \\ 
 &  &  &  & 
1.2 &  919.5 & 4455.6 & 4443.2 & 4431.0 & 4418.9 & 4406.8 & 4394.8 & 4382.9 \\ 
 &  &  &  & 
2.0 &  899.0 & 4472.8 & 4460.6 & 4448.5 & 4436.6 & 4424.7 & 4412.8 & 4401.0 \\ 
 
Sloan $g^\prime$
 & 1 &  5095.0 & 0.432 & 0.0 & 1320.7 & 4827.8 & 4811.9 & 4795.7 & 4779.4 & 4762.9 & 4746.4 & 4729.8 \\ 
 &  &  &  & 
1.2 & 1268.8 & 4849.8 & 4834.3 & 4818.7 & 4802.9 & 4786.9 & 4770.8 & 4754.6 \\ 
 &  &  &  & 
2.0 & 1226.7 & 4863.7 & 4848.6 & 4833.4 & 4817.9 & 4802.2 & 4786.4 & 4770.5 \\ 
 
Bessel $V$
 & 1 &  5171.0 & 0.423 & 0.0 &  560.7 & 5261.2 & 5256.8 & 5252.5 & 5248.3 & 5244.3 & 5240.3 & 5236.3 \\ 
 &  &  &  & 
1.2 &  562.0 & 5263.8 & 5259.4 & 5255.0 & 5250.8 & 5246.7 & 5242.7 & 5238.7 \\ 
 &  &  &  & 
2.0 &  562.4 & 5265.6 & 5261.1 & 5256.7 & 5252.5 & 5248.3 & 5244.3 & 5240.3 \\ 
 
Bessel $R$
 & 0 &  5929.0 & 0.386 & 0.0 & 1209.7 & 6490.8 & 6467.1 & 6444.1 & 6421.8 & 6400.1 & 6379.2 & 6359.0 \\ 
 &  &  &  & 
1.2 & 1228.2 & 6498.2 & 6474.6 & 6451.8 & 6429.6 & 6408.1 & 6387.3 & 6367.2 \\ 
 &  &  &  & 
2.0 & 1239.5 & 6503.9 & 6480.4 & 6457.6 & 6435.5 & 6414.1 & 6393.3 & 6373.2 \\ 
 
Sloan $r^\prime$
 & 0 &  6197.5 & 0.462 & 0.0 & 1498.8 & 6420.8 & 6387.5 & 6358.1 & 6331.9 & 6308.2 & 6286.5 & 6266.4 \\ 
 &  &  &  & 
1.2 & 1496.1 & 6436.2 & 6401.8 & 6371.6 & 6344.8 & 6320.6 & 6298.5 & 6278.1 \\ 
 &  &  &  & 
2.0 & 1493.7 & 6446.5 & 6411.4 & 6380.7 & 6353.4 & 6328.9 & 6306.5 & 6285.9 \\ 
 
Bessel $I$
 & 0 &  7617.0 & 0.359 & 0.0 & 2085.2 & 8506.1 & 8467.4 & 8429.4 & 8392.0 & 8355.4 & 8319.7 & 8284.8 \\ 
 &  &  &  & 
1.2 & 1886.7 & 8486.1 & 8447.8 & 8410.3 & 8373.6 & 8337.8 & 8302.9 & 8268.8 \\ 
 &  &  &  & 
2.0 & 1792.2 & 8475.1 & 8437.0 & 8399.8 & 8363.5 & 8328.0 & 8293.5 & 8259.9 \\ 
\enddata

\tablenotetext{a}{The index corresponds to the camera number for that
given filter. An index of ``0'' corresponds to the red side camera
(transmissive through the dichroic) and an index of ``1'' corresponds
to the blue side camera.}  

\tablenotetext{b}{The peak efficiency of the response curve
$R(\lambda)$ (col.~4) and the corresponding wavelength (col.~3)
computed before including a wavelength-dependent atmospheric effect of
the filter curves (i.e.,~airmass equal zero). The efficiency curves
are tabulated using laboratory transmission measures of the individual
filters and the dichroic.  The quantum efficiency of the CCDs are
taken from manufacturer data sheet for the SITe 512x512 24$\mu$m
chip.  In addition, to account for the 4 mirrors in the system, we use
the wavelength-dependent nominal reflectivity of aluminum as presented
on page {\bf 12}-117 of \citet{lide94}. We assume a wavelength
independent transmission (T = 98.5\%) for the remaining 9 optical
elements (field lens + collimator + 7-element lens). On-sky imaging
suggests that the peak response (col.~4) should be scaled downward by
3.170, 1.62, 1.53, 1.38, 1.75, 1.71, 3.54 in the filters $U$, $B$,
$V$, $g^\prime$, $R$, $r^\prime$, and $I$, respectively. See text for
an explanation.}

\tablenotetext{c}{The effective wavelength ($\lambda_{\rm eff}$) for
a given filter at the given airmass and input object spectrum.  An
average wavelength-dependent atmospheric transmission curve at Palomar
is assumed.  A power-law input object spectrum is assumed, with $f_\nu
\propto \nu^{-\beta}$.  The usual definition of $\lambda_{\rm eff}$
\citep[see][equation~A1]{fsi95} is for airmass = 0.0 and $\beta = -2$
(i.e.,~source flux $f_\lambda =$ constant).}

\end{deluxetable}

\subsubsection{Detectors}
\label{sec:detectors}

Both detectors are Apogee Ap-7b, with a SITe thinned,
backside-illuminated 512x512 CCD with 24 micron pixels.  The CCDs are
thermoelectrically cooled and, depending on the ambient temperature
inside the dome of the 200 inch, can typically reach temperatures of
$-$40 deg C after cooling for 15 minutes. Since JCAM is permanently
mounted, the self-contained cooling allows for a relatively
maintenance-free system which can, in turn, be brought to a ready
state by a remote observer (that is, without the need for on-site
filling of cryogenic dewars).  The total cost for both CCD systems was
\$16 k (as compared with $\age$ \$200 k for custom systems), greatly
reducing the overall cost of JCAM.
\begin{deluxetable}{lcccccl}
\tablecaption{Synthetic Magnitudes of Primary Standard Stars Through the JCAM Filter Set\label{tab:synth-mags}}
\tablewidth{0pt}
\tablecolumns{7} 
\tablehead{\colhead{Filter} & 
           \multicolumn{3}{c}{BD+26$^\circ$2606} & 
           \multicolumn{3}{c}{$\alpha$ Lyr} \\
           & \colhead{$m_{\rm AB,~JCAM}$} 
	   & \colhead{$\lambda_{\rm eff}$}
	   & \colhead{$\Delta m$} 
           & \colhead{$m_{\rm AB,~JCAM}$} 
	   & \colhead{$\lambda_{\rm eff}$}
	   & \colhead{$\Delta m$} \\
\colhead{} & \colhead{mag} & \colhead{\AA} & \colhead{mag} & \colhead{mag} 
           & \colhead{\AA} & \colhead{mag}}

\startdata

Bessel $U$\ldots         &   10.557 & 3735.812 &   &  0.596 & 3770.95 & \\
Bessel $B$\ldots         &   10.030 & 4416.799 &   & -0.119 & 4382.95 & 0.009 \\
Sloan $g^\prime$\ldots &    9.892 & 4796.175 & 0.002  & -0.092 & 4728.26 & -0.005 \\
Bessel $V$\ldots         &    9.750 & 5246.947 &   & -0.025 & 5232.20 & \\
Bessel $R$\ldots         &    9.584 & 6383.269 &   & 0.192  & 6324.35 & 0.007 \\
Sloan  $r^\prime$\ldots&    9.592 & 6302.615 & 0.018  & 0.176  & 6242.14 & 0.013 \\
Bessel $I$\ldots         &    9.504 & 8279.067 &   & 0.455  & 8233.87 & 0.001 \\

\enddata

\tablecomments{The synthetic AB magnitudes of the primary standards of
the Sloan system are calculated using the observed flux of the stars
(as given in table 6 of \citealt{fig+96}) convolved with the
calculated instrumental response. The effective wavelength
$\lambda_{\rm eff}$ of the filter (calculated following equation~3 of
\citealt{fig+96}) is also provided. The offset of the synthetic values
for some filters (cols.~4 and 7) are found using the magnitudes
presented in \citet{fig+96} and \citet{fsi95}. The good agreement
suggests that the JCAM Sloan filters should very well approximate the
same SDSS filters.  The absence of a detectable color term using
on-sky data corroborates this statement (see \S
\ref{sec:jcam-through}).}

\end{deluxetable}

The temperature and shutter control, as well as CCD binning modes, are
controlled through an ISA bus card which connects to the CCD housing
through a shielded cable.  The A/D converters reside on the CCD
housing and the ISA card provides a buffer for the data before they
are read to disk. A full-chip readout, including 30 extra bias
(``overscan'') lines, requires 9.5 sec per detector.  Though the
detectors may be readout ``simultaneously'' (13 sec total) such a
process introduces an apparent cross-talk which results in erratic and
high levels of read-noise on certain columns.  We were unable to
isolate and remove the source of the cross-talk but it likely occurs
in the interaction on the ISA bus since the effect can be mimicked by
forcing CPU interrupts during the readout.
\begin{deluxetable}{lcccc}
\tabletypesize{\footnotesize}
\tablecaption{Summary of GRB Triggers Observed to Date with 
              JCAM\label{tab:summary}} 
\tablewidth{0pt}
\tablecolumns{5} 
\tablehead{
\colhead{GRB/Trigger} & \colhead{$\Delta t$} & \colhead{Conditions} & \colhead{Result/} & \colhead{References} \\
\colhead{Name} & \colhead{} & \colhead{} & \colhead{Comments} & \colhead{}} 
\startdata
GRB 020124 & 1.88 day & thick cirrus & detection at $r^\prime \approx 24$ mag &
        \citet{bloom02} \\ 
GRB 011211 & 6.73 hr & very cloudy, 4\arcsec\ seeing & non-detection of candidate & \citet{bb01} \\
           &         &                              & solidifies transient \\
HETE \# 1793 & 1.62 hr & clear & false trigger from solar flare; & \\
             &         &       & not a GRB \\
GRB 010222 & 8.16 day & light cirrus & detection of OT in 4 bands &
           fig.~\ref{fig:jcam-example}; \citet{bloom02a} \\ & & & $R =
           23.19 \pm 0.15$ mag \\

GRB 001018 & 60.6 day  & thick cirrus, 2.8\arcsec\ seeing & non-detection of host & \citet{bdk+01} \\
           &         &                                 & to $R = 22.7$ mag \\
\enddata

\end{deluxetable}

The measured read noise and gain are given in table
\ref{tab:performance} and are in line with the manufacturer's
specifications. In table \ref{tab:performance} we give some relevant
characteristics of the JCAM CCDs and images.  Note that the gain in
both CCDs provide an adequate Nyquist sampling of the read noise. The
linearity of both CCDs is acceptable to about 55 k DN.

\subsection{Mechanical design and construction}

The design of the mechanical systems of JCAM dealt with three major
constraints.  First, the optics and physical dimensions of the
elements required the field lens to be $\sim$5 inches from the
entrance to the East Arm port and the CCDs to be at least $\sim$30
inch from the field lens.  Second, the entire system needed to rotate
so as to observe at a cardinal position
angle\footnotemark\footnotetext{We originally required the field
rotation so that we could be assured that the pick-off guide camera
could observe a bright enough guide star somewhere in the circular
annulus about the science field center.}\negafterfoot\ yet there is
only $\sim$13 inch clearance from the center of the optical axis to
the East Arm bench (see fig.~\ref{fig:jcampict}). Third, the East Arm
bench and the optical axis are aligned toward the pole and so the
entire system must operate at 33 deg angle from level without
significant flexure.

The bench clearance constraint and optical distances implied that the
two CCDs would have to be back-loaded at the end of the optical axis
if the instrument were to rigidly rotate.  This implied that the
rotator stage would have to carry a heavy load yet be capable of fine
angular positioning. We purchased a stepper-motor rotator stage from
Newport which offers a large load capacity, an 11" central clearance,
and angular positioning resolution of 0.001 degrees.  This stage is
mounted just before the field lens stage (fig.~\ref{fig:jcampict}).

The remainder of the stages were custom designed by us and constructed
out of aluminum at machine shops at the University of North Carolina
and Caltech.  The data-taking computer (see \S \ref{sec:software}),
rotator control module and filter-wheel staging electronics are all
housed in a rack which we mounted to the East Arm bench just below
JCAM.

\subsection{Electronics and software implementation}
\label{sec:software}

\begin{figure*}[tp]
\centerline{\psfig{file=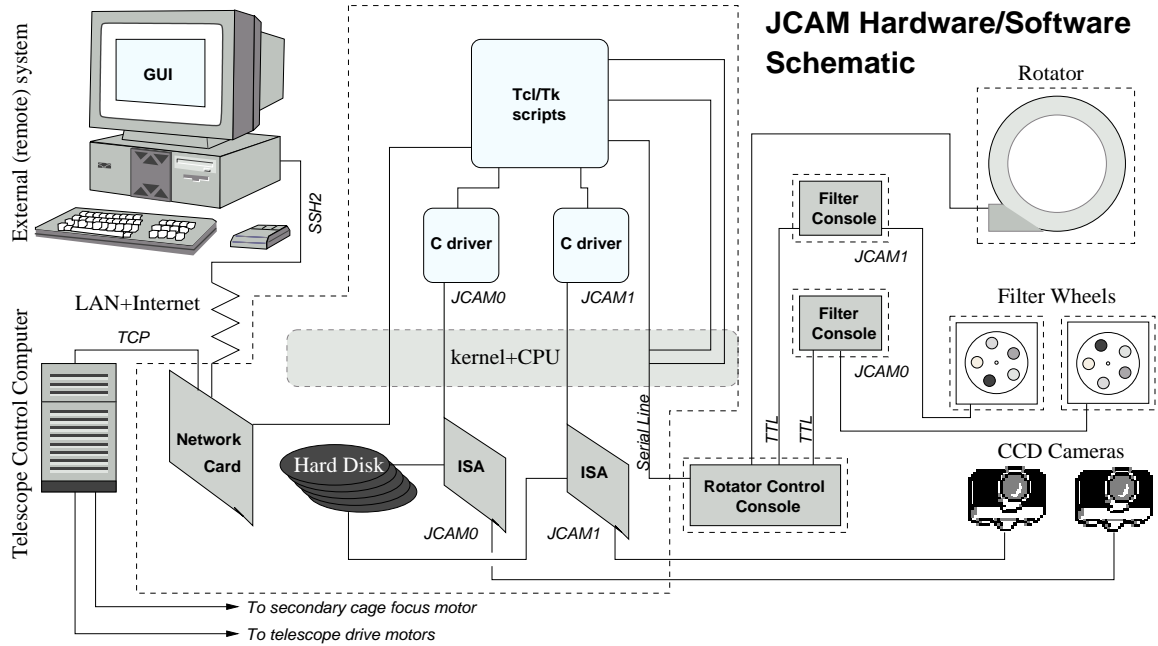,width=6.0in,angle=0}}
\caption[Schematic of the hardware and software configuration of
JCAM]{Schematic of the hardware and software configuration of JCAM.
The main computer, located inside the East Arm rack, is responsible
for the interaction with the Apogee CCD cameras, the control of the
rotator and filter wheels, data taking and storage, and communication
with the user via a secure SSH connection over the Internet.  The
interaction of the user (via a GUI on the remote machine; see
fig.~\ref{fig:jcam-screenshot}) with the instrument operations is
controlled though Tcl/Tk and C codes (see text) and routed through the
Linux kernel/CPU (depicted here as a filled gray block).  Separate
physical entities (e.g.,~the JCAM computer, filter wheels) are
encapsulated here in dashed lines.}
\label{fig:jcam-soft}
\end{figure*}

In the interest of simplicity, we decided that all camera, rotator,
and filter operations would be conducted through a single computer
which would be remotely controlled over the Internet. A schematic of
the basic software and electronics configuration is shown in figure
\ref{fig:jcam-soft}. The computer is currently running Linux RedHat
version 7.2 (kernel 2.4.2) and is accessed via a secure (SSH2)
connection. In the interest of security, we periodically update the
kernel and connection software.

The data acquisition and hardware are controlled by a single Dell
OptiPlex GX1 MiniTower 450 MHz Pentium III computer with 20 GB of IDE
disk space. A co-axial network cable inside the East Arm connects the
computer (automatically upon boot) to the Palomar network at data
transfer rates of $\sim$2 Mbps.  The specific model of the computer
was chosen since it allows up to 4 ISA card slots; JCAM uses only 2 of
these slots for the Apogee control cards but we wanted to keep the
machinery upgradeable.  The power for the two Apogee cameras, supplied
though the ISA cards, is provided from the computer; despite the
current load from the cameras (1.4 amp per camera) the computer power
supply performs well and there is no apparent pattern noise on the CCD
dark frames.

The interaction with the rotator and filter wheels is conducted
through a serial port interface with the rotator control stage from
Newport.  The control software sends short ASCII string commands to
this stage and can poll the stage for the status of the rotator.  The
control software also initiates the generation of TTL pulses from the
stage used to interact with the two filter wheels.

The low-level CCD control software---the device driver interface
between the kernel and the ISA cards---was written in C and purchased
as source code from the ClearSkyInstitute.  With each new kernel
update, we have modified this code to comply with the protocol for the
way that device drivers interact with the kernel.  In the future, we
may begin to use a new driver source, called Linux Apogee Instruments
camera drivers\footnotemark\footnotetext{See {\tt
http://www.randomfactory.com/apogee-lfa.html}}\periodafterfoot For the
control of the device drivers, installed at boot as two independent
kernel modules (one for each camera), we have written C-code wrappers
and a set of Tcl/Tk (current version 8.3) scripts.  This software,
amounting to about 10 k lines of code, is also used to interact with
the telescope control system (TCS) via a TCP connection, control the
filter wheels and rotator, and to write data acquired from the CCDs to
disk.  The user interacts with the code through a graphical user
interface (GUI) (also written in Tcl/Tk) which is displayed on the
remote (external) computer monitor though the SSH port.

\section{Operations and Performance}
\label{sec:jcam-ops}

JCAM is designed to operate in a ``Target of Opportunity" (ToO) mode,
temporarily interrupting the scheduled observer if a GRB afterglow or
other transient event is to be observed rapidly. Some reconfiguration
of the telescope may thus be necessary before on can start taking data
using JCAM.

\subsection{Initiating operations}

Here we outline the steps to begin observing with JCAM and the
approximate time required per step.  Notification of burst alerts
reaches our group members via cellular phone and email messaging
through the GRB Coordinates Network (GCN\footnotemark\footnotetext{See
{\tt http://gcn.gsfc.nasa.gov/gcn/}}\negafterfoot).  {\it Swift}
alerts will be nearly simultaneous with burst trigger, but the time
delay for {\it HETE-II} alerts depends on the precise location of the
satellite with respect to ground
stations\footnotemark\footnotetext{See {\tt
http://space.mit.edu/HETE/Bursts/}}\periodafterfoot Typically we
receive notice of an approximate position less than one minute after
the burst trigger and can then initiate a ToO by way of a phone call
to the observers within the next few minutes.

Running in ToO mode, the Coud\'e mirror may not yet be deployed and so
must to be swung into place by the telescope operator before JCAM
operations can commence.  To do so, the telescope first is stowed at
zenith; the slew to zenith can take between 1--4 minutes depending
upon the initial pointing. The telescope operator then operates the
Coud\'e crane from a console on the catwalk floor.  The Coud\'e crane
places the tertiary mirror atop the Cassegrain baffling tube and then
the tertiary is rotated by 90 deg about the Cassegrain tube so as to
point at the East Arm entrance hole.  This process requires about 22
minutes.  The secondary mirror is then swung into place, if it is not
in place already, and set to a nominal secondary focus distance 58mm.
This process requires about 4 minutes.

During the mechanical reconfiguration, the remote observer logs into
the JCAM computer and starts the JCAM software which then begins to
cool the CCDs and establish the connections to the TCS, rotator, and
filter wheels.  This process takes a total of $\sim$5 minutes with an
additional 10--20 minutes for the CCDs to cool fully. Once the mirrors
are in place, the telescope is slewed to a bright SAO star near the
science target to check pointing. We then perform a focus loop on the
SAO star requiring about 5--10 minutes. These operations can be
performed before the CCDs are fully cooled. In total, when the entire
telescope must be reconfigured, as in this example, the overhead time
from start to science target requires about 45 minutes.

In the case where the Coud\'e mirror has been put in place during the
daytime (i.e.,~when the observers are using only Prime focus as the
primary science instrument) the overhead ``time to science target'' can be
reduced substantially to $\sim$15 minutes. Here only the zenith slew
and secondary mirror flip, plus on-sky set up are required.

Though some science observations can likely commence within 15 minutes
increasing the likelihood that we will observe the transition from
reverse to forward shock (fig.~\ref{fig:revshock}), the 45 minute
timescale for longer ToO turn-around is comparable to this
timescale. We note, however, the transition time is a sensitive
function of the initial Lorentz factor of the explosion and the
redshift of the GRB so it may still be possible to observe this
transition. In either scenario, the observations of the peak of the
forward shock (fig.~\ref{fig:revshock}) and the small variability will
be possible.

\subsection{Data taking and observing procedures}

\begin{figure*}[p]
\centerline{\psfig{file=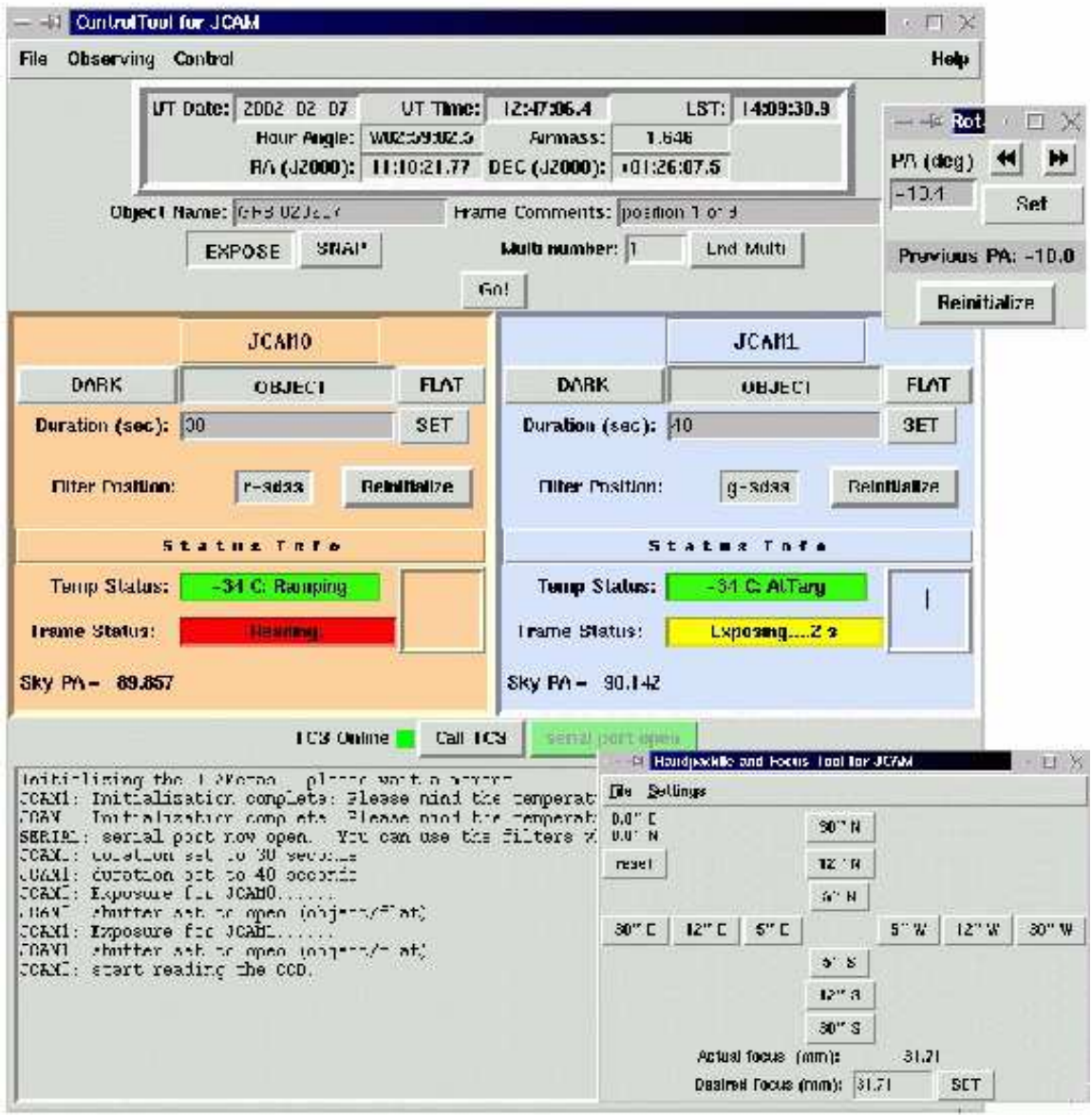,width=6.0in,angle=0}}
\caption[Screenshot of the JCAM graphical user interface]{Screenshot
of the JCAM graphical user interface (GUI).  In the main window of the
GUI, a real-time summary of the current pointing, time, etc.,~is
provided at top.  Object name and frame comments are entered by the
user and the observing mode (``EXPOSE'' for beginning exposures on
both CCDs or ``SNAP'' to begin the exposures separately).  The filter,
integration time, and frame type are chosen individually for each
camera.  A status display for each camera shows the CCD temperature,
exposure status, and current position angle of the exposure.  A small
compass rose shows the sky orientation of the CCD, with the longer
line pointing North and the shorter line indicating East.  Due to the
dichroic, there is a y-axis flip between the two CCDs. At the bottom
of the main window is a text log, which is also saved with time-stamps
to a log file. Offset from the main window are two smaller windows,
one to control the rotator angle (top) and the other to aid in small
changes in the telescope positioning (dithering) and secondary focus
(bottom).}
\label{fig:jcam-screenshot}
\end{figure*}

All data and instruments are controlled with a simple front-end GUI.
We have also written a hand-paddle and secondary focus tool to
facilitate dithering and focus changes that minimize the need to
communicate with the telescope operator.  A screenshot example of the
GUI is shown in figure \ref{fig:jcam-screenshot}.

All of the imaging data are stored on a data partition of the JCAM
computer hard drive.  Usually, after the data are written to disk, the
remote observer copies the data to the remote machine to inspect the
images; this can be done individually or by periodically synchronizing
the remote and JCAM data directories (using, for example, the
password-encrypted {\sc RSYNC} command).  As the images are rather
small (527 MB uncompressed; 150--400 MB compressed) the transfer times
require between 1--25 s per image, limited only by the bandwidth
between Palomar and the remote observer site.  The typical transfer
time between Palomar and Pasadena is 4 s for an uncompressed image.

We decided not to install the guider camera after realizing, by
comparison with the point-spread function of images obtained
contemporaneously in the same filter at the 60 inch telescope, that
200 inch Telescope tracks very well over times $\ale 300$ s.  Aside
from an occasional smearing due to telescope jump upon wind-shake
(which equally degrades guided-images) the image quality on
long exposures is comparable to the image quality of short exposures.
Nevertheless, to minimize tracking errors, we typically integrate for
shorter time periods than is usual with larger-format guided imaging.

The shorter exposure times, typically integration 100--250s, is
warranted by the short read times (9.5 s) and the fact that the frames
are sky-limited after $\ale$2 sec of integration time, except in
$U$-band where the sky dominates after about 30 s; the exact times
depend, of course, on the sky brightness contribution from the moon.
More frames per field, when dithered between exposures, also allow for
the construction of better supersky flat fields and the removal of
cosmic rays and CCD defects. When the seeing and/or transmission
changes rapidly, more images can also be useful in constructing a
higher signal-to-noise summed image of the field (that is,~by giving
lower weight to those frames with lower signal-to-noise detection of
point sources).  Note that thanks to the large collecting area of the
200 inch telescope, {\it JCAM imaging is never dark current limited}
despite being a thermoelectrically cooled system.

Depending on the science objectives, we typically observe at least 5
frames simultaneously in the Sloan $r^\prime$ and $g^\prime$ for 100 s
and 110 s, respectively.  The 10 s difference in exposure time allows
JCAM1 (Sloan $g^\prime$) to finish exposing just as JCAM0 (Sloan
$r^\prime$) finishes readout. For the other par focal set, Bessel $I$
and Bessel $U$, we typically acquired two 100 s frames in $I$ band
while exposing $U$ band for 200 s. (Note that the filters $B$/$V$ and $R$
are largely superseded in efficiency by the filters Sloan $g^\prime$
and $r^\prime$, respectively; see fig.~\ref{fig:filter-curves}).  We
have two modes of taking many frames automatically (``Multi'' in
fig.~\ref{fig:jcam-screenshot}), one which begins an exposure as soon
as the camera is finished reading out and the other which opens the
shutters of both cameras simultaneously. The latter mode, when the
exposure times of both cameras are equal, is particularly useful when
conditions are non-photometric since both images are exposed through
the same cloud pattern, preserving the relative flux of objects in the
two bandpasses.
 
As a result of shorter exposure times, a given field requires more
exposures for a given depth.  In a typical full night of science
imaging, we have typically generated between 600--800 images.
We found that the rate of frame acquisition is too high to
adequately log the frames by hand. As such, we wrote an electronic
logbook program in Tcl/Tk which automatically creates a new line for
each new image that is acquired.  The pertinent header information is
shown and the user can add comments to each line and then save the
logbook to text, Postscript, and graphical output.  Aided by the
existence of electronic logs, we have begun to archive each observing
run in a uniform set of web pages\footnotemark\footnotetext{{\tt
http://www.astro.caltech.edu/$\sim$jsb/Jcam/runsum/}}\periodafterfoot

\subsection{Preliminary results}

\begin{figure*}[tp]
\centerline{\psfig{figure=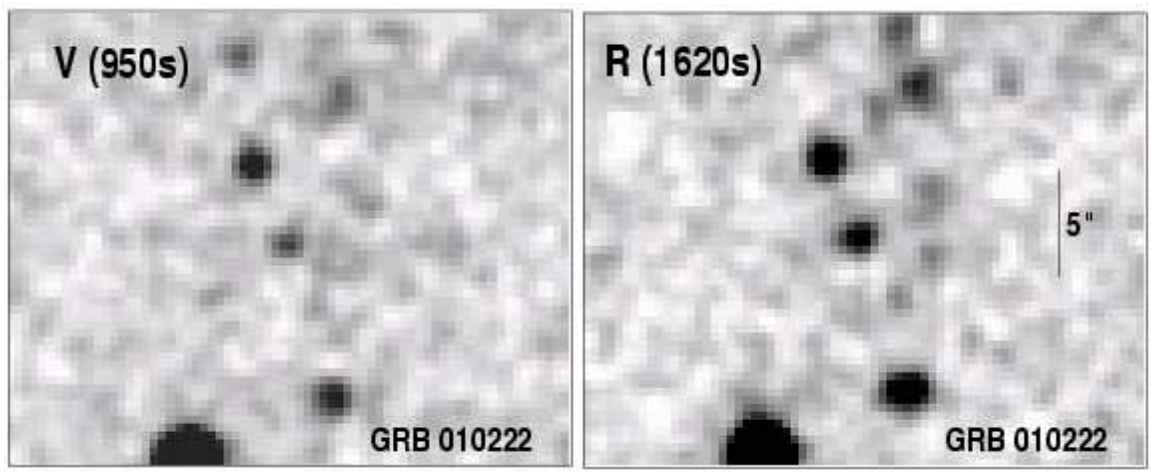,width=6.0in,angle=270}}
\caption[JCAM images of the afterglow of GRB 010222]{JCAM images of
the afterglow of GRB 010222 (object at center). The transient was
faint ($V$ = 23.54 mag, $R$ = 23.19 mag) at the time of observation (8
days after the GRB) but was detected at the 20-$\sigma$ level in
roughly 15 minutes of integration through light cirrus.}
\label{fig:jcam-example}
\end{figure*}

Despite poor observing conditions on all but one night of
commissioning over the past year (10 nights), we managed to observe a
number of GRB and GRB-related targets during the commissioning period
of JCAM. The results of some of these observations are summarized in
table \ref{tab:summary}; a JCAM dual-band image of GRB 010222 is shown
in fig.~\ref{fig:jcam-example}. On two occasions, observers have
successfully imaged a GRB position by operating JCAM from a remote
location, GRB 011211 (from Waimea, HI) and GRB 020124 (from Pasadena,
CA).

\begin{deluxetable}{lccl}
\footnotesize
\tablecaption{JCAM Detector Characteristics\label{tab:performance}}
\tablewidth{0pt}
\tablehead{ }\tablecolumns{2} 
\startdata
Pixel size\ldots                & 24 $\mu$m (0\arcsec .371) \\
CCD size\ldots                  & 512 $\times$ 512 pix$^2$ \\
~~~~including overscan regions  & 535 $\times$ 533 pix$^2$ \\
Field of View\ldots             & 3\arcmin.17 $\times$ 3\arcmin .17 \\
Avg.~Read time\ldots            & 9.5 s: single-frame mode\\
                                & 14 s: dual-frame mode \\
Inverse Gain\ldots              & 4.2 electron DN$^{-1}$ (JCAM0) \\
                                & 3.9 electron DN$^{-1}$ (JCAM1) \\
Read noise (r.m.s.) \ldots         & 8.5 electron (JCAM0) \\
                                & 9.0 electron (JCAM1) \\
Dark current  \ldots            & 0.16 electron pix$^{-1}$ s$^{-1}$ (JCAM0) \\
                                & 0.18 electron pix$^{-1}$ s$^{-1}$ (JCAM1) \\

\enddata

\end{deluxetable}

As can be seen from the table, we have unfortunately not yet observed
a {\it bona fide} GRB position on rapid turn-around timescale ($\ale
1$ hr). The only source which we followed-up rapidly (HETE \#1793)
turned out to be a particle event in the {\it HETE-II}\, detectors
generated after solar flare activity.  The disappointing absence of
rapid follow-up events is due to the low rate of GRBs with rapid
positional determination and the absence of localizations to
accuracies smaller than $\sim 5$ arcmin diameter.

Given the rapid readout and its ability to perform dual-band
simultaneous photometry, the instrument has also been in demand for
projects requiring deep color snapshots at a high duty-cycle (e.g.,
Kuiper Belt Object surveys, distant galaxy cluster surveys).  The
instrument would also be well-suited for photometric monitoring
programs of faint variable sources (e.g., gravitational lenses) which
require short pointed observations over a large number of nights.
Indeed, by reaching limiting magnitudes of 23--25 every 10 minutes, a
typical night with JCAM could be used to image $\sim$30 faint sources
in two colors with the same astrometric and photometric quality as
other larger-format optical imaging instruments.

\subsubsection{Observed System Throughput}
\label{sec:jcam-through}

On 4 Feb 2002, we observed several secondary standard stars through a
variety of airmasses in photometric conditions. The stars, spanning a
large range in colors, were selected from the calibrated Sloan
catalog of standards \citep{stk+02} as those near to the science
targets of the night. For all seven filters, we fit for the zeropoint,
color term(s), and airmass extinction to the SDSS filter system. The
transformations can be summarized as follows:

\footnotesize
\begin{singlespace}
\begin{eqnarray}
 u^\prime &=& (22.435 \pm 0.023) - (0.765 \pm 0.052) \times (X - 1) \nonumber \\
	  & & - (1.662 \pm 0.019) \times (u^\prime - g^\prime - 1) \nonumber \\
	  & & - 2.5 \log_{10} (R_{U, {\rm JCAM}}), \nonumber
\end{eqnarray}
\begin{eqnarray}
 g^\prime &=& (25.976 \pm 0.014) - (0.350 \pm 0.011) \times (X - 1) \nonumber \\
       	  &&  - (0.206 \pm 0.020) \times (g^\prime - r^\prime - 1) \nonumber \\
          &&  - (0.115 \pm 0.014) \times (u^\prime - g^\prime - 1) \nonumber \\
	  &&  - 2.5 \log_{10} (R_{B, {\rm JCAM}}),\nonumber
\end{eqnarray}
\begin{eqnarray}
 g^\prime &=& (26.971 \pm 0.030) - (0.323 \pm 0.042) \times (X - 1) \nonumber \\
	  &&  + (0.013 \pm 0.029) \times (g^\prime - r^\prime - 1) \nonumber \\
	  &&  - 2.5 \log_{10} (R_{g^\prime, {\rm JCAM}}),\nonumber 
\end{eqnarray}
\begin{eqnarray}
 g^\prime &=& (26.500 \pm 0.020) - (0.387 \pm 0.065) \times (X - 1) \nonumber \\
	  &&  + (0.370 \pm 0.008) \times (g^\prime - r^\prime - 1) \nonumber \\
	  &&  - 2.5 \log_{10} (R_{V, {\rm JCAM}}),\nonumber 
\end{eqnarray}
\begin{eqnarray}
 r^\prime &=& (26.903 \pm 0.004) - (0.181 \pm 0.007) \times (X - 1) \nonumber \\
	  &&  + (0.002 \pm 0.004) \times (g^\prime - r^\prime - 1) \nonumber \\
	  &&  - 2.5 \log_{10} (R_{r^\prime, {\rm JCAM}}),\nonumber
\end{eqnarray}
\begin{eqnarray}
 r^\prime &=& (26.576 \pm 0.003) - (0.134 \pm 0.004) \times (X - 1) \nonumber \\
	  &&  + (0.039 \pm 0.002) \times (g^\prime - r^\prime - 1) \nonumber \\
	  &&  - 2.5 \log_{10} (R_{R, {\rm JCAM}}),\nonumber
\end{eqnarray}
\begin{eqnarray}
 i^\prime &=& (25.893 \pm 0.003) - (0.101 \pm 0.006) \times (X - 1) \nonumber \\
	  &&  + (2.124 \pm 0.011) \times (r^\prime - i^\prime)\nonumber \\
	  &&  - 2.5 \log_{10} (R_{I, {\rm JCAM}}),\nonumber 
\end{eqnarray}
\end{singlespace}
\normalsize
\noindent with $X$ equal to the airmass and $R_{Y, {\rm JCAM}}$ equal to the
object flux rate as measured through the filter $Y$ in units of
electrons per second. The uncertainties do not include the systematic
uncertainties in the computed aperture correction (which are less than
0.01 mag). The zeropoints are expected, of course, to vary from night
to night but the color and extinction curves should remain fairly
stable. As expected, as there is no statistically significant color
term, the JCAM $r^\prime$ and $g^\prime$ filters very closely resemble
those of the SDSS filters.  This was also confirmed by the concordance
of the synthetic JCAM and SDSS magnitudes of primary standards (table
\ref{tab:synth-mags}).  The other filters do have statistically
significant color terms relative to the SDSS photometric system.  Note
that there is no need for a ``red leak'' correction to the $U$-band
magnitude (as required with Sloan), one extra benefit of using a
dichroic. We do not yet have enough photometric data to directly
compute the color terms for the JCAM filter set relative to other
photometric systems.

A sky brightness measurement was made at an airmass of $X=1.73$ and
77.6 deg from a half-illuminated moon.  The sky flux was 2.5, 61.1,
102.7, 66.8 electrons s$^{-1}$ pix$^{-1}$ in the $U$, $g^\prime$,
$r^\prime$, and $I$ filters, respectively. This is typical of a bright
moon-lit night, and represents a nominal upper limit to the expected
sky brightness levels.

Using the response curves and the measured fluxes of primary
standards, we compute the total instrument response as a scaling to
the calculated curves. First, we sample the standard star spectrum of
BD+26$^\circ$2606 \citep{fig+96} and the response curves at $\Delta_\lambda =
3$ \AA\ intervals through a spline interpolation, converting the
spectrum to flux, $S_\nu(\lambda_i)$, in units erg s$^{-1}$ cm$^{-2}$
Hz$^{-1}$. Here $i$ refers to the bin number of the sample. The total
expected count rate in the $Y$ filter is then,
\begin{equation}
R_{\rm Y, exp} = A_{\rm eff}\, \sum_i 
                 \frac{S_\nu(\lambda_i)\, R_Y(\lambda_i)\, \Delta_\lambda}
                      {h\, \lambda_i},
\end{equation}
\noindent where $R_Y(\lambda_i)$ is the tabulated response curve of filter $Y$
at an airmass of unity, $A_{\rm eff} = 1.93 \times 10^{5}$ cm$^{-2}$
is the effective collecting area of the 200 inch Telescope, and $h$ is
Planck's constant. If the calculated response curve is correct, we
expect $R_{U, {\rm exp}} = 8.606 \times 10^{5}$, $R_{B, {\rm exp}} =
5.114
\times 10^{6}$, $R_{V, {\rm exp}} = 5.269 \times 10^{6}$, $R_{g^\prime, {\rm exp}} = 1.039 \times 10^{7}$, $R_{R, {\rm exp}} = 1.040 \times
10^{7}$, $R_{r^\prime, {\rm exp}} = 1.419 \times 10^{7}$, and $R_{I,
{\rm exp}} = 1.281 \times 10^{7}$ electrons s$^{-1}$ from
BD+26$^\circ$2606. Using our observed data we actually observed
$R_{U, {\rm obs}} = 2.715 \times 10^{5}$, $R_{B, {\rm obs}} = 3.153 \times
10^{6}$, $R_{V, {\rm obs}} = 3.447 \times 10^{6}$, $R_{g^\prime, {\rm
obs}} = 7.528 \times 10^{6}$, $R_{R, {\rm exp}} = 5.959 \times 10^{6}$,
$R_{r^\prime, {\rm obs}} = 8.281 \times 10^{6}$, and $R_{I, {\rm obs}} =
3.620 \times 10^{6}$ electrons s$^{-1}$ from BD+26$^\circ$2606. This
implies that our response curves are overestimated by 3.170, 1.62,
1.53, 1.38, 1.75, 1.71, 3.54 in the filters $U$, $B$, $V$, $g^\prime$,
$R$, $r^\prime$, and $I$, respectively. The scale factors of $\sim
1.6$ in the $B$, $V$, $g^\prime$, $R$, $r^\prime$ filters are
reasonable and likely due to a lower throughput of the 9 optical JCAM
elements than assumed (98.5\%) and dusty telescope mirrors (these
observations were taken $\sim$10 months after re-aluminization).  For
example, the scale factor can be reproduced if each of these 13
elements are $\sim$4\% less efficient than assumed.  The large scale
factor in $I$-band is probably due to a less efficient CCD than
assumed. The large scale factor in $U$-band is probably due to a
combination of effects of poor UV response from the telescope mirrors
(for example, before re-aluminization, the $U$-band efficiency is
typically down by $\sim$20\% per mirror) and a lower UV throughput of
the optical elements in the Nikon lenses.

\subsection{Deficiencies}

In the construction and operation of JCAM we have realized a number of
deficiencies with the system, some expected and some not.  The
throughput in the near-UV, owing to the Nikon lens, is not as high as
we had required but we do not view this as a major impediment to the
science we hope to accomplish with the instrument.  Over the past
year, the T1-line Internet connection to the mountain has been largely
reliable but on a few occasions the network has been slow or down. Due
to weight balance restrictions, the Coud\'e mirror cannot be put in
place when the adaptive optics (AO) system is mounted at Cassegrain
focus.  Unfortunately, as the frequency of the AO system use has
increased, the availability of JCAM for science imaging is decreasing
(now about 90\% available).

Of greatest concern is the vignetting in the system which reduces the
effective throughput on the edges of the JCAM fields to as low as 40\%
from the peak throughput near the center of the field.  In practice,
though, since the vignetting pattern is stable, we have developed a
methodology for data reduction where the pattern may be found and
removed.  This leads to images where the noise is dependent upon
position on the chip.

\subsection{Future extensions}

We are looking into design changes to future minimize the vignetting
problem. We have also begun work to make JCAM fully scriptable; that
is, to be able to run JCAM from a command-line interface rather than
via a GUI.  We are also writing a number of other observing tools,
such as a graphical representation of the dither pattern on a given
field and an extension to the electronic log program which
periodically grabs water-vapor and IR weather maps from the Internet
and saves them to a local disk with timestamps.  The Tcl/Tk GUI is
currently run by the remote user on the JCAM computer, unnecessarily
taxing the CPU and memory of the computer.  We are, however, examining
software design changes that would allow the remote user to have a
local client which controls the GUI, which in turn sends and receives
small command packets to the JCAM computer (acting as a server).

For now, the bandwidth for image transfer is limited by the bandwidth
of the T1 line off the mountain. With future upgrades to the site
Internet connection, the current co-axial Ethernet connection of the
JCAM computer to the Palomar LAN will become the limiting factor.  As
such, we have purchased fiber optic electronics to put the JCAM
computer on the Palomar LAN at optical fiber connection rates, i.e.,~a
factor of 100 times faster than currently. We plan to install the
optical fiber system once the bandwidth for the Internet connection to
Palomar is increased by a factor of $\sim$10.

\acknowledgments

We are indebted to the financial and intellectual generosity of
M.~Jacobs without whom the project would have never been made
possible.  The staff and directors of Palomar Observatory are
applauded and thanked for their tireless effort to help the JCAM
project become as successful as possible: R.~Ellis, W.~Sargent,
R.~Brucato, R.~Thicksten, R.~Burress, H.~Petrie, J.~Henning, M.~Doyle,
J.~Mueller, D.~Tennent, S.~Kunsman, and J.~Phinney. At Caltech, we
thank T.~Small and J.~Yamasaki for their sagely advice at crucial
times during the construction of JCAM. We especially thank D.~Fox,
D.~Reichart, E.~Berger, and J.~Eisner for assistance during
commission.  JSB gratefully acknowledges the fellowship and financial
support from the Fannie and John Hertz Foundation.  SRK acknowledges
support from NASA and the NSF. AD was supported by a Millikan
Fellowship at Caltech.

\begin{subappendices}

\renewcommand{\arraystretch}{0.95}

\begin{deluxetable}{lccccccccccccccc}
\tabletypesize{\footnotesize}
\rotate
\tablewidth{0pt}
\tablecaption{JCAM Filter Response Curves\label{tab:filter-curves1}}
\tablecolumns{16}
\tablehead{
\colhead{$\lambda$} & 
\multicolumn{14}{c}{$R(\lambda)$} \\
\colhead{\AA}  
& \multicolumn{2}{c}{Bessel $U$}
& \multicolumn{2}{c}{Bessel $B$}
& \multicolumn{2}{c}{Sloan $g^\prime$}
& \multicolumn{2}{c}{Bessel $V$}
& \multicolumn{2}{c}{Bessel $R$}
& \multicolumn{2}{c}{Sloan $r^\prime$}
& \multicolumn{2}{c}{Bessel $I$}
 \\
  
& \colhead{0.0}
& \colhead{1.2}
& \colhead{0.0}
& \colhead{1.2}
& \colhead{0.0}
& \colhead{1.2}
& \colhead{0.0}
& \colhead{1.2}
& \colhead{0.0}
& \colhead{1.2}
& \colhead{0.0}
& \colhead{1.2}
& \colhead{0.0}
& \colhead{1.2}
 } \startdata
3080 &  \nodata &  \nodata &  \nodata &  \nodata &  \nodata &  \nodata &  \nodata &  \nodata &  \nodata &  \nodata &  \nodata &  \nodata &  \nodata &  \nodata \\ 
3160 &  5.57E-4 &  1.26E-4 &  \nodata &  \nodata &  \nodata &  \nodata &  \nodata &  \nodata &  \nodata &  \nodata &  \nodata &  \nodata &  \nodata &  \nodata \\ 
3240 &  8.20E-3 &  2.74E-3 &  \nodata &  \nodata &  \nodata &  \nodata &  \nodata &  \nodata &  \nodata &  \nodata &  \nodata &  \nodata &  \nodata &  \nodata \\ 
3320 &  2.89E-2 &  1.15E-2 &  \nodata &  \nodata &  \nodata &  \nodata &  \nodata &  \nodata &  \nodata &  \nodata &  \nodata &  \nodata &  \nodata &  \nodata \\ 
3400 &  5.75E-2 &  2.52E-2 &  \nodata &  \nodata &  \nodata &  \nodata &  \nodata &  \nodata &  \nodata &  \nodata &  \nodata &  \nodata &  \nodata &  \nodata \\ 
3480 &  8.66E-2 &  4.10E-2 &  \nodata &  \nodata &  \nodata &  \nodata &  \nodata &  \nodata &  \nodata &  \nodata &  \nodata &  \nodata &  \nodata &  \nodata \\ 
3560 &  1.18E-1 &  5.99E-2 &  1.01E-4 &  5.14E-5 &  1.37E-6 &  \nodata &  \nodata &  \nodata &  \nodata &  \nodata &  \nodata &  \nodata &  \nodata &  \nodata \\ 
3640 &  1.47E-1 &  7.91E-2 &  3.95E-3 &  2.12E-3 &  3.08E-6 &  1.66E-6 &  \nodata &  \nodata &  \nodata &  \nodata &  \nodata &  \nodata &  \nodata &  \nodata \\ 
3720 &  1.68E-1 &  9.45E-2 &  2.47E-2 &  1.39E-2 &  4.13E-6 &  2.33E-6 &  \nodata &  \nodata &  \nodata &  \nodata &  \nodata &  \nodata &  \nodata &  \nodata \\ 
3800 &  1.79E-1 &  1.05E-1 &  7.20E-2 &  4.21E-2 &  6.59E-6 &  3.85E-6 &  \nodata &  \nodata &  \nodata &  \nodata &  \nodata &  \nodata &  \nodata &  \nodata \\ 
3880 &  1.43E-1 &  8.70E-2 &  1.32E-1 &  8.06E-2 &  1.23E-3 &  7.46E-4 &  \nodata &  \nodata &  \nodata &  \nodata &  \nodata &  \nodata &  \nodata &  \nodata \\ 
3960 &  7.30E-2 &  4.66E-2 &  1.88E-1 &  1.20E-1 &  3.92E-2 &  2.50E-2 &  \nodata &  \nodata &  \nodata &  \nodata &  \nodata &  \nodata &  \nodata &  \nodata \\ 
4040 &  2.02E-2 &  1.34E-2 &  2.30E-1 &  1.53E-1 &  1.22E-1 &  8.14E-2 &  \nodata &  \nodata &  \nodata &  \nodata &  \nodata &  \nodata &  \nodata &  \nodata \\ 
4120 &  2.85E-3 &  1.96E-3 &  2.60E-1 &  1.78E-1 &  2.03E-1 &  1.39E-1 &  \nodata &  \nodata &  \nodata &  \nodata &  \nodata &  \nodata &  \nodata &  \nodata \\ 
4200 &  2.57E-4 &  1.81E-4 &  2.80E-1 &  1.97E-1 &  2.55E-1 &  1.80E-1 &  \nodata &  \nodata &  \nodata &  \nodata &  \nodata &  \nodata &  \nodata &  \nodata \\*
4280 &  2.00E-5 &  1.44E-5 &  2.92E-1 &  2.11E-1 &  2.94E-1 &  2.12E-1 &  \nodata &  \nodata &  \nodata &  \nodata &  \nodata &  \nodata &  \nodata &  \nodata \\* 
4360 &  \nodata &  \nodata &  2.97E-1 &  2.19E-1 &  3.28E-1 &  2.42E-1 &  1.29E-6 &  \nodata &  \nodata &  \nodata &  \nodata &  \nodata &  \nodata &  \nodata \\* 
4440 &  \nodata &  \nodata &  2.98E-1 &  2.24E-1 &  3.62E-1 &  2.72E-1 &  \nodata &  \nodata &  \nodata &  \nodata &  \nodata &  \nodata &  \nodata &  \nodata \\* 
4520 &  \nodata &  \nodata &  2.90E-1 &  2.23E-1 &  3.50E-1 &  2.68E-1 &  \nodata &  \nodata &  \nodata &  \nodata &  \nodata &  \nodata &  \nodata &  \nodata \\* 
4600 &  \nodata &  \nodata &  2.82E-1 &  2.20E-1 &  3.81E-1 &  2.97E-1 &  2.17E-6 &  1.69E-6 &  \nodata &  \nodata &  \nodata &  \nodata &  \nodata &  \nodata \\* 
4680 &  \nodata &  \nodata &  2.54E-1 &  2.01E-1 &  4.04E-1 &  3.19E-1 &  2.69E-6 &  2.13E-6 &  \nodata &  \nodata &  \nodata &  \nodata &  \nodata &  \nodata \\* 
4760 &  \nodata &  \nodata &  2.05E-1 &  1.64E-1 &  3.94E-1 &  3.15E-1 &  \nodata &  \nodata &  \nodata &  \nodata &  \nodata &  \nodata &  \nodata &  \nodata \\* 
4840 &  \nodata &  \nodata &  1.54E-1 &  1.24E-1 &  4.23E-1 &  3.42E-1 &  2.81E-3 &  2.27E-3 &  \nodata &  \nodata &  \nodata &  \nodata &  \nodata &  \nodata \\* 
4920 &  \nodata &  \nodata &  1.02E-1 &  8.32E-2 &  3.92E-1 &  3.20E-1 &  1.40E-1 &  1.14E-1 &  \nodata &  \nodata &  \nodata &  \nodata &  \nodata &  \nodata \\* 
5000 &  \nodata &  \nodata &  7.37E-2 &  6.05E-2 &  4.24E-1 &  3.48E-1 &  3.39E-1 &  2.79E-1 &  \nodata &  \nodata &  \nodata &  \nodata &  \nodata &  \nodata \\* 
5080 &  \nodata &  \nodata &  4.37E-2 &  3.61E-2 &  4.29E-1 &  3.54E-1 &  4.05E-1 &  3.34E-1 &  \nodata &  \nodata &  \nodata &  \nodata &  \nodata &  \nodata \\* 
5160 &  \nodata &  \nodata &  2.36E-2 &  1.96E-2 &  4.11E-1 &  3.41E-1 &  4.22E-1 &  3.50E-1 &  \nodata &  \nodata &  \nodata &  \nodata &  \nodata &  \nodata \\* 
5240 &  \nodata &  \nodata &  1.07E-2 &  8.91E-3 &  4.15E-1 &  3.46E-1 &  4.21E-1 &  3.51E-1 &  \nodata &  \nodata &  \nodata &  \nodata &  \nodata &  \nodata \\* 
5320 &  \nodata &  \nodata &  4.89E-3 &  4.10E-3 &  4.22E-1 &  3.54E-1 &  4.19E-1 &  3.51E-1 &  \nodata &  \nodata &  3.45E-6 &  2.89E-6 &  \nodata &  \nodata \\* 
5400 &  \nodata &  \nodata &  3.07E-3 &  2.58E-3 &  2.56E-1 &  2.16E-1 &  3.76E-1 &  3.16E-1 &  8.77E-6 &  7.38E-6 &  1.30E-3 &  1.10E-3 &  \nodata &  \nodata \\
5480 &  \nodata &  \nodata &  2.39E-3 &  2.01E-3 &  8.07E-2 &  6.81E-2 &  2.39E-1 &  2.01E-1 &  1.80E-4 &  1.52E-4 &  4.47E-2 &  3.77E-2 &  \nodata &  \nodata \\* 
5560 &  \nodata &  \nodata &  2.02E-3 &  1.71E-3 &  5.58E-4 &  4.72E-4 &  1.18E-1 &  9.97E-2 &  4.87E-3 &  4.12E-3 &  1.81E-1 &  1.53E-1 &  \nodata &  \nodata \\* 
5640 &  \nodata &  \nodata &  9.26E-4 &  7.85E-4 &  2.29E-6 &  1.95E-6 &  4.83E-2 &  4.10E-2 &  7.44E-2 &  6.31E-2 &  3.22E-1 &  2.73E-1 &  \nodata &  \nodata \\* 
5720 &  \nodata &  \nodata &  2.58E-4 &  2.19E-4 &  \nodata &  \nodata &  2.19E-2 &  1.87E-2 &  2.44E-1 &  2.08E-1 &  4.11E-1 &  3.49E-1 &  \nodata &  \nodata \\* 
5800 &  \nodata &  \nodata &  5.51E-5 &  4.70E-5 &  \nodata &  \nodata &  1.44E-2 &  1.23E-2 &  3.47E-1 &  2.96E-1 &  4.43E-1 &  3.77E-1 &  \nodata &  \nodata \\ 
5880 &  \nodata &  \nodata &  1.06E-5 &  9.14E-6 &  \nodata &  \nodata &  1.08E-2 &  9.26E-3 &  3.82E-1 &  3.28E-1 &  4.49E-1 &  3.86E-1 &  \nodata &  \nodata \\ 
5960 &  \nodata &  \nodata &  6.82E-6 &  5.90E-6 &  \nodata &  \nodata &  9.88E-3 &  8.55E-3 &  3.84E-1 &  3.33E-1 &  4.55E-1 &  3.94E-1 &  \nodata &  \nodata \\ 
6040 &  \nodata &  \nodata &  9.17E-6 &  8.02E-6 &  \nodata &  \nodata &  9.34E-3 &  8.17E-3 &  3.72E-1 &  3.26E-1 &  4.49E-1 &  3.93E-1 &  \nodata &  \nodata \\ 
6120 &  \nodata &  \nodata &  6.01E-6 &  5.32E-6 &  \nodata &  \nodata &  5.25E-3 &  4.65E-3 &  3.71E-1 &  3.28E-1 &  4.47E-1 &  3.95E-1 &  \nodata &  \nodata \\ 
6200 &  \nodata &  \nodata &  4.21E-6 &  3.76E-6 &  \nodata &  \nodata &  3.48E-3 &  3.11E-3 &  3.56E-1 &  3.18E-1 &  4.62E-1 &  4.12E-1 &  \nodata &  \nodata \\ 
6280 &  \nodata &  \nodata &  2.60E-6 &  2.33E-6 &  \nodata &  \nodata &  2.54E-3 &  2.28E-3 &  3.33E-1 &  2.99E-1 &  4.35E-1 &  3.91E-1 &  \nodata &  \nodata \\ 
6360 &  \nodata &  \nodata &  1.19E-6 &  1.07E-6 &  \nodata &  \nodata &  1.13E-3 &  1.02E-3 &  3.24E-1 &  2.92E-1 &  4.21E-1 &  3.80E-1 &  \nodata &  \nodata \\ 
6440 &  \nodata &  \nodata &  \nodata &  \nodata &  \nodata &  \nodata &  6.82E-4 &  6.19E-4 &  3.01E-1 &  2.73E-1 &  4.34E-1 &  3.93E-1 &  \nodata &  \nodata \\*
6520 &  \nodata &  \nodata &  \nodata &  \nodata &  \nodata &  \nodata &  4.85E-4 &  4.42E-4 &  2.65E-1 &  2.41E-1 &  3.92E-1 &  3.57E-1 &  \nodata &  \nodata \\*
6600 &  \nodata &  \nodata &  \nodata &  \nodata &  \nodata &  \nodata &  1.99E-4 &  1.81E-4 &  2.48E-1 &  2.26E-1 &  3.73E-1 &  3.40E-1 &  \nodata &  \nodata \\* 
6680 &  \nodata &  \nodata &  \nodata &  \nodata &  \nodata &  \nodata &  5.11E-5 &  4.68E-5 &  2.45E-1 &  2.24E-1 &  4.13E-1 &  3.78E-1 &  \nodata &  \nodata \\* 
6760 &  \nodata &  \nodata &  \nodata &  \nodata &  \nodata &  \nodata &  1.30E-5 &  1.19E-5 &  2.34E-1 &  2.14E-1 &  4.40E-1 &  4.04E-1 &  \nodata &  \nodata \\* 
6840 &  1.59E-6 &  1.42E-6 &  1.57E-6 &  1.40E-6 &  \nodata &  \nodata &  4.36E-6 &  3.88E-6 &  2.11E-1 &  1.88E-1 &  4.05E-1 &  3.60E-1 &  \nodata &  \nodata \\* 
6920 &  9.02E-6 &  7.86E-6 &  1.31E-6 &  1.14E-6 &  \nodata &  \nodata &  1.06E-6 &  \nodata &  1.90E-1 &  1.65E-1 &  3.67E-1 &  3.20E-1 &  1.90E-6 &  1.65E-6 \\* 
7000 &  3.06E-5 &  2.82E-5 &  1.04E-6 &  \nodata &  \nodata &  \nodata &  \nodata &  \nodata &  1.67E-1 &  1.54E-1 &  3.49E-1 &  3.22E-1 &  1.09E-3 &  1.01E-3 \\* 
7080 &  6.20E-5 &  5.73E-5 &  \nodata &  \nodata &  \nodata &  \nodata &  \nodata &  \nodata &  1.42E-1 &  1.31E-1 &  2.15E-1 &  1.99E-1 &  1.70E-2 &  1.57E-2 \\* 
7160 &  7.43E-5 &  6.55E-5 &  \nodata &  \nodata &  \nodata &  \nodata &  \nodata &  \nodata &  1.20E-1 &  1.06E-1 &  9.86E-3 &  8.68E-3 &  7.36E-2 &  6.48E-2 \\* 
7240 &  5.59E-5 &  4.91E-5 &  \nodata &  \nodata &  \nodata &  \nodata &  \nodata &  \nodata &  1.02E-1 &  9.00E-2 &  5.15E-4 &  4.52E-4 &  1.61E-1 &  1.41E-1 \\*
7320 &  3.44E-5 &  3.11E-5 &  \nodata &  \nodata &  \nodata &  \nodata &  \nodata &  \nodata &  9.02E-2 &  8.17E-2 &  5.80E-5 &  5.25E-5 &  2.48E-1 &  2.25E-1 \\*
7400 &  1.88E-5 &  1.74E-5 &  \nodata &  \nodata &  \nodata &  \nodata &  \nodata &  \nodata &  7.83E-2 &  7.28E-2 &  1.10E-5 &  1.02E-5 &  3.12E-1 &  2.90E-1 \\*
7480 &  1.09E-5 &  1.01E-5 &  \nodata &  \nodata &  \nodata &  \nodata &  \nodata &  \nodata &  6.63E-2 &  6.17E-2 &  2.76E-6 &  2.56E-6 &  3.47E-1 &  3.23E-1 \\* 
7560 &  5.55E-6 &  5.16E-6 &  \nodata &  \nodata &  \nodata &  \nodata &  \nodata &  \nodata &  5.45E-2 &  5.07E-2 &  1.17E-6 &  1.09E-6 &  3.58E-1 &  3.33E-1 \\* 
7640 &  2.66E-6 &  1.47E-6 &  \nodata &  \nodata &  \nodata &  \nodata &  \nodata &  \nodata &  4.44E-2 &  2.46E-2 &  \nodata &  \nodata &  3.58E-1 &  1.98E-1 \\* 
7720 &  1.17E-6 &  1.09E-6 &  \nodata &  \nodata &  \nodata &  \nodata &  \nodata &  \nodata &  3.58E-2 &  3.35E-2 &  \nodata &  \nodata &  3.53E-1 &  3.30E-1 \\* 
7800 &  \nodata &  \nodata &  \nodata &  \nodata &  \nodata &  \nodata &  \nodata &  \nodata &  2.86E-2 &  2.68E-2 &  \nodata &  \nodata &  3.46E-1 &  3.24E-1 \\ 
7880 &  \nodata &  \nodata &  \nodata &  \nodata &  \nodata &  \nodata &  \nodata &  \nodata &  2.28E-2 &  2.13E-2 &  \nodata &  \nodata &  3.39E-1 &  3.18E-1 \\* 
7960 &  \nodata &  \nodata &  \nodata &  \nodata &  \nodata &  \nodata &  \nodata &  \nodata &  1.79E-2 &  1.67E-2 &  \nodata &  \nodata &  3.31E-1 &  3.11E-1 \\* 
8040 &  \nodata &  \nodata &  \nodata &  \nodata &  \nodata &  \nodata &  \nodata &  \nodata &  1.39E-2 &  1.31E-2 &  \nodata &  \nodata &  3.23E-1 &  3.02E-1 \\* 
8120 &  \nodata &  \nodata &  \nodata &  \nodata &  \nodata &  \nodata &  \nodata &  \nodata &  1.07E-2 &  9.80E-3 &  \nodata &  \nodata &  3.12E-1 &  2.84E-1 \\* 
8200 &  \nodata &  \nodata &  \nodata &  \nodata &  \nodata &  \nodata &  \nodata &  \nodata &  8.16E-3 &  7.22E-3 &  \nodata &  \nodata &  3.00E-1 &  2.65E-1 \\* 
8280 &  \nodata &  \nodata &  \nodata &  \nodata &  \nodata &  \nodata &  \nodata &  \nodata &  6.19E-3 &  5.48E-3 &  \nodata &  \nodata &  2.90E-1 &  2.57E-1 \\* 
8360 &  \nodata &  \nodata &  \nodata &  \nodata &  \nodata &  \nodata &  \nodata &  \nodata &  4.75E-3 &  4.47E-3 &  1.14E-6 &  1.07E-6 &  2.85E-1 &  2.68E-1 \\* 
8440 &  \nodata &  \nodata &  \nodata &  \nodata &  \nodata &  \nodata &  \nodata &  \nodata &  3.59E-3 &  3.38E-3 &  \nodata &  \nodata &  2.78E-1 &  2.62E-1 \\* 
8520 &  2.13E-5 &  2.01E-5 &  3.41E-5 &  3.22E-5 &  \nodata &  \nodata &  2.60E-5 &  2.45E-5 &  2.94E-3 &  2.77E-3 &  1.34E-6 &  1.27E-6 &  2.71E-1 &  2.56E-1 \\* 
8600 &  1.06E-5 &  9.97E-6 &  2.73E-5 &  2.57E-5 &  1.98E-6 &  1.86E-6 &  1.64E-5 &  1.55E-5 &  2.14E-3 &  2.02E-3 &  2.51E-5 &  2.37E-5 &  2.64E-1 &  2.49E-1 \\* 
8760 &  1.18E-5 &  1.12E-5 &  2.18E-5 &  2.06E-5 &  1.93E-6 &  1.82E-6 &  1.90E-5 &  1.79E-5 &  1.23E-3 &  1.16E-3 &  2.01E-5 &  1.90E-5 &  2.45E-1 &  2.32E-1 \\
8840 &  1.78E-5 &  1.68E-5 &  1.67E-5 &  1.58E-5 &  2.60E-6 &  2.46E-6 &  2.04E-5 &  1.92E-5 &  9.46E-4 &  8.93E-4 &  2.32E-5 &  2.19E-5 &  2.36E-1 &  2.23E-1 \\ 
8920 &  7.39E-6 &  6.71E-6 &  1.45E-5 &  1.32E-5 &  1.89E-6 &  1.72E-6 &  1.60E-5 &  1.46E-5 &  7.12E-4 &  6.47E-4 &  2.19E-5 &  1.99E-5 &  2.29E-1 &  2.08E-1 \\ 
 
9080 &  1.58E-5 &  1.35E-5 &  2.23E-5 &  1.91E-5 &  2.09E-6 &  1.79E-6 &  1.31E-5 &  1.12E-5 &  4.36E-4 &  3.73E-4 &  2.02E-5 &  1.73E-5 &  2.13E-1 &  1.82E-1 \\ 
9160 &  1.25E-5 &  1.07E-5 &  9.32E-6 &  7.98E-6 &  1.29E-6 &  1.10E-6 &  1.34E-5 &  1.15E-5 &  3.45E-4 &  2.96E-4 &  1.60E-5 &  1.37E-5 &  2.05E-1 &  1.75E-1 \\* 
9320 &  9.84E-6 &  6.19E-6 &  7.33E-6 &  4.61E-6 &  1.65E-6 &  1.04E-6 &  1.12E-5 &  7.06E-6 &  2.20E-4 &  1.39E-4 &  1.75E-5 &  1.10E-5 &  1.82E-1 &  1.15E-1 \\* 
9400 &  1.63E-5 &  9.81E-6 &  1.44E-5 &  8.64E-6 &  1.80E-6 &  1.08E-6 &  8.83E-6 &  5.31E-6 &  1.89E-4 &  1.14E-4 &  1.71E-5 &  1.03E-5 &  1.71E-1 &  1.03E-1 \\* 
9480 &  1.26E-5 &  7.56E-6 &  7.12E-6 &  4.28E-6 &  1.50E-6 &  \nodata &  2.07E-5 &  1.25E-5 &  1.19E-4 &  7.17E-5 &  1.37E-5 &  8.25E-6 &  1.59E-1 &  9.57E-2 \\* 
9560 &  1.01E-5 &  6.47E-6 &  2.13E-5 &  1.37E-5 &  1.72E-6 &  1.10E-6 &  1.34E-5 &  8.60E-6 &  1.33E-4 &  8.57E-5 &  1.43E-5 &  9.16E-6 &  1.49E-1 &  9.55E-2 \\* 
9640 &  1.17E-5 &  9.11E-6 &  1.37E-5 &  1.06E-5 &  1.60E-6 &  1.24E-6 &  1.24E-5 &  9.65E-6 &  9.09E-5 &  7.07E-5 &  1.41E-5 &  1.10E-5 &  1.34E-1 &  1.04E-1 \\* 
9720 &  1.03E-5 &  8.99E-6 &  1.07E-5 &  9.37E-6 &  1.43E-6 &  1.25E-6 &  1.62E-5 &  1.42E-5 &  1.04E-4 &  9.10E-5 &  1.41E-5 &  1.23E-5 &  1.19E-1 &  1.05E-1 \\* 
9800 &  3.12E-6 &  2.82E-6 &  6.75E-6 &  6.09E-6 &  1.07E-6 &  \nodata &  1.27E-5 &  1.15E-5 &  6.30E-5 &  5.68E-5 &  1.06E-5 &  9.56E-6 &  1.06E-1 &  9.61E-2 \\* 
9880 &  1.48E-5 &  1.40E-5 &  8.37E-6 &  7.97E-6 &  1.34E-6 &  1.27E-6 &  1.29E-5 &  1.22E-5 &  7.13E-5 &  6.79E-5 &  1.35E-5 &  1.29E-5 &  9.38E-2 &  8.92E-2 \\* 
10040 &  9.42E-6 &  8.97E-6 &  4.73E-6 &  4.50E-6 &  \nodata &  \nodata &  8.57E-6 &  8.15E-6 &  5.85E-5 &  5.57E-5 &  2.78E-5 &  2.65E-5 &  6.82E-2 &  6.49E-2 \\* 
10120 &  3.66E-6 &  3.48E-6 &  1.20E-6 &  1.14E-6 &  \nodata &  \nodata &  8.53E-6 &  8.12E-6 &  3.80E-5 &  3.62E-5 &  6.37E-5 &  6.06E-5 &  5.77E-2 &  5.49E-2 \\* 
10200 &  6.36E-6 &  6.06E-6 &  5.55E-6 &  5.28E-6 &  \nodata &  \nodata &  2.73E-6 &  2.60E-6 &  4.75E-5 &  4.52E-5 &  2.50E-4 &  2.38E-4 &  4.78E-2 &  4.55E-2 \\* 
10360 &  3.14E-6 &  2.99E-6 &  4.95E-6 &  4.71E-6 &  \nodata &  \nodata &  5.93E-6 &  5.65E-6 &  2.28E-5 &  2.17E-5 &  1.76E-2 &  1.68E-2 &  3.16E-2 &  3.01E-2 \\ 
10440 &  1.45E-6 &  1.38E-6 &  4.54E-6 &  4.32E-6 &  \nodata &  \nodata &  4.63E-6 &  4.41E-6 &  2.42E-5 &  2.30E-5 &  3.96E-3 &  3.77E-3 &  2.44E-2 &  2.32E-2 \\ 
10600 &  2.46E-6 &  2.34E-6 &  2.68E-6 &  2.55E-6 &  \nodata &  \nodata &  \nodata &  \nodata &  8.08E-6 &  7.70E-6 &  9.43E-3 &  8.98E-3 &  1.32E-2 &  1.25E-2 \\ 
10680 &  1.86E-6 &  1.77E-6 &  1.86E-6 &  1.77E-6 &  \nodata &  \nodata &  1.87E-6 &  1.79E-6 &  1.44E-5 &  1.37E-5 &  3.05E-2 &  2.90E-2 &  9.99E-3 &  9.51E-3 \\ 
10760 &  1.20E-6 &  1.14E-6 &  1.58E-6 &  1.51E-6 &  \nodata &  \nodata &  2.09E-6 &  1.99E-6 &  1.24E-5 &  1.18E-5 &  9.46E-3 &  9.01E-3 &  7.25E-3 &  6.91E-3 \\ 
10920 &  1.12E-6 &  1.07E-6 &  1.47E-6 &  1.40E-6 &  \nodata &  \nodata &  1.22E-6 &  1.16E-6 &  4.10E-6 &  3.90E-6 &  2.87E-3 &  2.74E-3 &  3.87E-3 &  3.68E-3 \\ 
11000 &  \nodata &  \nodata &  1.10E-6 &  1.05E-6 &  \nodata &  \nodata &  1.15E-6 &  1.10E-6 &  3.91E-6 &  3.73E-6 &  2.90E-3 &  2.77E-3 &  2.80E-3 &  2.67E-3 \\* 
11080 &  1.01E-6 &  \nodata &  \nodata &  \nodata &  \nodata &  \nodata &  \nodata &  \nodata &  4.94E-6 &  4.71E-6 &  3.72E-3 &  3.55E-3 &  1.92E-3 &  1.83E-3 \\* 
11240 &  \nodata &  \nodata &  \nodata &  \nodata &  \nodata &  \nodata &  \nodata &  \nodata &  1.10E-6 &  1.05E-6 &  7.37E-3 &  7.03E-3 &  7.09E-4 &  6.77E-4 \\* 
11320 &  \nodata &  \nodata &  \nodata &  \nodata &  \nodata &  \nodata &  \nodata &  \nodata &  1.87E-6 &  1.79E-6 &  4.27E-3 &  4.07E-3 &  3.89E-4 &  3.71E-4 \\*
11480 &  \nodata &  \nodata &  \nodata &  \nodata &  \nodata &  \nodata &  \nodata &  \nodata &  \nodata &  \nodata &  1.53E-3 &  1.46E-3 &  1.84E-4 &  1.75E-4 \\* 
11560 &  \nodata &  \nodata &  \nodata &  \nodata &  \nodata &  \nodata &  \nodata &  \nodata &  \nodata &  \nodata &  1.35E-3 &  1.29E-3 &  1.38E-4 &  1.32E-4 \\* 
11640 &  \nodata &  \nodata &  \nodata &  \nodata &  \nodata &  \nodata &  \nodata &  \nodata &  \nodata &  \nodata &  1.40E-3 &  1.34E-3 &  1.04E-4 &  9.92E-5 \\* 
11720 &  \nodata &  \nodata &  \nodata &  \nodata &  \nodata &  \nodata &  \nodata &  \nodata &  1.49E-6 &  1.42E-6 &  1.66E-3 &  1.58E-3 &  7.52E-5 &  7.18E-5 \\* 
11800 &  \nodata &  \nodata &  \nodata &  \nodata &  \nodata &  \nodata &  \nodata &  \nodata &  1.12E-6 &  1.07E-6 &  2.20E-3 &  2.10E-3 &  5.66E-5 &  5.41E-5 \\* 
11880 &  \nodata &  \nodata &  \nodata &  \nodata &  \nodata &  \nodata &  \nodata &  \nodata &  \nodata &  \nodata &  3.04E-3 &  2.90E-3 &  4.15E-5 &  3.96E-5 \\* 
11960 &  \nodata &  \nodata &  \nodata &  \nodata &  \nodata &  \nodata &  \nodata &  \nodata &  2.19E-6 &  2.10E-6 &  3.86E-3 &  3.69E-3 &  3.05E-5 &  2.91E-5 
\enddata

\tablecomments{The calculated response curves are given for the 7 JCAM
filters for two different airmasses.  For clarity we suppress values
of $R(\lambda) < 1\times 10^{-6}$.  The wavelength coverage of the
response curves was chosen to coincide with the flux curves of the
spectroscopic standards $\alpha$ Lyr (Vega) and BD17$^\circ$4708 that
define the Sloan photometric system \citep[see][]{fig+96}.  To
reproduce the true total response on-sky imaging suggests that these
numbers should be scaled downward by 3.170, 1.62, 1.53, 1.38, 1.75,
1.71, 3.54 in the filters $U$, $B$, $V$, $g^\prime$, $R$, $r^\prime$,
and $I$, respectively. See text for an explanation.}

\end{deluxetable}

\end{subappendices}

\chapter{Epilogue and Future Steps}
\label{chap:epi}

\begin{figure*}[tp]
\centerline{
\psfig{file=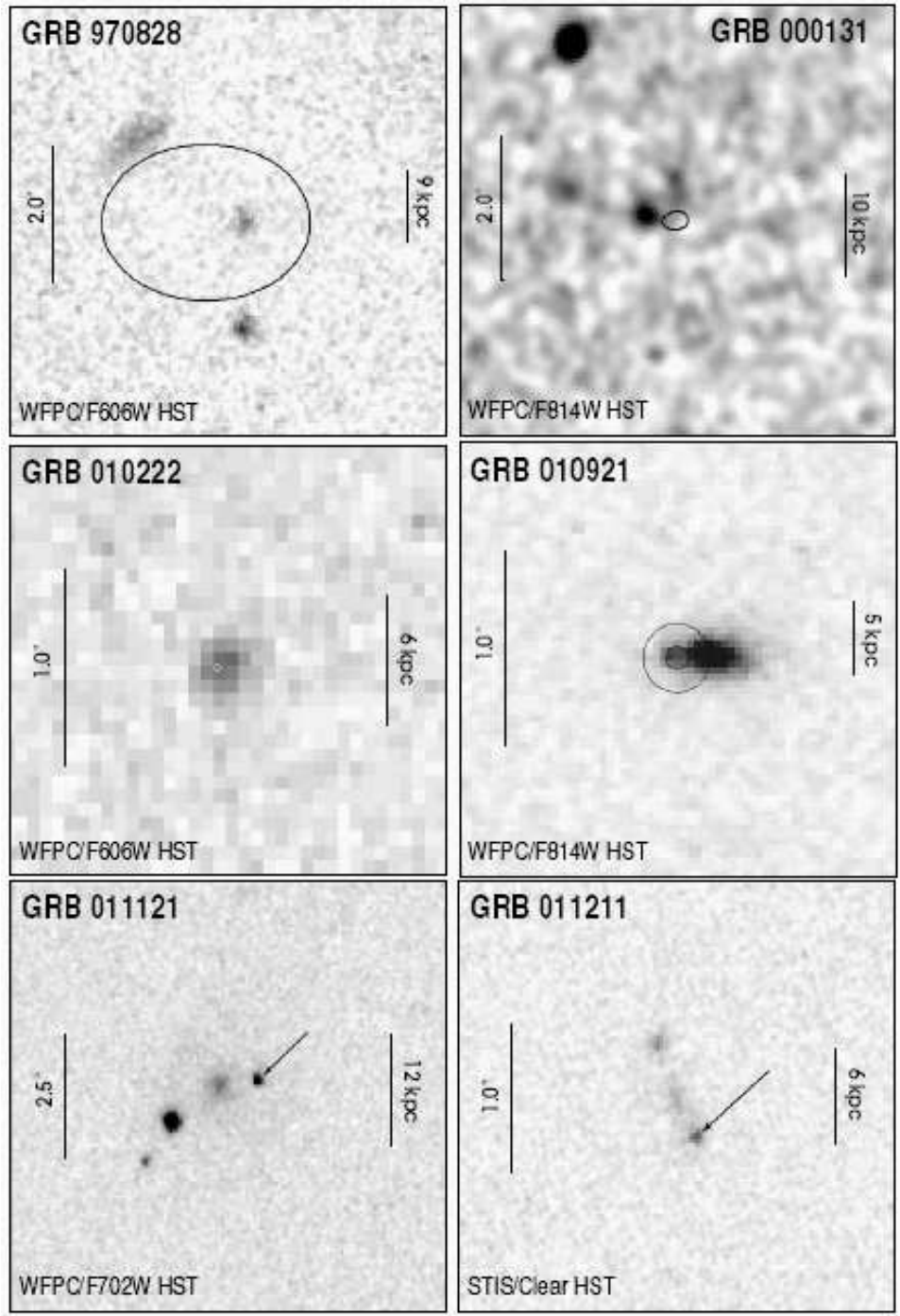,width=5in,angle=0}}
\caption[Update to chapter \ref{chap:offset} of new offsets and {\it HST} 
images of host galaxies.]{Update to chapter \ref{chap:offset} of new
offsets and {\it HST} images of host galaxies. (Top) 3 $\sigma$ error
contours are shown. (middle-left) The HST$\rightarrow$HST astrometric
tie places the OT very accurately ($\sigma_r = 6$ mas r.m.s.) on the
host galaxy image of GRB 010222; displayed is the 10 $\sigma$ error
contour. (middle-right) The afterglow is still visible as a point
source to the east of the apparent host galaxy; the error contours
correspond to the 1 and 3 $\sigma$ location predicted from an earlier
Palomar image. (bottom) Arrows point to the optical transients.}
\label{fig:offset-update}
\end{figure*}

\section{On the Offset Distribution of GRBs}

Since the offsets paper was completed (chapter \ref{chap:offset}),
there have been a few new afterglow discoveries and {\it HST}
observations of GRB hosts.  The offsets and hosts for these are
depicted in figure \ref{fig:offset-update}.  The new bursts, two of
which are the lowest-redshift GRBs measured, continue to show the
close connection of GRBs to the light of host galaxies.

The claim in my offsets work (chapters \ref{chap:nsns} and
\ref{chap:offset}) is that NS--NS and NS--BH mergers are inconsistent 
with the distribution of long-duration GRBs about their host
galaxies. This statement holds only under the caveat that the
calculated/expected radial models of such binaries are representative
of the true distribution. The populations synthesis work in chapter
\ref{chap:nsns} was the first attempt to produce such radial profiles
in a variety of realistic galactic potentials. Since then a number of
others \citep{fwh99,bbz99,bbz00} have followed suit using different
(and more sophisticated) population synthesis models for high-mass
binary evolution. For the ``standard'' NS--NS production channel
(\citealt{bv91}; see fig.~\ref{fig:prog-mod}), all of our studies
agree about offsets to within the uncertainties in input
parameters\footnotemark\footnotetext{Much of the early NS--NS and
NS--BH work was conducted primarily to predict the event rate for
LIGO; these proceeded on both observational \citepeg{phi91,nps91,kl00}
and theoretical (population synthesis) \citepeg{ty93,pzs96,vl96}
grounds. \citet{kns+01}, using the most up--to--date observations on
known Galactic NS--NS binaries and pulsar statistics, has recently
suggested that the birthrates of NS--NS binaries are still unknown by
at least a factor of $\sim$200.  While this may be true, it is
important to note that the uncertainty in the offset distribution
should be much smaller.  This is due to the somewhat counterintuitive
fact, as I discovered in chapter \ref{chap:nsns}, that different
supernovae kick distributions only effect the birthrate of merging
NS--NS, not the resulting distribution of systemic
velocities.}\commaafterfoot particularly the highly-uncertain
supernova kick distribution
\citep{fwh99}. The three later studies also roughly agree on BH--NS
offset distributions.

The comforting agreement between the different groups has recently
been upset by the potential ``discovery'' of new channels of NS--NS
production \citep{bk01,kkb02} that result in binaries which coalesce
rapidly after ZAMS ($\ale 10^6$ yr).  The existence of the dominant
new channel relies on the untested assumption that low-mass helium
stars can survive common-envelope evolution with a neutron star. If
true, then the expected distribution of NS--NS mergers may be
significantly more close to where massive stars are born \citep{bbr02}
and thus, by assumption, consistent with the observed offset
distribution\footnotemark\footnotetext{Note that BH--NS binaries are
still excluded by the measured offset distribution since $\sim$50\%
merger outside of 10 kpc.}\periodafterfoot Moreover, given the close
connection in time with star-formation, the redshift and the
host-galaxy properties of such NS--NS systems would be
indistinguishable from those properties expected from collapsar
progenitors.

Even if such channels dominate the rate of NS--NS production, since
neither NS is recycled and/or the binary merges quickly, it is
unlikely that we will ever observe such a system in the Galaxy.  We
are therefore left with the unsettling prospect of the existence of
new theoretical channels for NS--NS births without obviously testable
observations. A more detailed set of offset measurements might still
be able to discriminate between
progenitors\footnotemark\footnotetext{\citet{pb02}, using these new
channels, predicted the projected radial distribution of NS--NS and
found all but $\sim$5--10 percent of NS--NS merge within 10 kpc of
their host galaxy, depending on host mass. If the observed trend found
in chapter
\ref{chap:offset}, that all GRBs fall within 10 kpc of a galaxy, 
continues, then I calculate that we would need to observe between
26--72 (corresponding to a 10 and 5 percent extended population,
respectively) to rule out the new NS--NS scenarios at the 99\%
confidence level.}\nadaafterfoot and more detailed hydrodynamical
simulations should be employed of the common-envelope evolution phase
to test the assumptions that lead to short-lived NS--NS binaries.  The
required number of new offsets ($\sim$50) to discriminate between this
new NS--NS model and the collapsar model should be obtainable within
the first $\sim$six months of the {\it Swift} mission even if the
bursts occur in low-density environments \citep{pb02}.

Perhaps more fruitful to distinguish between progenitors will be the
use of high-quality early-time afterglow observations, afforded by
rapid {\it Swift} localizations and instruments such as {\it JCAM}, to
constrain the distribution of ambient density surrounding GRBs.  In
the rapid merger scenario, most bursts should still occur in
low-density environments ($n \ale 1$ cm$^{-3}$; \citealt{pb02})
whereas bursts from collapsars should occur with $n
\age 1$ cm$^{-3}$. Already, afterglow modeling of some of the more 
extensive datasets have yielded densities far in excess of unity
\citep{hys+01,pk01}.

Aside from the continuing potential for offsets to directly
discriminate between increasingly more sophisticated progenitors
models, the continued measurement of offsets will be useful in new
ways.  Already the offsets work has shown that most long-duration GRBs
probe the inner 10 kpc of their hosts.  This, coupled with an expanded
use of afterglow spectroscopic absorption line studies
\citep{bloom02a}, may shed new light on the enrichment history of
moderate- to high-redshift galaxies in a manner complementary to
quasar absorption-line studies (which probe the outer reaches of
galaxies).  Already we have seen that offsets plus spectroscopy can
yield a lower-limit to the dynamical mass of GRB hosts
\citepeg{cgh+02}.

\section{Re-examining the GRB--Supernova Connection}

\subsection{S-GRBs}

In chapter \ref{chap:sgrbs}, I suggested that there may be a sub-class
of GRBs that are associated with local SNe. As evidenced by the
exhaustive study undertaken by \citet{nbw99} to find archival examples
of S-GRBs in the {\it BATSE} catalog, the proposed sub-class has been
taken seriously as a legitimate class of GRBs.  In agreement with my
estimates, those authors too found about 1--2\% of the BATSE sample
that met our S-GRB criteria. As expected by these estimates, no new
examples of S-GRBs (as viewed purely from the $\gamma$-ray light curve
considerations in chapter \ref{chap:sgrbs}) have been found by {\it
BeppoSAX} or the IPN since 1998\footnotemark\footnotetext{In
retrospect, since supernovae remain bright (and detectable) for months
in the optical, even delayed follow-up of old IPN localizations of
GRBs might have uncovered a few supernova associations. Delayed
searches were performed in the radio \citep{fk95} but not
systematically at optical wavelengths}\periodafterfoot

\def\blo{1}
\def\pri{2}
\def\rei{3}
\def\gtv{4}
\def\reib{5}
\def\bdk{6}
\def\svb{7}
\def\fvh{8}
\def\bhj{9}
\def\lcg{10}
\def\csg{11} 
\def\sok{12}
\def\ddr{13}
\def\bdf{14}

\begin{deluxetable}{llllll}
\singlespace
\tablecolumns{4} 
\tablewidth{6in} 
\tablecaption{Summary of Proposed Cosmological GRBs with Associated Supernovae\label{tab:sn-summary}}
\tabletypesize{\scriptsize}
\tablehead{
\colhead{Light Curve \& Spectral} & \colhead{Viable Alter.}  & \colhead{Comments} &  \colhead{Refs.} \\
\colhead{Observations} & \colhead{Models} & \colhead{} & \colhead{}}

\startdata

\cutinhead{Good Candidates (definate bump detection)}

\sidehead{GRB 011121}
             Multi-color, multi-epoch observations
            &  ?
            &   Best case for a GRB--SN connection.
            & \blo, \pri    \\
             of a bump before, during, and after peak.
            &
            &   Not an exact fit to SN 1998bw,  but
            & \\
              Shows evidence for spectral 
            &
            &   differences probably reflective of
            & \\
              roll-over at $\sim 0.7 \mu$m.
            &
            &   the diversity in core-collapased SNe. \\

\cutinhead{Plausible Candidates (probable bump detection)}

\sidehead{GRB 970228}
             $V$, $I$, $R$, $K$-band observations
            &  dust echo?
            &   Good case for a GRB--SN connection,
            & \rei, \gtv, \reib    \\
             near peak. Broadband spectrum
            &
            &   but spectral-roll over found from 
            & \\
              shows evidence for spectral 
            &
            &   non-contemporaneous broadband obs.
            & \\
              roll-over at $\sim 0.9 \mu$m.
            &
            &   \& extrapolations are highly uncertain. \\

\sidehead{GRB 980326}

             One $R$-band detection, few $I$-band
            & dust echo
            &   Best detection of extra
            & \bdk \\
             upper-limits.  Crude spectrum 
            &
            &   emmission feature in $R$-band. Only 
            & \\
             reveals red colors relative to the
            &
            &   redshift upper-limit.
            & \\
             early afterglow.
            &
            &  First case for GRB--SN connection.
            \\

\sidehead{GRB 990712}
             Multiple $V$-, $R$-band observations 
           &     dust echo,
           &        Faint SN is at comparable
           &           \svb, \fvh, \bhj \\
             around SN peak, 2 HST epochs.
           &     thermal dust
           &         brightness to the host. \\
             No spectral info. Colors 
           & 
           &		 Wrong SN colors. \citet{fvh+00} \\
             indicate red OT at $t=48$ days.
           &
           &         finds no evidence for SN. \\

\cutinhead{Marginal Candidates (marginal bump detection)}

\sidehead{GRB 000911}

             $B$, $V$, $R$, $I$, $J$-band obs near SN
            & dust echo
            &   First claim of an IR brightening. Bright
            & \lcg \\
             peak. Rebrightening claimed in $R$-,
            &
            &   host yields $3\sigma$ detection of bump,

            & \\ 
             $I$-, $J$-band. Crude spectrum 
            &
            &   but claim hinges on one $J$-band
            & \\
             near SN peak.
            &
            &   point. Host light contiminates spectrum.
            \\

\cutinhead{Unlikely Candidates (improbable  bump detection)}

\sidehead{GRB 991208}
             Few $R$, $V$-band observations 
           &     \ldots
           &        some evidence for bump in $R$-band
           &           \csg \\
             around SN peak.
           &     
           &        but hinges on uncertain host flux. \\
             No spectral info. 
           & 
           &		   \\

\sidehead{GRB 970508}
             Few $B$, $V$, $R$, $I$-band observations 
           &     \ldots
           &        bump is a $\sim$2 $\sigma$ effect; only
           &           \sok \\
             around SN peak.
           &     
           &        marginally seen in $I$-band. \\
             No spectral info. 
           & 
           &               \\

\sidehead{GRB 000418}
             Few $R$-band observations 
           &     \ldots
           &        no apparent significance
           &           \ddr, \bdf \\
             around SN peak.
           &     
           &        to claimed $R$-band bump. See ref.~\bdf. \\
             No spectral info. 
           & 
           &               \\

\enddata

\tablecomments{Rank ordered list of possible GRBs with associated supernova
signatures, from most likely to least likely.  Ordering was based on
estimated significance of bump detection (photometrically and
spectroscopically) and number of viable alternative models to explain
the detection(s). }

\tablerefs{ 
\blo.~Chapter \ref{chap:grb011121}; \pri.~\citet{pbr+02}; 
\rei.~\citet{rei99}; \gtv.~\citet{gtv+00}; \reib.~\citet{rei01b};
\bdk.~\citet{bkd+99} (chapter \ref{chap:sn-grb});
\svb.~\citet{svb+00}; \fvh.~\citet{fvh+00}; \bhj.~\citet{bhj+01};
\lcg.~\citet{lcg+01};
\csg.~\citet{csg+01}; 
\sok.~\citet{sok01};
\ddr.~\citet{ddr02}; \bdf.~\citet{bdf+01}
}
\end{deluxetable}

There have been some weak constraints on the S-GRB population placed
by radio studies of Type Ib/Ic supernovae. None of the apparent
optical hypernovae (i.e., SN 2002bl, SN 2002ap, SN 2002ao, SN 2002J,
SN 2001bb, SN 1999as) were found to have prompt radio emission like
1998bw (though only 2002ap and 2002bl were promptly followed-up in the
radio\footnotemark\footnotetext{Four other non-hypernovae Type Ib/Ic
supernovae were also followed-up in the radio with no detections
(Berger, private communication).}\semiafterfoot Berger, private
communication).  One of these (SN 2002ap) was also determined to have
not produced a gamma-ray or X-ray burst \citep{hmg+02}.  These
non-detections are consistent with our suggestion that only a fraction
($\sim$10\%) of Type Ib/Ic supernovae could be truly the supernova
component of an S-GRB.

The predictions of the S-GRB hypothesis could be tested by {\it
Swift}.  At current predicted rates (about two localizations per
week), {\it Swift} should find about 1--3 S-GRBs per year down to the
{\it BATSE} fluence limit ($S \sim 8 \times 10^{-7}$ erg cm$^{-2}$).
Note that the {\it Swift} detection rate was based on extrapolated
detection rates from {\it BATSE}. Since {\it Swift} is $\sim$5 times
more sensitive than {\it BATSE}, it is possible that the expected rate
is a gross underestimate of the true rate---if the S-GRB hypothesis is
correct, then not only will {\it Swift} detect bursts at a higher rate
than presumed, but many of the faintest sources should be associated
with supernovae out to a distance of $\sim$225 $h_{65}^{-1}$ Mpc,
assuming that GRB 980425 was a standard candle. Even without decent
localizations from {\it Swift}, a local population could be manifest
in an upturn of the $\log N$--$\log S$ brightness distribution just
fainter than the sensitivity of {\it BATSE}.  Needless to say, if no
S-GRBs are found by {\it Swift} after hundreds have been
well-localized, the physical connection between 1998bw and GRB 980425
would be once again called in to question on statistical grounds.

Regardless of whether the S-GRB hypothesis is correct, there have been
some tentative/weak suggestions of the existence of other local GRBs
samples based upon gamma-ray properties:
\begin{itemize}
\item {\it Anisotropic Subsets}: The suggested angular
anisotropy---indicative perhaps of a clustering in the super-galactic
or Galactic plane---was found \citepeg{cws98,bmh98,mbh+00} in some
sub-classes that were created on purely phenomenological grounds
(i.e., by choosing a locus of ``intermediate'' duration,
soft-bursts). These findings are thus suspect because of the {\it ad
hoc} creation of sub-sets of the data which yield the desired results.

\item {\it Long-duration, long-lag bursts}: \citet{nor02} has recently suggested a phenomenological
sub-class of local GRBs motivated by the properties of GRB 980425 and
other bursts with known (higher redshifts).  The observed existence of
the so-called luminosity--pulse lag relation, on which the possible
local GRB sample is constructed, may even be motivated by a physical
origin
\citep{sal00}.

\item {\it X-ray Flashes (XRFs)}: Interestingly, XRFs detected 
by {\it BeppoSAX} are distributed such that $\langle V/V_{\rm max}
\rangle = 0.50 \pm 0.09$ \citep{hei01}, indicative of a distribution
in Euclidean space; this would suggest a relatively local population.

\end{itemize}

\subsection{Supernova bumps}
 
The supernova interpretation for intermediate-time deviations of GRB
afterglows from power-law behavior has become {\it en vogue} for new
and historical GRBs but, apart from the recent discoveries of GRB
011121 (chapter \ref{chap:grb011121}), the observational connection
has been largely disappointing.  In table \ref{tab:sn-summary}, I
provide a summary of the GRBs with a proposed detection of a supernova
component.  These bursts are ranked in order of the security of the
claimed detection as well as the number of plausible alternative
models. The only comprehensive study which did not find a SN component
(i.e.,~found an upper-limit on the peak brightness any associated SN)
was from
\citet{price02}, using GRB 010921.

The reasons for the lack of a convincing associations are many.
First, most historical GRBs were poorly sampled at crucial time
intervals: a GRB light curve must be well-sampled from about 1--3
months in order to conclusively see a SN signature; these time-scales
were often not probed as extensively pre-GRB 980326.  Second, there
are have been surprisingly few new GRBs since our discovery was
published; only 14 bursts since October 1999 until March 2002 had
optical or radio transients. This was due to a lower detection
efficiency of {\it BeppoSAX} and the IPN as well as a delayed start to
{\it HETE-II}.
Third, SN signatures from GRBs which arise from redshifts higher than
$z \sim 1$ are difficult to observe at optical wavelengths; only 5 of
the 14 bursts remained as possible candidates. Last, of the
low-redshift candidates, none of these bursts had a rapidly declining
afterglow that would have improved the chance of detecting of an extra
light curve component.  These primarily observational impediments were
ones which I anticipated in chapter \ref{chap:sn-grb}.

Though many of the alternative models for the ``red bump'' in GRB
980326 were refuted in chapter \ref{chap:sn-grb}, alternative
explanations for the red bump were later put forth. \citet{wd00}
suggested that the intermediate-time, red bumps could be produced by
the thermal reemission of afterglow light from dust.  While this
cannot be ruled out in the case of GRB 980326, I suggested at the 5th
Huntsville Conference in October 1999 that the spectral roll-over
inferred in the red bump of GRB 970228 \citep{rei99} excluded this
model. Though the broadband spectrum of GRB 970228 was approximately
thermal, the peak of the spectrum was at $\sim$0.8 $\mu$m, while the
peak of the thermal dust emission is expected at $\age 3 \mu$m for the
known redshift of GRB 970228. \citet{rei01b} and \citet{eb00} later
reiterated my point.

\citet{eb00} proposed that intermediate-time bumps could arise when dust 
around the GRB progenitor scatters the afterglow light into our
light--of--sight. Assuming that the initial flash due to a reverse
shock sublimates dust out to a radius $R$ ($\approx 0.1$--few pc;
\citealt{wd00}), the timescale for the onset of emission due 
to dust scattering emission from the explosion site is $t \approx R/c
\times (1 + z) \times (1 - \cos \theta_{\rm typ}) \approx 10^6$ s
(following eq.~6 of \citealt{eb00}). The typical scattering angle must
be less than $\theta_{\rm typ} \ale 30$ deg for a high probability of
scattering into the line--of--sight (see fig.~3 of
\citealt{eb00}). Dust should preferentially scatter blue light more
than red so that at low dust opacities, the intermediate-bump should
appear more blue than the afterglow.  At optical depths larger than
$\tau_{0.3 \mu{\rm m}} \approx 3$, however, the absorption dominates
and the bump appears more red than the afterglow \citep{eb00}. If the
latter is true, as is claimed by \citet{eb00}, then the
intermediate-time emission of GRB 980326 and GRB 970228 can be
explained as due to dust echoes and not an underlying supernova.

\citet{rei01b} has called into question some of the conclusions
of \citet{eb00}. First, he claims the assumption of a thin scattering
shell, left over by a wind-stratified medium after it has been
sublimated, may be unwarranted because the radius of the scattering
shell is required to be larger than the termination shock of the
stellar wind from the progenitor. Second, he points out that the
roll-over in the spectrum of GRB 970228 cannot be reproduced by dust
scattering unless the roll-over pre-exists in the early afterglow
itself (which it should not).  Instead, spectra of dust-scattered
light at a given time should be a monotonic function wavelength across
the optical/infrared spectrum.

The discovery of a supernova-like component in GRB 011121 appears to
disfavor the alternative models for red bumps. Though some may still
be skeptical of the association, the timescale, spectra, and light
curve of the GRB 011121 bump closely follows the simplistic
expectations of a core-collapsed SN. Regardless, this clearly bodes
well for future observational campaigns on low-redshift GRBs. What can
be done to strengthen (or refute) the associations? The clearest, most
unambiguous test for the GRB--SN connection will be obtaining a
spectrum of an intermediate-time bump. At optical wavelengths, one
expects to see redshifted broad metal-line absorption features from a
Type Ib/Ic supernova (see chap.~\ref{chap:sn-grb}).  Otherwise, in,
for example, the dust echo origin, a smooth continuum is
expected. This is a difficult observation from the
ground. Ground-based spectroscopy of $R \approx 25$ mag bumps require
a modest investment of observing time to see broad features from a
point source ($\age 2$ hr on a 10-m telescope in average seeing) but,
due to atmospheric smearing of the bump and host light, will likely be
hampered by the competing light of host galaxies. So unless the host
is significantly fainter than the bump (as in the case of GRB 980326),
even the detection of spectral features will lead to ambiguous
conclusions. In principle, a careful subtraction of the light of the
host (from a later-time spectrum after the bump has faded) could
remove the contamination problem, but in practice, even if the
observations were conducted using the same instrumental setup, it
would be hard to exactly match the same observing conditions as the
first epoch (e.g.,~seeing, airmass, slit location, etc.).

\begin{figure*}[tbp]
\centerline{\psfig{file=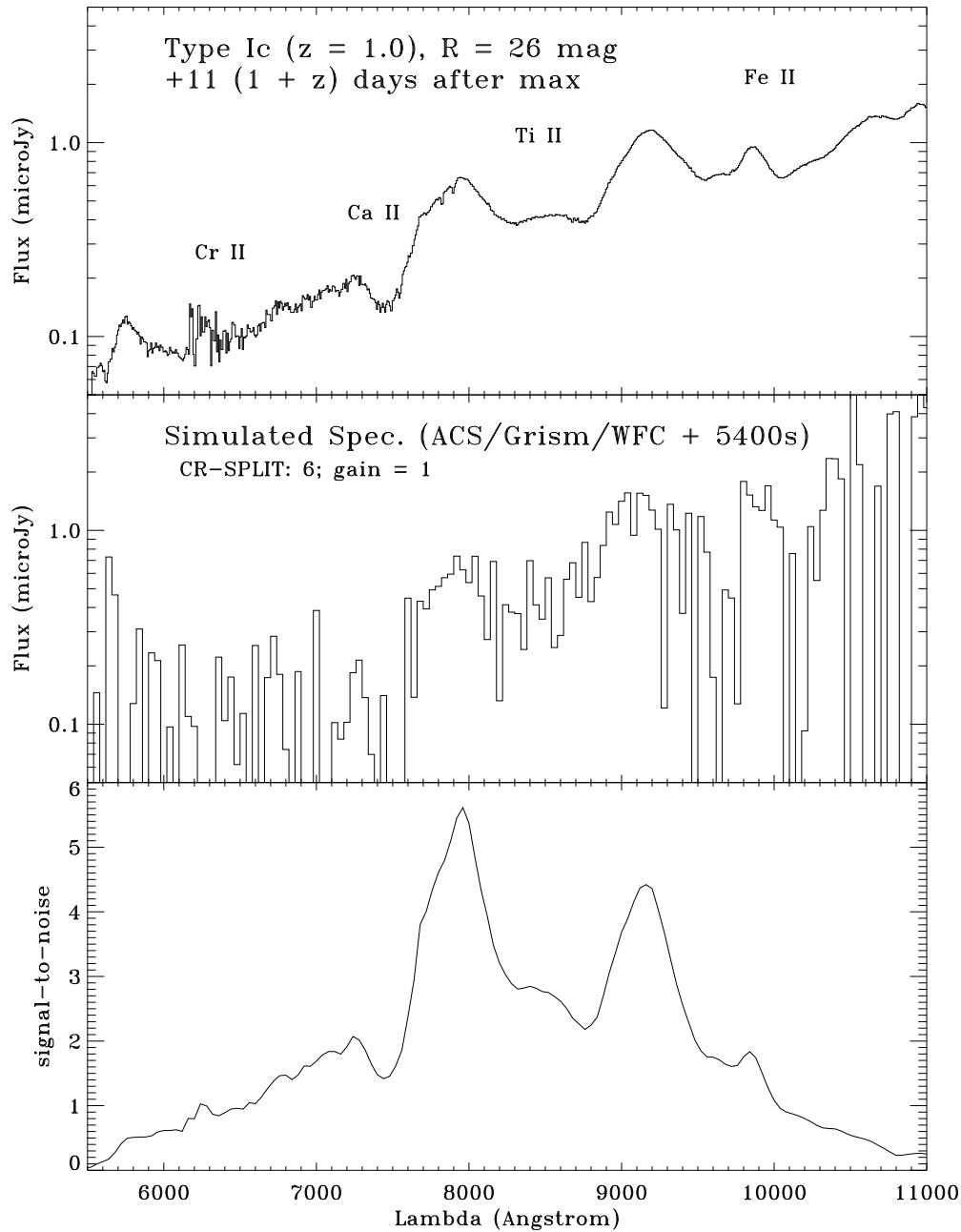,width=5.5in,angle=0}}
\caption[A future step toward resolving the progenitor question:
space-based spectroscopy of intermediate-time emission components]{A
future step toward resolving the progenitor question: space-based
spectroscopy of intermediate-time emission components. Here, I show a
simulated SNe spectrum at $R=26$ mag and $z=1.0$ as observed with the
{\it Advanced Camera for Surveys} on {\it HST}.  The template is SN
1994I (Type Ic) as reproduced in \citet{mbb+99}. The observing
parameters are given in the plot.  Clearly the slope of the continuum
is much steeper than $F_\nu \propto \nu^{-2}$ (as might be expected
from competing hypotheses such as dust echos) and, more importantly,
some broad spectral features are detectable even at such faint
magnitude levels. The dispersion scale is 39 \AA\ pixel$^{-1}$.  The
signal--to--noise on the bottom plot is given as the SN per resolution
element, which in this case consists of 5 (spatial) $\times$ 2
(dispersion) = 10 pixels.  The noise in the middle plot reflects the
noise per pixel, summed over 5 pixels in the spatial direction (this
encloses 88\% of the total flux of a point source).  }
\label{fig:sn10}
\end{figure*}

Space-based spectroscopy may be more productive and, just beginning in
March 2002, feasible. In figure \ref{fig:sn10}, I show a simulation of
a supernova spectrum from the recently-installed {\it Advanced Camera
for Surveys} (ACS) on the {\it HST}.  In a few orbits ($\sim4$), broad
SN features can be detected at better than $5 \sigma$ for a $R=26$ mag
SN at $z=1$.  The angular resolution of the grism spectrum will be far
superior than a ground-based spectrum, minimizing any host galaxy
contamination. And, if any ambiguity still exists, a repeat
observation at the same telescope roll-angle 6 months, or a 1 year
later should allow a precise subtraction of the host contribution
given the stability of the space-based instruments over time.  It is
not unreasonable to expect that such observations of bumps (and other
faint point sources; e.g.,~high-redshift SNe) will become one of the
great uses of ACS, and {\it The Next Generation Space
Telescope}. Another great hope for the next generation of sensitive
space-based GRB instrumentation is that, thanks to an order of
magnitude higher localization rate than currently, another apparently
rare burst like GRB 980326 may be localized.

\section{Conclusions: What Makes Gamma-ray Bursts?}

\begin{deluxetable}{|p{0.85in}||p{1.05in}|p{1.0in}|p{1.05in}|p{1.1in}|p{1.05in}|p{0.78in}|p{0.6in}|}
\rotate
\tabletypesize{\small}
\tablewidth{0pt}
\tablecaption{Summary Assessment of Progenitor Scenarios\label{tab:intro-sum}}
\tablehead{
\colhead{Issue/} &
\colhead{Double Neutron} &
\colhead{Black-hole--Neutron} &
\colhead{Black-hole--Helium} &
\colhead{Black-hole--White} &
\colhead{Collapsar} &
\colhead{DRACO} &
\colhead{AGN} \\
\colhead{Requirement} & \colhead{Star Merger}  & \colhead{Star Merger} &
\colhead{Star Merger}  & \colhead{Dwarf Merger} & \colhead{} &
\colhead{} & \colhead{}}

\startdata
\raggedright Energetics (10$^{51}$ of EM energy in a few secs)
           & \raggedright Accretion-driven \& spin energy driven possible.
           & \raggedright Accretion-driven \& spin energy driven possible.
           & \raggedright Only spin energy driven possible \citep{npk01}.
           & \raggedright Only spin energy driven possible \citep{npk01}.
           & \raggedright Accretion-driven \& spin energy driven possible.
           & \raggedright Spin energy driven possible.
           &  Not likely. \\
\hline
High $\Gamma$ outflow? 
               & Yes. 
               & Yes.
               & Yes.
               & Yes.
               & \raggedright Possibly. Baryon contamination a problem. 
               & Yes.
               & Yes \\
\hline
\parbox[t]{0.9in}{Burst \\ Durations} 
               & \raggedright Cannot explain long-duration bursts.
               & \raggedright Cannot explain long-duration bursts.
               & \raggedright Difficult to explain short bursts?
               & \raggedright Difficult to explain short bursts?
               & \raggedright Only long-duration bursts.
               & \raggedright Not well developed.
               & Unknown \\
\hline
\parbox[t]{0.9in}{Consistent with \\ burst \\ offsets?}
              & \raggedright No. Cannot explain observed distribution. See chapters \ref{chap:nsns} and \ref{chap:offset}.
              & \raggedright Probably no.
              & \raggedright Yes. Burst sites close to star-forming regions.
              & \raggedright Probably no.
              & \raggedright Yes. Burst sites inside star-forming regions.
              & \raggedright Probably yes: burst sites near star-forming regions.
              & {\raggedright No. Should be at centers of hosts.}\\
\hline
\parbox[t]{0.9in}{ \raggedright Consistent \\ with absence \\ of elliptical hosts?}
              & \raggedright Probably no. 
              & \raggedright Probably no. 
              & \raggedright Yes. 
              & \raggedright Yes. 
              & \raggedright Yes. 
              & \raggedright Yes. 
              & No.  \\
\hline
\parbox[t]{0.9in}{ \raggedright Explains transient iron lines?}
              & \raggedright No. 
              & \raggedright No. 
              & \raggedright Yes? 
              & \raggedright No. 
              & \raggedright Yes. 
              & \raggedright No. 
              & No. \\
\hline
\parbox[t]{0.9in}{ \raggedright Explains intermediate-time afterglow bumps?}
              & \raggedright No. 
              & \raggedright No. 
              & \raggedright Yes, but contrived.
              & \raggedright No. 
              & \raggedright Yes. A natural consequence of the model.
              & \raggedright ?
              & ? \\
\hline
\parbox[t]{0.9in}{ \raggedright Explains apparent homogeneous ISM?}
              & \raggedright Yes.
              & \raggedright Yes. 
              & \raggedright Yes, but requires kick outside of birthsite.
              & \raggedright Yes. 
              & \raggedright Yes, but contrived $\sim 1000$ yr wind turn-off before burst.
              & \raggedright ?
              & ? \\
\enddata
\end{deluxetable}

Table \ref{tab:intro-sum} depicts a critical (albeit cursory)
assessment of the current state of viable progenitor scenarios in the
face of existing observations. Dishearteningly, there is no obvious
progenitor scenario which can naturally explain all of the
observations of all of cosmological GRBs. Nevertheless, if we take the
bold step and suggest that the phenomenological sub-classification of
long-duration and short-duration bursts must actually represent a true
distinction in progenitor models, then we are free to ask the question
of what makes long-duration gamma-ray bursts.

The answer, which the reader should have seen growing clearer and
clearer throughout the progress of this thesis, is that collapsars
appear to be the only viable progenitor scenario which can explain
most, if not all, of the data to date. The issue that gives us most
pause is the disturbing lack of a detection of a wind-stratified
medium in a GRB afterglow.  Wind-stratified media are a natural
expectation of the massive stellar progenitor scenarios. One
suggestion is that there may be several different progenitors of
long-duration GRBs \citep{cl99,lw00} though it appears that almost all
proposed wind-stratified bursts can be adequately modeled by a jetted
burst in a constant-density medium \citepeg{fks+00,pk01}.  Instead, to
explain the observations, one could invoke the case where the
progenitor wind ``turns-off'' some $\ale$1000 yrs before the GRB so
that the dynamics of the afterglow are unaffected by the outwardly
flowing wind.  A wind-termination shock may also serve to homogenize
the ambient medium surrounding a massive star progenitor
\citep{wij01}. Scenarios where the wind is preferentially blown off
along the equator and the burst is jetted along the polar axis also
saves the collapsar.

Clearly, without an afterglow detection of a short burst, we have
almost no observational evidence which constrains the progenitors of
short-duration, hard-spectra GRBs (although see
\citealt{hbc+02}); however, the current theoretical picture, particularly
related to the timescales for energy release, fosters the ample
belief/hope that such bursts could be produced by merging remnants.
The progenitors of short bursts will almost certainly be a subject of
intensive study and discovery in the years to come.

\appendix				
\newpage				
\newpage				

\begin{singlespace}
\small
\bibpreamble{\addcontentsline{toc}{chapter}{Bibliography}}


\end{singlespace}
\end{document}